**The University of Reading**

# INSTRUMENTATION FOR ATMOSPHERIC ION MEASUREMENTS

**Karen Louise Aplin**

A thesis submitted for the degree of Doctor of Philosophy

Department of Meteorology

August 2000

# Abstract

Small ions are part of the atmospheric aerosol spectrum, and study of ion-aerosol interactions is fundamental in atmospheric physics. Air ion physics and instrumentation are reviewed, including the historical context. A miniaturised Gerdien condenser for ion measurement, operating *in situ* to minimise inlet errors, is described. Two operating modes using independent current and voltage decay measurements are employed. A more sophisticated self-calibrating and fully programmable ion mobility spectrometer (PIMS) based on the same principles, is also discussed. Detailed analysis of error terms and application of new technology is demonstrated to greatly improve its capability. A variety of self-consistent experimental approaches, including ionisation and ion concentration instruments, is used to calibrate the new PIMS instrument.

In developing and characterising the individual components of the PIMS, favourable and unfavourable operating régimes are identified: this approach can also be applied to other aspiration ion counting techniques. Use of a sophisticated programmable electrometer permits compensation for leakage terms. Electrically-charged aerosol particles have been found to complicate the ion measurements. Consequently, conventional ion-aerosol theory, which neglects the particulate concentration, is thought to be incomplete. The polymodal ion mobility spectrum is also found to influence the instrument's operation.

The classical theory for calculating air conductivity from voltage decay measurements is modified to take account of the ion mobility spectrum, and a new technique to obtain ion mobility spectra by inversion is presented. Ion mobility spectra are also derived using conventional ion current measurements, and the two methods of obtaining ion spectra from independent PIMS measurements are shown to be consistent. Development of the novel programmable ion instrumentation, in conjunction with consideration of the ion mobility spectrum yields an improved and flexible approach to *in situ* atmospheric ion measurements.

## Acknowledgements

This project was funded by the Natural Environment Research Council. I would like to thank Dr. Giles Harrison for his continual guidance and boundless enthusiasm for a subject that has also captured my imagination. The Meteorology Department technical team all deserve special thanks: Andrew Lomas for lots more than sorting out my PC problems and messy cables, and Stephen Gill for his excellent workmanship and coming to rescue me from Mace Head. Assistance in all matters electronic was provided by Roger Knight, who also helped to proof-read, and Steven Tames.

I have had extremely interesting and useful conversations with Dr. Charles Clement throughout the course of my studies. Dr. Jyrgi Mäkelä provided valuable insight during the long afternoon in Geneva when he was unable to escape my questioning.

My family and friends have been extremely supportive and encouraging, particularly during difficult times. My mother and brother Mark proof-read the majority of this Thesis. Mark's comments in particular were both helpful and hilarious during the long days of writing up. I thank Daniel Meacham for his friendship, humour and interest in my work. My late hamster, Hilary, was always an endearing distraction.

The musical life of the Meteorology Department has added an extra dimension to my time here. My devotion to one musical instrument has occasionally superseded my commitment to the scientific instruments I chose to make my career. It, and the people I meet with similar interests continue to inspire me.

# Contents

# Table of Figures

# Nomenclature

| | |
|---|---|
| $\mu$ | Electrical mobility |
| $a$ | Central electrode radius |
| $A$ | Gain |
| $b$ | Outer electrode radius |
| $B$ | Bandwidth |
| $C$ | Capacitance |
| $D$ | Absorbed dose rate |
| $E$ | Electric field |
| $E$ | Charge on an electron (1.6 x $10^{-19}$ C) |
| $G$ | Gain |
| | Chapter 2: Number of ions produced per 100 eV of absorbed energy |
| $i$ | Current |
| $i_B$ | Input bias current |
| $i_c$ | Compensated current |
| $i_F$ | Follower mode leakage current |
| $i_L$ | Leakage current |
| $i_{shot}$ | Current due to shot noise |
| $J$ | Current density |
| $j$ | Number of elementary charges on an aerosol particle |
| $J_{nuc}$ | Nucleation rate |
| $k$ | Constant for dimensions of Gerdien condenser |
| $K_B$ | Boltzmann's constant (1.38066 x $10^{-23}$ $JK^{-1}$) |
| $L$ | Length |
| $l$ | Mean eddy size |
| $l_s$ | Characteristic length scale |
| $M$ | Molecular mass (in SI units) |
| $m$ | Gradient of straight line |
| $n$ | Small ion concentration |
| | Chapter 5: DAC control code |
| $N$ | Total number of ions integrated across mobility spectrum |
| $n_i$ | Intermediate ion concentration |
| $n_L$ | Large ion concentration |
| $q$ | Volumetric ion production rate |
| $Q$ | Charge |
| $r$ | Ionic radius |
| | Chapter 2: a radius from the central axis of the outer electrode |
| $R$ | Resistance |
| $r$ | Correlation coefficient |
| $R^2$ | Coefficient of determination |
| $R_{cal}$ | Resistance of calibration resistor |
| $Re$ | Reynolds number |
| $R_F$ | Resistance of feedback resistor |
| $S_A$ | Surface area |
| $S$ | Supersaturation ratio |
| $t$ | Time |
| $T$ | Temperature |
| $t_d$ | Drift time |
| $u$ | Flow rate |
| $u_{ext}$ | External wind component |
| $u_{tube}$ | Flow rate in Gerdien condenser |
| $V$ | Voltage |
| $V_b$ | Bias voltage |
| $V_d$ | Drift velocity |
| $V_f$ | Voltage at central electrode |
| $V_{os}$ | Input offset voltage |
| $Z$ | Aerosol number concentration |
| $\varepsilon_0$ | Permittivity of free space (8.85 $pFm^{-1}$) |
| $\varepsilon_r$ | Relative permittivity |
| $\kappa$ | von Karman's constant (0.4) |
| $\alpha$ | Recombination coefficient |
| $\beta$ | Ion-aerosol attachment coefficient |
| $\gamma$ | Ion-assisted nucleation coefficient |
| $\gamma_t$ | Surface tension |
| $\eta$ | Viscosity |
| $\mu_c$ | Critical mobility |
| $\mu_I$ | Mobility of intermediate ions |
| $\mu_L$ | Mobility of large ions |
| $\rho$ | Density of a fluid<br>Chapter 7: space charge |
| $\sigma$ | Conductivity |
| $\tau$ | Time constant |

# 1 Ions in the Atmosphere

## 1.1 Introduction

The physics of electrically-charged clusters of molecules in the atmosphere, *air ions*, is inextricably entwined with the behaviour of other, larger particulates comprising the *atmospheric aerosol.* In one of the first publications in this area, Rutherford's (1897) comments encapsulated what is now known about ion-aerosol interactions with remarkable prescience,

> *"Later experiments on the influence of dust in the air led to the conclusion that it was due to the presence of finely divided matter, liquid or solid, in the freshly prepared gas… The presence of dust in the air was found to very greatly affect the ... conductivity… Since the dust-particles are very large compared to the ions, an ion is more likely to strike against dust-particle, and give up its charge to it or to adhere to the surface, than to collide with an ion of opposite sign. In this way, the rate of loss of conductivity is… rapid."*

Rutherford (1897) knew that the presence of aerosol particles reduced the ion concentration, and hence the electrical conductivity of the air[1], by attachment. This tenet has remained at the heart of ion-aerosol theory for over a hundred years[2]. Yet there is still a need to explore the physics of air ions and their interactions with atmospheric aerosol within and beyond Rutherford's framework.

*Aerosol* is a collective term for the myriad of particles present in the atmosphere. The size spectrum of these suspended particulates ranges from the smallest cluster ions[3] to relatively large organic matter with radii of order $10^{-4}$ m (Pruppacher and Klett, 1998). The significance of these particles for climate and health is a strong motivation for understanding them further. Rutherford's (1897) comments suggest the behaviour of aerosol can be inferred from observing its interaction with ions. The electrical properties of air ions permit application of experimental techniques that could not be used to measure the bulk properties of aerosol, which are not sufficiently charged to be deflected by modest electrical fields. Ions are the smallest form of atmospheric particulates, therefore studying them can give insight into the formation and growth of larger aerosol particles, which have a greater direct atmospheric impact.

[1] Air conductivity and air ion concentration are directly proportional.
[2] There is a further discussion of some historical aspects of this Thesis in Appendix E.
[3] Atmospheric ions are frequently classified into *large ions* (r > 3 nm), *intermediate ions* (1 > r > 3 nm), and *small ions* which are typically 0.5 nm in radius. Large ions are often classified as charged aerosol particles and have a distribution of electrical charges, whereas intermediate and small ions have unit charge. (*e.g.* Hõrrak *et al,* 2000)

## 1.2 Atmospheric aerosol

The increased political significance of environmental science has necessitated investigation and characterisation of atmospheric aerosol particles. Atmospheric aerosol absorbs infrared radiation and is significant in climate forcing. Knowledge of aerosol concentrations is crucial for climate models, and this entails an understanding of aerosol production and removal mechanisms. Although the science of climate change is largely based on computer modelling, real measurements of aerosol are vitally important to both support and corroborate them.

On a smaller scale, the emission of aerosol from vehicles and industry is also a pressing issue. Aerosols of radius less than 10 μm are small enough to penetrate deep into the human lungs and have become classified as PM10 and PM2.5, with PM2.5 being smaller than 2.5 μm. (The distinction arises because PM2.5 can penetrate within the alveoli of the lungs, whereas PM10 cannot). Contemporary studies have recognised their effects on the body, and have identified aerosol particles as an important type of pollution. The emission and dispersion of such pollutants needs to be better understood to improve public health. Yet such essential, and apparently simple, scientific problems like measuring aerosol pollutants in different size ranges have not been completely solved. Gravimetric and optical instruments commonly used for measurement of PM10 and PM2.5 only operate within certain (often poorly defined) size ranges. Furthermore, there is increasing concern about the health effects of the very smallest particles which are often missed by common measurement methods, despite making up the main body of the aerosol number concentration. There is therefore a clear need to continue increasing our understanding of the whole spectrum of atmospheric aerosol.

### 1.2.1 Aerosol nucleation

An important mechanism of aerosol formation is *gas-to-particle conversion* (GPC) (Pruppacher and Klett, 1998), where gas molecules become clustered together to create a macroscopic particle. *Homogeneous nucleation* is typically a catalysed gas phase chemical reaction, by which sulphates and some ammonium salts can be produced. Homogeneous nucleation of water vapour can only occur spontaneously in highly supersaturated vapour. Such supersaturations are not found in the atmosphere. Tropospheric water vapour clouds are observed to form readily at supersaturations of 2 %, which cannot be accounted for by the theory of homogeneous nucleation. Therefore, another mechanism nucleating water vapour into droplets must be occurring: this is *heterogeneous nucleation*, which requires the presence of some pre-existing particle to reduce the vapour pressure at which condensation occurs. The particles which can *potentially* act as nucleii for cloud droplet growth are known as *condensation nucleii* or *CN*[4], and are part of the aerosol continuum. CN concentrations are typically a few thousand per cubic centimetre; concentrations are higher in urban areas, and depleted in marine environments.

*Ion-induced nucleation,* the growth of an aerosol particle by vapour condensing onto an ion, has been shown to be theoretically possible by Castleman (1982). This effect has yet to be observed in the atmosphere, although it has been measured in the laboratory on several occasions (*e.g.* Bricard *et al,* 1968). There is also provocative evidence to suggest that ion-assisted nucleation is an aerosol-forming process, particularly in areas where CN may be depleted. Rapid "bursts" of particle growth are commonly observed at Mace Head on the west coast of Ireland (O'Dowd *et al,* 1996), and have not been explained. Slower ionic growth has also been reported in Estonia, and it has been suggested that this is the first stage of a nucleation process (Hõrrak *et al*, 1998a). A mechanism for ion-induced nucleation in the atmosphere has been proposed (Turco *et al*, 1998; Yu and Turco, 2000), but not observed, and there is some disagreement whether the ionisation rates used in Yu and Turco's (2000) simulation are appropriate (Harrison and Aplin, 2000b). Therefore the potential rôle of ions in climate processes remains controversial and uncertain. Whatever ions' precise importance, it is clear that they need to be characterised in detail in order to gain any understanding of physical mechanisms in which they could be implicated.

[4] CN which do nucleate water vapour at atmospheric supersaturations are known as cloud condensation nucleii (CCN).

### 1.2.2 The solar cycle and climate

The possibility of solar modulation of the Earth's climate has long been a contentious issue, with many correlations reported which are frequently dubious or short-lived. Ney (1959) speculated that changes in cosmic ray intensity would cause increased storminess, cloudiness and affect the earth's weather systems. Svensmark and Friis-Christensen (1997) have observed a correlation between the cosmic ray flux and the cloud cover on the Earth. Short-term fluctuations in cosmic ray activity, known as *Forbush decreases*, have also been associated with changing cloud cover (Pudovkin and Veretenko, 1995). The cosmic ray flux is modulated by the solar cycle because when solar activity is at a maximum, the sun's magnetic field is sufficiently large to deflect the least energetic (sometimes called "soft") cosmic rays away from the Earth. Cosmic rays are primarily made up of energetic protons and alpha particles (Svensmark and Friis-Christensen, 1997) and so have an ionising effect, including generating air ions. During solar minima, more cosmic rays reach the earth's atmosphere where they can ionise atmospheric molecules: variations in ionisation result.

The influence of cosmic rays on ionisation is well accepted, but recent controversy concerns the proposition that ions help to form CN, which changes the cloud cover over the earth. There is no existing atmospheric experimental evidence to support or refute hypotheses associating ionisation and particle formation, so increased observations of ionic processes are essential. Ion mobility spectra, which can resolve ionic growth, are one such set of necessary data, but the instrumentation to make routine atmospheric ion measurements is lacking. Until such instrumentation has been developed, testing the hypothesis that ionisation affects CN in the atmosphere would remain unfeasible.

## 1.3 Atmospheric small ions

*Atmospheric small ions* are small molecular clusters carrying a net electric charge. They are produced by ionisation of molecules in the air, and these initial ions are quickly clustered by water molecules to produce a central, singly charged, ion surrounded by 4-10 water molecules. Air ions exist at typical ground level concentrations over land of, on average, a few hundred per cubic centimetre. They are subject to considerable variability from atmospheric turbulence and transport effects. This influences the ion concentration both directly and indirectly *via* the aerosol population, as identified by Rutherford (1897).

### 1.3.1 Production of atmospheric ions

Radon-222 decay products emitted from the soil are important contributors to ionisation at the land surface. One alpha-particle from radon typically has an energy of 4 MeV, and since the average ionisation energy is around 35 eV, each $\alpha$-particle will produce about $10^5$ ion pairs (Israël, 1971). The differing chemical composition of negative ions reduces the mean number of water molecules attached to the central cation. Consequently, negative ions are slightly smaller than positive, and can move faster in an electric field. A schematic diagram of the atmospheric ion production mechanism is given in Figure 1.1.

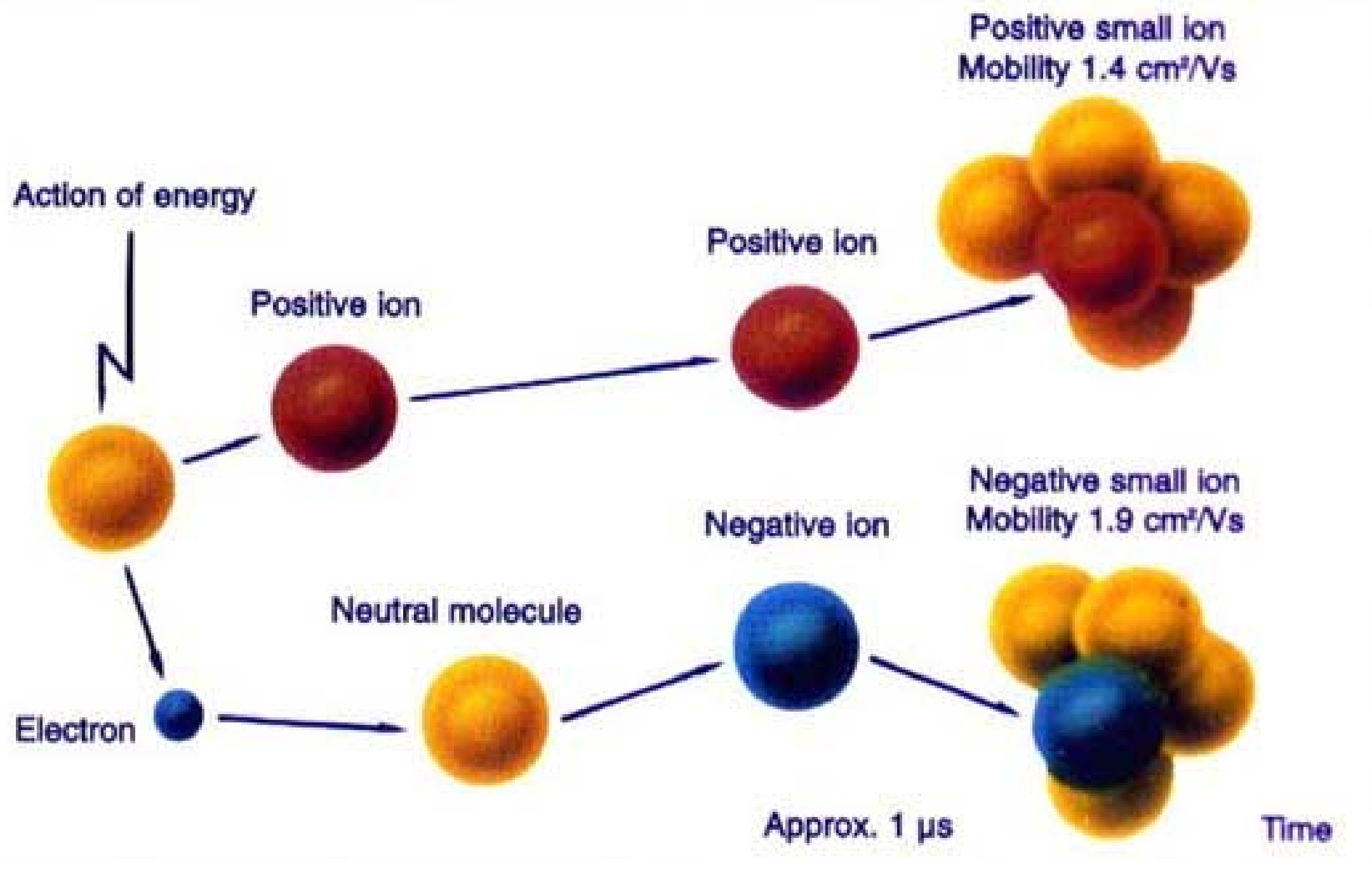

*Figure 1.1 Schematic illustrating the production of atmospheric small ions from neutral molecules.*

Cosmic rays are high-energy particles from outside the solar system which cause about 20 % of the ionisation at ground level. (However the ionising potential of one such particle, typically moving at a sub-relativistic velocity, will clearly vastly exceed that of a single alpha-particle resulting from terrestrial radioactive decay.) The mean ion production rate is subject to considerable variability, but an accepted value for the long-term mean is 10 pairs $cm^{-3}$ $s^{-1}$ (Chalmers, 1967). The number of ions increases with altitude as the relative contribution of cosmic rays to ionisation increases, and their intensity. Other, less significant sources of atmospheric ions are corona ions from large electrical fields (*e.g.* those found below high voltage power lines). Ions can also be produced from the breaking of water droplets, leading to larger concentrations near to waterfalls and at the seashore (Chalmers, 1967).

### 1.3.2 Size and composition of atmospheric small ions

$N_2^+$, $O_2^+$, $N^+$ and $O^+$ are the main primary ions produced from ionisation of the most common gases in the atmosphere. It is energetically favourable for the ions to react very quickly with water. The time constant of this reaction, when water molecules complex around the primary ions, is rapid[5] and is proportional to the humidity of the air (Keesee and Castleman, 1985). The most common anions and cations in the troposphere are $H^+(H_2O)_n$ and $NO_3^-(HNO_3)_n$ where $n < 10$ (Keesee and Castleman, 1985), although mass spectrometric studies have shown that a wide variety of ions can exist, including organic species such as amines and pyridines (Eisele, 1988, 1989). Typically, these molecular cluster ions are about 0.5 nm in radius, with a mass of a few hundred atomic mass units (Hõrrak *et al*, 1999). Pruppacher and Klett (1998) state that the mean velocity attained by a particle with one elementary charge, and radius 1 μm in a typical atmospheric electrical field of 100 $Vm^{-1}$ is $10^{-7}$ $ms^{-1}$, which is insignificant in comparison to ambient air motion. Air ions carry one unit charge concentrated over a smaller volume, and the electrical and mechanical forces acting on them are comparable. Therefore, in an electric field, ions are influenced by electrical forces at least as much as mechanical ones.

### 1.3.3 Mobility of atmospheric small ions

The concept of *electrical mobility*, $\mu$ is useful to describe the behaviour of atmospheric small ions, because there is a linear relationship between it and the

[5] The timescale is typically nanoseconds.

magnitude of the electric force acting on the ions. It was first defined by Thomson (1928) as

$$\mu = \frac{v_d}{E} \qquad \textit{Eq. 1.1}$$

(here written in scalar form) where $E$ is the magnitude of the electric field and $v_d$ the associated drift velocity attained by the charged particle when in the Stokes régime, (*i.e.* when the electrostatic forces acting on the particle balance the drag forces). Small ions have a relatively high mobility (typically 1 $cm^2V^{-1}s^{-1}$)[6]. Other, larger ions exist, but their electrical mobilities are several orders of magnitude less than the typical clusters comprising a small ion, so their contribution to the bulk electrical properties of air ions is negligible. It is helpful to be able to relate ion mobility to radius, as shown in Figure 1.2. This problem is non-trivial because of the size-dependent interplay between different forces and effects.

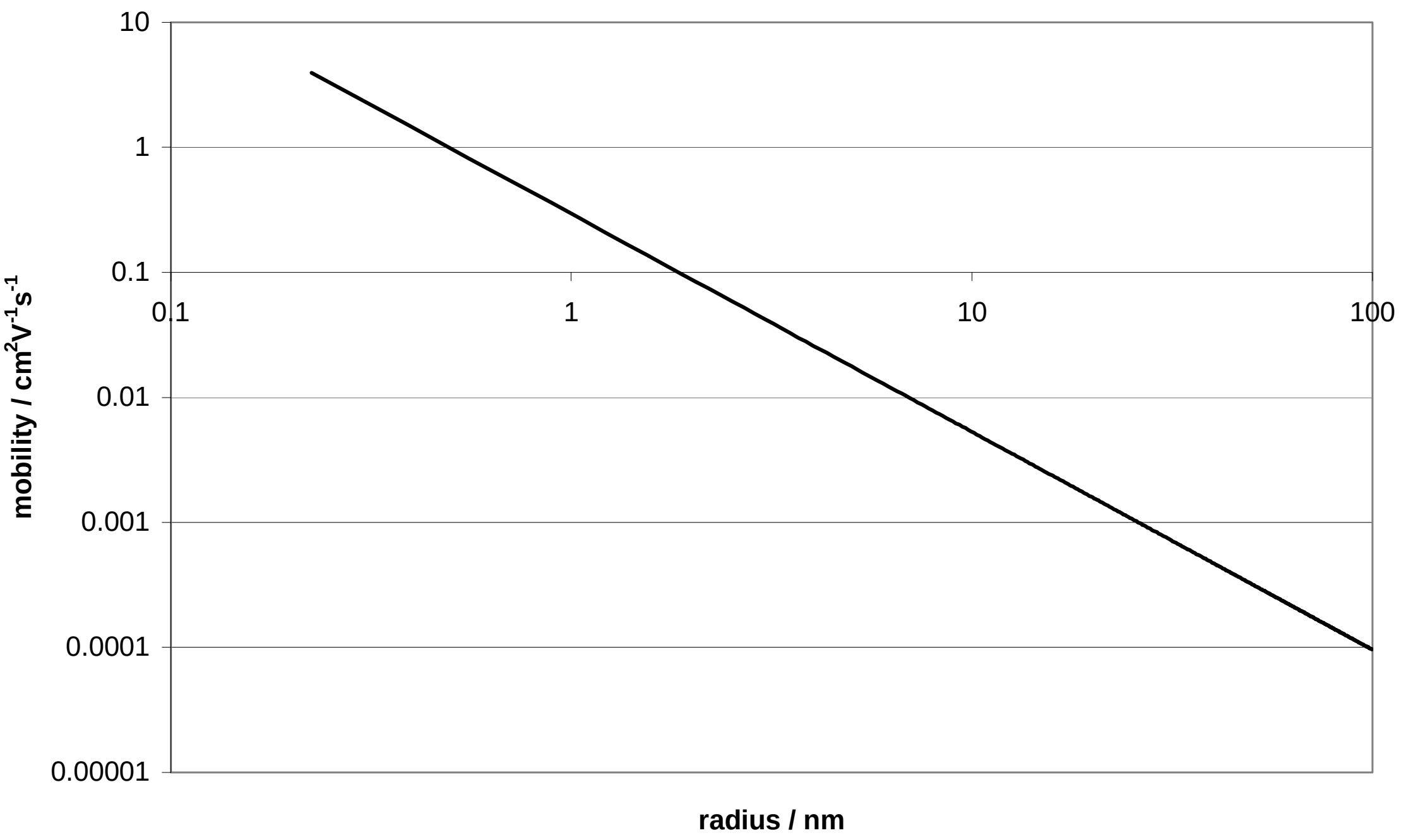

*Figure 1.2 Ionic mobility as a function of radius, from the expression derived by Tammet (1995).*

### 1.3.4 The ion balance equation

Ions recombine with oppositely charged ions, and, as Rutherford (1897) observed, they also attach to larger aerosol particles. The ion balance equation describes the fluctuations in ion concentration. When all the aerosol particles have the same radius, this is known as a *monodisperse* population, and the following equation applies where $q$ is the volumetric ion production rate, $\alpha$ the recombination coefficient, $\beta$ the ion-aerosol attachment coefficient, and $Z$ the aerosol number concentration.

[6] The units of mobility can be written as $cm^2V^{-1}s^{-1}$ , which is convenient for atmospheric ions because the mobility of a typical ion is $1 cm^2V^{-1}s^{-1} = 1x\ 10^{-4}\ m^2V^{-1}s^{-1}$.

$$\frac{dn}{dt} = q - \alpha n^2 - \beta n Z \qquad \textit{Eq. 1.2}$$

## 1.4 Ions in the global atmospheric electrical circuit

Ions are important in the maintenance of the global atmospheric electric circuit, a schematic diagram of which is shown in Figure 1.3. At the upper levels of the atmosphere, ionisation is extensive and there is a layer of conductive air known as the ionosphere[7]. Strictly, the *atmospheric current density J* is defined by

$$J = \sigma E \qquad \textit{Eq. 1.3}$$

where $\sigma$ is the air conductivity and $E$ the atmospheric electric field. The conduction current is calculated over the entire planet, and directly results from the electrical conductivity due to air ions. The magnitude of the fair weather current density is small compared to the typical current of 1 A delivered by each thunderstorm. The global circuit exists due to the approximate equality between the global fair weather current and the thunderstorm current, when compared on the planetary scale.

[7] In the context of the global atmospheric electrical circuit, the ionosphere is frequently referred to as the *electrosphere*. The differences between them are discussed by MacGorman and Rust (1998).

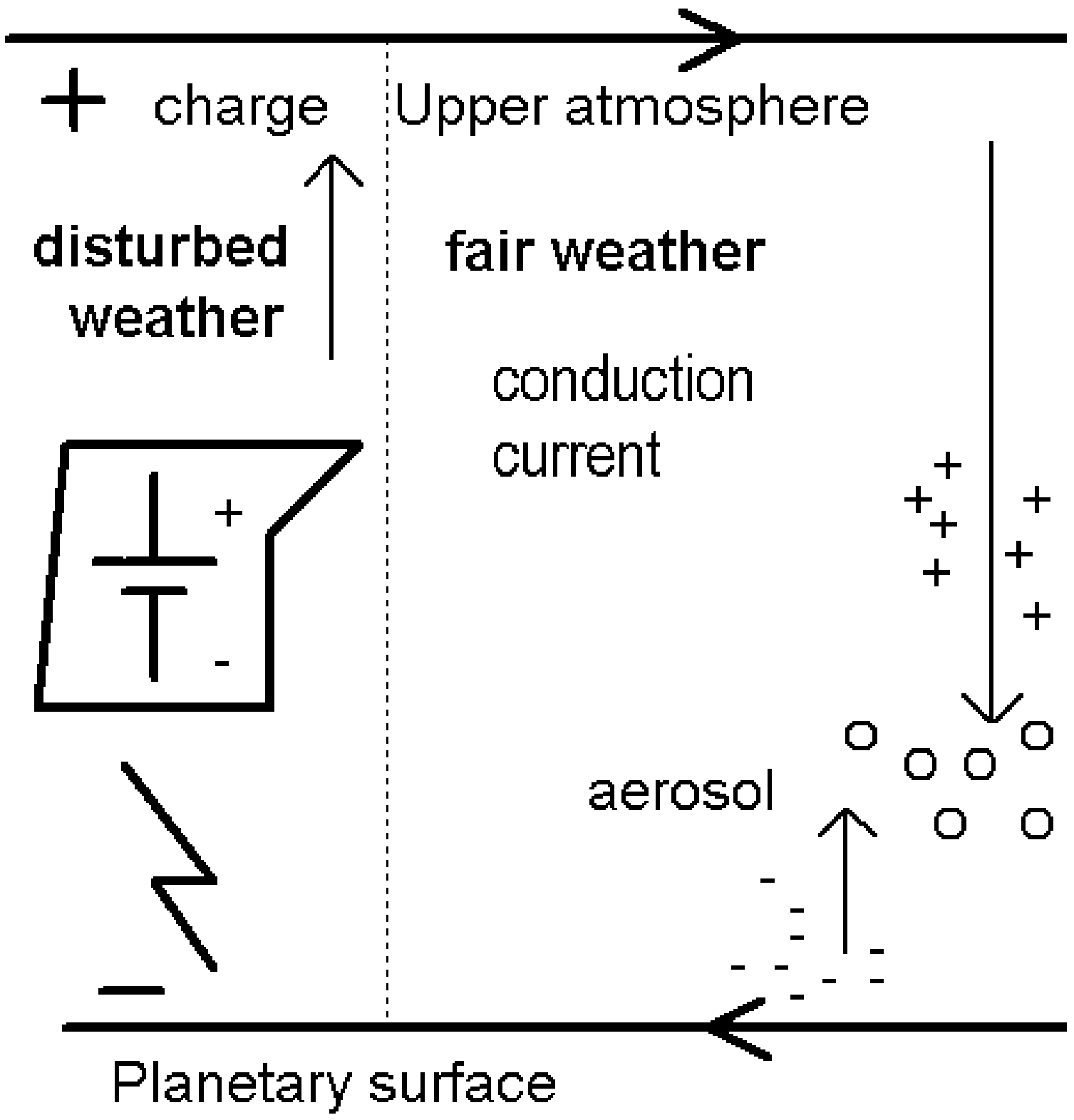

*Figure 1.3 The global atmospheric electrical circuit (Harrison and Aplin, 2001) showing the combination of charge separation from thunderstorms, and maintenance of the circuit by air ions.*

## 1.5 Measurement of atmospheric small ions

Small ions can be measured by exploiting their electrical properties such as electrostatic attraction. The simplest way of counting atmospheric ions is to subject them to an electric field, for example by blowing air between two metallic plates or into a conducting cylinder with a central electrode. If an electric field is applied across the electrodes, ions will be electrostatically attracted to them. Therefore this configuration can be considered to store charge, and is referred to as a *capacitor* or *condenser*. If the air-spaced capacitor is charged up to some voltage and the charge allowed to decay, the rate of decay through air will be related to the ion concentration. Alternatively, a current can be measured which is proportional to the concentration of air ions and their electrical mobility. As small ions are singly charged, this enables the air conductivity to be determined.

Gerdien (1905) first used such a cylindrical capacitor to measure air conductivity. The *Gerdien condenser* has become the classic instrument for air ion measurement, albeit in a different configuration to the one Gerdien (1905) originally proposed. Initially, the conductivity was inferred from the rate of decay of the voltage across the capacitor's electrodes, but this technique has fallen almost completely out of use in modern implementations. This is unfortunate, because measuring a rate of voltage decay is far simpler than attempting to resolve the very small ion current at the central electrode (of order $10^{-13}$ A). It is also difficult to understand why one of the two methods of ion counting with the Gerdien condenser should be so favoured, when corroboratory measurements can be obtained with the same instrument operating in both modes.

An additional application of the Gerdien condenser is that varying the electric field in the condenser allows selection of different mobilities of ions. This enables resolution of spectral information about the ion population; although relatively simple in principle, such measurements are rare. Modern electronic and computer technology is infrequently applied to exploit the self-corroborating and spectral measurement properties of the Gerdien instrument. There is clearly scope for modernisation of this classical instrument to measure ion mobility spectra, and to utilise both methods of conductivity measurement.

## 1.6 Motivation

Many of the primary motivations to improve instrumentation for investigating air ions can be traced back to the earlier summary of Rutherford (1897). Rutherford's observation that small ions to attach to aerosol particles has developed into a hypothesis that air conductivity can be used as a pollution indicator, both on a secular and local scale. Although theoretically sound, the dependency of air conductivity on the aerosol concentration has not been shown to be universally valid. One of the motivations for improving air ion measurements is to test the precept that conductivity is generally useful as a pollution indicator.

Rutherford's comments summarised early perceptions of atmospheric particulates. His conceptualisations have been developed throughout the century into viewing ions as part of the atmospheric aerosol spectrum[8]; this provokes further study of ion-aerosol interactions. If ions are viewed as part of the aerosol spectrum, then the study of atmospheric aerosol encompasses the study of atmospheric ions. All the motivations to measure aerosol (discussed above) apply directly to air ions and motivate the study of air ions in their own right.

More specifically, ions have been implicated in aerosol particle formation, but this has not yet been observed directly in the atmosphere, though there is increasingly a theoretical framework providing a plausible atmospheric mechanism (Turco *et al*, 1998; Yu and Turco, 2000). There is a controversy about the rôle of ions in aerosol formation, with two rival theories, one of which does not involve atmospheric ions[9]. Measuring atmospheric ion growth processes will clearly be expedient in resolving these issues.

The Gerdien condenser is a classical instrument, which can measure both air conductivity and atmospheric ion spectra. It is capable of air ion measurements by two methods, although few attempts have been made to compare the two operating modes. A combination of the two ion measurement methodologies using the same sampling instrument would be a powerful, and novel, way to check its self-consistency. In particular, it is hypothesised that the measurement of voltage decays from the Gerdien instrument can be mechanised and used to make systematic measurements at the ground in conjunction with the direct measurement of current at the central electrode. This, in combination with ion mobility spectra, would provide reason to trust the detailed measurements which will be required to detect ion-assisted particle formation.

## 1.7 Thesis Structure

This Thesis develops an experimental methodology to investigate air ion properties in the atmosphere, with particular emphasis on surface measurements. Chapter 2 discusses existing instrumentation and previous ion measurements, with some of the theory relating them to other aspects of atmospheric physics. The design and components of a modernised Gerdien ion counter are investigated in Chapter 3, and it is developed into a programmable device in Chapter 4. Chapter 5 describes further refinements of the instrument and the implementation of a self-calibrating feature. In Chapter 6 a programmable ion mobility spectrometer is tested in the atmospheric surface layer for self-consistency, and against other instruments. Chapter 7 discusses some of the variability in the measurements, and Chapter 8 presents some new atmospheric observations made with the spectrometer. Finally, Chapter 9 summarises the principal findings of the Thesis and suggests some directions for further research.

[8] This idea is explored directly in Appendix E.

Appendix A contains a detailed description of the multimode electrometer used in the Thesis, and Appendix B discusses some aspects of current amplifier calibration. Appendix C shows the source code for the most important computer programs. Appendix D contains engineering diagrams for the new Gerdien condenser used in Chapters 5 onwards, and Appendix E is a philosophical discussion of the history of air ion measurement.

[9] These theories will be discussed in Section 2.7

*"The picture presented here is that the language and practice of experimentation, instrumentation and theory are distinct, but linked – and in interesting ways" Peter Galison*[10]

# 2 The significance of atmospheric ion measurements

In this section the scientific rôle of atmospheric ion measurements is discussed. The historical motivation and development of instrumentation, with typical results are described in Sections 2.1 to 2.4. The remainder of the chapter discusses modern applications of air ion measurements, in particular air ion mobility spectra and their relevance. Section 2.7 describes a recent and important motivation for further study of air ions.

## 2.1 Historical introduction

The first publication directly related to air ions was by Zeleny (1898). Research into the subject became popular in the early years of this century: it was predominantly concentrated in the Cavendish Laboratory at Cambridge but also in the US and Germany. Initially, work on air ions was not initially motivated by questions in atmospheric science, (*e.g.* Rutherford, 1897; McClelland, 1898), but a tendency has developed during this century towards viewing ions as part of the atmospheric aerosol spectrum.

Ions were first identified by Faraday in his nineteenth century studies of electrochemistry, and were known to be produced by some sort of breakdown of molecules in an electric field. Helmholtz developed this work to define the electric charge on the atom as the finite quantity of electricity carried by all ions. Stoney then used the word "electron" for the first time to describe a fundamental amount of electricity (Robotti, 1995). J.J. Thomson identified the constituents of cathode rays as the first sub-atomic particles, which he called "corpuscles", at the Cavendish Laboratory in 1897[11]. Ernest Rutherford and C.T.R. Wilson were taken on as research students in the 1890s. Although they ultimately became famous outside this field, both men produced valuable work related to air ions during this period at the Cavendish Laboratory. In particular, whilst still a student, C.T.R. Wilson (1897) developed his cloud chamber, in order to investigate particle formation from air ions[12]. Roentgen's discovery of X-rays (1896) was timely for research into ions in gases, since "Roentgen rays" were soon found to make gases electrically conductive by ionisation. The conductivity of atmospheric air was explicitly attributed to the presence of molecular cluster-ions by Elster and Geitel in 1901 (Torreson, 1949; Schonland, 1953).

[10] Galison P. (1997), *Image and Logic : A Material Culture of Microphysics*, University of Chicago Press

[11] "Corpuscles" were later renamed "electrons".

## 2.2 Overview of the Gerdien method

### 2.2.1 Principles of the Gerdien Condenser

Zeleny (1898) measured the mobilities of air ions using a ventilated cylinder, as shown in Figure 2.1. A section of the airflow was ionised with a radioactive source, and the ion present were attracted to wire gauze by an electric field. The wire gauze was connected to an electrometer, to measure the ions. Zeleny (1898) found that negative ions needed a larger potential difference to be deflected by the same amount as the positive ions. This implied that the negative ions had a higher velocity, and hence mobility (see Eq. 1.1) than positive ions.

[12] This won Wilson the Nobel Prize in 1927 after the cloud chamber had been applied to measure sub-atomic particles.

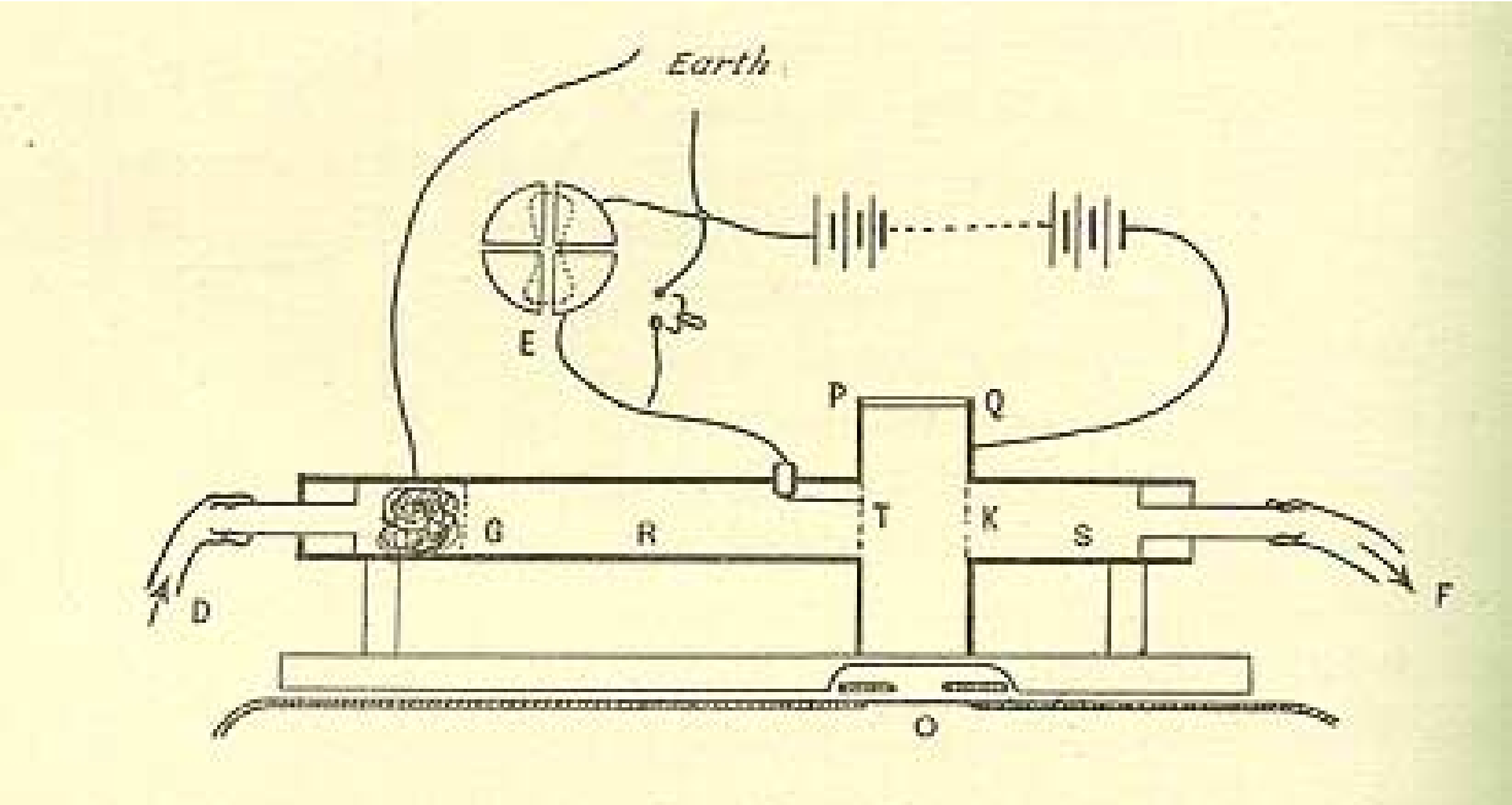

*Figure 2.1 The apparatus used by Zeleny to measure ionic mobility. The gases enter at D, pass through a plug of wire wool, to remove "dust", and leave at F. O is an ionising X-ray source. Q and P are brass plates with a voltage across them. T and K are wire gauze. If the ions produced hit P, Q or K they are conducted to earth, but ions hitting T modify its potential, which is measured relative to the battery by the electrometer E (Thomson, 1928).*

In 1905 Gerdien developed a device which has become the standard instrument for measuring air conductivity. The eponymous *Gerdien condenser* will be referred to as a "Gerdien" in this chapter. It is made of two coaxial electrodes, a hollow cylinder known as the *outer electrode* containing a thinner *central electrode* (which is frequently a solid wire). This configuration has a finite capacitance, which can be theoretically derived from Gauss' Law, and is therefore frequently referred to as a "condenser" (Swann, 1914). If a potential is applied across the electrodes and the tube is ventilated, then air ions of the same sign as the voltage are repelled from the outer electrode and attracted to the central electrode. If they meet it, a small current flows, which is proportional to the ion concentration and electrical conductivity of the air.

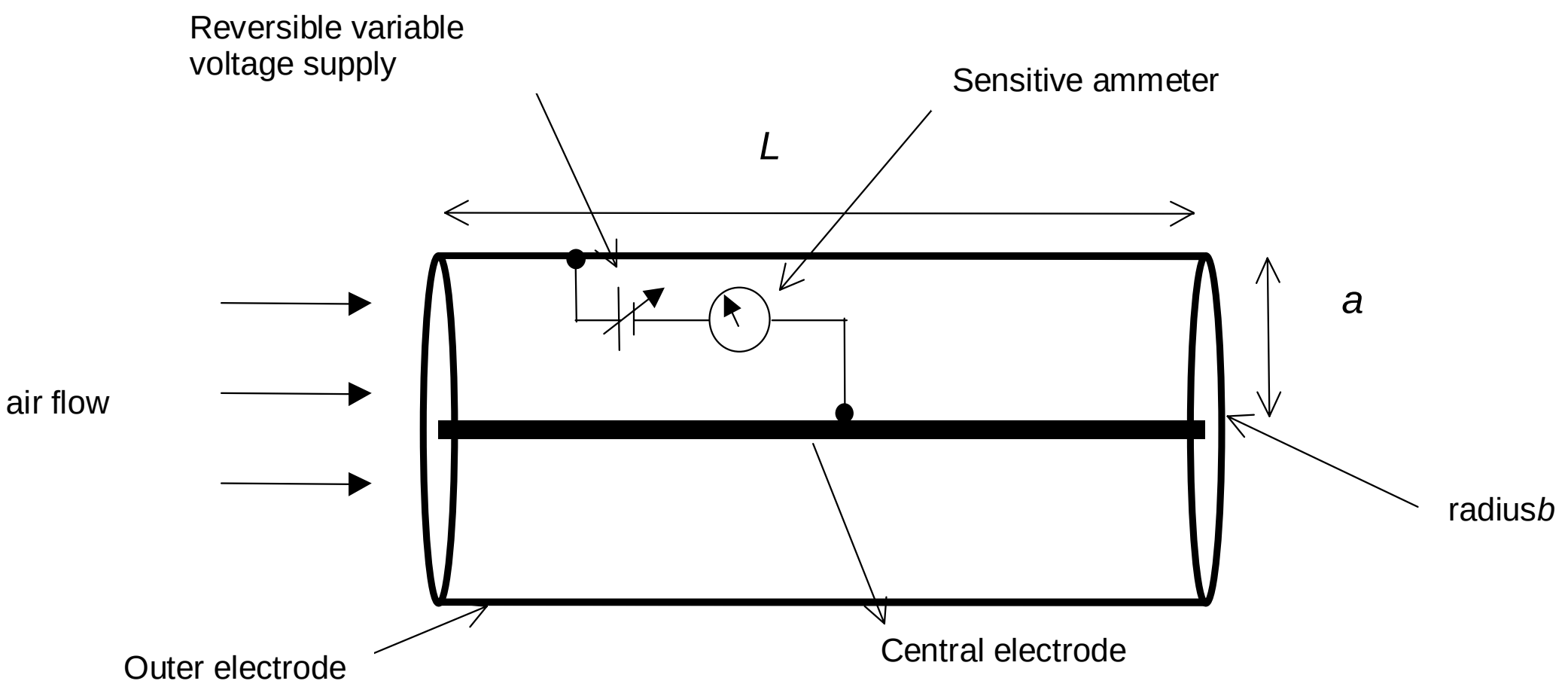

*Figure 2.2 Schematic of a Gerdien condenser*

Early Gerdiens were substantially proportioned, with linear dimensions of about 8 cm by 50 cm (Torreson, 1949). This was necessary in order for the output current to be large enough to be resolved by contemporary electrometers. In the 1950s smaller tubes were developed for radiosonde measurements, which were typically of order 5 cm in diameter and 30 cm in length. Attaching the Gerdien to a fixed potential, and allowing the voltage to decay through air permits a voltage to be measured rather than the current from the central electrode. This is preferential for conductivity measurement on radiosondes, because it is more straightforward to measure voltage than small currents, and immensely simplifies the issues associated with electrometry. The smaller sized tube was not problematic when ion currents did not have to be directly measured. Recent improvements in electronics have improved measurements of very small currents in the femtoampere range (Harrison, 1997a; Harrison and Aplin, 2000a).

The most common of the varied materials used to make Gerdien condensers has been brass (*e.g.* Higazi and Chalmers, 1966; Brownlee, 1973) probably following successful use of this material by early investigators (*e.g.* Rutherford, 1897). Gerdiens designed for use on radiosondes have necessarily been made of less dense materials such as aluminium or aluminised paper (Hatakayema *et al,* 1958; Venkiteshwaran, 1958). The tube material could be significant for two reasons: firstly that of its durability when used in the field, and secondly that of the effect of contact potential.

Wåhlin (1986) discusses the effect of the material of the tube, opining that contact potentials cause offsets in the i-V response. This is credible, as two electrodes touching a fluid containing ions do make an electrochemical cell (*e.g.* Atkins, 1989), even if the electrolyte (air) is dilute. According to Wåhlin (1986), a Gerdien will measure a non-zero current when the bias voltage is zero, due to electrochemical potential at the surface of the tube. Wåhlin (1986) suggests that a steel tube has to be biased at about 0.4 V and the aluminium at 1 V to counteract this. However, he does not explain how the effects of tube material alone were distinguished in his experiments from the many other sources of offset associated with Gerdien condenser measurements. The effect of different tube materials has not been systematically investigated, but Hatakayema *et al* (1958) did apply an offset of -140 mV corresponding to the contact potential for their aluminium tube, though they do not explain how this quantity was obtained.

### 2.2.2 Calculating conductivity from Gerdien condenser measurements

Chalmers (1967) explains the fundamentals of conductivity ($\sigma$) measurement from the rate of decay of charge $Q$, carried by a capacitor of surface area $S_A$, through a medium. Gauss' Law states that the charge per unit area is proportional to the electric field, which is (from Eq. 1.3) proportional to the current density $J$, given by

$$J = \frac{\sigma Q}{S_A \varepsilon_0} \qquad \textit{Eq. 2.1.}$$

hence the current $i$, for a flow of negative ions from a positively charged body is

$$i = -\frac{\sigma Q}{\varepsilon_0} \qquad \textit{Eq. 2.2.}$$

Since current $i$ is equal to the rate of change of charge $dQ/dt$, Eq. 2.2 is a differential equation, the solution of which is

$$Q = Q_0 \exp(-\frac{\sigma t}{\varepsilon_0}) \qquad \textit{Eq. 2.3.}$$

This indicates that the charge will decay exponentially with time $t$ from the original value $Q_0$. In terms of conductivity measurement, the rate of decay is related to the ion concentration. By analogy with the expression for exponential decay of charge from a capacitor (*e.g.* Duffin, 1980), it is clear that conductivity is inversely proportional to the time constant $\tau$ of the decay,

$$\sigma = \frac{\varepsilon_0}{\tau}$$ *Eq. 2.4.*

For a fixed capacitance C,

$$Q = CV$$ *Eq. 2.5*

(*e.g.* Duffin, 1980). It is clear from Eq. 2.4. and Eq. 2.5 that the conductivity can be determined by measuring the time constant of the voltage decay from the capacitor using

$$V = V_0 \exp(-\frac{t}{\tau})$$ *Eq. 2.6.*

As in the direct ion current measurement method, the ions entering the Gerdien condenser which are the same sign as the voltage are repelled towards the central electrode. The sign of the ions measured in this way is the same as the applied voltage (Venkiteshwaran, 1958; MacGorman and Rust, 1998[13]). The measurements differ in the instrumentation used to sense the motion of the ions, and are therefore called the *Voltage Decay* and *Current Measurement* modes in this Thesis.

The Voltage Decay mode was the method originally proposed for conductivity measurement by Gerdien (1905), and although subsequently discussed (Swann, 1914), it seems to have been largely neglected until conductivity measurements on radiosondes were made by Hatakayema *et al* (1958), Venkiteshwaran (1958) and Rosen *et al* (1982). This neglect is surprising, given that conductivity measurements made by the two methods would be extremely useful to check self-consistency, especially as there is no absolute calibration technique for Gerdien condensers.

Conductivity by the Current Measurement mode can be derived by considering the electric field in the Gerdien condenser, as in MR (1998). If *a* and *b* are the radii of the outer and central electrodes of a Gerdien, *V* the voltage across the electrodes, and *r* a radius from the central axis of the outer electrode as in Figure 2.2, then the electric field for a cylinder of infinite length is calculated from Gauss' Law

$$E = \frac{V}{r \ln\left(\frac{b}{a}\right)}$$ *Eq. 2.7*

[13] MacGorman and Rust will be referred to as MR.

(*e.g.* Duffin, 1980). An ion of mobility $\mu$ will move an incremental distance in the electric field given by

$$dr = \mu E dt \qquad \textit{Eq. 2.8}$$

so ions moving inwards in the tube take a time *t* to move the total distance from the outer to the central electrode, given by

$$t = \frac{\ln\left(\frac{b}{a}\right)}{\mu V}\int_b^a r dr \qquad \textit{Eq. 2.9}$$

In this time, if the air is moving at a speed *u*, then the distance it moves is *ut*. If *ut* is less than the length of the tube *L* then all the ions are collected when

$$u < \frac{2\mu VL}{(b^2 - a^2)\ln\left(\frac{b}{a}\right)} \qquad \textit{Eq. 2.10.}$$

From this a minimum mobility of ion can be defined, the *critical mobility* which is collected by the central electrode whilst in the tube. This definition is subject to error since Eq. 2.7 for electric field in the tube assumed it was infinite in length, and the Gerdien is too short for end effects to be ignored. Substituting in the expression for mobility (Eq. 1.1) then the critical mobility $\mu_c$ can be defined as

$$\mu_c = \frac{(a^2 - b^2)\ln\left(\frac{a}{b}\right)u}{2VL} \qquad \textit{Eq. 2.11.}$$

The theory detailed above assumes that ions take a parabolic path in the tube. In reality the ions enter the tube at a distribution of angles, and a few can thus be collected even if they are slightly below the critical mobility. The error from this source is likely to be much smaller than those introduced by measurement errors.

The current arriving at the central electrode per unit time is defined in MR (1998) as

$$i = \frac{2\pi n \mu VLe}{\ln\left(\frac{b}{a}\right)} \qquad \textit{Eq. 2.12}$$

Since the conductivity of air $\sigma$ is related to the total number of ions and their mobilities,

$$\sigma = e\int n d\mu \qquad \textit{Eq. 2.13}$$

assuming that all small ions have the same mobility, *a unimodal ion spectrum*, then ionic conductivity can be approximated as

$$\sigma = ne\mu \qquad \textit{Eq. 2.14}$$

which is derived in *e.g.* Harrison (1992). Substituting Eq. 2.12 into Eq. 2.9, the ion current is proportional to the conductivity so that

$$i = \frac{2\pi VL\sigma}{\ln\left(\frac{b}{a}\right)} \qquad \textit{Eq. 2.15}$$

The constants in this equation have the same units as the theoretical expression for the capacitance, derived from Gauss's Law (*e.g* Duffin, 1980),

$$C = \frac{2\pi\varepsilon_0 L}{\ln\left(\frac{b}{a}\right)} \qquad \textit{Eq. 2.16}$$

so if the *measured* capacitance is substituted into Eq. 2.15, conductivity can be written as,

$$\sigma_{\pm} = \frac{i\varepsilon_0}{CV_{\pm}} \qquad \textit{Eq. 2.17.}$$

This removes some of the error ensuing from the Gauss's Law assumption, due to the end effects of the Gerdien tube (Chalmers, 1967). Repulsion of ions of the same sign as the outer electrode towards the central electrode ensures that bipolar conductivity of the same sign as the bias voltage is measured. This equation is used to calibrate current measurements from a Gerdien condenser to obtain bipolar conductivity. *Total conductivity* is defined as the sum of the bipolar conductivities

$$\sigma_t = \sigma_+ + \sigma_- \qquad \textit{Eq. 2.18.}$$

When conductivity measurements are made with a Gerdien condenser they will vary with the bias voltage used and the dimensions of the tube according to Ohm's law. However, different bias voltages and physically different tubes will have different critical mobilities, and therefore measure different fractions of the ion spectrum. The variety of atmospheric ion species (*e.g.* Keesee and Castleman, 1985) suggests that there is a non-unimodal ion spectrum, so measurements made at one bias voltage might be affected by the whole ion spectrum. Conductivity measurements made at other bias voltages may have only a small fraction of ions contributing to the measurement. Caution must therefore be exercised when comparing conductivity measurements made with different instruments.

There is a need to redefine conductivity as an *average* property of air ions, with the number concentration as the average over the whole mobility spectrum. If the ion spectrum is not measured at the same time as the conductivity, the best estimate is possibly to take the critical mobility to be the average for small ions, so that a significant fraction of the spectrum is observed.

The concept of a Gerdien condenser as an instrument to measure air ions is well-established, but the theory of its operation is explored in inadequate detail, resting on many assumptions about the nature of the ions it measures. The capacity of the Gerdien to measure air ions by two independent methods could easily have been exploited, particularly in the absence of absolute calibration techniques. Yet this has been profoundly neglected in the hundred-year history of the instrument.

### 2.2.3 Fair weather measurements of conductivity at the ground

| **Reference** | **Year of Expt.** | **Location** | **$\sigma_t$ (fSm$^{-1}$)** |
|---|---|---|---|
| Hogg (1939) | 1939 | Kew, UK | 3 |
| Misaki (1964) | 1963 | Socorro, USA | 60 |
| Higazi and Chalmers (1966) | 1965 | Durham, UK | 11 |
| Brownlee (1973) | 1970 | Auckland, NZ | 8 |
| Retalis *et al* (1991) | 1968-1980 | Athens, Greece | 3 |
| Guo *et al* (1996) | 1973-4 | Toronto, Canada | 8 |
| Tammet *et al* (1992) | 1985-6 | Tahkuse, Estonia | 18 |
| Pauletti and Schirripo Spagnolo (1989) | 1987 | Apennnines, Italy | 8 |
| Kamra and Deshpande (1995) | 1990 | Bay of Bengal | 9 |
| Arathoon (1991) | 1990 | Lancaster, UK | 8 |
| Dhanorka & Kamra (1992) | 1991 | Indian Ocean | 9 |

*Table 2.1 Summary of ground level fair weather conductivity measurements, averaged over time periods from one day to twelve years. Where bipolar measurements have been made, their averages have been summed to give the total conductivity.*

Several sets of measurements of the diurnal variation of conductivity have been made (Paoletti and Schirripa Spagnolo, 1989; Misaki, 1964; Retalis *et al,* 1991; Dhanorkar and Kamra, 1992; Guo *et al* 1996; Kamra and Deshpande, 1995). All the measurements over land show a peak just before sunrise, caused by the nocturnal inversion trapping air near to the surface in which natural radioactive gases such as radon can build up and increase the ionisation rate (Wait and Parkinson, 1945), shown in Figure 2.3. The conductivity decreases mid-morning, because turbulent mixing becomes more significant and disperses the radioactive gases that ionise the air. For the rest of the day the conductivity measurements are more dependent on the local conditions.

*Figure 2.3 Typical diurnal variation of conductivity measured at Kew, from Chalmers (1967)*

In clean-air regions of the world, the conductivity trace tends to match the diurnal variation of ionisation rate well. Other factors, primarily the presence of aerosol, can also influence the local ion concentration. Insight into factors governing the ion concentration $n$ (and hence conductivity by Eq. 2.14) can be obtained from the ion balance equation in its simplest, steady-state form, where $q$ is the ionisation rate, $Z$ the monodisperse aerosol number concentration, and $\alpha$ and $\beta$ are the ionic and self-recombination and attachment coefficients:

$$q = \alpha n^2 + \beta n Z \qquad \textit{Eq. 2.19}$$

When conductivity varies with the ionisation rate, this implies that the aerosol term must be negligible, as it would otherwise modulate the ion concentration. This is known as the *recombination limit* (Clement and Harrison, 1992) when

$$n = \sqrt{\frac{q}{\alpha}} \qquad \textit{Eq. 2.20}$$

If the aerosol term dominates, this is referred to as the *attachment limit*, because of the effect of small ions attaching to larger aerosol particles. In this limit conductivity is less clearly related to the ionisation rate as in Retalis *et al* (1991) and Guo *et al* (1996). These measurements were made in polluted areas, and also showed the lowest mean values of conductivity with $\sigma_t \approx 8$ fSm$^{-1}$. Conductivity can also vary with local meteorology affecting dispersion of aerosol, as in Figure 2.3. These measurements were made in the 1930's when there were frequent London smogs, when aerosol emissions became trapped near the ground in winter, and insolation was insufficient to substantially mix the air and disperse the aerosol[14].

Conductivity measurements over the ocean (Kamra and Deshpande, 1995) show a much smaller diurnal variation. Over the ocean, *q* is low, since there are no natural radioactive emissions from the sea. However, ion concentrations remain similar to the land because of the reduced aerosol concentration, which attenuates the conductivity over land. The effects of human activity on conductivity over the ocean are also minimal, though detectable and are discussed further in Section 2.5.

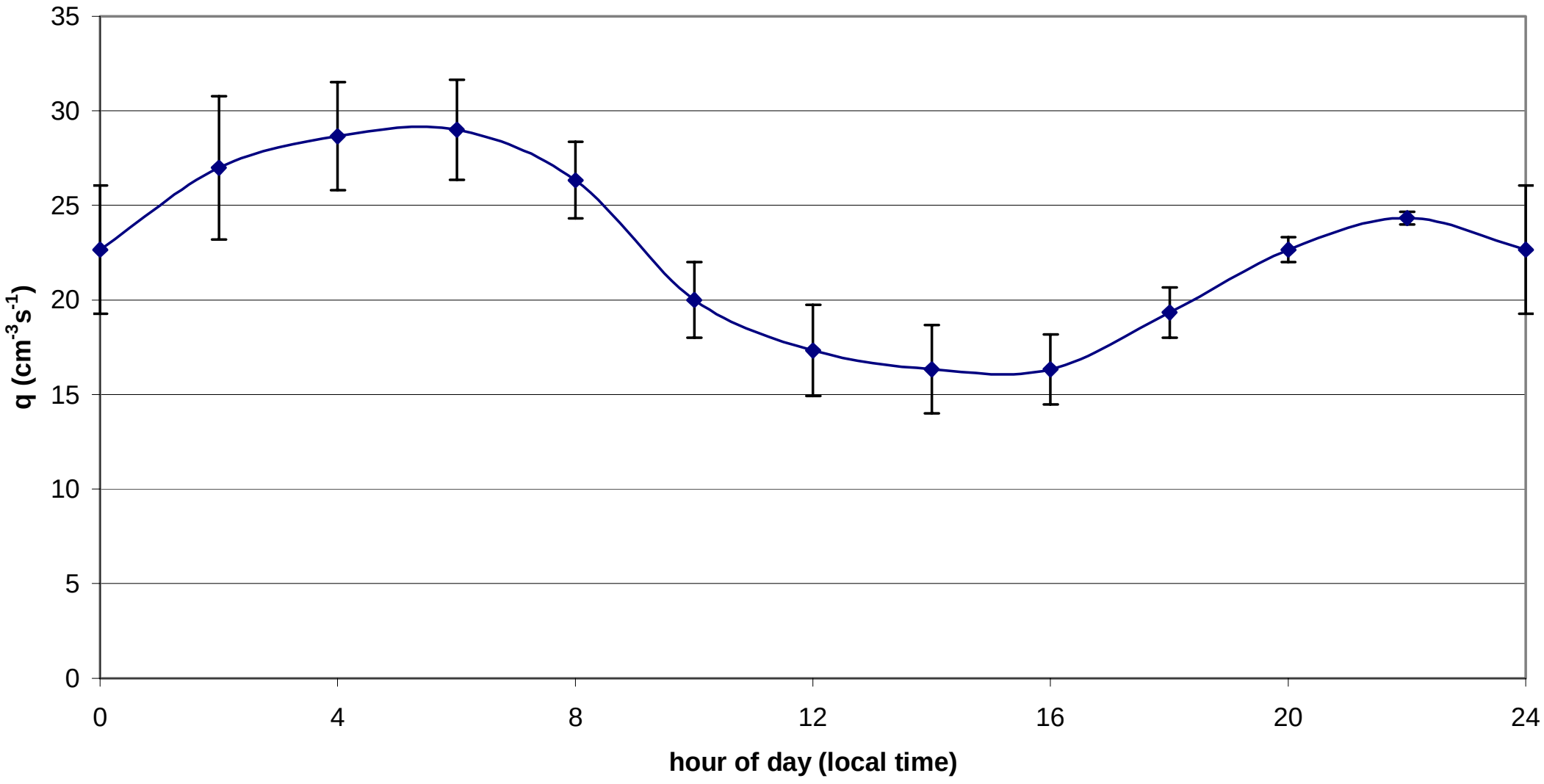

*Figure 2.4 The diurnal variation of ionisation rate q, from an average of three data sets in Wait (1945). The measurements were made at Canberra, Australia, Huancayo, Peru and Washington DC, USA. Error bars are the standard error of the mean.*

The *ion asymmetry ratio, x* is positive at the surface because of the *electrode effect*. This is an excess of positive charge near the surface resulting from the global electrical circuit; in fair weather negative ions move upwards and positive downwards. Near to the ground the motion of the positive ions is constrained, so there is a build-up of positive charged particles (MR, 1998). Although the existence of the electrode effect was posited from purely electrical arguments, micrometeorological factors are also vital in understanding the motion of charged particles in the surface layer (Barlow, 2000). $x$ is known to tend to unity as the wind speed increases, owing to increased turbulent mixing (Higazi and Chalmers, 1966; Retalis *et al,* 1991) reducing the asymmetry from the electrode effect.

[14] A special meeting of the Royal Meteorological Society in 1954 chronicled this problem *(Q. J. Roy. Met.Soc.*, **80**, 344).

An increase in the relative humidity is thought to cause a decrease in the conductivity resulting from the increased hydration of air ions with water molecules, which decreases their mobility and hence the conductivity. In extreme cases, ions attach to water droplets, resulting in a more pronounced decrease in conductivity. This has been suggested as a technique to predict fog formation, and is reviewed by van der Hage and de Bruin (1999). According to van der Hage and de Bruin (1999), this method is successful in clean marine environments, but in urban environments the presence of aerosol tends to perturb the expected relationship[15]. Kamra *et al* (1997) measured conductivity over the ocean, where there is little variability in the ionisation rate relative to over land, and concluded that an increase in the relative humidity of 10-20 % caused the conductivity to decrease by a factor of two. In a thorough laboratory investigation, Moore and Vonnegut (1988) showed that the conductivity over boiling water was half its value compared to dry air with a relative humidity of 20 %. Harrison (1992) calculated relative concentrations of different hydrations of the common atmospheric ion $H^+(H_2O)_n$ by extrapolating thermodynamic data obtained from mass spectrometry. According to these calculations, at 300 K for a change in relative humidity from 20 % to 100 % the mean mobility should decrease from $2.05 \times 10^{-4}$ to $1.74 \times 10^{-4}$ $m^2V^{-1}s^{-1}$, *i.e.* a factor of 17 %. The discrepancy between the theoretical and experimental values is to be expected, since the mean mobility of atmospheric ions (Mohnen, 1974) implies that other, more massive, air ions exist as well as the hydronium species, and extrapolation of thermodynamic data beyond the experimental conditions may be problematic. However, the trend of the experimental measurements is in agreement with the theory that increased humidity decreases the mean mobility.

[15] It is difficult, however, to understand how such a traditionally "fair weather" instrument such as the Gerdien, which is susceptible to stray currents from condensation and across insulators, could be trusted to measure conductivity accurately in fog.

## 2.3 Fair weather measurements at altitude

Many balloon-borne measurements of the conductivity profile have been made (*e.g.* Woessner *et al,* 1958; Rosen *et al,* 1982; Gringel, 1979; Gringel *et al,* 1983). However these studies can be viewed critically for their lack of a systematic approach. For example, the synoptic situation may affect the ion concentration significantly, since a marine air mass may contain more ions and fewer aerosol, and continental air masses from different sources may contain varying amounts of aerosol. In the above papers, no attempt was made to analyse the data with respect to the meteorological conditions and hence only general conclusions can be drawn.

The over-riding characteristic is an increase in conductivity with altitude for two significant reasons: in the first 6 km of the atmosphere, the aerosol concentration drops exponentially with height, and in the troposphere it is significantly lower than in the boundary layer (Pruppacher and Klett, 1997). Incoming cosmic rays are also more intense at higher altitudes, so the ion production increases with height. The latter point explains the greater rate of increase, and lower variability, of conductivity observed above about 13 km. The classic expression for the variation of conductivity $\sigma$ with height $z$ (in km) is quoted in MKS units in MR (1998), but was originally derived by Woessner *et al* (1958):

$$\sigma_+(z) = 3.33\times10^{-14}\exp\left(0.254z - 0.00309z^2\right) \qquad \textit{Eq. 2.21}$$

$$\sigma_-(z) = 5.34\times10^{-14}\exp\left(0.222z - 0.00255z^2\right) \qquad \textit{Eq. 2.22}$$

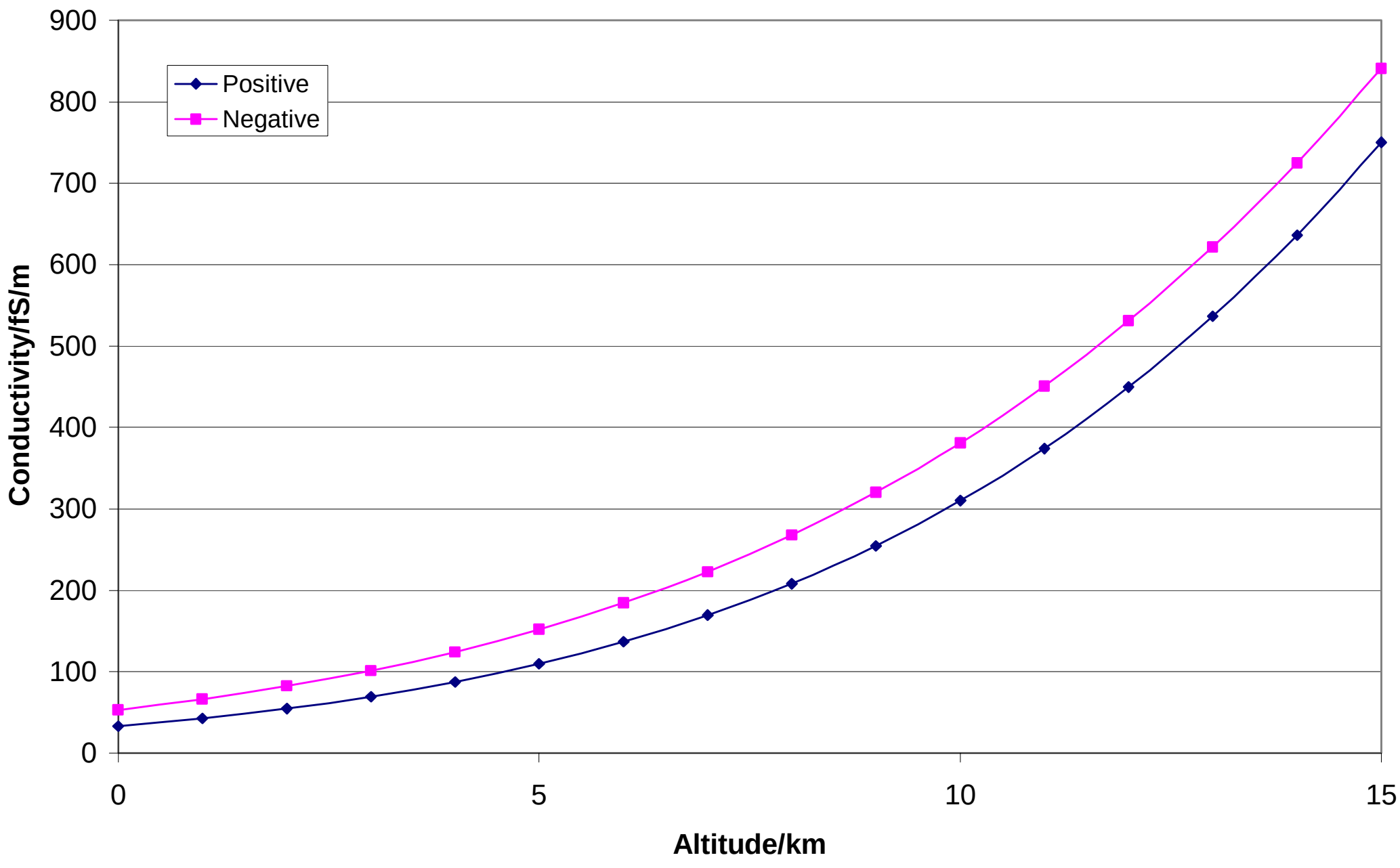

*Figure 2.5 Variation of conductivity with altitude from Eq. 2.21 and Eq. 2.22, derived by Woessner et al (1958).*

Woessner *et al* (1958), measured conductivity with Gerdien condensers suspended below balloons and operating in the Voltage Decay mode, as shown in Figure 2.5. Woessner *et al*'s (1958) expressions (Eq. 2.21 and Eq. 2.22) are derived from a theoretical consideration of ionisation by cosmic rays only, and the coefficients are obtained empirically from measurements. Positive conductivity at ground level is predicted to be 42 $fSm^{-1}$, which is about a factor of four higher than the average value. If only cosmic rays were being considered, conductivity would be under-predicted at ground level, however Woessner *et al's* model (1958) does not include the effect of aerosol, which significantly attenuates the ion concentration. In the boundary layer, natural radioactivity contributes to much of the ionisation, and the aerosol concentration is variable; these factors preclude accurate prediction of conductivity based on Woessner *et al's* (1958) criteria. Gringel (1978) also measured conductivity, by current measurement, and compared results between solar maximum and solar minimum. He calculated empirical expressions for the variation of positive conductivity with height, which were of the form

$$\sigma_+(z) = 1\times10^{-14}\exp(a + bz + cz^2 + dz^3) \qquad \textit{Eq. 2.23}$$

where the coefficients change depending on the solar cycle. Gringel's (1978) expression is plotted in Figure 2.6 below. Since Gringel's expressions are completely empirical and based on radiosonde ascents from the ground, they include the effect of aerosol and therefore predict conductivity more effectively at ground level than Woessner *et al* (1958). The conductivity is lower at solar maximum because the Sun's increased magnetic field is more effective at deflecting the lower energy cosmic rays away from the Earth's atmosphere, hence there is less ionisation, the difference being most pronounced at higher altitudes. Positive conductivity averaged over both solar cycles at z = 1 m is 27 $fSm^{-1}$, which concurs with other ground-based measurements. Woessner *et al* (1958) made their measurements between September 1956 and June 1957 which was at a period of solar maximum, which agree well with Gringel's (1978) for the same part of the solar cycle .

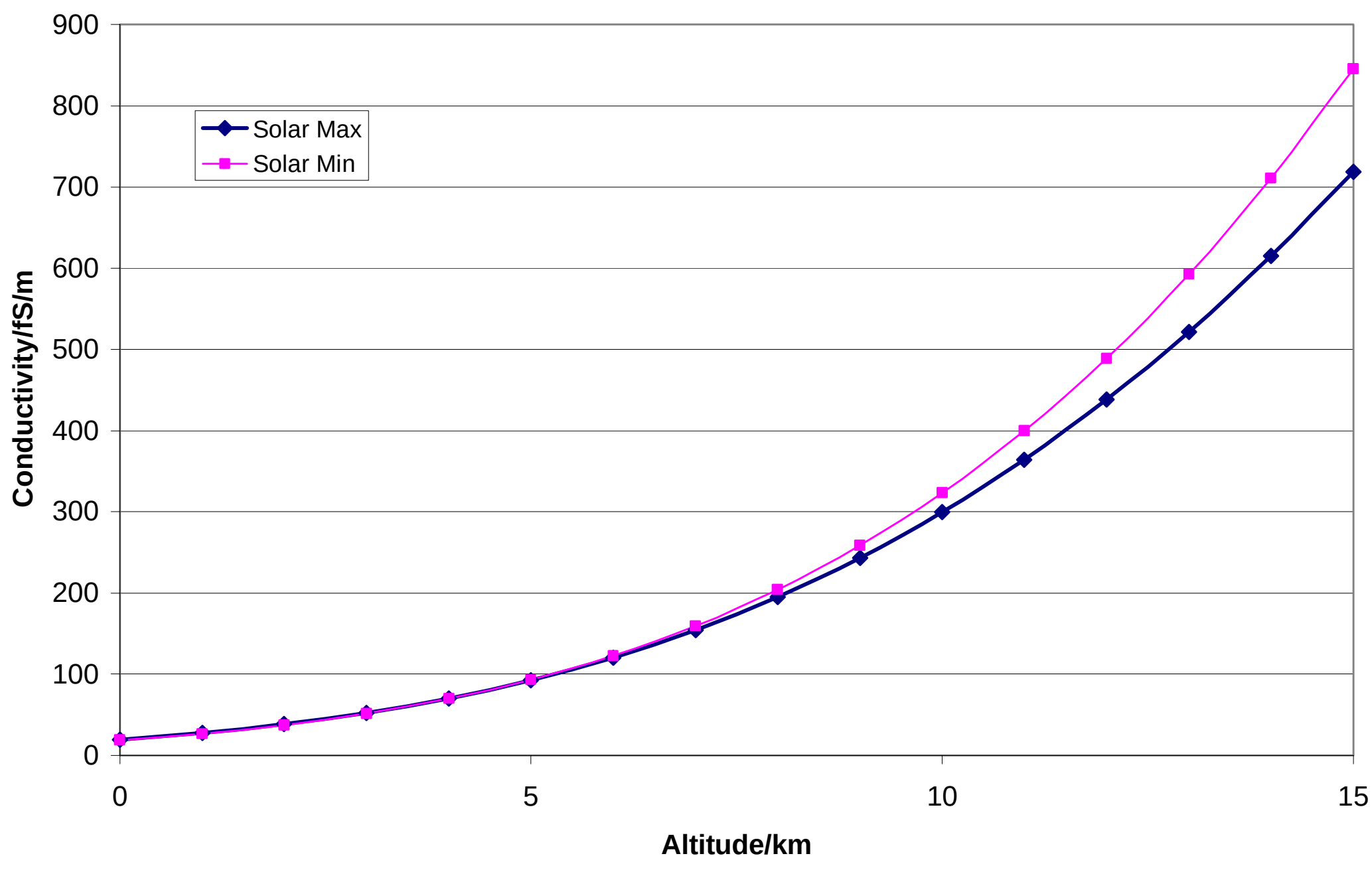

*Figure 2.6 Variation of positive conductivity with altitude and solar cycle, calculated using Eq. 2.23 from Gringel (1978).*

## 2.4 Errors in the Gerdien method

Errors in the collected fraction of ions result from diffusional losses to inlets and walls, and electrically produced errors (Anderson and Bailey, 1991). Repulsion at the inlet of ions of a suitable mobility to contribute to the measurement, due to the disturbance caused by the tube in the local electric field, has been recognised. This can be removed by keeping the inlet at local atmospheric potential to minimise such perturbations (Higazi and Chalmers, 1966; Anderson and Bailey, 1991). This technique is known as *guarding,* (Horowitz and Hill, 1994) and whilst worthwhile to reduce leakage currents and field perturbations, it requires additional driving and measurement circuitry.

Anderson and Bailey (1991) carried out a detailed study of the losses of ions by diffusion. They calculated expressions for the losses to the tube walls from theoretical expressions and previous conductivity measurements. The fraction of ions $n/n_o$ collected depends on the tube diameter $d$, flow rate $\Phi$ and tube length $L$:

$$\frac{n}{n_0} = \exp\left(cd^{-7/8}\Phi^{-1/8}L\right) \qquad \textit{Eq. 2.24}$$

($c$ is a dimensionless constant: for positive ions $c = -3.8 \times 10^{-3}$, for negative ions $c = -4.5 \times 10^{-3}$). Anderson and Bailey (1991) did not find a bend in the inlet pipe to have a noticeable effect, but Higazi and Chalmers (1966) reported that this produced a 15-20 % error. Another expression is cited in Hõrrak *et al* (2000) specifically for one instrument[16], is in a considerably different form ($c$ is a constant which varies with the dimensions of the instrument):

$$\frac{n}{n_0} = \frac{1}{(1 - c\mu^{0.67})} \qquad \textit{Eq. 2.25}$$

This suggests that Eq. 2.24 may not be directly applicable for all geometries, which may explain the differing results of Higazi and Chalmers (1966). Anderson and Bailey (1991) unfortunately neglect to state the dependence of Eq. 2.24 on geometry.

Since the determination of the conductivity from Gerdiens by current measurements clearly depends on the tube capacitance (Eq. 2.16), the value of the capacitance used and its accuracy are important, but not easy to determine due to stray capacitances *e.g.* in connecting leads. Calculation of the capacitance from Gauss's Law for an infinitely long cylindrical condenser is not appropriate because end effects introduce a substantial error. Smith (1953) describes a method to find the capacitance of a Gerdien by comparing measurements made with and without the central electrode present. It is of concern that the method of capacitance determination and the uncertainty are rarely cited, but it is apparent that Smith's (1953) method has become fairly common (MR, 1998). Blakeslee (1984) used a similar technique to Smith by applying a 20 kHz signal across the outer electrode and ground, with an alternating current (ac) voltmeter between the central electrode and ground, and tuned the circuit with a variable capacitor to give the same output after the central electrode had been removed. Although Blakeslee (1984) proves that the change in the variable capacitor is equivalent to the capacitance with the central electrode present, he gives no indication of the uncertainty. Blakeslee and Krider (1992) state that the capacitance of their Gerdien is "about 6.2 pF", suggesting that their technique for measuring it was approximate. The capacitance is usually the largest source of error in the calibration process. In Aplin and Harrison (2000) the relaxation time technique used has an associated error of about 20 %. However, since the development of a new calibrator based on a precision voltage ramp delivered to a capacitor (Harrison and Aplin, 2000a), this problem has considerably diminished[17].

[16] The Estonian Air Ion Spectrometer, which is discussed in detail in Section 2.6.2.

Moody (1981) believes that "ion-induced noise" is a significant problem with Gerdien condenser measurements. This apparently results from the impaction of oppositely charged particles to those being measured on to the Gerdien's electrodes, which can cause noise if the timescale of the fluctuations greatly exceeds the time constant $\tau$ of the electrometer

$$\tau = RC \qquad \textit{Eq. 2.26}$$

where $R$ is the resistance and $C$ the capacitance across the electrometer. For the values cited by Moody (1981) $\tau$ is 0.3-3 s. Moody (1981, 1984) designed a tri-electrode Gerdien with a grid electrode between the outer and central electrodes to solve the problem. However, he does not state how he was able to separate the effects of "ion-induced noise" from the other sources of noise or uncertainty in his measurements, or indeed speculate on what they might be. "Ion-induced noise" is only thought to be significant if the concentration of one sign of ion exceeds the other by a few orders of magnitude. The maximum measured atmospheric ion asymmetry ratio is only of order 10 (MR, 1998), so if Moody's "ion-induced noise" does exist, it is unlikely to be relevant to measurements in atmospheric air.

[17] This is discussed in more detail in Chapter 5 and Appendix B.

## 2.5 Air conductivity and aerosol pollution

Rutherford (1897) was the first to propose the idea that larger particles reduce the number of small ions by attachment. This behaviour explains the difference between the conductivity in clean air and urban locations, and the lack of dependency of conductivity on the ionisation rate in polluted regions. Since the new charged aerosol particle produced by attachment is much bigger than the original small ion, its mobility is orders of magnitude smaller, which reduces air's electrical conductivity. This effect was first observed in the atmosphere by Wait (1946), who formalised Rutherford's ideas following oceanic conductivity measurements on the cruises of the research ship *Carnegie* (Chalmers, 1967) from 1912 onwards. The fifty-year delay between Rutherford (1897) and Wait (1946) may have been because the concept of secular change of air conductivity resulting from a global increase in pollution would, by definition, require many years to be established. It is arguably not established now, a further fifty years on. To prove its existence, measurements need to be made over the ocean where the ionisation rate is not controlled by local radioactivity, and local effects do not dominate the ion concentration. These are rare, but recent results (Kamra and Deshpande, 1995) from a cruise in the Bay of Bengal, confirm that there has been a secular decrease in air conductivity this century. Kamra and Deshpande (1995) noted that there was a strong relationship between distance from the land, wind direction, and conductivity. Measurements of conductivity over the oceans are plotted below in Figure 2.7, although it is perhaps naïve to directly compare many sets of data made with different instruments under different conditions, particularly since the previous measurements were made at different distances from landmasses. In the light of this, the concept of a secular decrease in conductivity this century has been neither corroborated nor refuted.

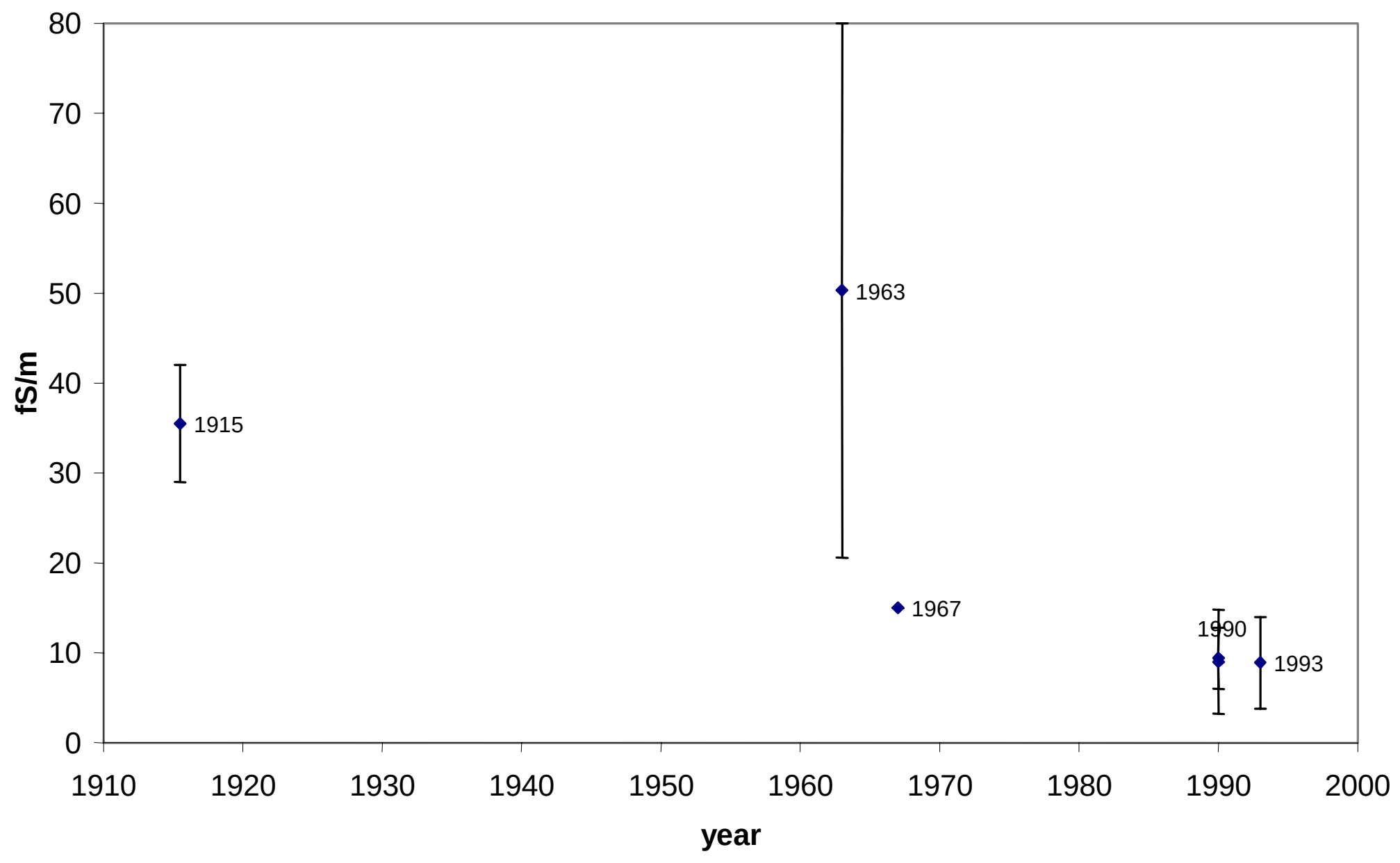

*Figure 2.7 Comparison of total conductivity measurements made over tropical oceans, from data presented in Kamra and Deshpande (1995). Where a range of conductivities has been given, the average has been plotted and the error bars are the range of values. If measurements were made over a period of more than one year (e.g. the Carnegie measurements were from 1911-1920) the average has been plotted.*

Attempts have also been made to quantify the relationship between air pollution and conductivity in urban areas. Retalis *et al* (1991) show there is a statistically significant negative trend in air conductivity at Athens from 1968-1980, and opine that there is a connection between this and increasing pollution; but no attempt is made to compare the results with any other measurements, this can therefore remain no more than speculation. Guo *et al* (1996) used two years' data from Toronto in 1973-4 in order to analyse correlations between total conductivity and aerosol, carbon monoxide, sulphur dioxide and nitrogen dioxide. They found negative linear correlations between conductivity and all the pollutants, but the best correlation (-0.46) was between conductivity and aerosol.

The theoretical relationship between conductivity and aerosol concentration can be simply derived from the steady-state solution of the ion balance equation (Eq. 2.19), assuming a monodisperse aerosol population

$$\sigma = e\mu\left(\frac{\sqrt{(\beta Z)^2 + 4\alpha q} - \beta Z}{2\alpha}\right) \qquad \textit{Eq. 2.27.}$$

Eq. 2.27 shows that Guo *et al's* (1996) assumption of a linear model between conductivity and aerosol number concentration has no physical basis unless the aerosol concentration is very high, so that $\beta Z >> 4\alpha q$ in the ion balance equation (Eq. 2.19). Although Guo *et al* (1996) did see an inverse correlation between conductivity and aerosol, it is regrettable that their paper was accepted for publication despite many of the plots having no units. It is therefore impossible to determine any quantitative link between conductivity and aerosol concentration from their work.

Guo *et al* (1996) were motivated to investigate the correlation of conductivity and gaseous pollutants because they "reduce the ion productivity of the air" with respect to the recombination term $\alpha n^2$. The recombination coefficient $\alpha$ is generally calculated from Thomson's theory (*e.g.* Harrison, 1992), which is based on the precept that two oppositely charged molecular ions will recombine if they are within a certain distance from each other, and when a third molecule enters the sphere of influence, it will catalyse the recombination process. So the implication of Guo *et al* (1996) must be understood as the presence of extra gaseous pollutants can increase ionic recombination. This is an unconventional viewpoint, but as some negative correlation was seen between ions and gaseous pollutants the idea may warrant further investigation. There could be many other explanations for the correlations, for instance if aerosol particle formation onto ions was catalysed by these trace gases there might be a similar relationship between the trace gases and conductivity. Perhaps the most likely explanation is that gases are acting as a proxy for particles, since sulphate aerosol is initially formed from sulphur dioxide (*via* homogeneous nucleation, which will be explained in Section 2.7). If this is so then the correlation may depend on the timescale of the measurement, if comparable with the particle formation processes. Guo *et al* (1996) use long-term average measurements, so their results should be unaffected.

Eq. 2.27 strongly suggests that there should be a quantifiable link between air pollution and the electrical conductivity of air. It is commonly proposed that this could potentially be useful as a proxy for aerosol measurement. However, Eq. 2.27 is reliant on the assumption that air conductivity is only influenced by atmospheric small ions. If intermediate or large ions contribute to the conductivity, as suggested by Dhanorkar and Kamra (1997), then there is scope for a positive $\sigma$-$Z$ correlation. There is a dearth of rigorous work investigating the relationship between $\sigma$ and $Z$, but systematic measurements should be capable of at least clarifying the issue.

## 2.6 Ion mobility spectrometry with a cylindrical condenser

An ionic mobility spectrum describes the distribution of ions across different electrical mobility categories. Since mobility is inversely related to ionic mass and radius (Tammet, 1995), then mobility spectra can be a powerful way to investigate ionic properties because they respond to number concentrations in the whole ion population, and can resolve changes in individual classes of small ions. Depending on time resolution, changes in the spectrum can be related to ionic growth processes. Ionic radii and masses can be calculated directly from mobility, thus providing size-resolved data with the additional benefit of electrical information. Mobility spectra of atmospheric ions can also be thought of as linking the electrical parameters of the atmosphere to the chemical ones (Nagato and Ogawa, 1998). Ion mobility spectra, with their potentially detailed information about ion evolution (primarily limited in time resolution), can complement mass spectra, which require discrete sampling but are able to give more precise chemical details. Hõrrak *et al* (1998b) believe a full air ion mobility spectrum gives a thorough description of the electrical state of air. This facilitates comprehension of the characteristics of both the global electrical circuit and electrically charged atmospheric aerosol, which are emerging as increasingly important in the climate system.

In the Gerdien condenser, changing the magnitude of the electrical field or the dimensions of the capacitor varies the fraction of ions that can pass through the tube without being drawn to the central electrode and contributing to the current. This is described by Eq. 2.11. Varying the voltage across the tube, or the flow rate through it, therefore provides a way to study the characteristics of ions of different sizes. If current measurements are made with different voltages across the capacitor, the ion concentration $n$ can be found if the conductivity and critical mobility are known.

### 2.6.1 Theory

Gerdien condensers and other ion spectrometers provide discrete measurements of number concentration in separate mobility categories. Ion spectra from such data can be calculated by a number of methods, discussed in this section.

The method can vary depending on which part of the ion spectrum is under consideration. For example, Hõrrak *et al* (1998b) have obtained spectra of the whole ion size range. For large ions, they calculate a least squares fit to a distribution similar to the log-normal distribution representing the broader aerosol spectrum. Brownlee (1975a) also fitted distributions by least squares regression to his ion spectrum measurements, choosing to use a log-normal distribution for the small ions and a delta function for the large ions. The difficulty with this method is that it assumes a form for the ion spectrum, which to some extent requires pre-solution of the problem before measurements have been attempted. If physical processes occur that are incommensurate with the previously selected model, they will reduce the "goodness of fit" to the model[18]. Consequently, events that might be worthwhile to study may be viewed as undesirable perturbations, or may not even be noticed if "goodness of fit" parameters are not quoted. In their comparison of spectral inversion methods, Hõrrak *et al* (1998b) observe that the technique of calculating the spectrum from the aerosol charge distribution over-predicted the number of particles of radius below 1.3 nm. They suggested that this was due to non-equilibrium charging because of the presence of very fresh charged particles, *i.e.* that a mechanism was occurring which had not been accounted for in their presupposition of the aerosol distribution. This exemplifies the difficulties of assuming a model for the ion spectrum, but data inversion would be impossible without a predetermined model to fit to.

Brownlee (1975a) also discusses a more analytical approach, deriving the "inverse analytic spectrometer transform" for measurements made with a Gerdien condenser, where $i$ is the current, $\mu$ mobility, $u$ flow rate, $n$ the $n$th mobility category and $f(\mu)d\mu$ the spectrum integral, as

$$\frac{di}{d\mu} = \frac{ue}{\mu_n}\int_{\mu_n}^{\mu_{n-1}} f(\mu)d\mu - \frac{i_n}{\mu_n} \qquad \textit{Eq. 2.28.}$$

[18] Identification and explanation of such events has, however, been described as a vital part of scientific progress (Kuhn, 1968).

The spectrum can be calculated from a series of measurements of ion currents $i_n$ for a range of mobilities $\mu_n$. Since the spectrum integral is only the number of ions in the range between each pair of mobility steps, the mobility spectrum obtained is in the form of a histogram with each bar width equivalent to the mobility interval and the height the number of ions within that mobility step. This method sacrifices mobility resolution but does not require *prima facie* assumptions about the nature of the ion spectrum in order to interpolate between ion concentrations obtained at discrete mobilities.

### 2.6.2 Instrumentation

There are two types of mobility spectrometer based on the Gerdien-like concept of a capacitor collecting an ion current. The *drift tube* is not ventilated, whilst the ventilated spectrometer is known as an *aspiration spectrometer*. The drift tube is preferable when there is a limited volume of air to be sampled, a relatively high ion concentration and small (high mobility) ions only are to be measured (Matisen *et al,* 1992). The aspiration spectrometer is probably more suited to measurements in atmospheric air, where air ions exist in a wide mobility range and varying concentrations.

Measurements of air ions can of course be made with other mobility spectrometers. But since by definition, the drift tube requires a particular sample of air to be analysed, they are poorly suited to *in situ*, high time resolution atmospheric applications. Drift tube spectrometers operate by measuring ion current as a function of distance away from the source, and Nagato and Ogawa's (1998) instrument is depicted in Figure 2.8. Since mobility is a measure of ionic velocity in an electric field, the currents at a certain time and distance from the source allow mobility selection. In this case the drift time $t_d$ is given by Nagato and Ogawa (1998) as

$$t_d = \frac{L^2}{\mu V} \qquad \textit{Eq. 2.29}$$

where $\mu$ is the mobility, $L$ distance from the source and $V$ voltage in the chamber. A further disadvantage of this spectrometer for air ion measurements, is that artificial ionisation with a radioactive source is usually necessary to provide big enough currents from the small volume of air sampled. However, drift tube spectrometers can give high mobility resolution and estimates of the times of evolution of different ions.

Air ion counters have been designed and made at the University of Tartu in Estonia for fifty years, but the first air ion *spectrometer* was made in 1972 (Matisen *et al,* 1992). The Estonian small ion spectrometer is depicted in Figure 2.9. It is a multi-electrode cylindrical capacitor, which measures currents from corona charging of the particles with MOSFET electrometers and can detect intermediate and small ions at concentrations down to 10 $cm^{-3}$ (Matisen *et al,* 1992). In Figure 2.9, the high voltage supply is used to produce the corona, and the voltage supply is the bias voltage across the electrodes. Since there are several electrometers measuring currents at different points along the tube, each current measurement comes from ions in a mobility range defined by the position of the connection to the electrometer. The instrument is located inside a building with ions entering through a 2m inlet made of aluminium tube, and three measuring capacitors of the same design are used to detect small, intermediate and large ions in a total of 20 different mobility categories simultaneously (Hõrrak *et al* 1998a). The mobility[19] range covered is 0.00032-3.2 $cm^2V^{-1}s^{-1}$ , once every five minutes. The results are corrected for losses to the walls of the inlet pipe using a correction derived specifically for the instrument (see Section 2.4).

Gerdien condensers have been used to measure ionic mobility spectra by Brownlee (1973, 1975a, 1975b), Misaki (1964) and Dhanorkar and Kamra (1993). Brownlee had an aspirated Gerdien sectioned into insulated collecting rings of different lengths, which measured naturally occurring air ions in a mobility range 0.96-0.23 $cm^2V^{-1}s^{-1}$. His aim was to develop a fast response instrument but it was limited to twenty minutes per spectrum by the availability of only one electrometer, which had to be used to store charge from each of the rings in turn, with reed relay switching. Dhanorkar and Kamra (1993) made simultaneous measurements of the currents from three Gerdiens of different sizes with one air inlet. Their instrument permitted variation of both the bias voltage and flow rate, but they chose to vary the flow rate to avoid transients. The mobility range covered was wide (3.37 – 6.91 x $10^{-4}$ $cm^2V^{-1}s^{-1}$), but it took four hours to scan through the whole size range. Dhanorkar and Kamra (1993) calculated mobility spectra by a differencing method from measurements made at different flow rates, but they do not explain their procedure. Calculating the spectrum every four hours assumes that the ionic properties of the air remained constant over this period. This is questionable since synoptic-scale meteorological phenomena can vary over a few hours, let alone the very rapid turbulent fluctuations of the air. Ion variations are not exclusively synoptic-scale features, but medium-scale variations which can happen in a few hours, like a change of air mass from clean maritime to polluted continental, will modify the electrical characteristics of the air. If Dhanorkar and Kamra (1993) did difference over a four hour period, then the local meteorological conditions will definitely have changed because of factors like the angle of the sun, and this alone is enough to cast doubt on their procedure.

[19] Since 1 $cm^2V^{-1}s^{-1}$ = 1 x $10^{-4}$ $m^2V^{-1}s^{-1}$, it is more convenient to refer to mobilities in units of $cm^2V^{-1}s^{-1}$, and this convention is adopted from here onwards.

*Figure 2.8 Ion drift spectrometer, from Nagato and Ogawa (1998). A and B are the ion source (radioactive), C 500V cell, D 1MΩ, E shutter, F drift ring, G aperture grid, H collector. Insulating spacers between drift rings are made of Teflon.*

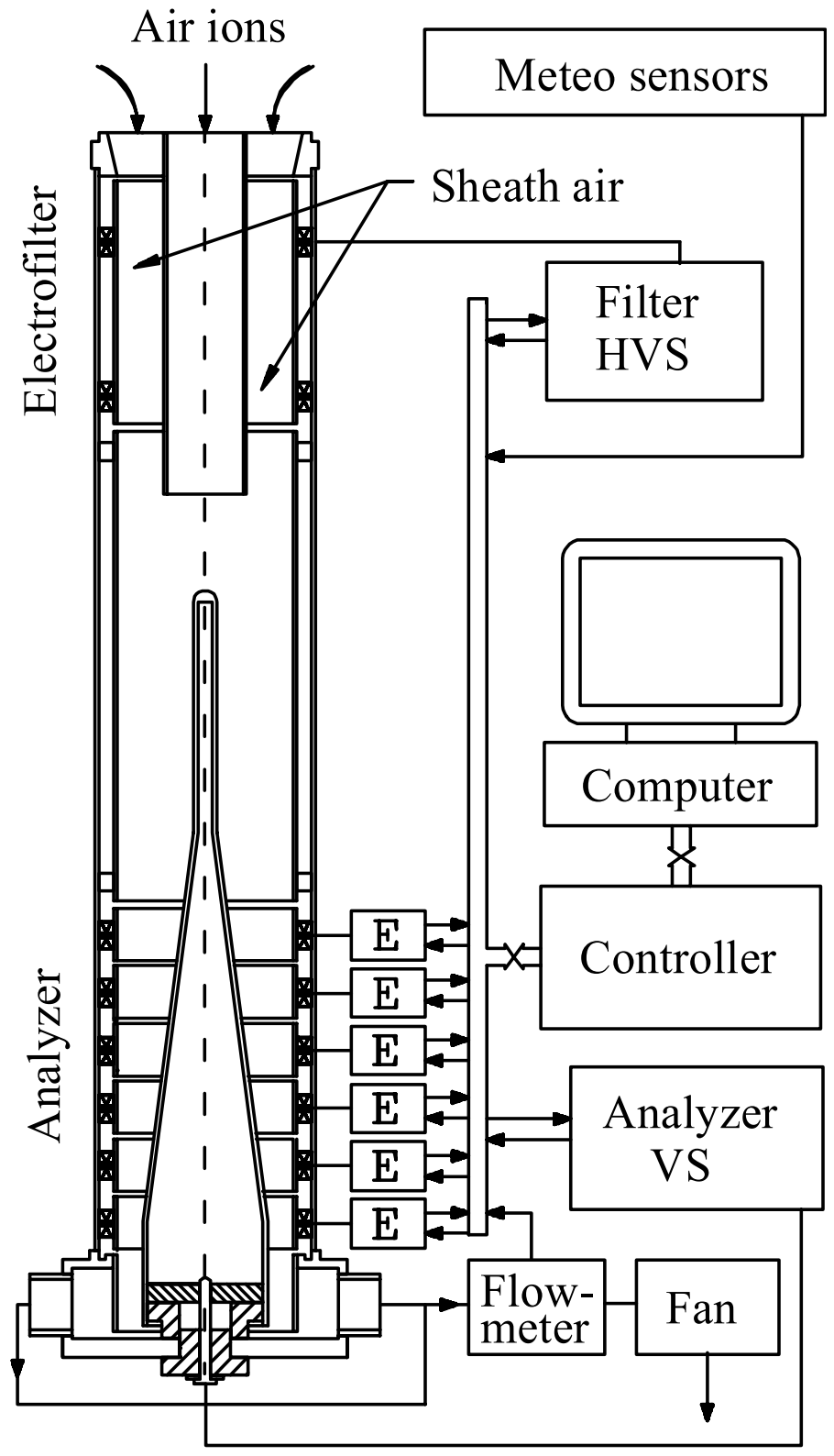

*Figure 2.9 The Estonian small air ion spectrometer, from Hõrrak et al (2000). E is an electrometer, HVS high voltage supply, and VS voltage supply. The height of the spectrometer is 69.5 cm, and the diameter is 12.2 cm.*

### 2.6.3 Air ion mobility spectra

Ion mobility spectra have often been measured in laboratories with the ions produced artificially. Whilst this is a valuable tool for analysing ions under controlled conditions, the behaviour of ions in the atmosphere is very different, since the atmospheric aerosol continually interacts with the ion population. As natural ion concentrations are lower than artificially produced populations, recombination is often exaggerated in ionisation chambers. Since this Thesis describes the development of an instrument to measure natural atmospheric ions, only measurements of ions in atmospheric air are considered here. Brownlee (1975a,b) ultimately only partitioned his own ion measurements into two mobility categories; therefore his results are not discussed here as mobility spectra.

Misaki (1964) obtained several hundred large ion spectra in the New Mexico semidesert from a Gerdien condenser with varying bias voltage and two central electrodes of different lengths. This arrangement enabled a wide mobility range to be studied, covering ions in the size range 7-44 nm. He also compared the measurements on one day to conductivity, space charge and residual current, and found an inverse relationship between large ions and electrical conductivity. This is because "large ions" are charged aerosol particles, so oppositely charged ions are even more likely to attach to them than neutral aerosol particles, enhancing the inverse relationship described in Section 2.2. The mobility spectra were rather smooth, with a tendency to shift to lower mobilities throughout the day. On several days peaks in the spectrum appeared mid-morning which shifted to very low mobilities ($10^{-4}$ $cm^2V^{-1}s^{-1}$) over the next few hours. Misaki (1964) thought this was due to a coagulation process, and suggested that future studies of ion growth measure neutral particles as well. However, Misaki (1964) did not make measurements of smaller ions, without which it is difficult to postulate a growth mechanism for large ions.

Dhanorkar and Kamra (1993) measured ion mobility spectra at Pune in India on two fair weather days at four-hour intervals. Interestingly, their data showed that there was an inexplicable excess of negative ions and space charge at 0400 and 0900, which is opposite to what is expected, given the electrode effect (MR, 1998). On both days the mobility spectra showed a slight shift towards lower mobilities throughout the day, coinciding with an increase in the population of the smallest ions in the afternoon.

Photochemical activity was suggested as an explanation for the new particle production later in the day. Natural radioactivity trapped in the nocturnal inversion produced high concentrations of new ions in the early mornings, but there was no associated peak of slower ions. This evidence suggests that the ionic growth processes observed are not associated with higher concentrations of radioactive gases, although the idea about photochemical activity was not investigated further by Dhanorkar and Kamra (1993).

Air ions have been measured at Tahkuse Laboratory, a rural site in Estonia, since 1988 (Hõrrak *et al,* 2000). This has permitted long-term average spectra to be produced, and other detailed analyses of air ion characteristics. Figure 2.10 is a summary of the measurements made at Tahkuse Observatory, which have been used to characterise and classify air ions in detail. Small ions are defined as having mobilities above 0.5 $cm^2V^{-1}s^{-1}$, after which there is a depletion of ions down to mobilities of about 0.05 $cm^2V^{-1}s^{-1}$. Figure 2.11 shows the average small ion spectrum, which is the longest average ion mobility spectrum available. In the air ion literature the average mobility is defined as the most frequently occurring mobility (Mohnen, 1974) and in Hõrrak *et al* (2000) the average mobilities are both in the range 1.3-1.6 $cm^2V^{-1}s^{-1}$, implying that the ion asymmetry ratio is almost 1. This is ascribed to an absence of the electrode effect because of the height of the inlet: about 2 m. According to Hõrrak *et al* (2000), the growth of small ions is usually thermodynamically hindered at $\mu < 0.5$ $cm^2V^{-1}s^{-1}$, so at this point there is a natural depletion of ions, defined as “intermediate ions”. Still slower ions are effectively charged aerosol particles, also known as large ions. Hõrrak *et al* (2000) state that the characteristics of the large ion spectrum concur with theoretical calculations for aerosol charging.

*Figure 2.10 Average air ion mobility spectrum, with classification categories, from Hõrrak et al (2000). Measurements were made at Tahkuse Observatory in 1993-1994.*

*Figure 2.11 Average small ion mobility spectrum, from Hõrrak et al (2000). Measurements were made at Tahkuse Observatory in 1993-1994.*

The standard deviation in the intermediate ion range is quite large, owing to regular "bursts" of enhanced intermediate ion production. Typical concentrations are about 50 $cm^{-3}$, but during the bursts, the concentration can increase to 900 $cm^{-3}$ over the day. Hõrrak *et al* (1998a), describe slow increases in the fraction of ions in the intermediate size range, peaking at about 10 nm, occurring over a period of about eight hours during daylight. This is indicative of an ionic growth process, and has occurred consistently in measurements made from 1988-1994. The events occur in fair weather periods, typically associated with anticyclonic, clean air masses from the Baltic Sea (Hõrrak *et al,* 2000). This behaviour is very similar to the observations by Misaki (1964) and Dhanorkar and Kamra (1993). Hõrrak *et al* (1998a) also suggest that these intermediate ions continue growing into large ions due to a high time-lagged correlation between intermediate and large ions. Two mechanisms were suggested, firstly that new particles were formed by homogeneous nucleation and then charged by small ion attachment. The second idea was that the particles were growing by an ion-induced nucleation process, following sudden bursts of some precursor gas, which was not nitrogen dioxide or water vapour, but was otherwise unknown.

There exist relatively few measurements of ion mobility spectra, and the only set of long-term averages are from the clean air measurements at Tahkuse Observatory. In a classic paper, Mohnen (1974) pictured the atmospheric ion spectrum as shifting and elusive, from which only instantaneous snapshots can be captured. With increasing observations of ion growth processes, there is even greater contemporary need to develop instrumentation which can focus on the areas clearly demanding further study.

## 2.7 Nucleation

A further motivation is the possibility that ions could be involved in the production of particles acting as condensation nucleii (CN) in the atmosphere, which is currently attracting increased attention. C.T.R. Wilson (1897, 1899) reported that new particles were made in the presence of ultraviolet radiation and radioactive sources. He was certainly the first person to suggest that ions might be involved in particle formation, and also to realise the importance this would have for climate. Apart from some sporadic studies, (Verzar and Evans, 1953; Bricard *et al,* 1968) it is only in approximately the last ten years that the importance of new particles in the atmosphere, and the potential rôle of ions in their formation, has begun to be recognised widely. Harrison (2000) has even suggested a modification to the ion balance equation (Eq. 2.19) to include a coefficient for ion-induced nucleation, $\gamma$, where $\gamma$ is proportional to the ion concentration raised to some power $p$ where $p$ is between 0 and 1, depending on the nucleation mechanism. If $\gamma$ does exist, then the ion balance equation can be rewritten as

$$\frac{dn}{dt} = q - \alpha n^2 - \beta n Z - \gamma n^p \qquad \text{Eq. 2.30.}$$

The recent surge of interest in this subject could have been provoked by several factors, for example unexplained observations of bursts of new particles in the atmosphere (O'Dowd *et al,* 1996, Mäkelä *et al*, 1997). Theories of binary homogenous nucleation, which is the classically assumed aerosol formation mechanism (O'Dowd *et al,* 1996), also predict rates that are too slow to account for observed aerosol concentrations (Yu and Turco, 2000; Kulmala *et al,* 2000). A number of laboratory studies have shown that particles can be produced in the presence of ionising radiation, although the findings are not concordant, and the experiments often involved so much radiation that their results cannot be extrapolated to atmospheric conditions. Climate models need to include aerosol particle source and loss terms accurately to parameterise the radiative effects of the atmospheric aerosol burden. Finally, the controversial theory relating the solar cycle to global cloud cover suggests that ionisation contributes to particle formation without postulating a microphysical mechanism (*e.g.* Svensmark and Friis-Christensen, 1997). Although this has attracted considerable attention to the topic, in particular to rebutting their theory, (a particularly recent example is Jorgensen and Hansen (2000)) little effort has been made to prove or refute the microphysical aspects.

### 2.7.1 Nucleation onto ions

*Heterogeneous nucleation* is the term used to describe the condensation of some vapour onto a particle of a different composition to the vapour. When the particle is an ion, the physics is quantified by the Thomson equation, which describes the variation in the Gibbs free energy of molecules clustering around a nucleus. When the Gibbs free energy $\Delta G$ is at a saddle point, particle growth becomes energetically feasible. If $r$ is particle radius, $\rho$ the fluid density, $S$ is supersaturation ratio, $M$ the mass of the molecule (in SI units), $q$ particle charge, $\gamma_T$ the surface tension, $k_B$ Boltzmann's constant, $T$ temperature, $\varepsilon_r$ the relative permittivity and $r_o$ the initial particle radius, the equilibrium condition for this situation can be written as:

$$\ln S = \frac{M}{k_B T \rho}\left[\frac{2\gamma_T}{r} - \frac{q^2}{32\pi^2\varepsilon_0 r^4}\left(1 - \frac{1}{\varepsilon_r}\right)\right] \qquad \textit{Eq. 2.31}$$

which expresses the necessary conditions for nucleation in terms of a supersaturation $S$ (Mason, 1971). The presence of a charged central particle lowers the supersaturation needed for condensation to occur. Wilson (1899) used his cloud chamber to measure the supersaturations required for nucleation in the presence of radioactive sources (*i.e.*

almost certainly onto ions), which were in the range 250-370%, which is too high to occur in the atmosphere. However, comparison with the greater supersaturations needed for homogenous nucleation illustrates the way in which electrical charge can stabilise cluster growth.

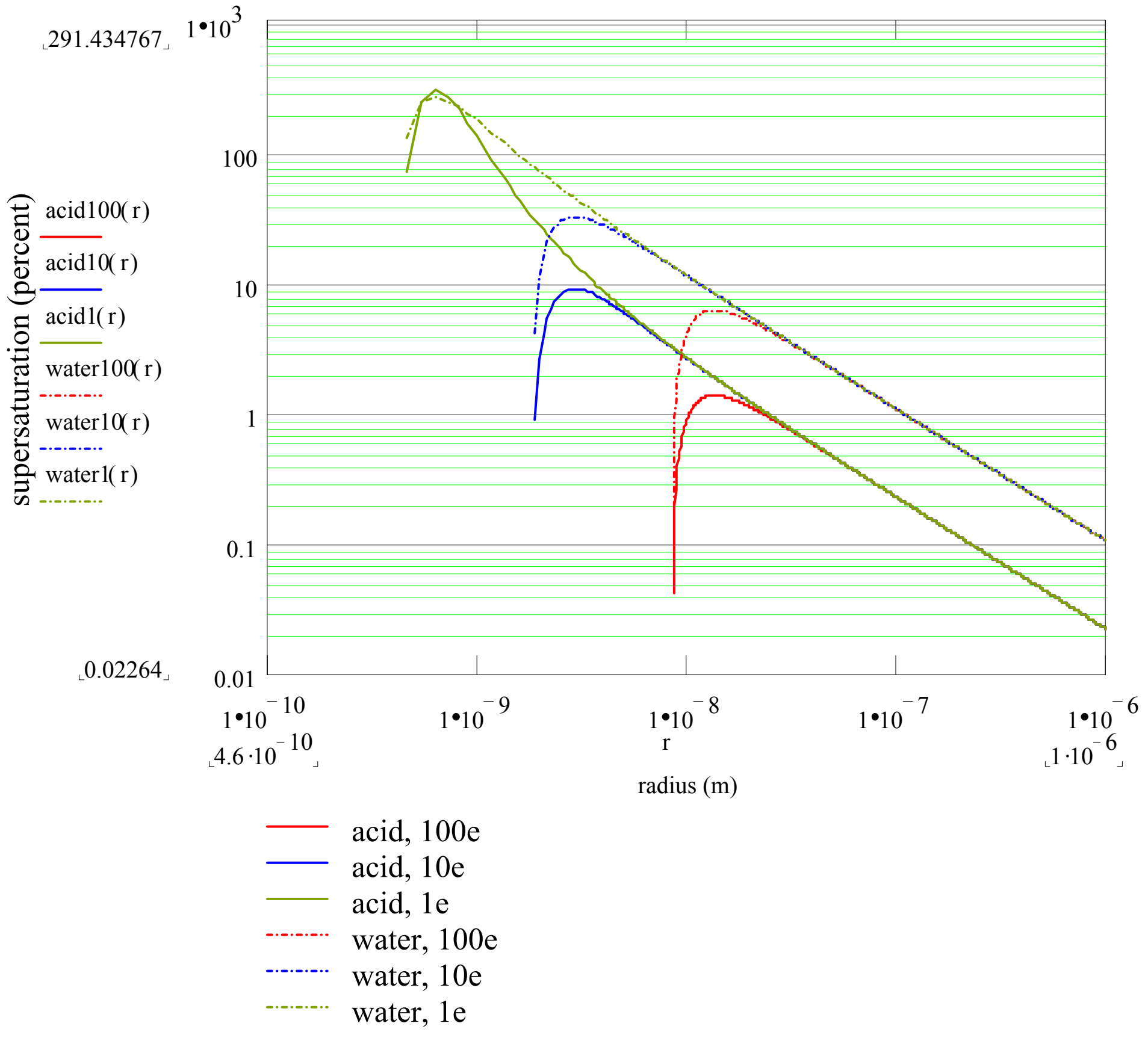

*Figure 2.12 Equilibrium vapour pressure as a function of particle radius from Eq. 2.31, for droplets of water and sulphuric acid growing onto ions with different elementary charges (at 300K). The saturation vapour pressure of sulphuric acid is calculated from the expression in Laaksonen and Kulmala (1991). The surface tension of sulphuric acid was assumed to be the same as water.*

Figure 2.12 shows that increasing the charge stabilises droplets at lower supersaturations. A water droplet can grow at atmospheric supersaturation levels when it has a critical radius of about 5 nm and a charge exceeding about ten elementary charges. This suggests that ions can act as nucleation sites for water droplet growth, but since few particles in the atmosphere hold the necessary charge, there is a low probability of this occurring. Ion-assisted nucleation of sulphuric acid is possibly more likely because of its lower vapour pressure, and Figure 2.12 shows that for greater than one elementary charge, sulphuric acid can be nucleated at lower supersaturations than water. Typically, sulphuric acid exists in the atmosphere at 300K with a partial pressure of $10^{-4}$ mbar (Warneck, 1987) and in concentrations of order 0.04 pptv (Kirkby *et al*, 2000). The saturation vapour pressure of sulphuric acid at 300K is around 400 mbar, so any supersaturation in these conditions is highly unlikely.

### 2.7.2 Laboratory observations of nucleation

Bricard *et al* (1968) measured the number of new particles formed in clean air in a chamber under the influence of ionising radiation, different trace gases and illumination. They found the CN concentration could be made to cycle up and down when regularly adding a short-lived source of α-particles (0.5 $Bqm^{-3}$)[20] on a five-minute timescale, when the concentration of new aerosol varied by approximately a factor of two (Figure 2.13). The radii of the new particles were estimated to be 2 nm in the presence of the radiation, and 12 nm out of it. The smaller particles correspond to the size of atmospheric small ions, which are produced in abundance by the radioactive source. However, further measurements showed that only about 10% of the total particle number was charged, indicating that in the presence of ionising radiation an excess of neutral particles were produced. This suggests an ion-ion recombination mechanism; but Bricard *et al* (1968) hypothesised that the coagulation rate depended on the sulphur dioxide concentration, in the presence of some other gaseous impurity.

Bricard *et al* (1968) also measured ion mobility spectra during the nucleation process, finding a number of well-defined ion mobilities, which they matched to certain species. Addition of $SO_2$ caused the mean mobility to decrease, and the characteristics of the mobility spectrum changed to a continuous shape, indicating ionic growth processes. Application of Thomson's theory for heterogeneous nucleation gave a *critical radius*, (the minimum radius of stable particles) of 0.7 nm if the mobility was 1 $cm^2V^{-1}s^{-1}$ (corresponding to a typical atmospheric small ion of radius 0.5 nm (Tammet, 1995)). It was suggested that the presence of sulphuric acid in the vicinity of the droplets promoted nucleation by increasing the supersaturation. The hypothesised physical growth mechanism (Bricard *et al,* 1968) was that the particles grew because neutral nucleii of below the critical radius captured a charged particle to stabilise and then grow. However, this does not explain Bricard *et al's* observations (1968) of production of neutral particles in a radioactive environment.

[20] 1 Becquerel = 1 decay per second

*Figure 2.13 The increase in particle concentration every time 10 rads*[21] *of thoron, a short-lived radon isotope, is introduced into a reaction chamber containing filtered urban air likely to be high in sulphur dioxide and other trace pollutant gases (from Bricard, 1968).*

In the 1980s there was a resurgence of interest in the formation of aerosol particles in the atmosphere. Increasing concentrations of the radioactive gas Krypton-85 from nuclear fuel reprocessing generally encouraged research into the atmospheric effects of ionising radiation. Observations of the environmental effects of acidic rainfall, and the depletion of forests near to nuclear installations, motivated studies of whether radioactivity could be implicated in the production of sulphuric acid aerosol in polluted (particularly $SO_2$ laden) environments (Raes *et al*, 1985).

[21] The rad is defined by Handloser (1959) as the amount of radiation which results in the absorption of 100 ergs/gram of any material. Therefore it represents the energy imparted to the medium by the radioactive source. In SI units, 1 rad = 0.01 J/kg.

Consequently there were several laboratory studies attempting to simulate aerosol production from ionising radiation, usually in air containing $SO_2$, and a radioactive source. Raes *et al* (1985) carried out experiments in synthetic air containing typically 0.015 ppm $NO_2$ and 0.5 ppm $SO_2$, with cobalt-60 (a γ-source) and ultra-violet irradiation. They found that at the higher dose rates, γ-radiation could produce an extra $10^5$ particles $cm^{-3}$, but at low dose rates, ultra-violet radiation was needed in addition to the ionising radiation to produce any new aerosol particles. The threshold activity was found to be $10^{-2}$ μ$Gys^{-1}$ [22], which is about three orders of magnitude higher than natural background activity. This study is unrealistic for the atmosphere, since γ-radiation is poor at ionising air and only contributes slightly to atmospheric ionisation, and Raes *et al* (1986) conceded that their results were not applicable to the atmosphere.

More recent work (Emi *et al*, 1998) contradicts both Bricard *et al* (1968) and Diamond *et al* (1985) who reported that only neutral particles were produced from ions. Emi *et al* (1998) describe results obtained with an ion mobility analyser, with an americium-241 source in air containing 5 ppm $SO_2$ and 3500 ppm water. They observed a mobility peak at 0.1 $cm^{-2}V^{-1}s^{-1}$, which clearly shows the presence of charged particles of 1 nm radius (Tammet, 1995). This is suggestive of an ion growth process, but since the particles formed are only intermediate ions (Hõrrak *et al*, 1998a) this provides no evidence of nucleation to larger aerosol particles. Adachi *et al* (1998) measured the charged fraction in a mixture of $NO_2$, ammonia (0.68 ppm), water (3000 ppm) and air, in the presence of two 3.03 MBq americium-241 sources. About $10^3$ $cm^{-3}$ new particles were produced after three minutes in the presence of small concentrations of $NO_2$, 80% of which were charged. This again suggests ion growth is occurring, but no information regarding the size of the particles produced was available.

[22] 1 Gray =1 J/kg

Diamond *et al* (1985) suggest a mechanism whereby ions are not directly involved in a chemical reaction to make sulphuric acid. Their mechanism involves an excited ion transferring its energy to assist in $SO_2$ oxidation, which then leads to nucleation. Although they argue relatively convincingly for this mechanism given their observations of the chemistry of different energy ions from different sources, they do not address the fundamental question of whether the precursor ions are likely to meet the energetic requirements for the reaction. Further research into the basic chemical properties of air ions is clearly required for this line of enquiry to progress further.

The experiments of Vohra *et al* (1984) appear to be the only laboratory studies of particle formation with levels of radioactivity naturally occurring in the atmosphere. They used a radon source of 100-500 pCi $m^{-3}$ (atmospheric concentrations are cited by Vohra *et al* (1984) as 90-400 pCi $m^{-3}$)[23], and found that 7 x $10^4$ $m^{-3}$ particles were formed in their 4:1 nitrogen-oxygen mixture in the presence of $SO_2$, ozone and ethene, with radon present at the lowest atmospheric concentrations. Since Vohra *et al* (1984) did not even mention ions in their discussions, it would be helpful for these experiments to be repeated in conjunction with detailed ion measurements. A similar experiment has been proposed at CERN to simulate atmospheric ionisation in a controlled environment (Kirkby *et al*, 2000). This will be the first of its kind, since the ionisation will be similar to that caused by cosmic rays, and may help to explain the observations of Vohra *et al* (1984).

It would be useful to have an empirical expression for ion (or particle) production as a function of radiation based on all these experiments. This is difficult because of the different radioactive sources and units used. In terms of ionisation, the units of Roentgen and Becquerel appear favourable to describe radiation, since Roentgens are defined explicitly in terms of ionisation in air: 1 Roentgen is the dose of X or $\gamma$-radiation such that the emission per 0.001293g of air produces 1 e.s.u. of electric charge (Handloser, 1959). In modern units, 1 Roentgen is equivalent to 208230 ions $m^{-3}$, or 8.7 mGy in air. The Roentgen is rarely used now because of its irrelevance to health physics, and the ionisation in air is only defined in terms of X and $\gamma$-radiation, which are not the major sources of air ions. The Becquerel is defined as one decay per second; a single radioactive source has a total activity, often defined in Bq, whereas for a fluid a volume must also be quoted. The experiments described above use a wide variety of sources, units and chamber volumes, not all of which are quoted to facilitate the calculations necessary for comparisons.

[23] 1 Curie = 3.7 x $10^{10}$ Bq

The majority of ionisation at ground level is from $\alpha$-radiation, which is effective at ionisation in air (hence its short range), so all $\alpha$-decays can be assumed to produce ions. As the energy required to make an air ion pair is constant, the ion production rate $q$ should be directly proportional to the activity expressed in Becquerels at ground level, because of the dominance by $\alpha$-radiation. However, the relationship is likely to vary with the type of ionising radiation, since the amount of energy given up by different radioactive particles to ionisation varies greatly. Mäkelä (1992) discusses a more generally applicable measure of ionisation, the *G-value*, which is the number of ion pairs produced per 100eV of absorbed energy. This varies depending on the species ionised, because of their varying ionisation energy.

| **Product** | **G-value (number of pairs/100 eV)** |
|---|---|
| $N_2^+$ | 2.24 |
| $O_2^+$ | 2.07 |
| $H_2O^+$ | 1.99 |

*Table 2.2 Number of ion pairs produced per 100 eV of absorbed energy for common atmospheric species, from Mäkelä (1992).*

A plot of the dose in Grays against $q$ should also be linear, independent of the source, since it is related the energy used in ionisation. The effectiveness (or not) of the radioactive species in ionisation is accounted for, because by measuring absorbed energy in a Geiger tube, this energy must have also been used in ionisation. As about 35.6 eV is required to ionise one molecule of air, an average value of G in air is 2.81 ions/100 eV.

### 2.7.3 Observations of nucleation in the atmosphere

Observations of new particle production in clean air, or air otherwise depleted of CN, for example in the marine boundary layer, have recently become available. These typically show bursts of particles in the nanometre size range, which are first observed in the morning and grow throughout the day. O'Dowd *et al* (1996) have observed sudden increases in the CN concentration, in marine environments in Antarctica and in the Outer Hebrides. During a particularly sudden nucleation burst in the Weddell Sea, the CN concentration increased from 550 to 3500 $cm^{-3}$ in 15 minutes, corresponding to a nucleation rate of about 3 particles $cm^{-3}s^{-1}$ (though Antarctic conditions are fairly unique in meteorological and biological terms). The CN concentration was reported to decay in 3-4 hours after the bursts due to rapid loss from mixing, advection and coagulation. This time scale is less than the time scale of CN growing to CCN ($\approx$ 50 nm). There were also regular observations of nucleation bursts in the Outer Hebrides, where CN grew from $<$ 8 nm to about 40 nm in approximately an hour, then decayed rapidly in half that time. O'Dowd *et al* (1996) were unable to suggest a mechanism for these events, but noted that at the Scottish site, there were large kelp beds near the coast leading to enhanced levels of dimethyl sulphide (DMS), a compound known to be significant in aerosol formation.

Mäkelä *et al* (1997) measured particles with radii from 1.5-250 nm with differential mobility particle sizers at Hyytiälä, in Finnish pine forest. Unlike O'Dowd *et al* (1996), particles were observed growing from 5-6 nm at a few nanometres per hour continuously until the evening, when they were big enough to be subsumed into the background aerosol. These events occurred during sunny days in spring and summer when there was a large temperature difference between the morning and early afternoon, and the relative humidity decreased consistently. On the days when nucleation was observed, there was often a reversal of the vertical temperature gradient when the stable nocturnal boundary layer broke down at 0800-0900. This suggested that the particle formation events were associated with vertical mixing and could have been produced at a different height to the measurements, although Mäkelä *et al* (1997) were confident that the smallest particles were produced *in situ*, probably because of their short lifetime. The origin of the particles could not be ascertained and was not directly related to any measured soil conditions. Photochemical or biogenic activity producing trace species was a possible cause, but photosynthesis was discounted since it does not occur in Finland in early spring.

### 2.7.4 Models

Mäkelä (1992) modelled the production of sulphuric acid aerosol in an irradiated environment. His first step was to model the formation of ions from radioactivity (assuming an aerosol-free environment, and that the dose rate distribution is Gaussian). He expressed the change of ion concentration *dn/dt* in unit volume as

$$\frac{dn}{dt} = GD(t)\rho - \alpha n^2 \qquad \textit{Eq. 2.32}$$

where $n$ is the ion concentration, $G$ the number of ion pairs produced per unit of absorbed energy, $D$ the absorbed dose rate, $\rho$ the gas density and $\alpha$ the self-recombination coefficient.

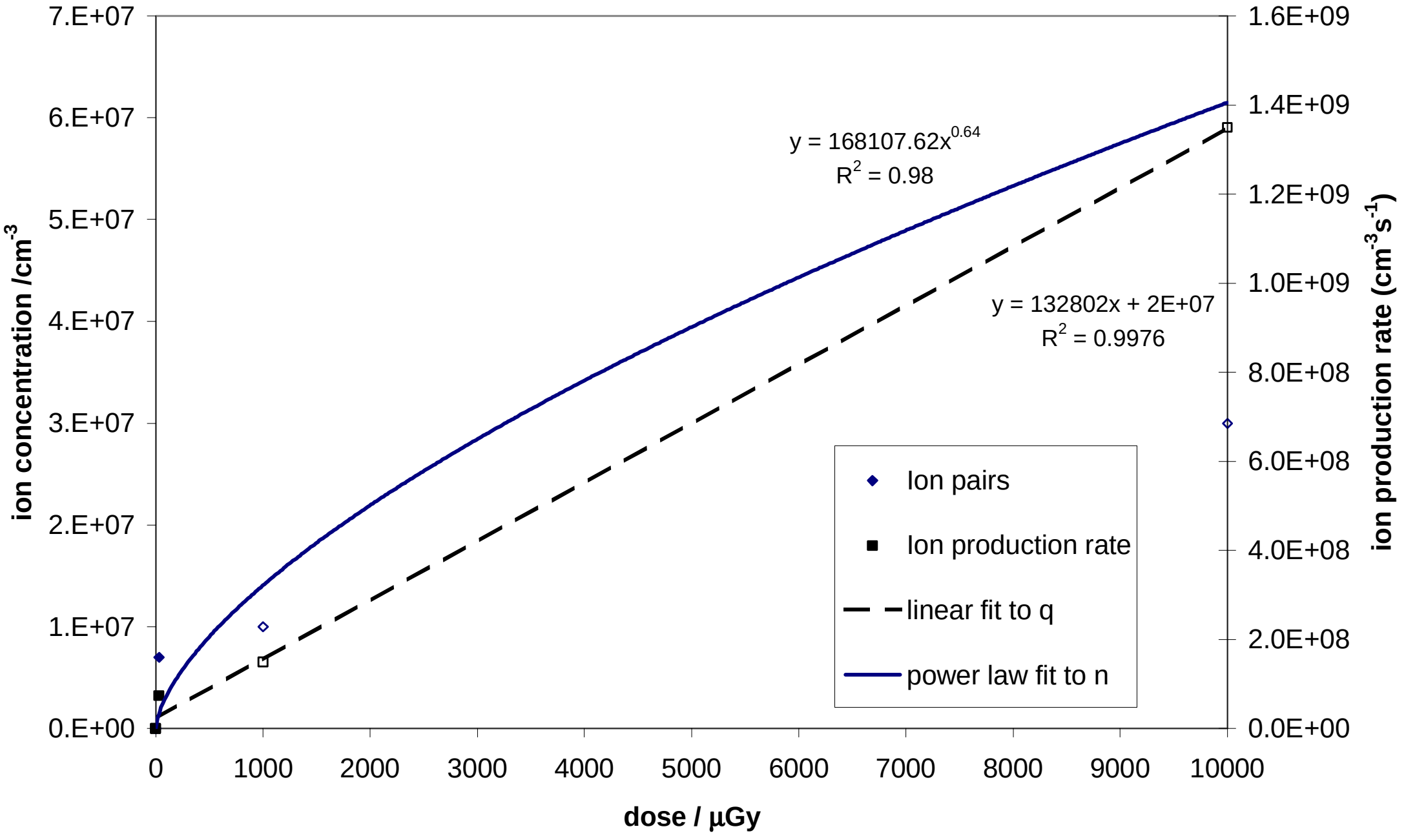

*Figure 2.14 Ion concentration (n) (left axis) and ion production rate (q) (right axis) as a function of dose rate. Shaded points are from Raes et al (1985, 1986), unshaded points are from Mäkelä (1992).*

Figure 2.14 shows the relationship between dose rate and ion concentration and production rate, from values of $n$ quoted by Raes *et al* (1985, 1986) and Mäkelä (1992), which they calculated assuming the recombination limit (Eq. 2.20). The ion production rate $q$, also calculated assuming the recombination limit (Eq. 2.20), shows a good ($R^2 = 0.998$) linear relationship with radioactive dose $D$. In the recombination limit, $D$ should be proportional to $n^2$. The best power law fit between $D$ and $n$ gives $D \propto n^{0.64}$, which shows that the recombination limit is a good approximation. The recombination limit is not a perfect fit to the data, because Raes *et al* (1985, 1986) reported aerosol production in their experiments (described in Section 2.7.2) so there would have been attachment effects. This plot is not applicable to atmospheric air, since the dose and ionisation rates are much higher than typical atmospheric ionisation rates (2-10 ion pairs $cm^{-3}s^{-1}$).

Mäkelä (1992), empirically estimated sulphuric acid aerosol production as a function of irradiation. From two sets of existing measurements, he was able to state that for a total dose of 1mGy, if one ion is produced per 100eV this gives 8 x $10^{13}$ particles $m^{-3}$. He then measured the production of ultra-fine aerosol particles in a water-sulphuric acid system in a chamber irradiated with $^{241}$Am. His results are shown in Figure 2.15. Above a dose of about 0.4 mGy, the number of particles formed varied linearly with respect to dose rate, with a gradient equivalent to the square root of the dose. This implies that the particles are behaving as ions under the recombination limit, which Mäkelä (1992) interprets as ion-induced nucleation of particles provoked by the very high ion concentration. This hypothesis is supported by his observations of shifts in the ion mobility spectrum towards larger sizes. The “critical dose rate” above which the experimental results become linear is probably when there are enough ions present for them to condense onto every sulphuric acid particle. Mäkelä (1992) also explains that at high dose rates, the particle concentration becomes steady because of attachment. These results are clearly inapplicable to the atmosphere, since the chamber is so highly ionised that the recombination limit affects even the new particles that are produced, and in the atmosphere the background aerosol would modulate the ion production rate, probably to the extent that it never gets high enough for ion-induced nucleation.

*Figure 2.15 Results obtained by Mäkelä (1992). $N_{i,max}$ (dotted line) is the ion concentration calculated from the dose using the G-value and assuming the recombination limit. $J_{ion}$ (filled circles) are model predictions for particles produced from ion-induced nucleation. $J_{hom}$ (crosses) are predictions for homogenous nucleation. EMS (empty squares) are the results of Mäkelä's (1992) experimental investigations, made at 20 °C, 15 % RH and 10ppm $SO_2$.*

A microphysical model of ions and aerosol including detailed electrostatic effects has been used by Turco *et al* (1998) and Yu and Turco (2000) to show that ions in atmospheric concentrations can facilitate production of sulphate aerosol by nucleation. Sulphate was chosen for use in the model because from observations, it is the most likely nucleating agent, and its thermodynamics is better characterised than other potential nucleating agents such as organic compounds (Turco *et al*, 1998). The microphysical model had previously been used to accurately predict particle formation from ions in aircraft plumes (Yu and Turco, 1997). Though under these conditions ion and sulphate concentrations are much higher than in the atmosphere, these results indicate that the microphysics of ion-mediated nucleation is an existing effect, which has been accurately parameterised in the aerosol model. The mechanism suggested by Yu and Turco (2000) is one whereby the recombination of a hydronium positive ion and a sulphate negative ion can produce a cluster that is larger than the critical size for nucleation. This effect is strongly dependent on the sulphuric acid concentration, with the degree of dependency varying non-linearly with the various feedback mechanisms. Turco *et al* (1998) calculate that the nucleation rate $J_{nuc}$ depends on the sulphuric acid concentration $[H_2SO_4]$ raised to some power $n$, where the constant of proportionality accounts for the ionisation rate

$$J_{nuc} \propto [H_2SO_4]^n \qquad \text{Eq. 2.33.}$$

The magnitude of $n$ is related to the number of ligands that are required to stabilise a neutral recombination product, and will vary according to the temperature and relative humidity. A secondary mechanism is that the electrical charge held by an ion cluster can also enhance the probability of a collision with a molecule of sulphuric acid by an order of magnitude (Yu and Turco, 2000).

In Yu and Turco (2000) the model is tested by applying similar code to the initial conditions for observations of particle formation from air ion mobility spectra at Tahkuse Observatory (discussed above), and also measurements of ultrafine aerosol and sulphuric acid made in Colorado. In both cases the model accurately predicts similar results to the experiments. However, the effects of background aerosol are ignored which meant that to get ion concentrations comparable with those measured at Tahkuse, the surface ionisation rate had to be set at a value which is comparable to the cosmic ray contribution only. Rather than scaling $q$ to 2 $cm^{-3}s^{-1}$ to arrive at $n = 1000cm^{-3}$, it might be better to retain a more typical atmospheric value of $q$ = $10cm^{-3}$ $s^{-1}$ (Chalmers, 1967). In this case, assuming a steady-state variation and using

$$Z = \frac{q - \alpha n^2}{\beta n} \qquad \text{Eq. 2.34.}$$

(Clement and Harrison, 1992) there should actually be an associated pre-existing monodisperse aerosol concentration of $Z \approx 1000cm^{-3}$, using an ion-aerosol attachment coefficient $\beta$ typical of a 0.2 μm radius aerosol. This presence of the background aerosol particles is particularly important because they can also remove the necessary condensable vapour, cutting off the nucleation (Harrison and Aplin, 2000b).

It is interesting that Yu and Turco's (2000) model should match the Estonian results so well, since their ion-ion recombination mechanism suggests that neutral particles should be produced, and in this case there would be no progressive shift in the mobility spectrum. This implies one of two things; either that the enhanced coagulation onto charged particles is actually more significant than their model suggests, or secondly that the sulphuric acid concentrations (which were not measured at Tahkuse) were selected so as to give results which fitted the observations best. The second set of results the model was compared are more appropriate since neutral particles were measured as well as sulphuric acid.

Kulmala *et al* (2000) propose a different mechanism, which has also had some success in explaining experimental observations. They have modelled a ternary nucleation system, of sulphuric acid, ammonia and water, which predicts nucleation under typical tropospheric sulphuric acid concentrations. Other suggested mechanisms tend to rely on transiently high supersaturations of sulphuric acid for nucleation, which are both unproven and difficult to measure. They suggest a model where there are reservoirs of thermodynamically stable clusters (TSCs) which are too small to detect (< 1nm), and certain conditions will trigger their growth to resolvable sizes, which is when a "nucleation burst" occurs. The detectable TSC concentration is dependent on the existing aerosol concentration; when this is high the TSCs are lost by coagulation to larger aerosol particles. The sulphuric acid concentration is also important and nucleation rates are about a thousand times slower than in binary nucleation because of the lower concentrations required. If, however, air containing TSCs is advected over a mass of a source of condensable species (for example, photochemically produced organics) this can produce a burst of larger particles which can continue to grow to the sizes detected by Mäkelä *et al* (1997). In Kulmala *et al* (2000), their ternary nucleation model is applied to the preliminary conditions for nucleation at Hyytiälä in Finland and matches the observations well. It does not appear to be possible to test this model properly until aerosol instrumentation that can measure particles in the 1-3 nm size range is available. This nucleation mechanism, however, does not invoke ions, unlike Yu and Turco (2000) and Hõrrak *et al* (1998a, 2000), since the TSCs are neutral particles and for there to be charge equilibrium with the background small ion concentration, only a tiny fraction could be charged. Kulmala *et al* (2000) firmly rule out ion-induced nucleation because their ternary nucleation mechanism (in conjunction with advection of condensable vapour) occurs at rates of 50-100 $cm^{-3}s^{-1}$, which considerably exceeds average atmospheric ionisation rates. However, the ionisation rate can vary considerably (Harrison, 2000) and it is possible that short-lived transients caused by highly energetic particles could cause the ionisation rate to peak for long enough to push the TSCs into the detectable size range.

There is a well-established body of evidence that ionising radiation can result in the production of fine particles. Additionally, it seems probable that ions are implicated in this process. However, the paucity of experimental atmospheric observations of ions and particle formation, and lack of a plausible theoretical basis has meant that experimentation has not been directed towards investigating ions and atmospheric nucleation. Only very recently have microphysical simulations suggested nucleation mechanisms feasible in atmospheric conditions.

## 2.8 Conclusions

The two differing theories of atmospheric aerosol formation, both of which have successfully explained aspects of existing experimental data, now provide a very strong motivation to measure air ion mobility spectra. In one of these theories, ions are crucial, and accurate measurements of the atmospheric ion spectrum are clearly required to corroborate or refute. There is no reason why both the mechanisms proposed by Kulmala *et al* (2000) and Yu and Turco (2000) cannot apply under different conditions, but further investigations of ultrafine aerosol particles, in conjunction with meteorological, electrical and chemical measurements will be required to differentiate between them.

At the moment, the instrumentation required to test the hypothesis that ions are involved (by whatever mechanism) in CN production is not available. Instruments to measure ion mobility spectra have been developed and deployed, but there are often play-offs between mobility and time resolution, and fundamental assumptions about the nature of the ion spectrum. Instrumentation developed to date has usually had its time and mobility resolution rigidly fixed at the outset as a consequence of its design, although suitable operating conditions may only be known after observations have been made. The classical Gerdien instrumentation also fundamentally lacks a viable calibration technique, despite the availability of an alternative mode to check self-consistency. There is a pressing need for a flexible and portable instrument that can be utilised for a variety of purposes.

*"In order to make it work, you had to spit on the wire in some Friday evening in Lent" Patrick Blackett*[24]

# 3 A miniaturised Gerdien device

In this chapter, an ion counter based on the classical cylindrical capacitor designed by Gerdien (1905) is analysed in terms of its separate components. Firstly, some semantics are required to differentiate the components of the ion counter. The term "Gerdien tube" eponymously refers to the part of the instrument which replicates Gerdien's, *i.e.* the central and outer cylindrical electrodes. "Gerdien condenser" is used to describe the capacitative properties of the Gerdien tube. The instrument described here differs from the classical Gerdien device in that it includes a third, screening electrode, and here this configuration is simply called a "Gerdien". "Gerdien system" refers to the complete instrumentation necessary to measure ions, which consists of the Gerdien, associated electrometer and bias voltage supply.

The objectives of the chapter are to attain a detailed knowledge of the Gerdien system components, and to test the instrument in atmospheric air. The physical properties of the tube and the flow through it, and the measurement systems are all examined in terms of setting design criteria (Section 3.1), and their direct application to making measurements in the atmosphere (Section 3.2). Some measurements are made with the Gerdien system in air in Section 3.3. They are tested against theoretical predictions, and for self-consistency.

## 3.1 Design criteria

### 3.1.1 Tube material and dimensions

The critical mobility defines the minimum mobility of ion which contributes to the measurement (see Eq. 2.11) and is the single most important Gerdien design specification. The instrument dimensions need to be chosen so that the critical mobility, given the tube voltage and airflow rate, is an appropriate value for measuring atmospheric ions. Since typical ion mobilities ≈ $10^{-4}$ $m^2V^{-1}s^{-1}$, the critical mobility needs to be of this order of magnitude for conductivity measurement (see

[24] (about the Geiger tube). This quotation was on display at the Science Museum in April 2000.

Section 2.2.2). If the Gerdien is to be used as an ion mobility spectrometer, the tube dimensions need to be chosen so that a realistic voltage range will investigate the desired part of the mobility spectrum. Exact details will depend on the context, for example the ion size range to be investigated. Limitations to the bias voltage supply are also significant: large voltages are difficult to switch and control accurately, being prone to ripple (Harrison, 1997b).

If the critical mobility is low enough so that the vast majority of the air ions are collected, there is no change in conduction current with voltage. This is the flat region on Figure 3.1, known as *saturation*, when the Gerdien only counts the total number of ions. Outside the saturation region there should be an approximately Ohmic relationship between bias voltage and measured current, depicted as the straight line in Figure 3.1. To see a true Ohmic relationship there would need to be a unimodal ion mobility spectrum; in reality, changes in the mobility spectrum cause slight variations in the conductivity at different voltages (as mentioned in Section 2.2.2). The best way to maintain Ohm's law is to ensure the Gerdien is sufficiently ventilated, since the critical mobility is directly proportional to the flow rate (MR, 1998).

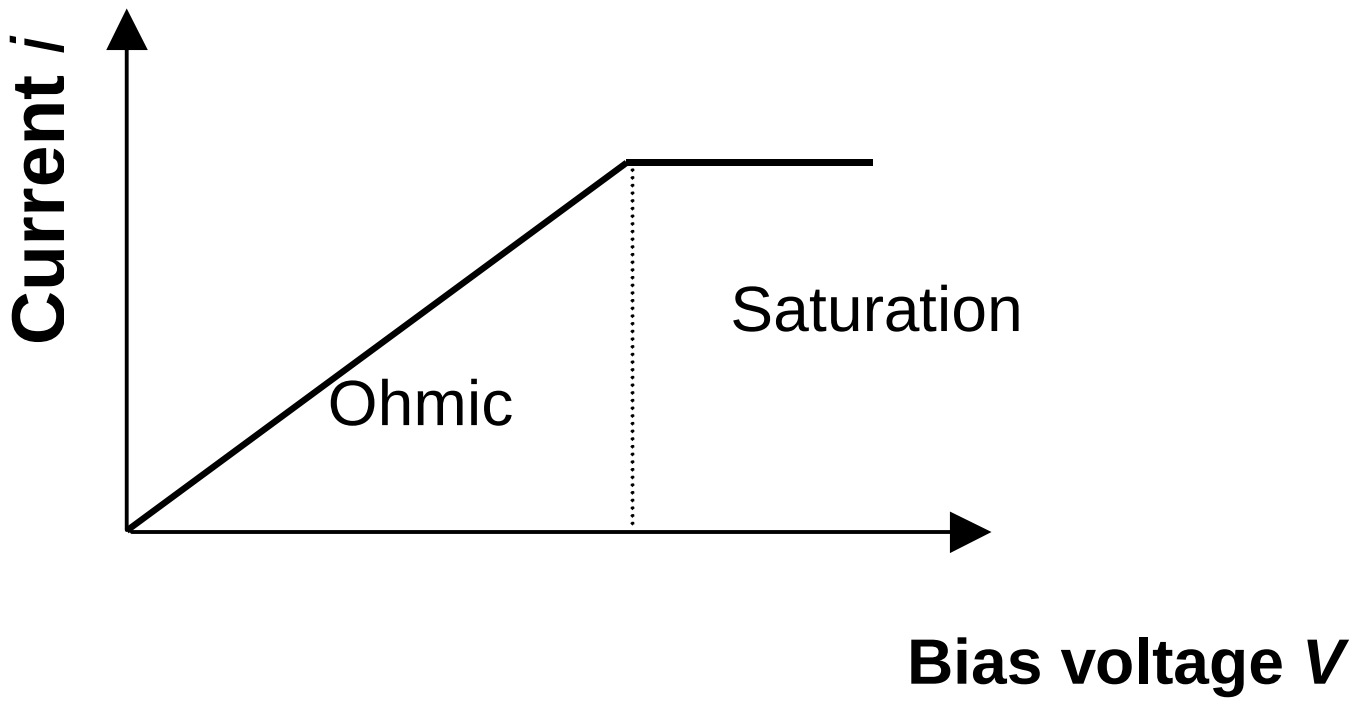

*Figure 3.1 Idealised i-V relationship for a Gerdien condenser, showing the two operating régimes (after Chalmers, 1967).*

In Chapter 2, tube material was discussed briefly. The literature suggests no scientific reasons to prefer one metal in particular, although it seems sensible to start with a material that has previously been used successfully. Aluminium tubes were used by Hatakayema *et al* (1958) and Venkiteshwaran (1958), and for the Gerdien tested in this chapter, aluminium was initially chosen for the outer electrode because of its workability. The central electrode was made of stainless steel wire. A second aluminium cylinder, which was earthed, was used as a screen to reduce electromagnetic interference. The outer electrode and screen were insulated from each other with a PVC separator. The tube dimensions are tabulated below, and Figure 3.2 is a simplified diagram of the system.

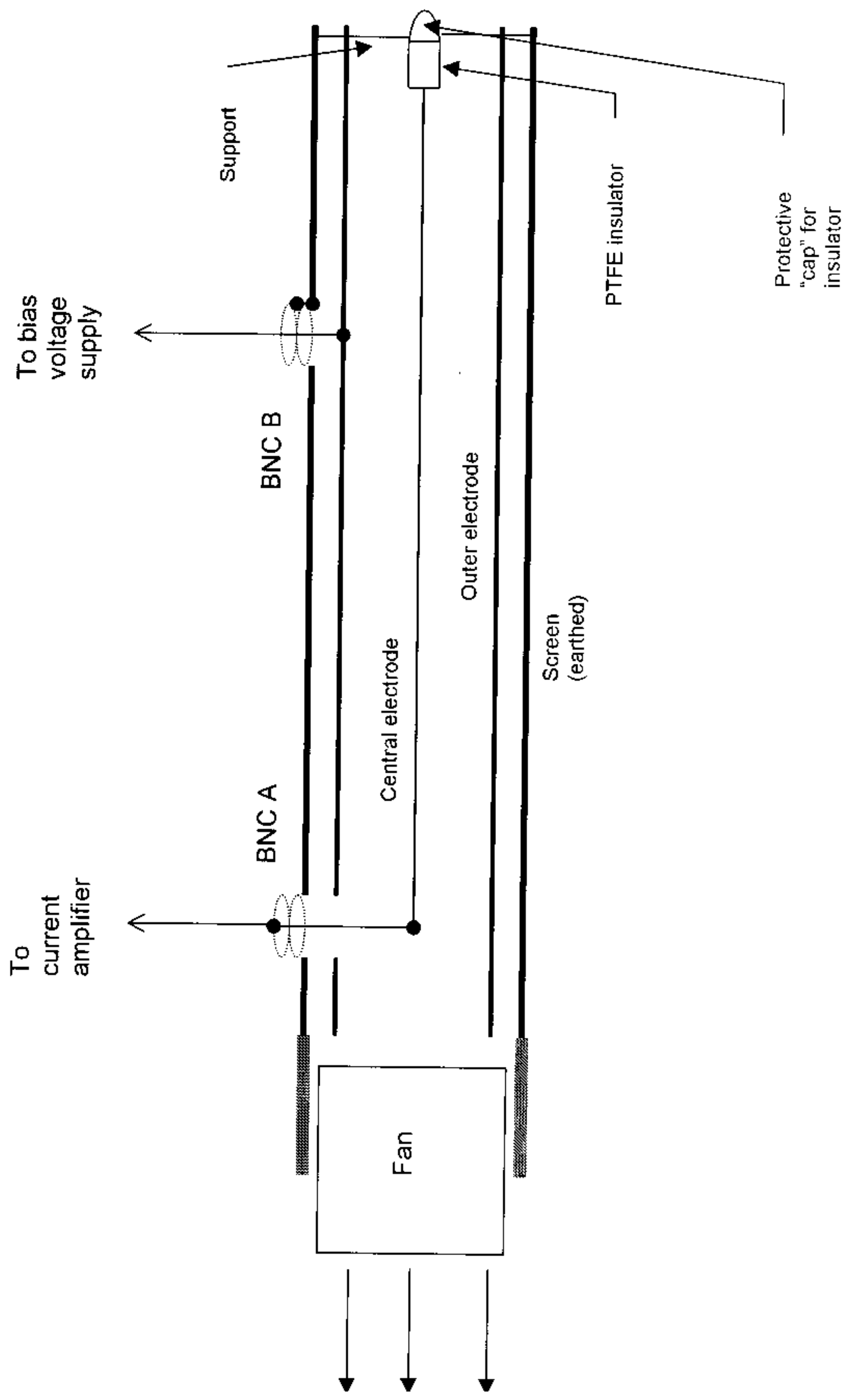

*Figure 3.2 Schematic diagram of the Gerdien system in Current Measurement mode. BNC B supplies the bias voltage. BNC A measures the ion current, and its outer connection is driven by the current amplifier.*

| | |
|---|---|
| Outer electrode radius *a* (m) | 0.0125 |
| Central electrode radius *b* (m) | $7.94 \times 10^{-4}$ |
| Length of central electrode *L* (m) | 0.258 |

*Table 3.1 Summary of the dimensions of the Gerdien condenser described in this chapter*

There were two BNC connectors mounted on the outside of the aluminium screen, with their casing secured to the outer electrode with nylon screws.

*a) current amplifier connection – BNC A in Figure 3.2*

This connection was located level with the end of the central electrode, which was bent at right angles and soldered directly to the inner of the BNC. The outer of the BNC was earthed to the screen.

*b) bias voltage connection – BNC B in Figure 3.2*

This BNC connects to the bias voltage supply to provide a potential across the outer and central electrodes.

The central electrode has to be particularly well insulated from other possible current paths, because it conveys the ion current, which is so small that great care has to be taken to avoid leakage. One of the highest quality insulators available today is Teflon[25], with a volume resistivity exceeding $10^{18}$ Ωcm (Keithley, 1992). Additional beneficial properties are that it is hydrophobic, repels surface contaminants and, when compared to other high performance insulators such as sapphire, it is inexpensive (Keithley, 1992). For these reasons, Teflon has been deployed for high quality insulation in the Gerdien tube. Insulating the end of the central electrode to the current amplifier connection, where there is no electrical connection but the electrode needs to be supported, is crucial. A small cylinder of Teflon with a hole drilled in it, to wedge the central electrode in, was used at the inlet end, with a cross-support of thin wire wrapped around the insulator and wedged between the PVC supports.

[25] Teflon is the trademark (DuPont Corporation) of polytetrafluoroethene (PTFE)

The insulated end of the central electrode is at the Gerdien inlet. It is important that the disturbance to the air entering the tube is minimised to reduce turbulence, whilst ensuring charge does not accumulate on the insulator and perturb the electric field at the inlet. For these reasons, an electrically floating metallic cap with rounded edges in a “bullet” shape was attached to the insulator at the end of the central electrode to protect it, whilst keeping the flow into the tube as smooth as possible.

### 3.1.2 Flow rates

The critical mobility (derived in Section 2.2.2) is determined by the ratio of flow rate $u$ and bias voltage $V$ such that

$$\mu_c = k\frac{u}{V} \qquad \textit{Eq. 3.1.}$$

where $k$ is a constant (with dimensions of length) related to the physical dimensions of the tube: outer electrode radius $a$, central electrode radius $b$ and tube length $L$, as

$$k = \frac{\left(a^2 - b^2\right)\ln\left(\frac{a}{b}\right)}{2L} \qquad \textit{Eq. 3.2.}$$

As can be seen from Figure 3.3, the critical mobility varies considerably with $V$ and $u$. The flow rate and bias voltage therefore need to be selected in order to measure the required critical mobilities. It is clear that some source of external ventilation is required to prevent the tube saturating and operating in the ion counter mode for periods of low wind speeds, for example during testing in the laboratory. The bias voltage has to be chosen once the average flow in the tube has been estimated.

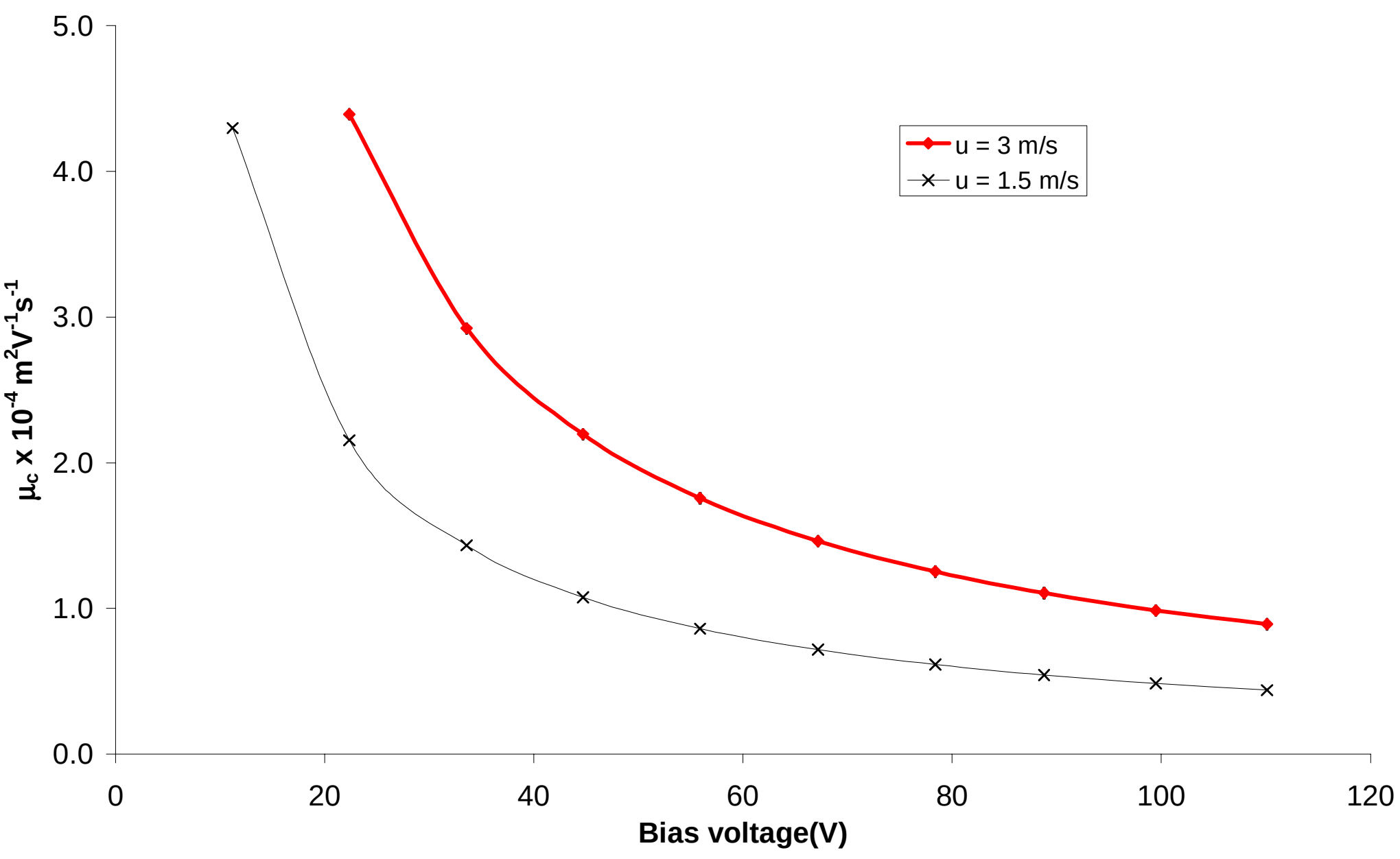

*Figure 3.3 Critical mobility $\mu_c$ against bias voltage at two flow rates for k = 0.00178 m, the value used in the Gerdien experiments discussed in this chapter.*

For the Gerdien used here, a small electric fan in a cylindrical plastic casing[26] was inserted in one end of the tube to ventilate it. If the polarity is reversed when connecting it to the supply, air can alternatively be sucked through the tube, passing through the fan after the electrodes. This is preferable to measurements being made from air that has already been disturbed by the fan. Previous conductivity measurement do not always state the direction of ventilation, but since many of them had the Gerdien located inside a building or at a window, they would have needed to suck air through the tube (*e.g.* Higazi and Chalmers, 1966).

The flow rate in the Gerdien was measured with a hot wire anemometer[27] inside a cardboard tube of similar dimensions to the Gerdien, ventilated with the fan. A 12V supply gave a flow rate of about 2.2 ms$^{-1}$. The bias voltages were selected based on this, using a reversible, variable voltage supply constructed from ten 12 V batteries to provide Gerdien voltages in the range 12-120 V with a ten-way switch. This was connected to the outer electrode *via* a BNC connector on the tube. When smaller voltages were needed, a battery or millivolt calibrator could be used.

[26] 12 V dc, ~ 40 mA, RS components part no. 600-904
[27] Testo 435 meter, measures to ± 0.1 ms$^{-1}$

Chalmers (1967) explains the need for Ohm's Law to apply in conductivity measurements, as a way of ensuring that the ion current never totally depletes the ion concentration in the air near the electrodes. Chalmers (1967) defines a criterion to maintain Ohm's Law: the streamlines of the fluid flow (in this case, the fluid referred to is ionised air) in a ventilated tube must be parallel to the electrodes, which implies that the flow must be laminar (Tritton, 1988). Otherwise, Chalmers (1967) believes the decay of charge from the condenser may not follow the classical exponential formation in Eq. 2.3, from which the conductivity by the Voltage Decay method and direct current measurement can be derived[28]. Perfectly laminar flow will be difficult to maintain in a Gerdien tube, because of the ventilation requirement. The transition to turbulent flow (*i.e.* random and chaotic fluctuations of turbulent eddies in the fluid) can be defined in terms of the Reynolds number, *Re*, which is given by

$$\mathrm{Re} = \frac{\rho u L_s}{\eta} \qquad \textit{Eq. 3.3}$$

(*e.g.* Tritton, 1988) where $\rho$ is the fluid density, $\eta$ the fluid viscosity, $L_s$ a characteristic length scale and $u$ the mean flow speed. Since turbulence starts to occur at Re ~ $10^3$, then for a Gerdien of comparable dimensions to the one used here (length 25 cm, radius 2 cm), the flow will be turbulent for speeds exceeding about 0.1 $\mathrm{ms}^{-1}$, which are likely to be insufficient to measure conductivity. Chalmers' (1967) criterion therefore seems approximate.

Consider the motion of ions in the tube; in a perfectly laminar flow they will be subject to two forces only. An electrical force will act towards the central electrode, and the mean flow will blow the ions along the tube. In turbulent flow there is a third effect from the action of turbulent eddies, which introduce a chaotic aspect into the ionic motion. As long as the mean flow exceeds the scale of the turbulent eddies, the ionic motion predicted by theory (resulting most importantly in the critical mobility Eq. 2.11) should still hold. In terms of the design of a Gerdien condenser, the condition of laminar flow imposed by Chalmers (1967) can be relaxed, as long as it can be proved that the mean flow exceeds the motion due to turbulent eddies.

[28] An alternative and perhaps more intuitive derivation is usually used for the Current Method, described in Section 2.2.2

### 3.1.3 Measuring the ion signal

Eq. 2.17 gives the relationship between ion current *i*, Gerdien bias voltage *V* and capacitance *C* as

$$\sigma = \frac{\varepsilon_o i}{CV} \qquad \textit{Eq. 3.4.}$$

If typical values of atmospheric conductivity are 10 $fSm^{-1}$, the Gerdien bias voltage is 20 V and its capacitance is 10 pF, then substituting into Eq. 2.17 suggests that typical Gerdien currents from an instrument of that specification will be ~ 200 fA in magnitude[29]. The small magnitude of these currents implies that a specialist electrometer is necessary. Commercially available instruments exist with resolution of 1 fA, such as the Keithley 6512, but these are bulky and expensive calibration devices unsuited to fieldwork. When measuring such small currents, effects often considered negligible in other circumstances are comparable to the signal, such as leakage currents, and also the 50 Hz interference caused by the ac mains supply. Great care therefore has to be taken to minimise these problems whilst maximising the time response, by careful design and component selection, as in the current amplifier developed by Harrison (1997a).

Measurement of the change in voltage across the Gerdien electrodes when the charge on them decays through air is more straightforward, and a simple follower circuit can minimise leakage currents if a suitable op-amp is chosen. However, the currents flowing are clearly no larger than when they are measured directly at the central electrode, and in the Voltage Decay mode, particular attention has to be paid to the connections from the tube to the follower input to avoid leakage. This can be avoided by driving the screen of the input at the output potential, a method known as *guarding* (Horowitz and Hill, 1994).

[29] Although Gerdiens used in the literature have varied considerably in size, their capacitance is usually of order 10 pF (*e.g.* Arathoon, 1991; Blakeslee and Krider, 1992). Consequently, the currents are usually dependent on the magnitude of the bias voltage, which is chosen based on critical mobility considerations, indirectly related to the size of the instrument through the constant *k* defined in Eq. 3.4. Since the Gerdien used here is smaller than average, this results in smaller output currents than measurements with larger Gerdien condensers.

A *current amplifier* is an instrument to measure a small current (Keithley, 1992) by converting it into a voltage, which is usually linearly related to the input current[30]. The simplest form of *i*-to-*V* converter is a resistor: for example, currents of 1 pA could be converted into 1 V by passing them through a high value resistor of 1 TΩ. However, a resistor is not ideal for current measurement because its time response is limited by the time constant of the resistor and its own capacitance (typically a few pF). With the high value resistors of order 1 TΩ required to measure small ion currents, this immediately implies a time response of order seconds.

An alternative approach is to utilise the properties of operational amplifiers. The output voltage of an op-amp is proportional to the potential difference at the inputs, with the constant of proportionality given by the gain (Keithley, 1992). In conjunction with a fixed resistance, operational amplifiers can also be used to convert current to voltage. There are two configurations in which op-amps can act as i-to-V converters, shunt and feedback[31].

a) Shunt ammeter b) Feedback ammeter

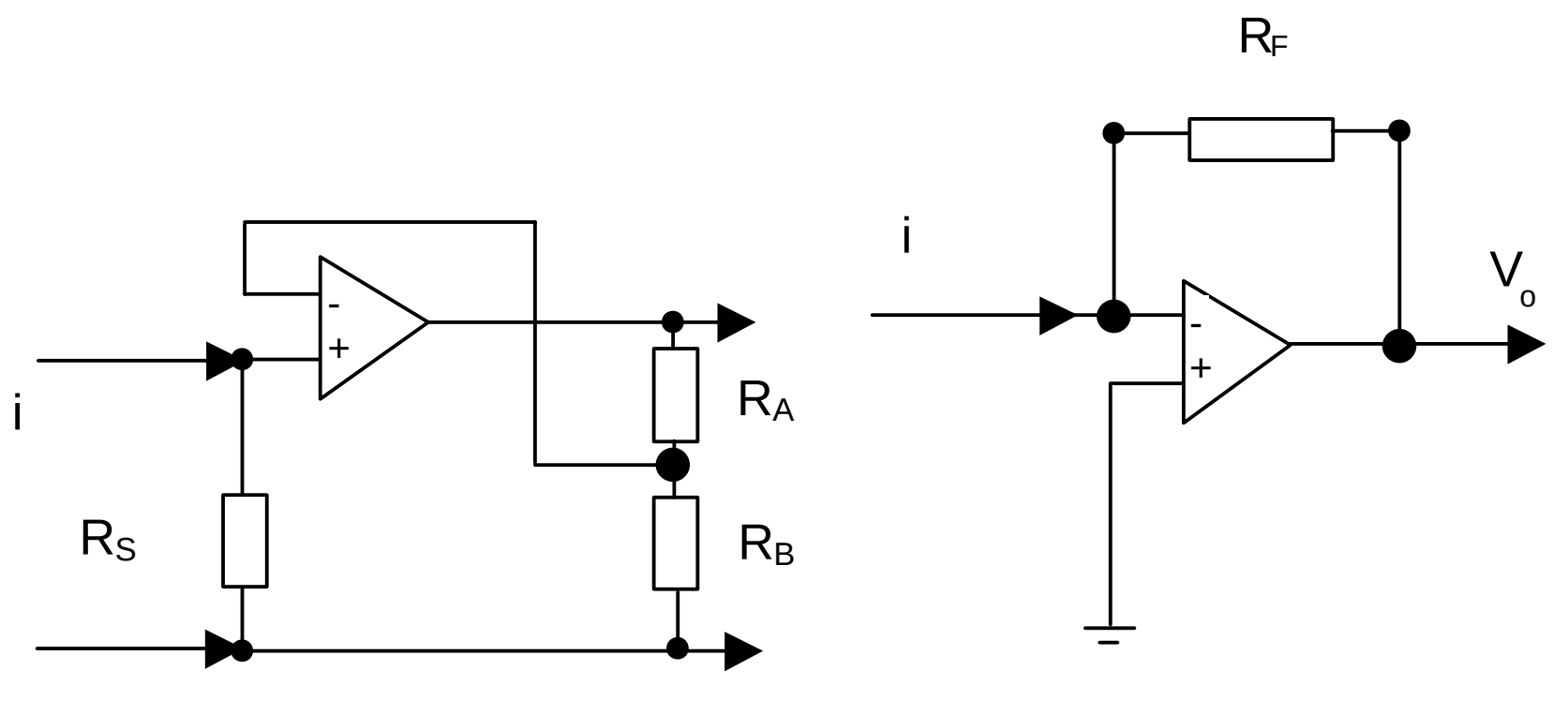

*Figure 3.4 Ammeters using op-amps, after Keithley (1992). $R_F$ and $R_S$ are the large value resistors. i is input current, $V_o$ is output. Note that for the shunt ammeter (a), the signal joins the non-inverting input, and for the feedback ammeter (b), the inverting input.*

In the shunt ammeter (Figure 3.4a), the output voltage $V_o$ is given by the expression for a non-inverting amplifier with an input voltage $iR_s$, *i.e.*:

[30] It is also possible to make a logarithmic current amplifier (Keithley, 1992) by using a logarithmic feedback element. They are well suited to measuring currents in a wide range, for example atmospheric point discharge currents in fair weather and thunderstorms.

$$V_o = iR_s(1+\frac{R_A}{R_B}) \qquad \textit{Eq. 3.5.}$$

In the feedback ammeter (Figure 3.4b), the output voltage is related to the current *i* by

$$V_o = -iR_F \qquad \textit{Eq. 3.6.}$$

(Keithley, 1992). The shunt ammeter suffers from a similarly limited time response to the resistor described above because the input capacitance and any stray capacitance are in parallel with the shunt resistor $R_S$ (Arathoon, 1991). The feedback ammeter is preferable because the op-amp's high gain means that negative feedback keeps the inverting input as a virtual earth. As the non-inverting input is grounded, the voltage across the inputs is very small so there is virtually no input capacitance, and the time constant is only limited by the feedback resistor $R_F$ and its stray capacitance.

The feedback current amplifier used in this application was developed by Harrison (1997a). It was designed specifically for atmospheric current measurements, and incorporates a 30dB 50 Hz noise-rejection feature which uses a tuned filter to extract the 50 Hz signal, phase-shift it by 180°, and feed it back to the input to cancel it out. It also includes several features to minimise leakage currents; the input connection is soldered directly to the op-amp inverting input.

Harrison's (1997a) original current amplifier design had a feedback resistor of 1 GΩ. Initial measurements showed that the Gerdien currents measured in the laboratory were smaller than expected, probably due to ion-deficient indoor air. One possible solution would be to add a further op-amp with gain $A$ to increase the total sensitivity of the system

$$V = A(-iR_F) \qquad \textit{Eq. 3.7.}$$

depicted in Figure 3.5. If $A = -100$, and $R_F = 1$ GΩ, then the output is nominally 1V/pA. Another approach, which has the same overall effect on the gain, would be to increase the feedback resistor to 1 TΩ, omitting the second gain stage or having $A = -1$. "Gain trimming" is the process of adjusting $A$ to get the desired response.

[31] The feedback amplifier is sometimes also called the “transresistance amplifier”.

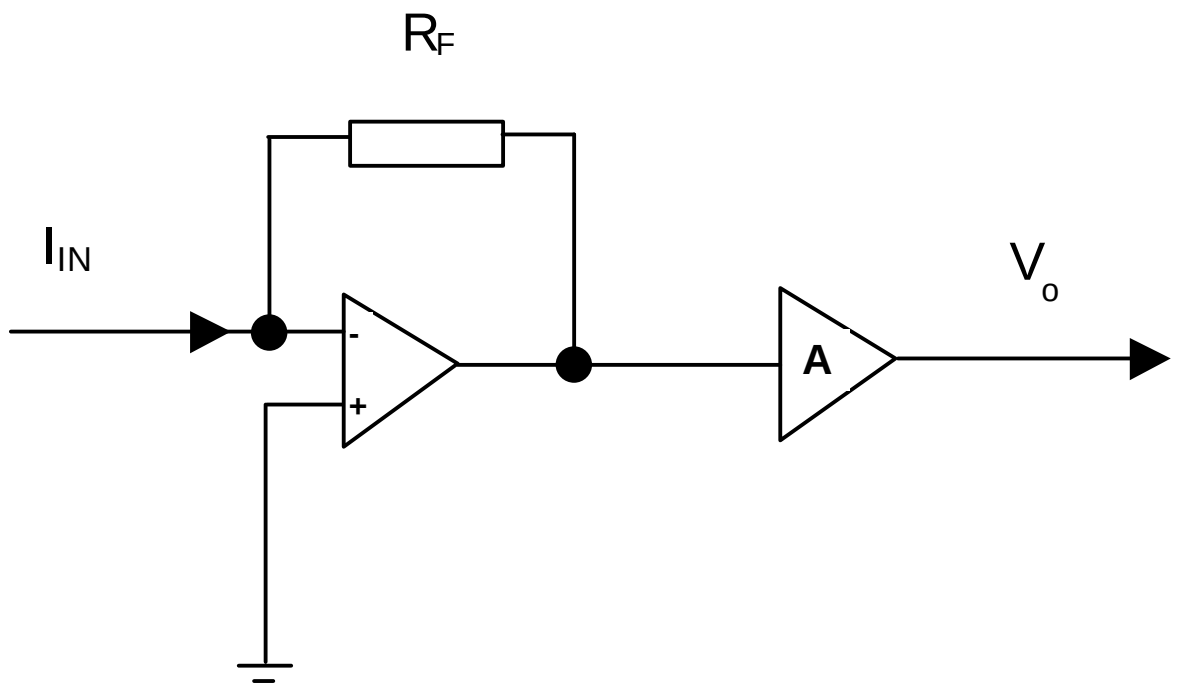

*Figure 3.5 Feedback current amplifier including gain stage A*

The principal difference between the two options is one of temperature stability. In practice, there exists an *input offset voltage* $V_{os}$, which is a consequence of the potential across the terminals of the input amplifier not being exactly zero; $V_{os}$ is defined as the voltage required at the input to cancel it at the output (Keithley, 1992). There is also an *input bias current* $i_B$ which is defined as the average of the input currents with the inputs shorted together. Physically, the input currents result from the base leakage currents of the input transistors (Horowitz and Hill, 1994). The effects of input offset voltage and input bias current are both magnified by the second op-amp's gain $A$, thus the total response $V_0$ of the system is:

$$V_0 = A[-\{(i + i_B)R_F + V_{os}\}] \qquad \textit{Eq. 3.8}$$

$R_F$, $i_B$ and $V_{os}$ all have a temperature dependence, with higher value resistors being more variable with temperature. (Resistor temperature dependencies are shown in Table 3.2 below, where it can be seen that a 1 TΩ resistor is 70 % more temperature sensitive than a 1 GΩ resistor.) The change in the output voltage $\Delta V$ for a finite change in temperature $\Delta T$ can be written as:

$$\frac{\Delta V}{\Delta T} = A\left\{\left[\frac{\Delta R_f}{\Delta T}(i + i_B)\right] + \frac{\Delta V_{os}}{\Delta T}\right\} \qquad \textit{Eq. 3.9}$$

There is thus a compromise between using a 1 GΩ resistor with a larger gain stage magnifying the input offset voltage, and the 1 TΩ resistor with its greater temperature dependence but no multiplication of $V_{os}$. The change of input bias current with temperature has exactly the same effect on the output voltage in both cases, which is discussed separately below. For the MAX406 operational amplifier, $\Delta V_{os}/\Delta T$ = 10 μV/ºC. The temperature coefficient of resistance for resistors from $10^9$-$10^{12}$ Ω was obtained from the manufacturer's information[32], shown in Table 3.2.

| **$R_f$ (Ω)** | **Temp coeff of resistance (/ºC)** | **ΔRf/ΔT (Ω /ºC)** | **Gain *A* to keep output 1V/pA** |
|---|---|---|---|
| $10^9$ | 0.0010 | $10^6$ | 1000 |
| $10^{10}$ | 0.0012 | $1.2 \times 10^7$ | 100 |
| $10^{11}$ | 0.0015 | $1.5 \times 10^8$ | 10 |
| $10^{12}$ | 0.0017 | $1.7 \times 10^9$ | 1 |

*Table 3.2 Resistance values, temperature coefficients and gains*

The temperature sensitivity of the output voltage was calculated for different values of feedback resistance, varying the gain to keep the response nominally 1V/pA, by substituting data from Table 3.2 into Eq. 3.9 above.

Figure 3.6 shows that increasing the value of the feedback resistor decreases the overall temperature sensitivity of the current amplifier. With the 1 GΩ feedback resistor, the term where the input offset voltage is multiplied by the gain in Eq. 3.9 dominates, with the error increasing for smaller currents. For larger resistors, the first term in Eq. 3.9 dominates, improving the overall thermal stability despite the greater sensitivity of the larger resistors to temperature changes.

[32]ELTEC Type 104, L.G. Products Limited, Slough (1994). (Now distributed through LGP Electro-optics, Chobham, Surrey.)

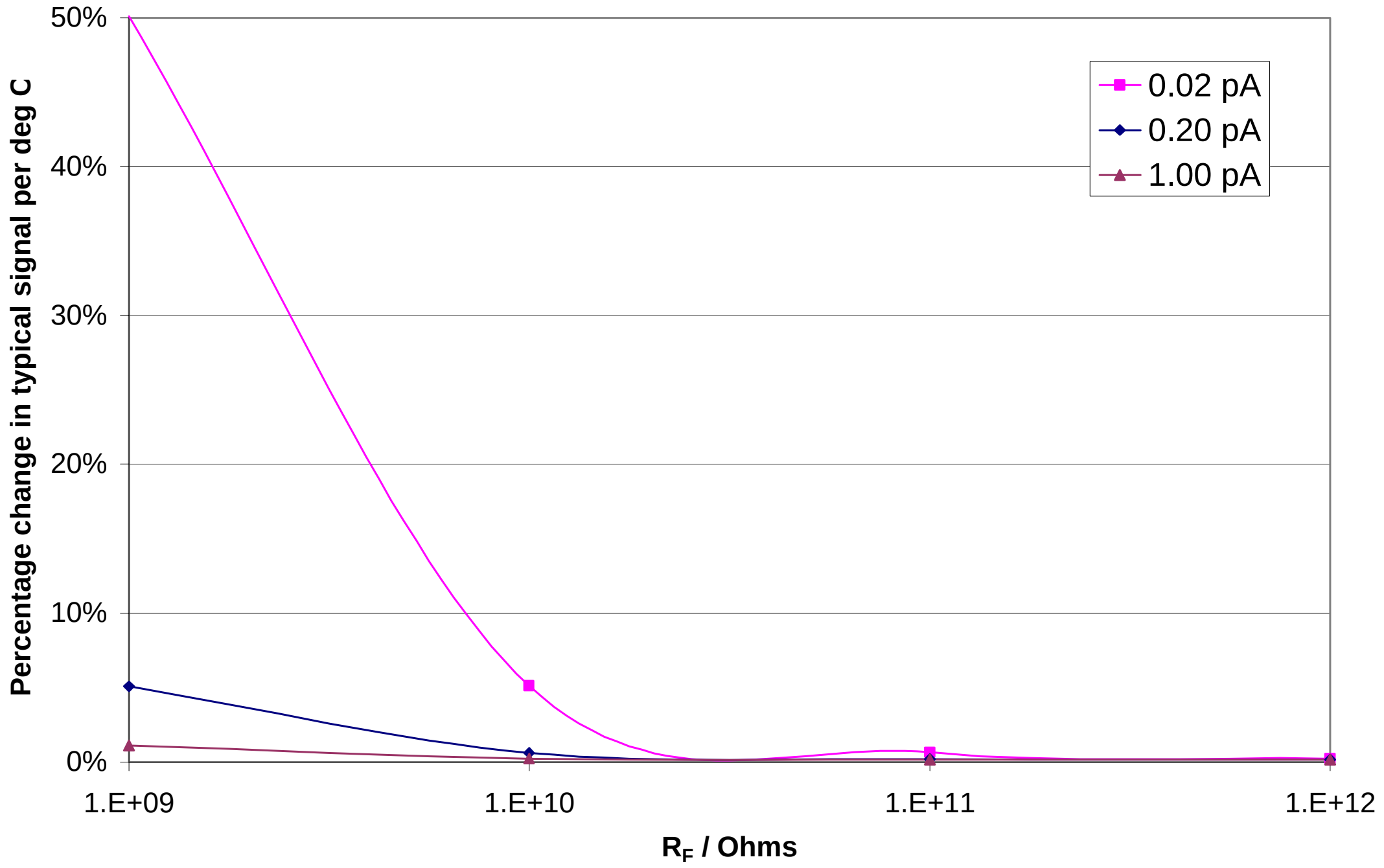

*Figure 3.6 Temperature sensitivity of the output voltage with $R_f$ for typical input currents from atmospheric ions (shown in pA in the legend), calculated from Eq. 3.9.*

Comparisons were made of the temperature stability of a current amplifier using a 1 TΩ feedback resistor with two different electrometer grade op-amps:

- AD549JH (which has the $\Delta V_{os}/\Delta T$ = 10 µV/ºC as the MAX406), which was used successfully by Arathoon (1991) in a feedback current amplifier with a 100GΩ feedback resistor.
- MAX406, used in the Harrison (1997a) current amplifier.

Using the AD549JH compared to the MAX406 would reduce the input bias current at 0 °C by an order of magnitude, but increase it by a factor of two at 30 °C. Since the two op-amps are otherwise similar, the more consistent value of $i_B$ is preferable as it is easier to quantify.

Following the analysis of temperature dependencies of input bias current and input offset voltage, the MAX406 op-amp was unchanged from the original design (Harrison, 1997a), but the feedback resistor was changed to a 1 TΩ with the second stage having a gain of -1.

The thermal stability of the current amplifier's leakage terms were tested by logging its output when its input was capped, at the Reading University Meteorology Field Site. In this configuration, the output is from the combination of the input bias current and leakage currents from the current amplifier, and shows a clear temperature dependence which can be seen in Figure 3.7. The predicted effect of changes in temperature on the input bias current alone is plotted in Figure 3.8. Although the MAX406 has a small input bias current compared to other op-amps, if typical ion currents are about 200 fA, then the input bias current introduces an error of about 5 %, and in indoor tests this will probably be worse because of the higher temperatures and smaller ion concentration.

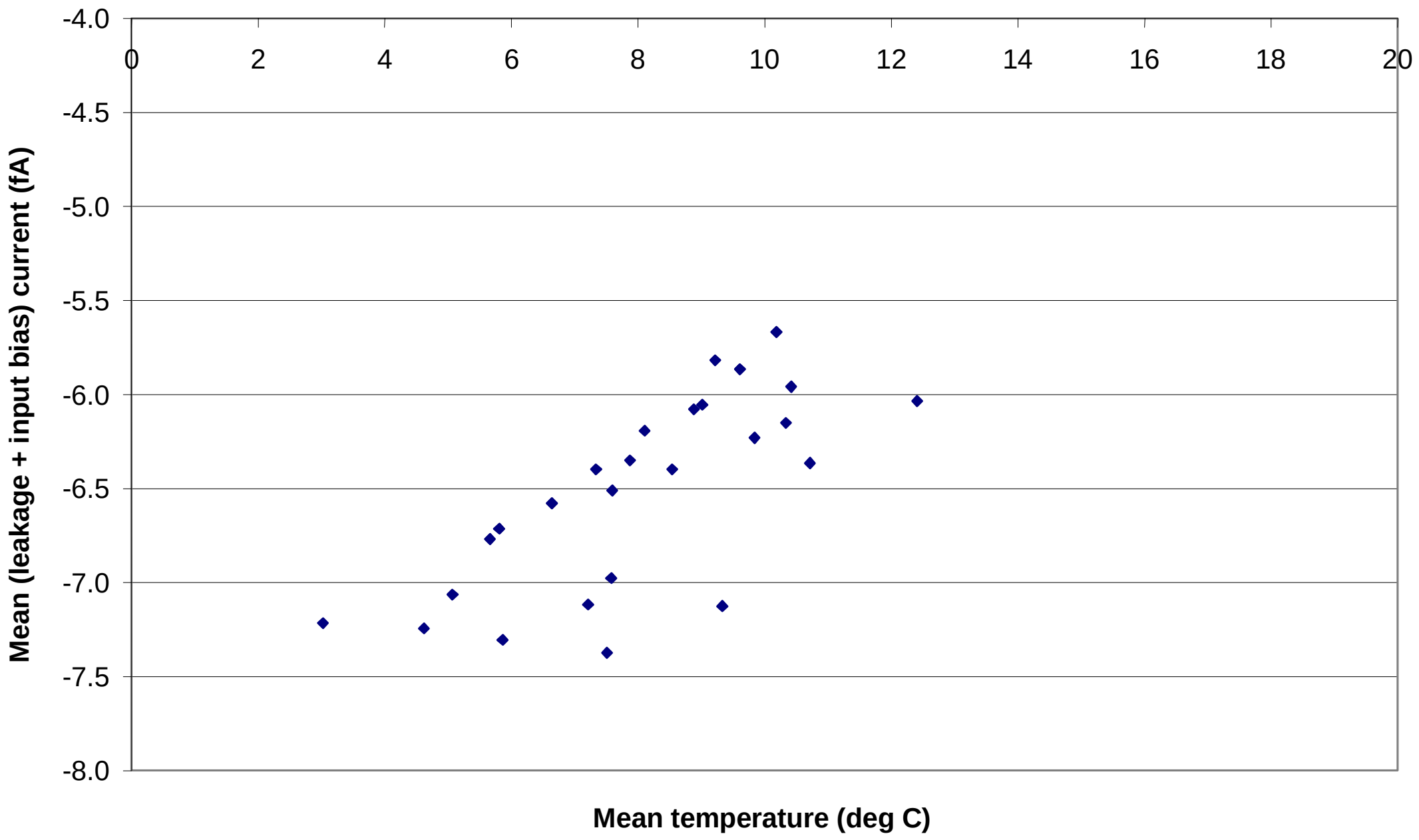

*Figure 3.7 Daily mean leakage current and temperature for March 1998, calculated from automatically logged five minute averages of capped current amplifier output (in volts) and temperature.*

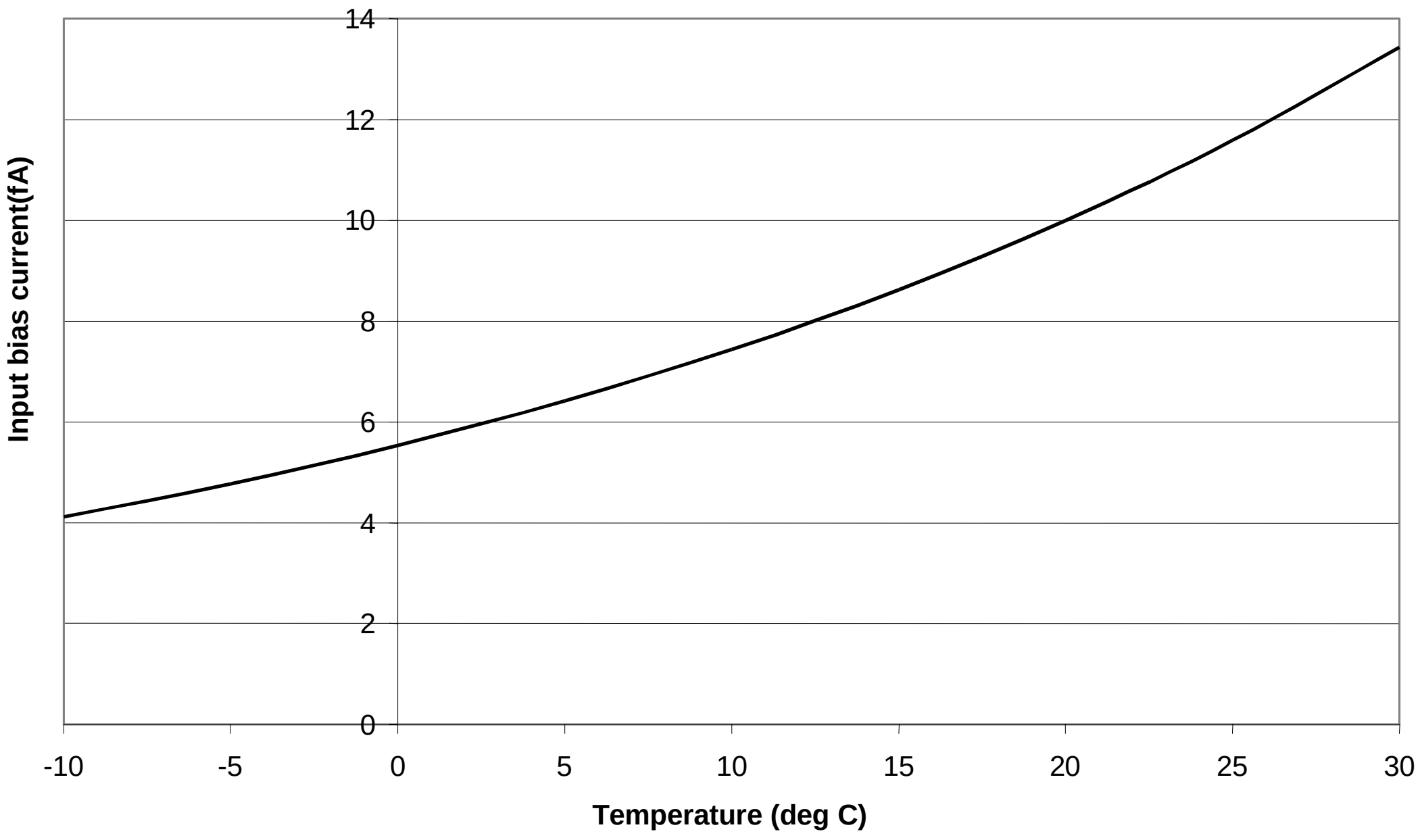

*Figure 3.8 Variation of the typical MAX406 input bias current with temperature (from manufacturer's data sheet).*

Both these effects will vary with the individual op-amp used, and Figure 3.8 in particular is likely to be the worst case scenario quoted by the manufacturer. However, it is clear that the temperature dependence of both the input bias current and the electrometer leakage current are significant effects when measuring at the femtoampere level. The observed current amplifier resolution of ± 3 fA at 20 ºC [33], concurs with the trend in the sum of leakage and input bias currents shown in Figure 3.7, but the results suggest the leakage currents will be greater at lower temperatures. The atmospheric results (measured at 2 – 12 ºC) support this; because $i_b$ is positive, then the leakage currents in the circuit must be dominant to make the overall response negative.

### 3.1.4 Switching devices in electrometry

When a switch is required, for example in the Voltage Decay method to control the bias voltage, great care has again to be taken to avoid leakage currents by choosing well-insulated components. For this reason, sub-miniature glass encapsulated reed switches[34] in a activating electromagnetic coil[35] were used for low leakage switching applications. For the Gerdien system, the major applications were shorting out the current amplifier's feedback resistor, and controlling the voltage which charges the electrodes.

[33] This will be discussed further in section 3.2.3

This reed relay, cycling at 0.1 Hz, was used to short out the feedback resistor in the current amplifier, with its input floating and screened. When the relay opened there was a voltage transient of about 30 mV (corresponding to a negative charge injection of ~ 6 fC) into the system. This exhibited a RC decay with a settling time of approximately 5 s. Brownlee (1973) observed a similar 20 mV transient with the ERG type MA/GO/J reed relays he was using to switch between connections to his electrometer. To test whether this effect was a property of the relay he compared the voltage transients using different capacitor values across the reed relay. From Eq. (2.5), if $q$ does not vary with time, then doubling the capacitance should halve the voltage transient. Brownlee (1973) concluded from this that his relays injected a charge of about 20 fC into the system. This confirms suspicions that the reed has a small capacitance, and discharges when closed, which can cause transients[36]. To avoid this charge injection affecting the current measurements, a suitable decay time should be allowed to elapse after the relay switching, before making a measurement.

Another test of the reed relay, when connected across the current amplifier feedback resistor, was to subject it to step changes in input current when it was fixed either on or off, and observe the effect on its switching capabilities. When the relay was open, changes in the input current greater than from +1 to –1 pA could cause it to jam in the off (open) position. A possible explanation for this could be that as the op-amp is inverting, the feedback voltage is of the opposite sign to the input current, but at the instant when the sign of the input current is changed, the input and the feedback potentials on the reeds are the same sign. The repulsion between the reed and its input means that there is nowhere for the charge on the relay to dissipate to, which could cause it to stick open. This would probably be an insignificant problem in atmospheric measurements, because step changes of order 2 pA with a change of sign, caused by ions are extremely unlikely. However, this effect might compromise the long-term reliability of the relay switching.

[34] RS type 229-3658
[35] RS type 349-844
[36] An example of this is described in Section 3.3.1.

## 3.2 Characterisation of individual components of the Gerdien system

In this section, the individual components of the Gerdien system are investigated. Some wind tunnel experiments to identify the flow régime in the tube were carried out and are described in Section 3.2.1. The findings of the flow investigation suggest that fluctuations in the external ventilation may be important, necessitating further consideration of its effects on the size of ions contributing to the measurement in Section 3.2.2. The measurement system has also been studied in detail, and the findings are presented in Section 3.2.3.

### 3.2.1 Flow properties of air in the Gerdien tube

The aim of the experiments described in this section were to determine some characteristics of the flow in the Gerdien tube, in particular how the flow in the tube varied with the nature of the external flow. It is also necessary to confirm that flow in the tube is approximately laminar, so classical Gerdien theory (Chalmers, 1967) would apply (as discussed in Section 3.1.2).

The flow properties of the tube were investigated by making measurements with a Disa type 56C01 hot-wire anemometer in a wind tunnel. The tube used was physically identical in dimensions and composition to a Gerdien condenser, but lacked the electrical connections. Measurements were made when the external ventilation was sucking air through the Gerdien in the same direction as the fan, and also when the external and induced ventilation (from the fan) were opposed. The probe was inserted 33.5 mm into the Gerdien, to measure the flow adjacent to the central electrode.

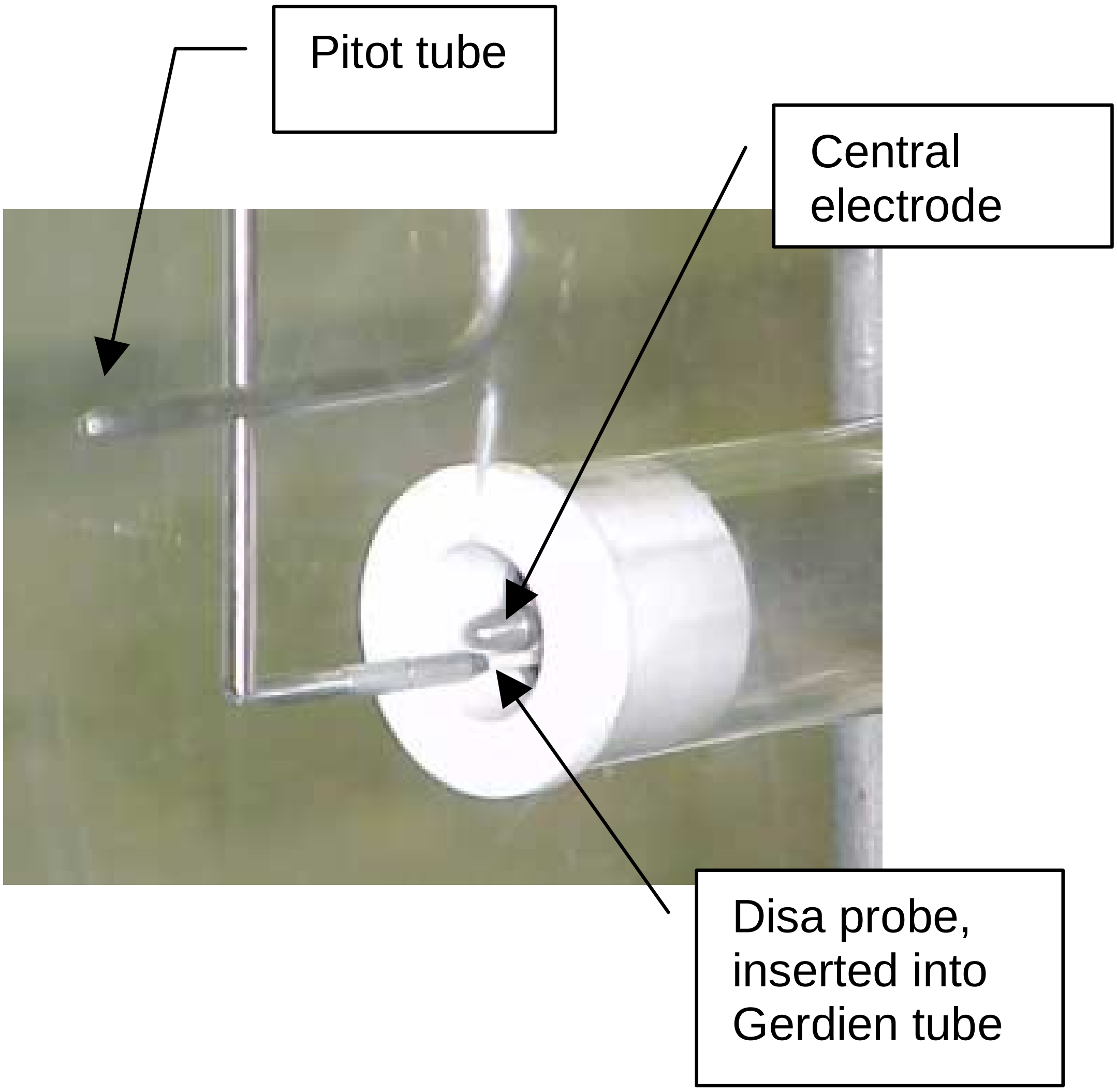

*Figure 3.9 Photograph of the Disa probe inserted into the end of the Gerdien, and showing the Pitot tube which measured the external flow.*

The relationship between the flow in the tube and the external ventilation can be used to empirically determine $u_{tube}$ from the measured wind component. Three régimes were investigated, which are summarised below in Table 3.3. The first case was when the external ventilation exceeded the fan speed so that it dominated the flow in the tube, the second when the external ventilation was low so that the fan was the principal component of the flow, and the third when the flows were approximately equal and opposite. Only wind speeds in the typical atmospheric range were chosen, so that the study would be as applicable as possible.

| Case | External wind velocity ($ms^{-1}$) | Tube flow speed ($ms^{-1}$) | Reynolds number |
|---|---|---|---|
| **a) Velocities opposed, fan dominating** | 0.8 | 1.9 | 5000 |
| **b) $u_{ext} \approx u_{tube}$** | 0.8 | 1.3 | 3000 |
| **c) Velocities aligned, external ventilation dominating** | 2.3 | 1.6 | 4000 |

*Table 3.3 Conditions for measurements in the Gerdien in a wind tunnel, and the Reynolds number estimated from Eq. 3.3. External wind velocity was measured using a Pitot tube positioned near the Gerdien in the tunnel. The tube flow speed was measured with the hot-wire probe positioned inside the tube.*

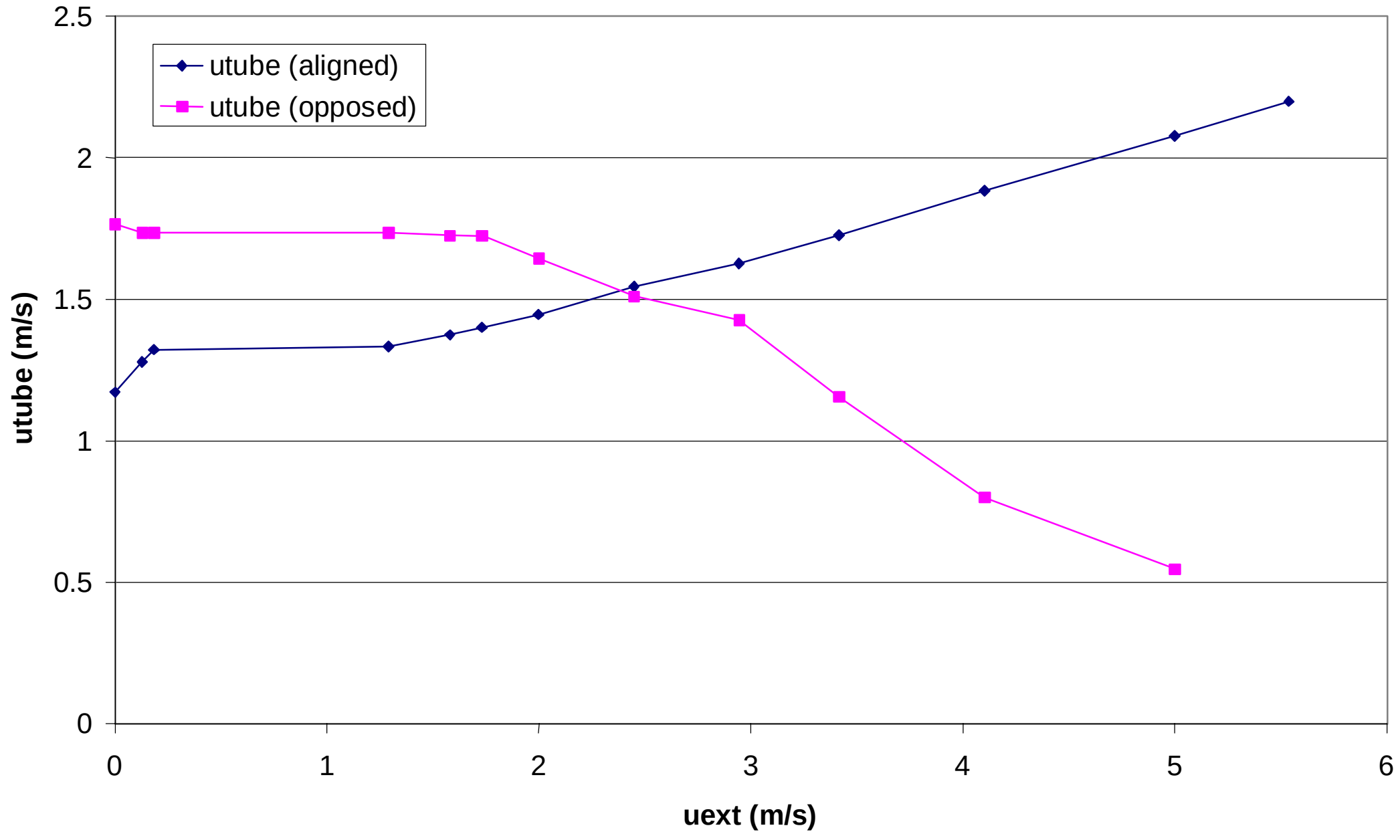

*Figure 3.10 Relationship between the flow in the tube $u_{tube}$ and the external ventilation $u_{ext}$, for the two cases where the external and fan flow are in the same and opposite directions.*

If curves are fitted to Figure 3.10, they can be used to calculate $u_{tube}$ for all $u_{ext}$ in the measured range by interpolation. Two polynomials were fitted for the flow in the tube when for the two cases when the external flow is opposed ($u_o$ in Eq. 3.10) and aligned ($u_a$ in Eq. 3.11) with the fan flow.

$$u_{tube} = 1.81 - 0.771u_o + 0.873u_o^2 - 0.325u_o^3 + 0.034u_o^4 \qquad \textit{Eq. 3.10}$$

$$u_{tube} = 1.25 + 0.021u_a + 0.047u_a^2 - 0.004u_a^3 \qquad \textit{Eq. 3.11}$$

To calculate $u_{tube}$ for atmospheric air, the external wind component is calculated from the wind speed and direction. If it is negative, then the flow is opposed to the direction of the fan flow, and Eq. 3.10 is used to give $u_{tube} \pm 0.1$ ms$^{-1}$. If it is positive then the aligned wind speed is substituted into Eq. 3.11, with a slight associated error in $u_{tube} \pm 0.04$ ms$^{-1}$.

The Reynolds number *Re* (Eq. 3.3) was calculated for the length scale of the tube diameter (0.04 m), since this indicates the scale of motion of the ions when they enter the tube. Turbulence of this length scale is important, because if it is too vigorous the motion of the ions towards the central electrode will be significantly perturbed. The results, in Table 3.3, show that the flow in the tube is at the lower limit for the onset of turbulence, which starts when $Re \approx 10^3$, depending on the exact geometry and flow characteristics (Tritton, 1988). This suggests that the flow in the tube is only slightly turbulent, if at all, so the "relaxed Chalmers criterion" discussed in section 3.1.2 holds, although some noise may be introduced into the measurements from fluctuations in the flow. From the values of *Re* for different flow conditions, it appears as if the fan is primarily responsible for introducing some more turbulent flow (with higher *Re*), but when the fan and external wind are equal and opposite, then this effect is reduced. The difference, however, is insufficient to justify eliminating data based on wind speed variations alone.

### 3.2.2 A simplified relationship between critical mobility and maximum radius

It is clear from Figure 3.10 that the flow inside the tube is not linearly related to the external ventilation. This has consequences for the critical mobility of ion contributing to the measurement, which is a function of the flow in the tube, as shown in Eq. 3.1. The non-linear relationship between the external and internal ventilation implies it is important to calculate the critical mobility in the tube, because small changes in the external wind speed or direction may cause dramatic fluctuations in the critical mobility.

Some flow rates may be inappropriate because the critical mobility is not in a suitable range for conductivity measurement (see Section 2.2.2). Theoretical predictions (Dhanorkar and Kamra, 1997) show that if intermediate ions are contributing to the conductivity, then classical ion-aerosol theory (outlined in Section 2.2.3 and described in more detail in Clement and Harrison (1992)) does not hold. Ion mobility varies inversely, but non-linearly with ion mass and radius and intermediate ions are defined as having radii ~ 1 nm (Hõrrak *et al*, 1999). Since mobility is not the most intuitive of quantities to relate to other physical properties of an ion, it is desirable to find a simple way to convert between ion mass and radius. Tammet (1995) has derived a complicated expression to relate the two quantities, which has been simplified for use in this Thesis. Ion size and mobility data (calculated using Tammet's (1995) expression) have been given in Hõrrak *et al* (1999), and are plotted below. A power law fit has been applied which gives a simple way to relate ion mobility to radius.

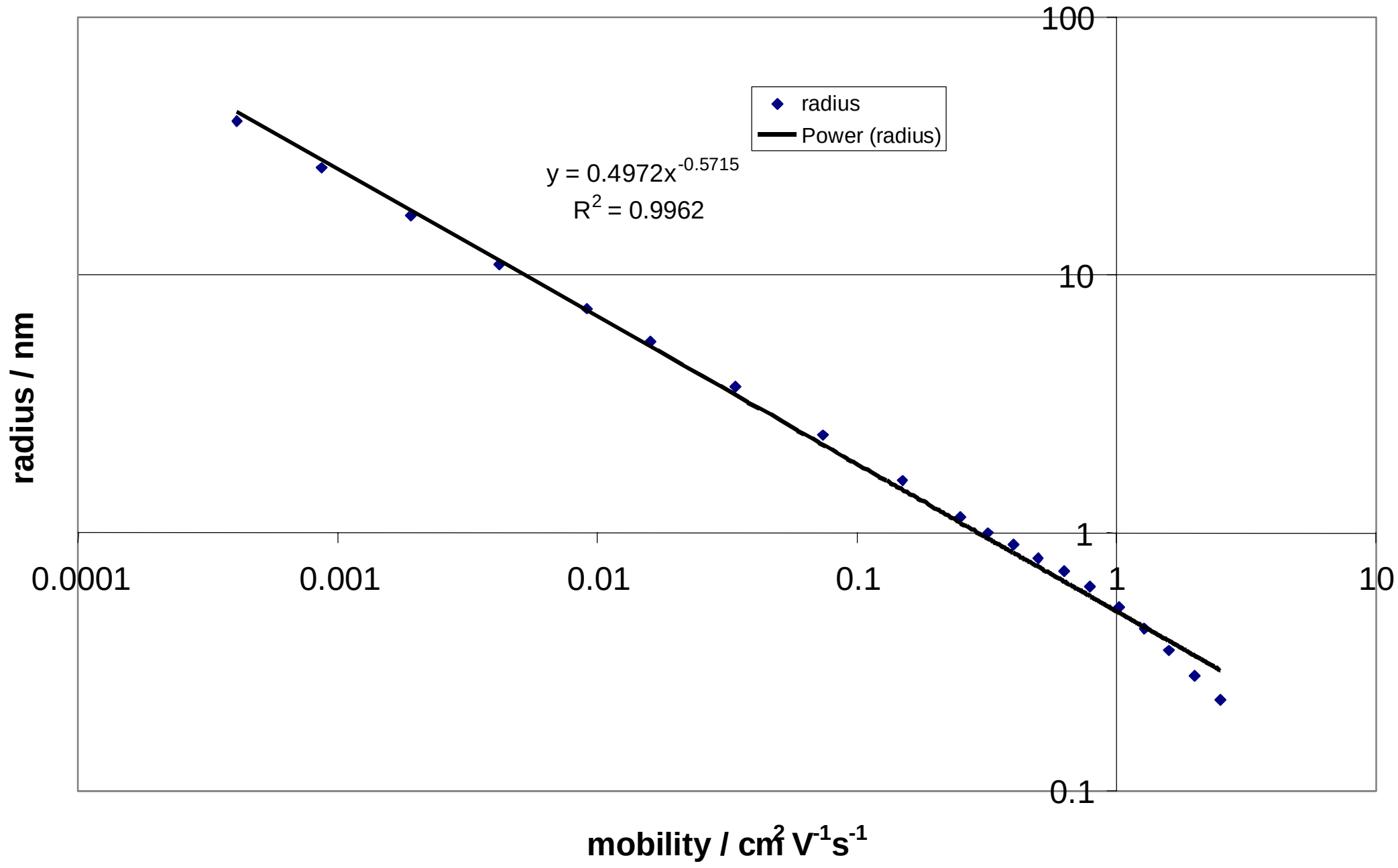

*Figure 3.11 Relationship between ionic mobility and radius, originally calculated using Tammet's (1995) expression and here reduced to a simple power law.*

Critical mobility is defined as the minimum mobility of ion contributing to the Gerdien measurement. The inverse relationship between it and radius means that the maximum radius of ion contributing to the measurement can be known. This permits easy conversion between ionic mobility and radius, so there can be continual awareness of more intuitively understood physical properties. It is therefore easier to select filtering criteria, for example, based on flow considerations, to exclude the effect of intermediate ions on the conductivity measurement.

### 3.2.3 Measurement System

The current amplifier was calibrated in the laboratory using a resistive method of generating small currents with a millivolt calibrator[37] and a 1 TΩ resistor. From Figure 3.12 it is clear that for this current amplifier, the dc response is linear to 99% in the range $-2.5 < i_{in} < +3$ pA.

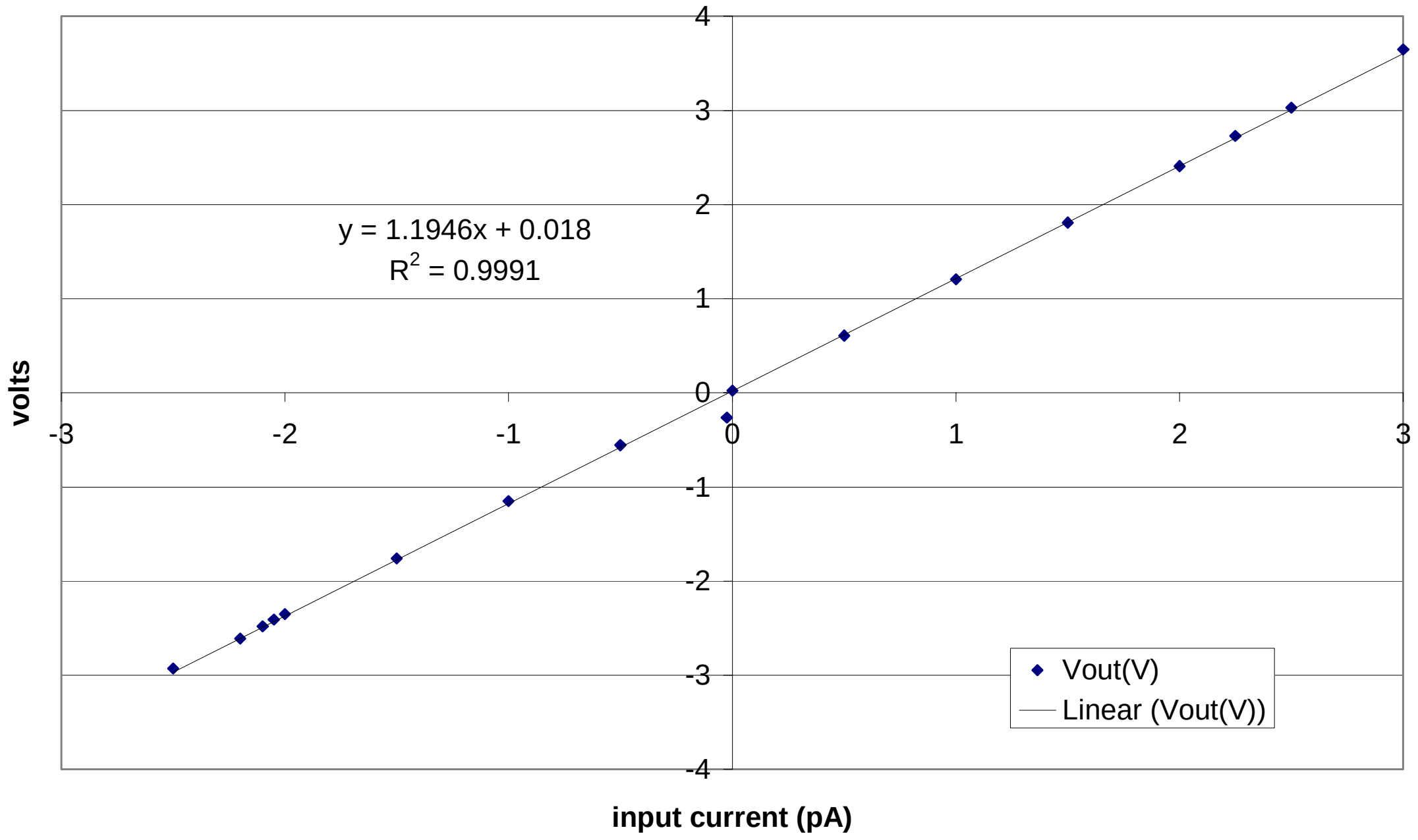

*Figure 3.12 Resistive current amplifier calibration. The input current was generated using a series 1 TΩ calibration resistor (± 10 %) driven using a dc millivolt calibrator (± 0.02 %). The output voltage was measured using the Keithley 2000 digital voltmeter (± 35 ppm).*

[37] The millivolt calibrator used in this Thesis is a Time Electronics 2003S (± 0.02 %)

The current amplifier resolution was ± 3fA, and a Fourier power spectrum of its output when connected to a current source showed that the dominant component of the noise was mains at 50Hz, although attenuation resulted from the 50 Hz reject feature (Harrison, 1997a). A RC low pass filter with $\tau$ = 2 Hz (1 MΩ, 220 nF) across the output was used to eliminate high frequency variation without attenuating the mean signal, although when computerised data logging was used in the field, averaging reduced the need for this filter.

Initial atmospheric tests were hindered by an over-range response from the current amplifier op-amp saturating, probably due to an excessive leakage current. The saturation periods did not relate to any of the meteorological and other atmospheric electrical measurements made at the same time. The persistence of this response also increased with time, indicating a progressively developing problem. Possible causes were investigated, and are discussed below.

<u>ac effects</u>

The 1TΩ feedback resistor used is at the upper limit of commercially available resistors, and such a value is infrequently used, except in electrometry. If there is feedback through the circuit involving a 180° phase shift, the circuit becomes an ac amplifier, with changes in the values of the capacitors in the circuit affecting the gain. If this occurs, it could cause an amplification of any resonance, which might submerge the true ionic conduction current in oscillation.

To investigate this, the equivalent circuit of the Gerdien and current amplifier was constructed, and is shown in Figure 3.13. The Gerdien condenser was sealed with foil to act as a capacitor with very high (but unknown) resistance. Its resistance should remain roughly constant since the only current in the tube would be from cosmic and gamma background ionisation. Step changes in currents were induced, and the capacitance across the Gerdien was also varied. Oscillations were clearly visible with step changes of the orders of magnitude that could be expected from atmospheric ions (*e.g.* 100-200 fA). This "ringing" indicated a resonance and hence an instability. The experiment was repeated with lower value resistors instead of the Gerdien, and there was no ringing. However, the magnitude of the ringing was not large enough to cause saturation of the op-amp, which occurs at about 3 pA.

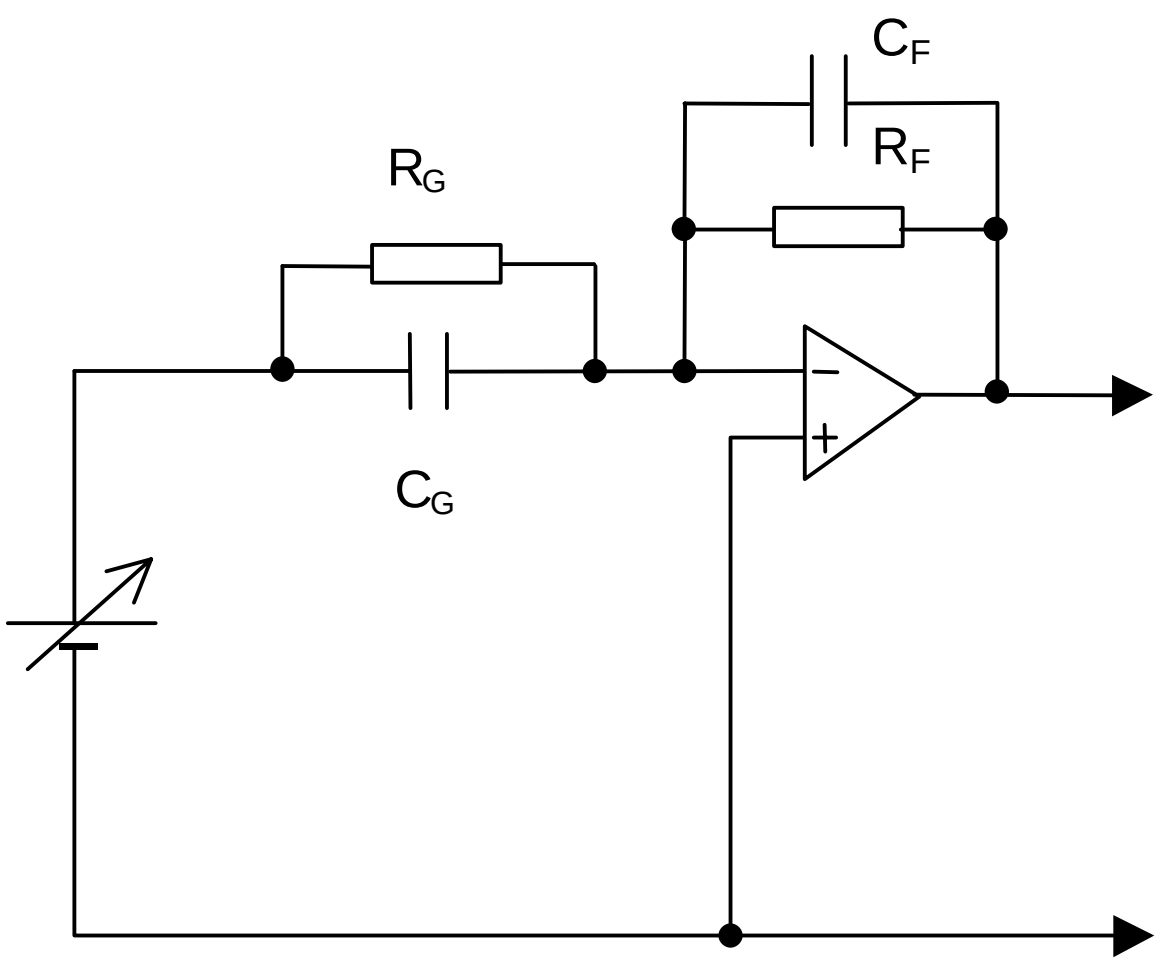

*Figure 3.13 The Gerdien system's analogous electrical circuit. The Gerdien is a capacitor $C_G$ ~ 8pF, and $R_G$ is effectively the resistance of air (of order 1 PΩ). The op-amp and feedback resistor $R_F$ (1 TΩ) are the current amplifier. $C_F$ is parasitic capacitance ~ 1pF. The cell was a millivolt calibrator, and provides a bias voltage from ± 0-10 V.*

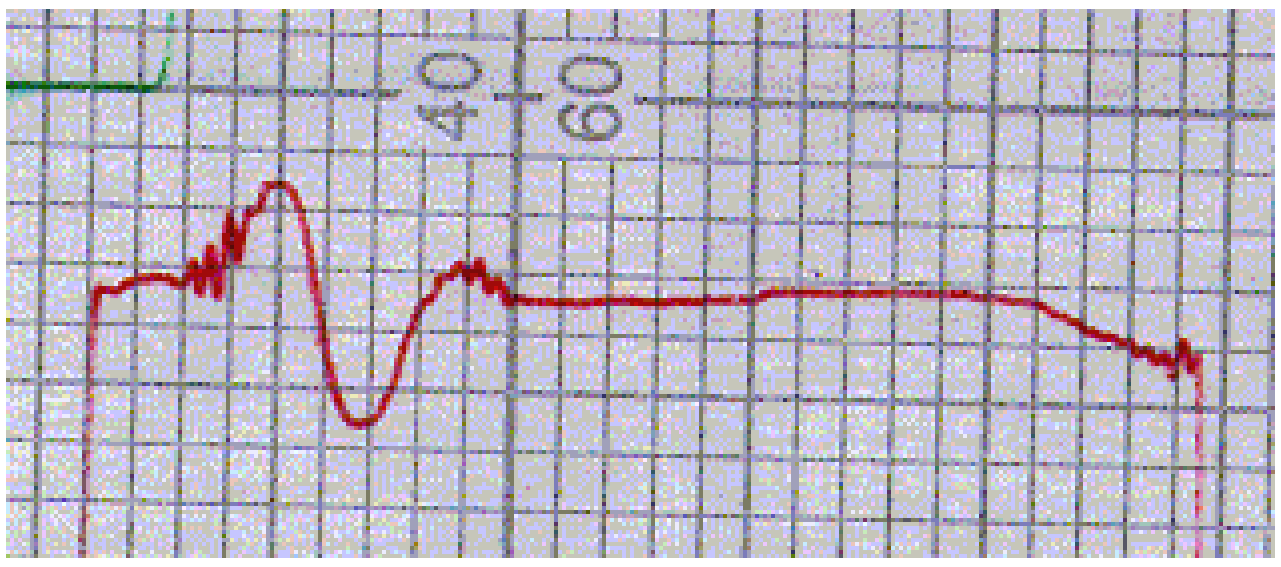

*Figure 3.14 Example of the oscillations in the current amplifier output ($C_G$ ~ 8 pF) when subjected to a step change from 100 to 200 fA. Oscillations are clearly visible each time a transient is induced. The x-axis scaling is 1 small square = 0.2 s. y-axis scaling is approximately 4 mV/small square (the total amplitude of the oscillation was about 10 % of the signal)*

The oscillations could be caused by the parasitic capacitance across the op-amp, or some effect due to the particular (and unusual) combination of low capacitances and very high resistances in the circuit. Keithley (1992) describes a circuit which can neutralise the capacitance across the op-amp. It adds another capacitor and resistor to the feedback loop, so that if their time constant is made to match the time constant of the feedback resistor and its capacitance, then the effect is cancelled out. This circuit (known as a "T-network") was added, and removed the instability, appearing stable in the long term. However, the step changes imposed in the laboratory to cause this effect are unlikely to occur in practice, because they would require an extremely sudden and rapid change in the ionic composition of the air. Fluctuations in the ion production rate may be rapid, but are unlikely to cause a doubling of the ion concentration because the effects of self-recombination and attachment act to damp the changes. The ventilation of the Gerdien tube also acts to smooth the effect of ionic fluctuations. Other ac effects which might have been responsible for the over-range output were investigated in detail, but this was the most conclusive finding.

Effect of tube material

The Gerdiens were regularly cleaned with isopropanol, but this process was inadequate at the end of the tube with the fan sealed watertight into it. Accumulation of dirt on the electrodes and insulators could provide an effective current leakage path and result in the progressive degradation observed. Aluminium surfaces, like the Gerdien tube's outer electrode, become permanently coated with a layer of $Al_2O_3$ on exposure to air, so chemical processes occurring on the electrodes are not inconceivable. Previous aluminium Gerdiens were only used for radiosonde ascents (Hatakayema *et al,* 1958; Venkiteshwaran, 1958), which typically last for a few hours, so it is difficult to compare them to long-term deterioration of Gerdiens running in the field.

As Wåhlin (1986) suggests that contact potentials cause differences in the offsets of the Ohmic response plot for Gerdien tubes (discussed in section 2.2.1), the Ohmic response of several Gerdien tubes of different materials was tested in the laboratory. The same current amplifier was used throughout.

| Tube material | Length (m) | Radius of outer electrode (m) | Radius of central electrode (m) | Flow rate (m/s) |
|---|---|---|---|---|
| Aluminium | 0.258 | 0.0125 | 0.0008 | 2.1 ± 0.1 |
| Graphite | 0.5 | 0.0405 | 0.002 | 0.8 (estimated) |
| Steel | 0.3 | 0.011 | 0.002 | 2.1 ± 0.1 |
| Brass | 0.258 | 0.0115 | 0.0008 | 2.1 ± 0.1 |

*Table 3.4 Dimensions of the different tubes tested. The same fan was used for three of the tubes. The flow rate is obtained from a calibration of fan supply voltage against flow rate measured inside a cardboard tube of the same dimensions. The flow rate through the graphite tube was estimated with a hot wire anemometer at the tube inlet.*

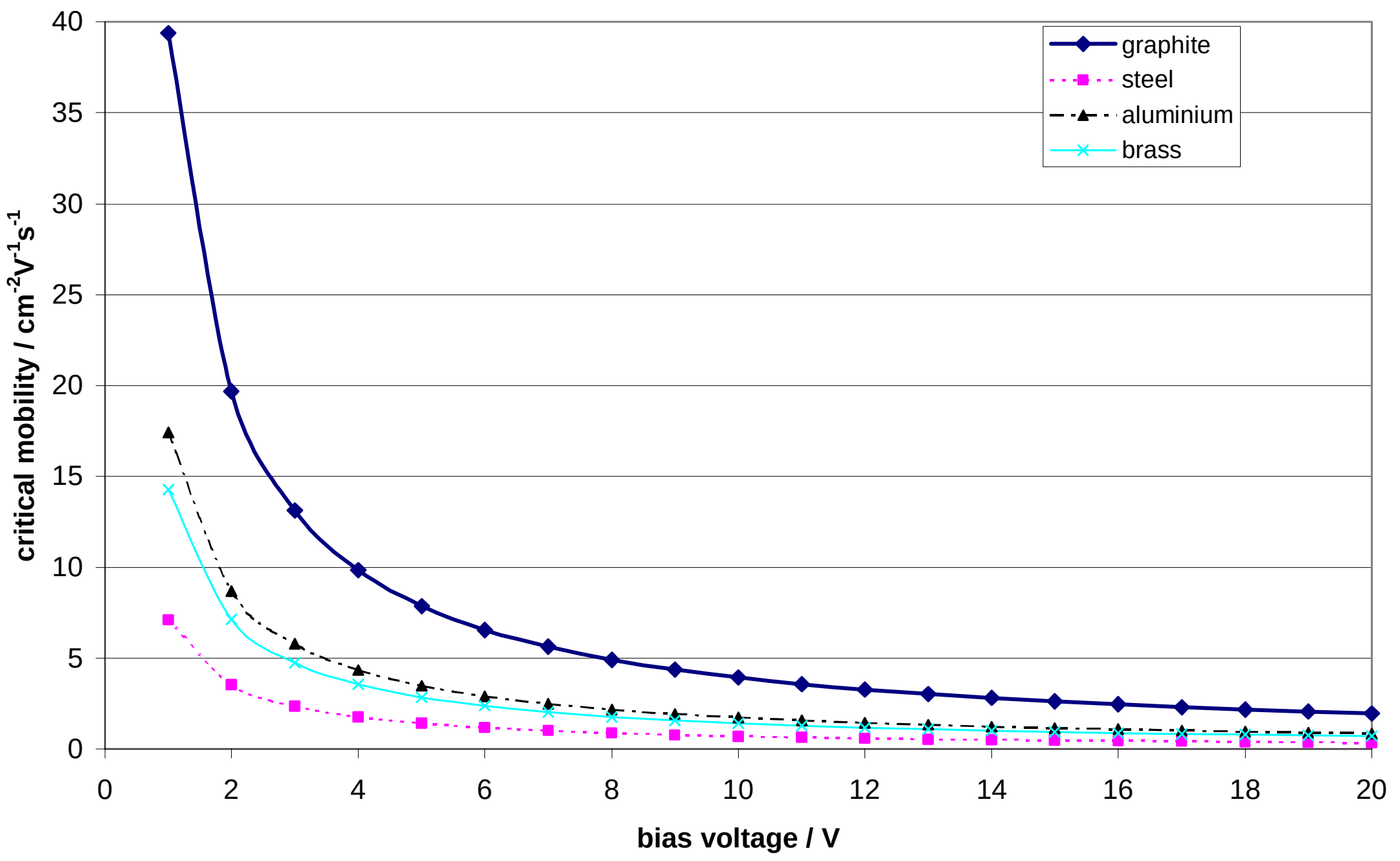

*Figure 3.15 Variation of critical mobility ($cm^2s^{-1}V^{-1}$) with voltage (V) for the four different tubes tested.*

Table 3.4 and Figure 3.15 show that small differences in the dimensions of the tube can lead to a large variation in the critical mobility. Any comparison of the i-V relationship for these tubes is clearly dominated by the critical mobility criterion, and material effects are indistinguishable.

Even when the Gerdien was clean, the current amplifier would frequently record a current exceeding 2 pA. This behaviour was repeatable, typically starting in the early evening and lasting until the next morning, and a typical day showing this pattern is depicted in Figure 3.16. These periods correlated well with high relative humidity, and it was possible to predict their onset when the relative humidity exceeded about 75 %. This was thought to be due to condensation onto the insulators or electrodes causing an excessive leakage current.

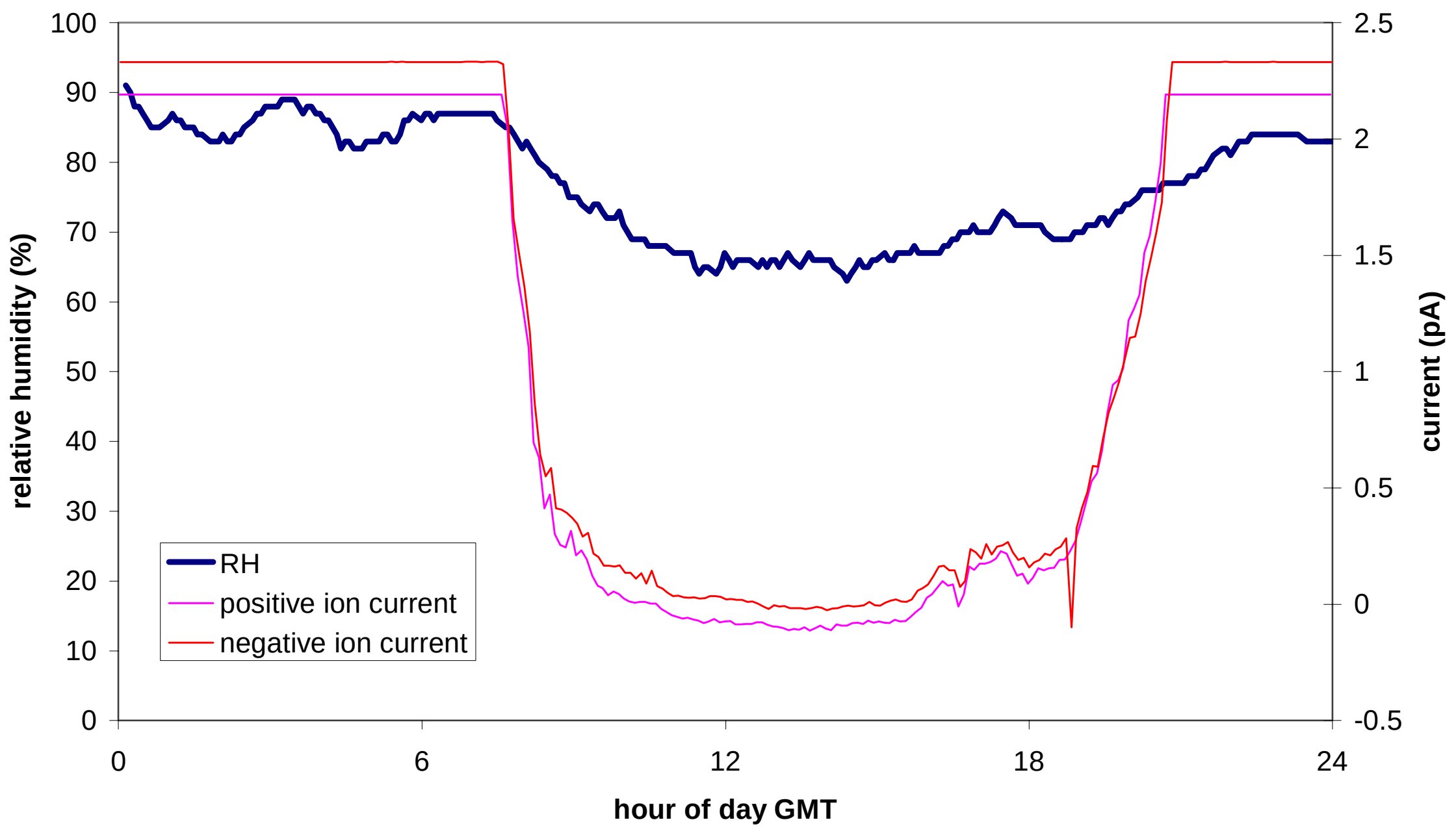

*Figure 3.16 Example of high Gerdien currents at high relative humidities at Reading on 20th February 1999. A current exceeding 2 pA is the op-amp's over-range saturation response; currents << 1 pA are expected to be caused by atmospheric ions.*

## 3.3 Implementation of the Gerdien system

The Gerdien system (in Current Measurement mode) has been used for two test applications: measurement of the conductivity variation with aerosol concentration at a semi-rural site near Reading (Section 3.3.1), and calibration against the ion production rate in clean air (Section 3.3.2). Finally, (in Section 3.3.3) air conductivity is measured by the Current Measurement and Voltage Decay modes with the same Gerdien condenser, and the results are compared.

### 3.3.1 Measuring physical properties of ions with the Gerdien system

In an aerosol-ion system such as the atmosphere at Reading, the concentration of atmospheric small ions is related to aerosol concentrations, as discussed in Section 2.2.2. The values of $\alpha, \beta$ and $q$ in Eq. 2.27 were estimated manually using a least squares method, to minimise residuals within a sensible range. This enables the aerosol concentration to be calculated from conductivity measurements. The aerosol mass concentration was measured using an optical instrument, the TSI 8520 DustTrak[38] and converted to a number concentration. The direct conversion of aerosol mass to number concentration can only be an approximation, since it assumes a monodisperse aerosol concentration of spherical particles of constant density.

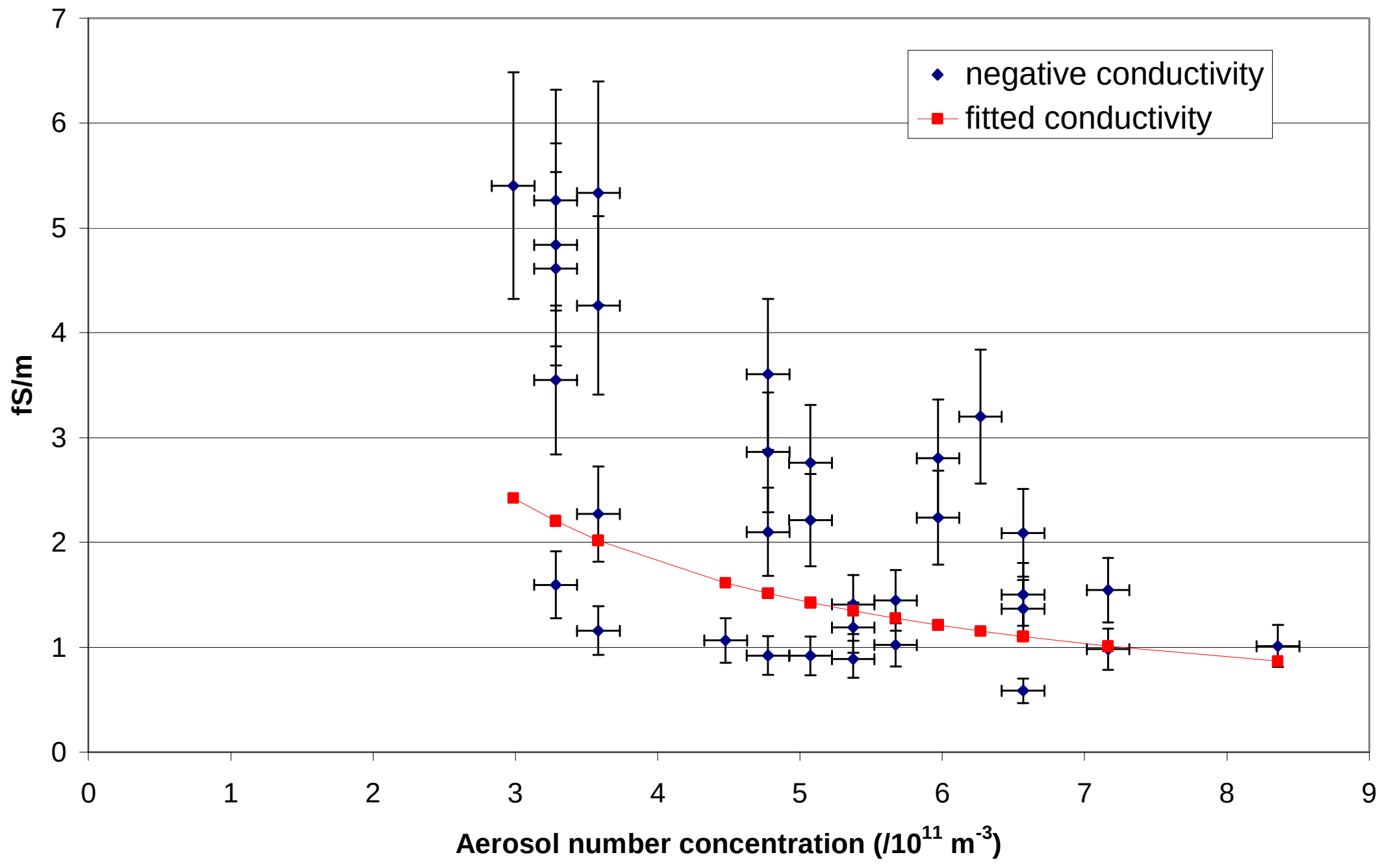

*Figure 3.17 Five minute averages of negative conductivity and aerosol number concentration at Sonning-on-Thames, a semi-rural site NE of Reading (described in detail by Barlow (2000)), on 12th August 1998. The fitted values are indicated in red. The x-axis error bars are calculated from the error in the DustTrak mass concentration measurement, converted to aerosol number concentration. Error bars in the ordinate are ±20%, from the error in determining the Gerdien capacitance.*

The conductivity curve calculated from Eq. 2.27, (having fitted for $\alpha, \beta$ and $q$) is a tolerable match to the measurements, but the paucity of conductivity data and its large spread of values indicate that significant improvements to the Gerdien system were needed. The fitted values of the coefficients were $\alpha \approx 10^{-14}$ m$^3$s$^{-1}$, $\beta \approx 10^{-13}$ m$^3$s$^{-1}$, and $q \approx 5$ ion pairs cm$^{-3}$s$^{-1}$, which are up to two orders of magnitude away from the textbook values of $\alpha \approx 1.6 \times 10^{-12}$ m$^3$s$^{-1}$, $\beta \approx 1.4 \times 10^{-12}$ m$^3$s$^{-1}$ (Clement and Harrison, 1992) and $q \approx 10$ ion pairs cm$^{-3}$s$^{-1}$ (Chalmers, 1967), again suggesting problematic measurements. There may be other explanations for the difficulties encountered with quantifying the $\sigma$-$Z$ relationship. For example, the DustTrak is unlikely to give a true picture of the aerosol concentration because of the approximation required to convert mass into number concentrations, and its cutoff size (10 nm) may exclude smaller particles participating in ion-aerosol attachment. There are also suggestions (Dhanorkar and Kamra, 1997*)* that aerosol concentrations exceeding $10^{10}$ m$^{-3}$, as seen here, may invalidate the expected inverse relationship. Other experiments measuring aerosol and conductivity (Guo *et al*, 1996) showed a large scatter in the fit, which may also be expected for this case. However, these effects can not be considered until the Gerdien data is improved.

### 3.3.2 Calibration in clean air

A further test on the Gerdien system is to compare its readings with the source rate of ions, which is principally from local radioactivity. The complicating effect of ion removal by aerosol can be reduced if these measurements are made in clean air, (see Section 2.2.3, particularly Eq. 2.20) for which reason the measurements were made at Mace Head, W. Ireland, in summer 1999. In the recombination limit, when the aerosol concentration is negligible, the ion concentration depends only on the square root of the ion production rate, $q$, as in Eq. 2.20. Classical work (Chalmers, 1967) suggests that alpha-radiation due to Radon isotopes measurable with a Geiger counter, comprises a significant fraction of $q$ at the surface.

A ZP1410 Geiger tube operated at a bias voltage of 550 ± 5 V, in the plateau of its sensitivity curve, was located approximately 5 m away from the Gerdien condenser, at the same height. The Gerdien system made ten samples, which were subsequently averaged, at 0.66 Hz on a nominally three-minute cycle. Figure 3.18 shows data from the early morning of a fine day, where the square root of the Geiger count rate is correlated with the ion concentration, in broad accordance with theory.

[38] This measures aerosol particles with a radius greater than 0.01μm to ± 0.001 μgm$^{-3}$ or ± 1% (TSI, 1994).

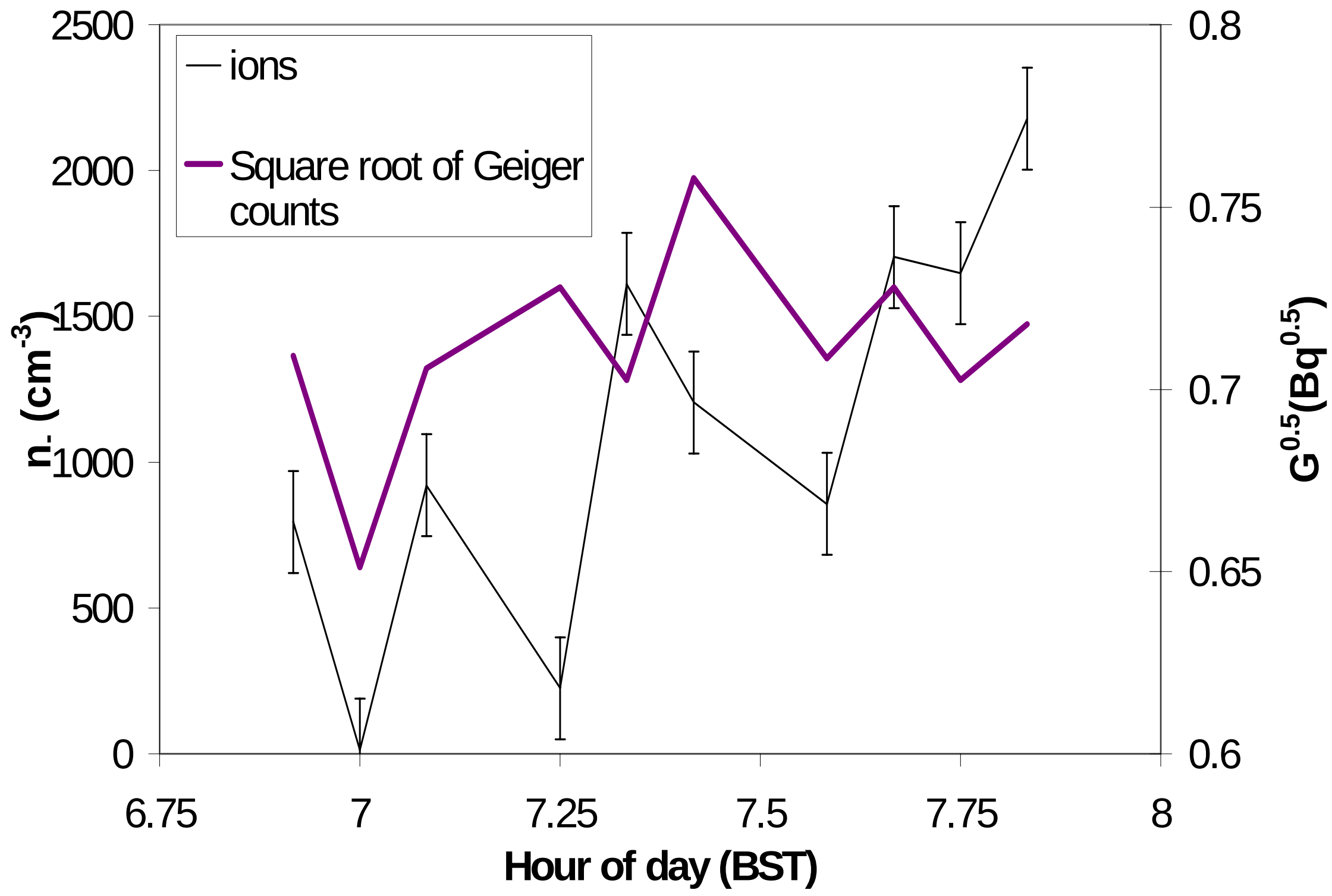

*Figure 3.18 Negative ion concentration $n_-$ (obtained as an average of ten 0.66 Hz samples made approximately every 3 min on 27 June 1999 at Mace Head, Ireland) against $G^{0.5}$ where the Geiger counter output rate is G. The negative ion concentration is indicated on the left-hand axis; error bars are the standard error of the 0.66 Hz samples. The Geiger counter output is shown on the right-hand axis, sampled at 1 Hz and averaged every 5 minutes. The mean wind component into the Gerdien was 5 $ms^{-1}$, and the approximate aerosol mass concentration was 1 $\mu gm^{-3}$. The correlation coefficient is 0.38. From Aplin and Harrison (2000).*

### 3.3.3 Measurements in both Voltage Decay and current modes

Simultaneous measurements of positive and negative conductivity were made using the two different Gerdien measurement modes for three weeks. Atmospheric air was sampled in the laboratory using adjacent Gerdien tubes by an open window at 1.5 m measuring conductivity by the two methods. A circuit (shown in Figure 3.19) was set up to apply a voltage across the electrodes of the Gerdien condenser, turn it off and measure the Voltage Decay as the charge on them dissipated. A manually operated reed relay was used for the switching, and the potential was measured with a unity gain electrometer buffer. The case of the buffer was guarded to the system ground (the earthed screen) to minimise possible leakage paths. The buffer was connected to a chart recorder[39] to measure the change of voltage with time.

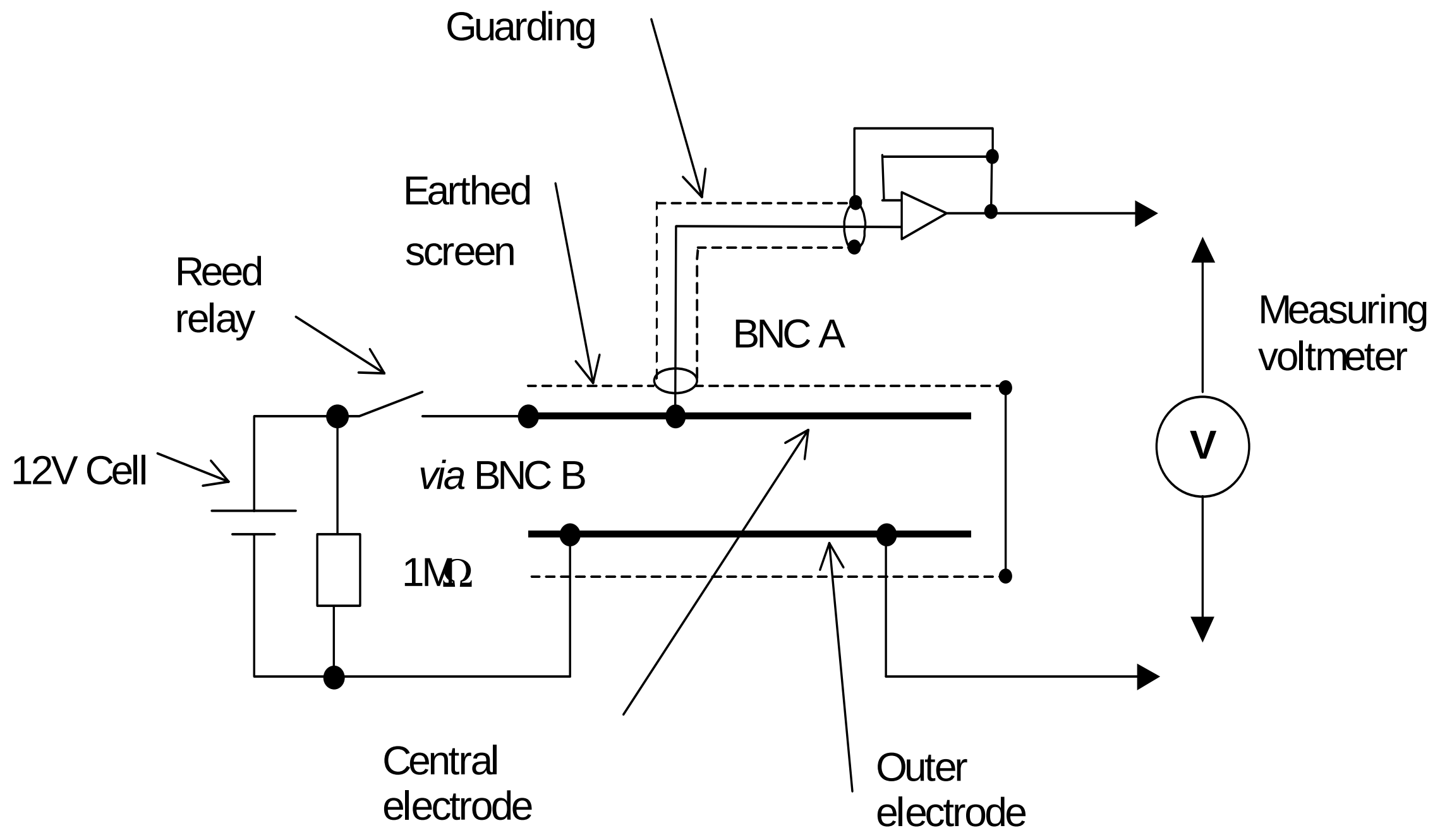

*Figure 3.19 Voltage Decay measurement configuration. The 1MΩ resistor in the reed relay part of the circuit was needed to protect from transients. The cell supplies the bias voltage to the electrodes, controlled by the relay.*

The results shown in Figure 3.19 were calculated by recording the voltage decay on a chart recorder, calculating $\sigma$ as an average from $V$ and $t$ values every 30 s during the decay, which typically becomes unidentifiable after ten minutes. If the exponential decay approximation is assumed, conductivity $\sigma$ can be calculated from Eq.s 2.4 and 2.6:

[39] Gould BS-272

$$\sigma = \frac{\varepsilon_o \ln(\frac{V}{V_o})}{-t}$$ *Eq. 3.12*

where $V$ is the instantaneous measured voltage across the electrodes at a time $t$ after $V_o$ was applied to charge up the electrodes. Conductivity from the current method was calculated from Eq. 3.4.

Equal times every day were spent measuring positive and negative conductivity by the current method, and the daily averages summed to give a total conductivity value for comparison with measurements made by the alternative method. The daily average values of conductivity are shown in Figure 3.20, along with the ratio of positive to negative conductivities. This shows an average value of about 1.4, which is typical for polluted urban air (Higazi and Chalmers, 1965; Retalis *et al,* 1991).

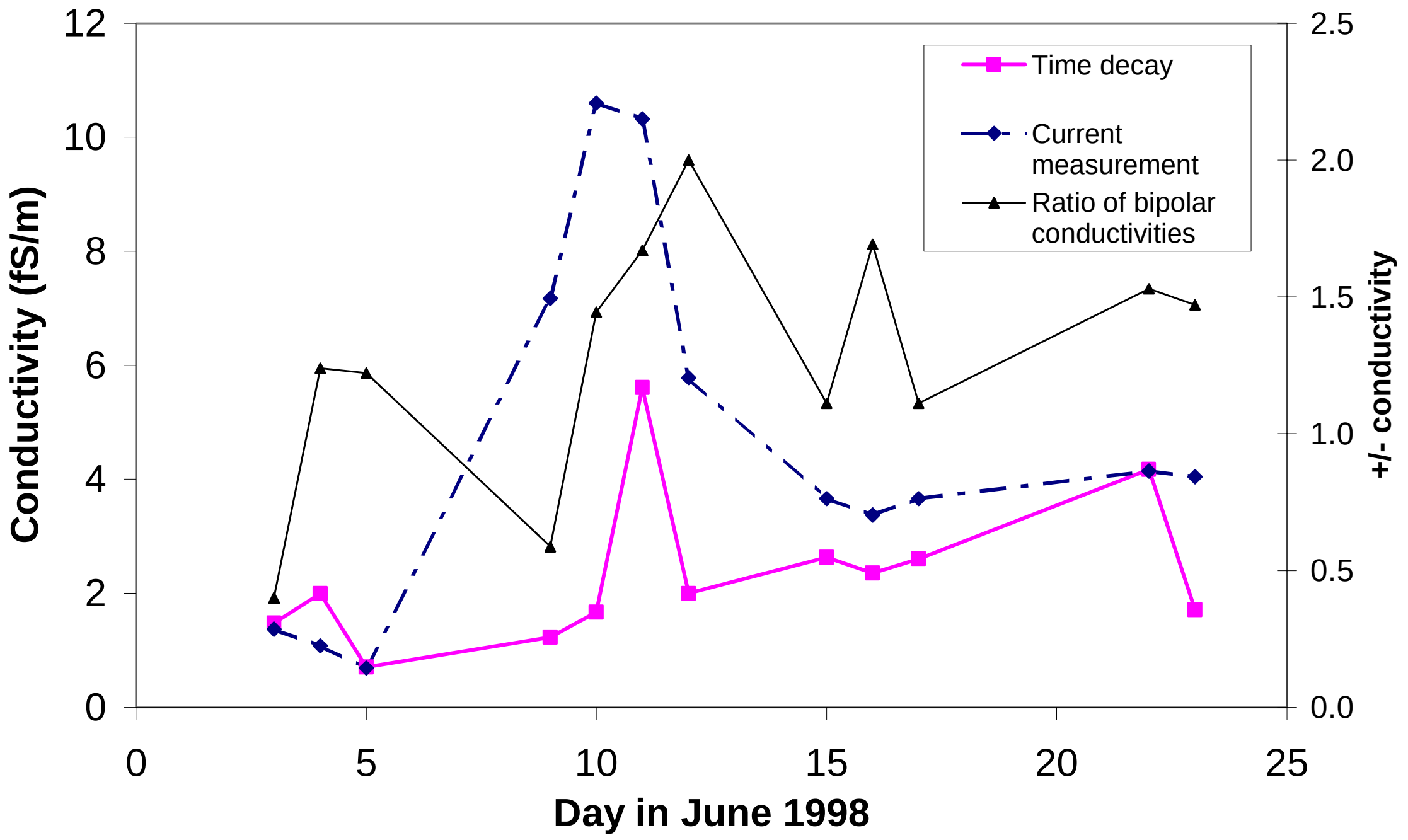

*Figure 3.20 A comparison of daily averages of air conductivity measured by the Current Measurement and time decay methods (left axis). The ratio of positive to negative conductivity is also shown (right axis). From Aplin and Harrison (2000).*

Results for the air conductivity obtained from both methods are consistent, with a positive correlation of 0.43. Experimental error in the Current Measurement method is ± 20 %, due to the error in the capacitance determination of the Gerdien condenser. The error in the Voltage Decay method is more difficult to quantify, because the validity of the exponential assumption is not known. The voltage and time values were obtained manually from the chart recorder trace, which may contribute further error; therefore the uncertainty is likely to be comparable to or greater than the error in the Current Measurement results. The conductivities calculated by the two methods agree to within the errors in the measurements, except when there is a slightly greater discrepancy between 9 and 12 June 1998. Much of this is probably accounted for by variations in flow speeds within the tubes, and the effects of wind direction on the sampling orifice. Average conductivity from the Current Measurement mode was 5 $fSm^{-1}$, which is in agreement with similar measurements (*e.g.* Blakeslee and Krider, 1992, see also Table 2.1).

## 3.4 Conclusions

In this chapter each of the Gerdien system components has been analysed. The current amplifier's response has been optimised for low errors due to leakage, and some disadvantages of the original tube material were found. Investigation of leakage currents and thermal stability show some temperature-dependent effects. The Voltage Decay method as an alternative mode of conductivity measurement has also been successfully deployed and results from the two methods agreed. Measurements of the relationship between conductivity and aerosol were successful to an order of magnitude, and calibration in clean air accords with theoretical predictions. This suggests the techniques used can be independently validated, but their accuracy and reliability would benefit from refinement.

*'this would have obvious meteorological applications if atmospheric air were ionised to even a very small extent' C.T.R. Wilson*[40]

# 4 A switched mobility Gerdien ion counter

In this chapter, a microcontroller is integrated with the Gerdien device described in Chapter 3. This permits the ion counter to measure currents from ions at two different critical mobilities. The reasoning behind this particular development strategy is explained in Section 4.1. An overview of the computer-controlled, bimobility Gerdien system is described in Section 4.2.

## 4.1 Justification and strategy

One of the conclusions of Chapter 3 was that the Gerdien device can measure ions in two modes of operation, a configuration which is clearly advantageous for self-consistency checking. However, the tests were labour-intensive, and required manual operation with results obtained from a chart recorder trace. Introducing some sort of computer control would reduce these data logging difficulties, and other benefits of such a system would greatly improve it. For example, it would be relatively easy to switch between voltages to measure two mobilities of ion with some binary control capability. Running the Gerdien switching in this way would also circumvent the over-range problems, which took some hours to start. Computer control was preferred because it would allow great flexibility in different modes of operation, for example running the Gerdien in the voltage decay mode, carrying out basic data processing, logging to a PC and switching between ion polarities and mobilities. A microcontroller was chosen because of its low cost, ease of incorporation into electronic circuits and ability to support serial communication with the PC.

## 4.2 The microcontrolled system

The system used, pictured below in Figure 4.1 is a commercial module, the BASIC Stamp[41] based on the PIC 16C56 processor, with eight I/O pins and an integral

[40] From Wilson's notebooks, referring to the cloud chamber (quoted in Galison P. (1997), Image and Logic, Chicago University Press)
[41] Parallax Instruments Inc., 3805 Atherton Road, Suite 102, Rocklin, California 95765, USA

EEPROM for program storage. Once the microcontroller is switched on, and a program downloaded to the EEPROM *via* the parallel port of a PC, it runs until the power is disconnected, and if the power is reconnected, the program will restart. This makes it ideal for remote logging applications.

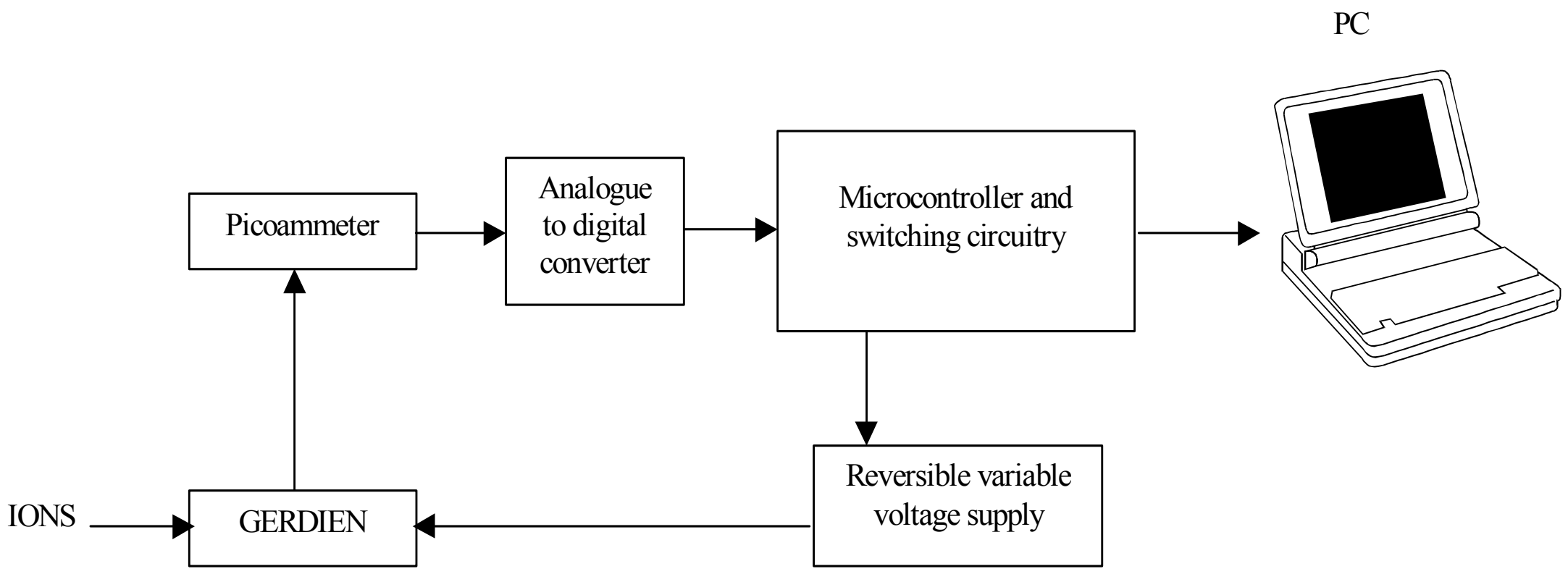

*Figure 4.1 Schematic showing the integrated air ion measurement system, from Aplin and Harrison (1999)*

### 4.2.1 Analogue to digital converter

An analogue to digital converter (ADC) was necessary to convert the output voltage from the picoammeter into a digital format suitable for serial communication. The LTC1298 12-bit ADC (IC1 in Figure 4.2) was used in single-ended mode, controlled by two serial lines from pins 2 and 3 of the microcontroller. A stable 2.5 V voltage reference (measured at 2.51995 ± 0.00005V) is added to the current amplifier output voltage before the ADC input, to ensure that bipolar ion currents can be recorded.

The level shift was tested by measuring the picoammeter output voltage accurately with a Keithley 6512 electrometer before and after the shift stages. It was constant to ± 0.05mV for $-2 < i_{in} < +2$ pA. For larger positive voltages the level shift decreases and for smaller negative voltages the level shift increases, such that the ADC input voltage tends to constant values of approximately +5 V or – 75 mV. The computer logging system records over range values of 4095 or 0 respectively when the ADC receives these inputs. The ADC was calibrated by supplying it with known voltages from a millivolt calibrator, and the output counts were serially sent to a PC; the calibration graph is Figure 4.3 below.

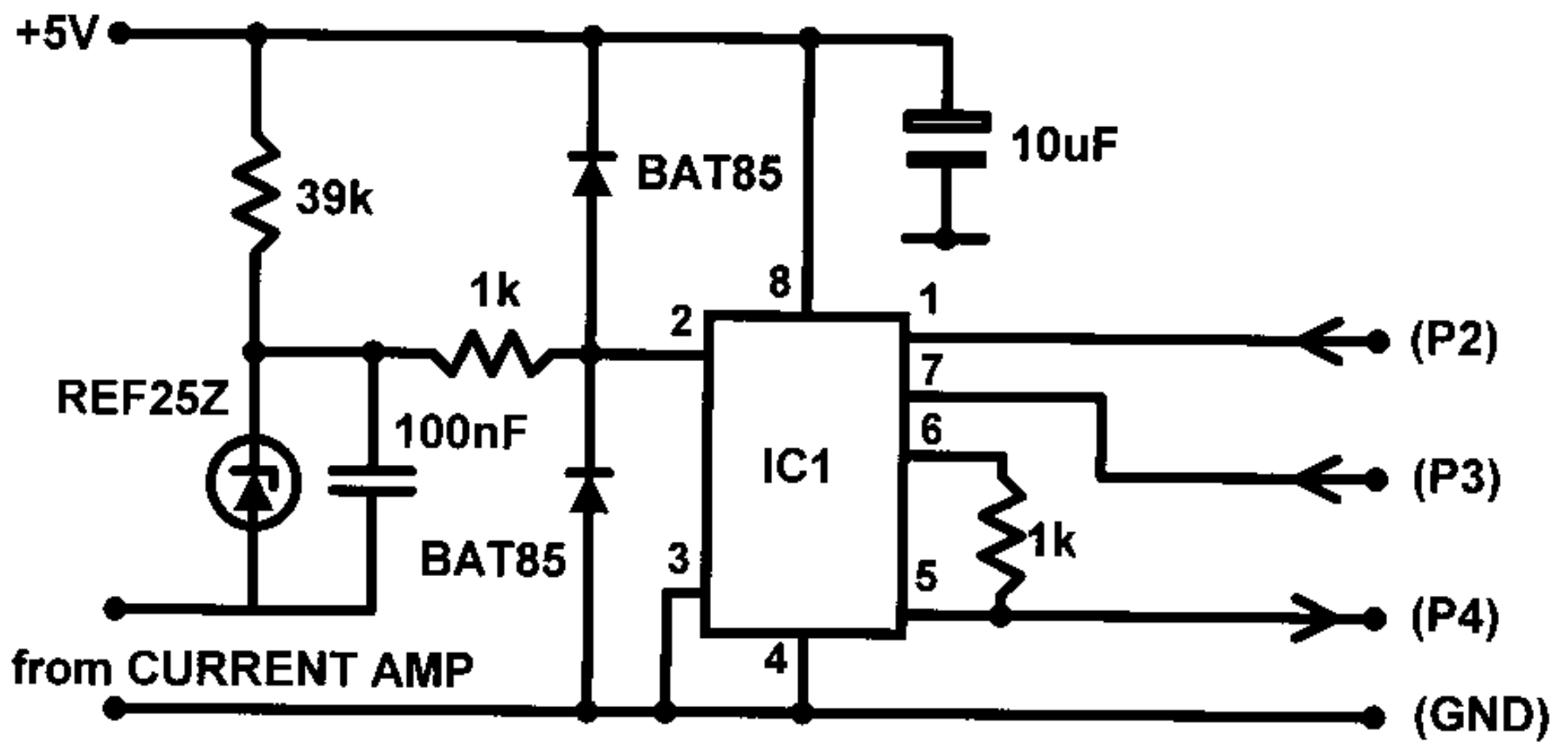

*Figure 4.2 Schematic diagram of the ADC (IC1) and level-shifting circuitry, showing the power supply and connections to the microcontroller I/O pins. From Aplin and Harrison (2000).*

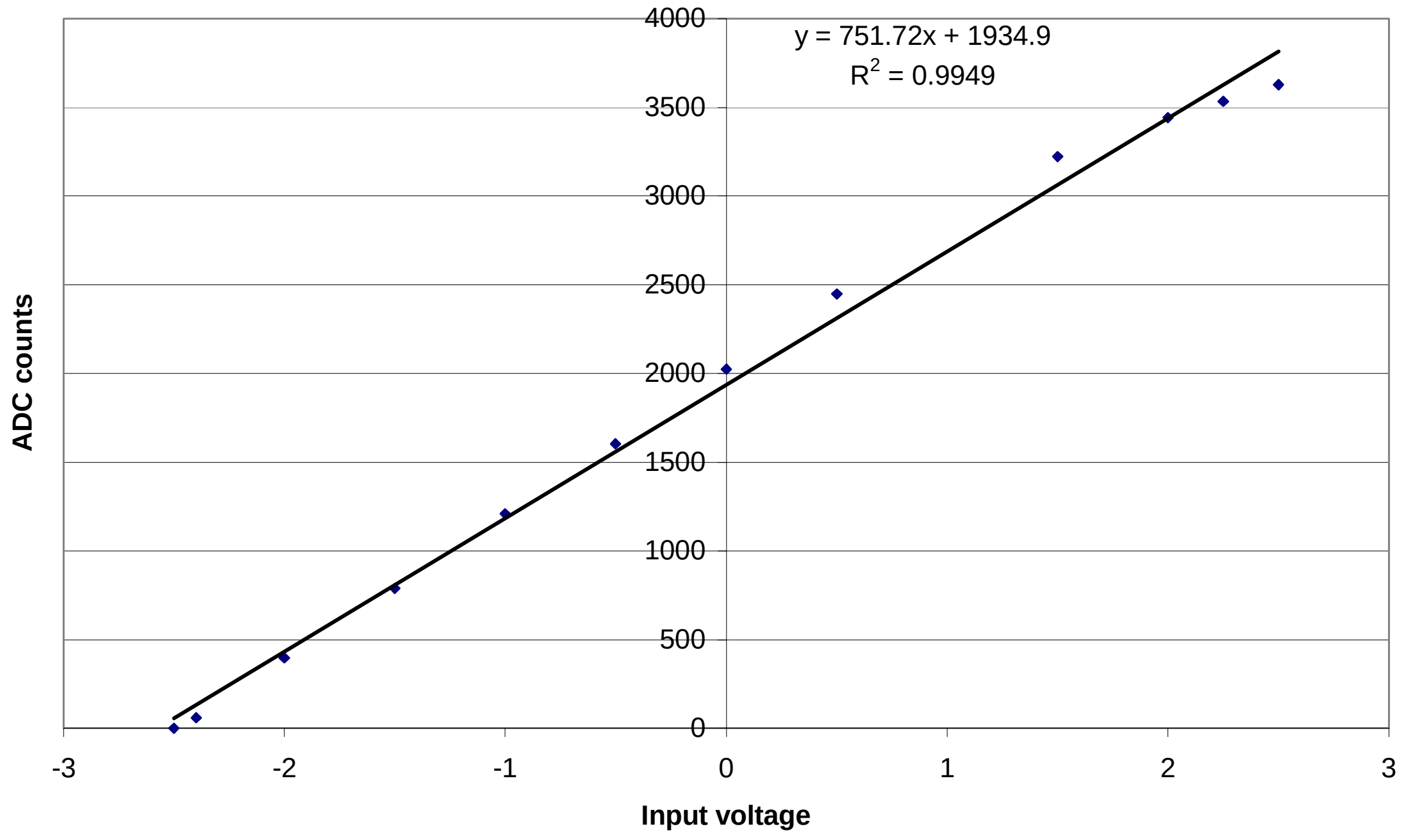

*Figure 4.3 ADC calibration, with bipolar voltages from a millivolt calibrator. The linear response range is shown here.*

### 4.2.2 Switching circuitry

The simple switching circuit employed essentially comprised two low leakage current relays, driven by VN10KM MOSFET transistors controlled by digital outputs from the microcontroller. One relay (a changeover device) was connected to two batteries to provide positive (+9 V) and negative (-6 V) bias voltages for the Gerdien condenser. The other relay was connected across the feedback resistor of the picoammeter to short out the picoammeter feedback resistor. When activated, this relay protects the picoammeter from the transients which occur immediately after the voltage is switched.

### 4.2.3 Software

Programs were written to control and test each component individually before the system was tested as a whole. The program to control the ADC was from Parallax (1998) with some modification. The separate programs controlling the ADC, serial communication and relays were combined to make the overall control program, *GNOCLOCK*, the operation of which is outlined in Table 4.1. The source code is supplied in Appendix C.

| Name | Actions |
|---|---|
| Zero | Turns reed relay on to protect current amplifier<br>Sets other relay to “0” to reduce power consumption<br>Waits for 6s |
| + | Turns reed relay off to allow ion current measurements<br>Sets other relay to bias voltage 1<br>Waits 3 minutes<br>Takes 10 readings of ion current at 1Hz<br>Sends readings to PC via serial protocol |
| - | Sets relay to bias voltage 2<br>As + above and repeats forever |

*Table 4.1 Outlines the actions performed by the program running on the microcontroller. The delay of three minutes after the change of bias voltage is required to allow complete recovery from the transient that occurs on switching.*

### 4.2.4 Serial communication

Data was sent serially to a PC at 300 baud at the following settings: no parity, eight data bits, one stop bit, at negative polarity, and with the output mode always driven.

There was a permanent connection from the microcontroller to the serial port of a Toshiba 1800 laptop computer, which was mains powered and stored in a waterproof metal box at the base of the Gerdien mast assembly. A simple data-logging program, *DATAGRAB*[42] was used to receive data via the serial port and store it as a text file on the hard drive of the computer. *DATAGRAB* also added the time of each reading from the PC clock (which was frequently corrected) to the file.

## 4.3 Results with the microcontrolled system

The dual mobility microcontrolled ion counter was run at Reading University Meteorology Field Site, with the aim of obtaining measurements of two ion size categories, and testing the new ADC logging capability. Measurements of negative ion concentrations at different mobilities were made by using the microcontroller system to switch the Gerdien between two voltages of the same sign, –6 V and –9 V. The number concentration of ions with a mobility exceeding the critical mobility, $n(\mu > \mu_c)$ is approximated by

$$n(\mu > \mu_c) = \frac{\sigma}{e\mu_c} \qquad \textit{Eq. 4.1},$$

where $e$ is the electronic charge and $\mu_c$ the critical mobility. The critical mobility was calculated using Eq. 2.11. Flow inside the tube, which is needed to calculate critical mobility, was determined from a fourth-order polynomial fit to the wind tunnel measurements of flow in the tube against external ventilation (Figure 3.10). During the measurement period, the wind component into the Gerdien tube was positive, so the polynomial was fitted to the measurements for the case when the external wind was aligned with the flow. The results are shown in Figure 4.4.

[42] N. Ellis, Department of Meteorology, University of Reading, 1993

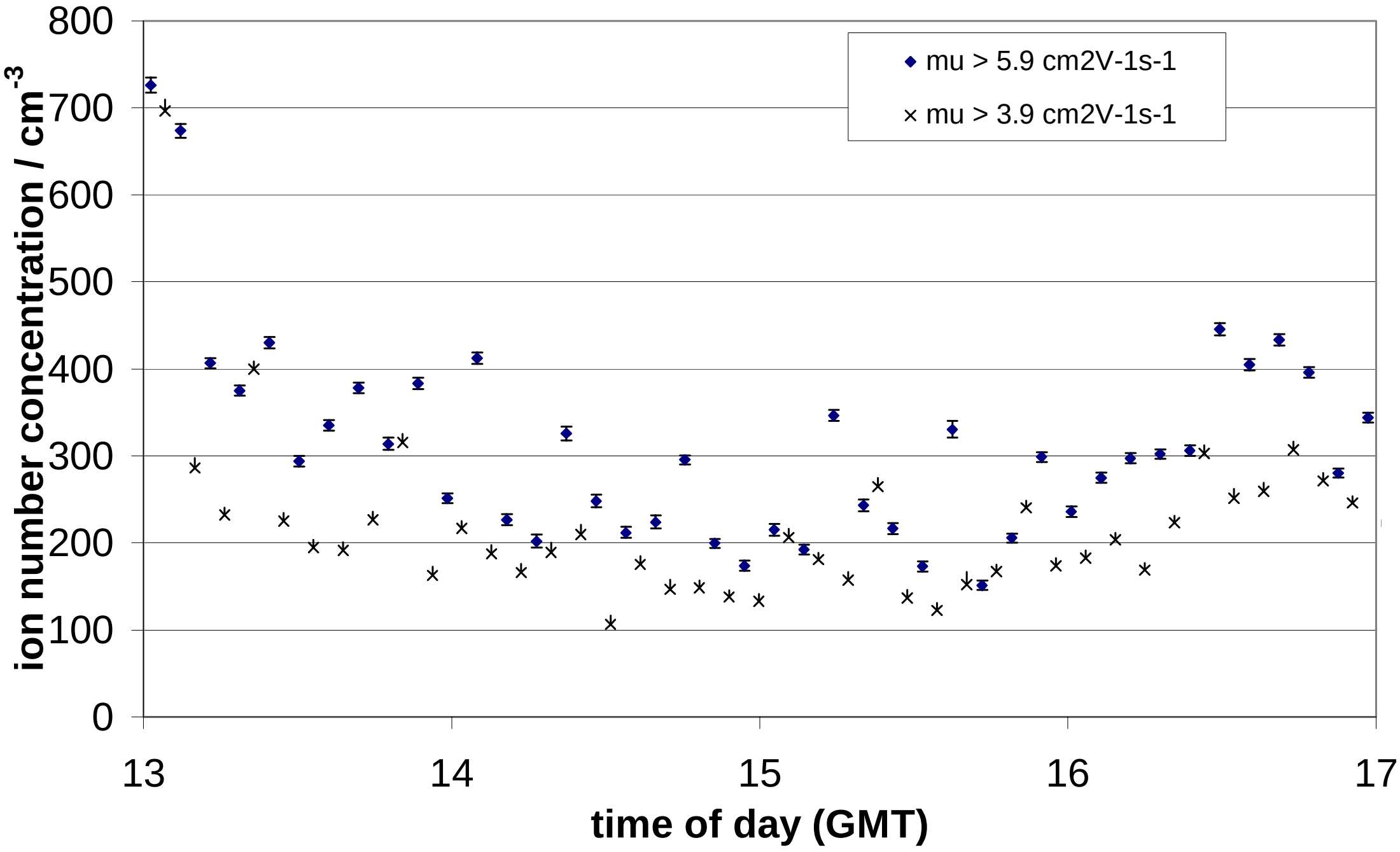

*Figure 4.4 The negative ion number concentration at critical mobilities of 5.9 and 3.9 $cm^2V^{-1}s^{-1}$ for the afternoon of 23rd February 1999.*

## 4.4 Conclusions

A simple microcontrolled system has been tested to measure ions by switching between two critical mobilities and serially output the data to a PC. This is a considerable improvement on the manual measurements described in Chapter 3, and means the instrument can be effectively deployed for remote measurements in the field. Encouraging results also suggest that running the instrument at more than one critical mobility is advantageous. The success of this initial microcontrolled system means that it can be applied to the more complex problems of controlling the voltage decay mode and running both modes on one instrument.

*"To make a real, significant impact on Science, you need to invent a new instrument" James Lovelock, 15th November 1999*[43]

# 5 The Programmable Ion Mobility Spectrometer

Chapter 4 discussed an ion measuring device using an elementary switching arrangement to select different critical mobilities. This chapter covers the development and testing of a Programmable Ion Mobility Spectrometer (PIMS), to permit a greater flexibility in the ion measurements. The major part is the development of a new multimode electrometer (MME), which in addition to measuring small currents[44], can act as a voltage follower, and compensate its own error terms in software. This allows the use of Current and Voltage Decay modes in ion measurements. A further addition to the system described in Chapter 3 is digital-to-analogue converter circuitry to add a range of programmable critical mobilities, by changing the bias voltage, in the Current Mode[45].

The equations used to calculate conductivity from Gerdien devices in Current and Voltage measurement modes are:

*Current Measurement method*

$$\sigma_{\pm} = \frac{i\varepsilon_0}{CV_{\pm}} \qquad \textit{Eq. 5.1}$$

*Voltage Decay method*

$$\sigma = \frac{\varepsilon_0}{\tau} \qquad \textit{Eq. 5.2}$$

for which voltage, current and time measurements are all required, which are discussed in this chapter. I have adopted a more rigorous approach to improving and characterising the accuracy and uncertainty in each of the quantities required in Eq. 5.1 and Eq. 5.2. Section 5.3 explains the generation of the programmable critical mobilities by varying the bias voltage, $V$ in Eq. 5.1. Current and voltage measurements are discussed in the context of the two measurement modes of the multimode electrometer, in Section 5.4, which raises questions of calibration (Section 5.5). Section 5.6 discusses the two other MME modes, in which it checks its own error terms. Section 5.7 describes the integration of the Gerdien, electronics and microcontroller to comprise the PIMS device. Using the integrated system, the Gerdien condenser capacitance $C$ is measured, using a precision voltage ramp initially developed to calibrate current measurements, and microcontroller software is developed. Finally, in section 5.7.4, a preliminary atmospheric test compares $\sigma$ measured by the two methods.

[43] This quotation is paraphrased from a lecture given by James Lovelock at the Department of Meteorology, University of Reading, to a packed audience consisting almost entirely of scientists working with numerical atmospheric models.

[44] In this chapter, the mode of the multimode electrometer which measures small currents will be referred to as the "picoammeter mode".

## 5.1 Motivation

### 5.1.1 Limitations of other systems

The microcontrolled Gerdien condenser system described in Chapter 3 has measured air ion concentrations as predicted by ion-aerosol theory (*e.g.* Clement and Harrison, 1992), and all its components have been thoroughly tested and investigated. However, there were clearly some aspects in need of improvement. The system was unreliable at high relative humidities. The reason for this effect was not identified, but many possible causes were ruled out. Aluminium as a Gerdien tube material was not thought to be suitable for extended field measurements due to its chemical reactivity, and the previous Gerdien design was also difficult to clean. Preliminary investigations of leakage currents and their characteristics showed that their magnitudes and temperature dependencies were clearly significant, but could not be tested at the same time as the ion measurements. A further difficulty, relevant to Gerdien condenser measurements in general, is that the instrument lacks an absolute calibration technique. It is difficult to generate sources of ions with a well-defined spectrum and concentration; for the concentration to be known accurately, they would have to be confined in some defined volume, which by definition introduces problems of unknown losses to the walls.

## 5.2 Objectives

The primary objective is to produce a flexible instrument that can measure atmospheric ions reliably and consistently, by further development of the microcontrolled system described in Chapter 3. To meet the reliability criterion, there needs to be some way to assess the quality of the data. Direct calibration of a Gerdien-type instrument is impractical with air ions for the reasons outlined in Section 5.1.1 above, therefore the only means of "quality assurance" is to include some degree of self-calibration. This is possible with a Gerdien condenser by comparing ion measurements made in the voltage decay and current measurement modes. Further tests of the reliability of the data can be made by quantifying, and compensating for, as many leakage terms as it is possible to determine. Several PIMS instruments will be made to further improve the accuracy of the measurements.

[45] The nomenclature conventions adopted in the introduction to Chapter 3 are continued here.

A secondary objective of developing the programmable ion mobility spectrometer (PIMS) is to make an instrument that is well-suited to making measurements of ionic mobility spectra.

### 5.2.1 New features of the system

The new system is fully automated, and controlled by the original 16C56 microcontroller, with more features added in software and hardware. It measures ions both by either the Voltage Decay or Current Measurement techniques, allowing self-consistency checks. A programmable range of bipolar bias voltages for spectral measurements is attained by adding a digital bias voltage generator based on a DAC.

A current amplifier is required in the Current Measurement mode, as before. However, voltage decay measurements require incorporation of a unity gain electrometer follower into the signal processing system. Separate op-amps could be used to measure voltage and current, with the inputs switched by reed relays controlled by the microcontroller. However, this configuration would further complicate assessing the errors due to leakage currents, which vary with the op-amp. A new electrometer was therefore developed which uses Reed Relays to switch between the voltage follower and picoammeter modes.

A new tube design with a stainless steel outer electrode was intended to improve the reliability of outdoor measurements. The dimensions of the tube were chosen specifically to measure small to intermediate ions with $\mu > 0.2\ cm^2V^{-1}s^{-1}$ (at an average flow rate of $2\ ms^{-1}$), given the practical limitations on bias voltage supply.

This corresponds to a maximum particle radius of ~ 3 nm (Tammet, 1995), which is approximately equivalent to the lower limit of presently-available neutral particle detectors such as the Pollak counter (Metnieks and Pollak, 1969).

New tube design

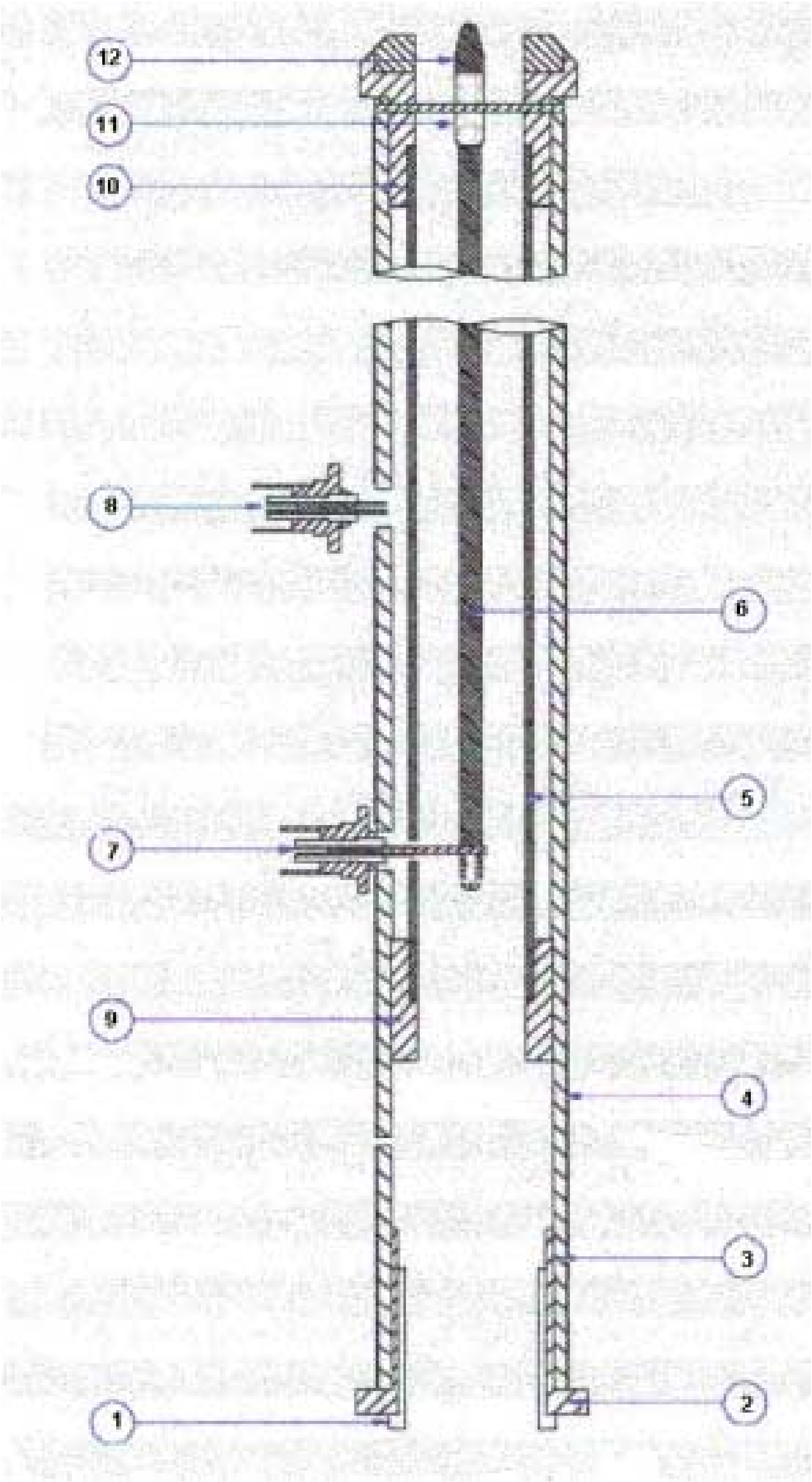

*Figure 5.1 The principal components of the PIMS Gerdien. 1 indicates the fan position, 2 fan housing, 3 acrylic outer screen, 4 electrical outer screen (conductive tape), 5 outer electrode, 6 central electrode, 7 central electrode BNC connector, 8 outer electrode BNC connector, 9 outer electrode rear support, 10 outer electrode front support, 11 central electrode insulator, 12 inlet bullet-shaped cap.*

Both electrodes of the new Gerdien device (5, 6) were constructed of stainless steel, and the tube was constructed so that it could be completely dismantled for cleaning or repair. These design decisions were based on the problems encountered with an aluminium tube with the components sealed in, discussed in Chapter 3. Approximate dimensions are: central electrode length 25 cm, tube radius 11 mm, central electrode radius 2 mm. The outer screen was constructed of acrylic (3) wrapped in a sleeve of conducting tape (4) to provide the necessary electrical screening. At the ends of the Perspex sleeve were Teflon insulating supports (3, 10), into which the fan (1) was housed at one end (2) (a small hole was drilled in the Perspex casing to thread the fan power cables through). The outer screen was longer than the central and outer electrodes, so a further insulating support (9) was required to support the outer electrode. At the other end of the tube, the bullet-insulator arrangement (11, 12) was similar to that discussed in Section 3.1.1, except that the ends of the cross-supports were wedged into the Teflon sleeve (10), rather than into the earthed screen to provide better electrical insulation.

The central electrode (6) was again connected directly to the centre of the BNC electrometer connection. However, this was effected by attaching another piece of wire perpendicular to it with a small screw (7), rather than bending the central electrode at right angles. This was so that the connection could be taken apart for cleaning or repair. The connections to both electrodes (7,8) were carefully positioned so that the whole system could be placed in a box for field measurements, with its apertures exposed to the air, but the electrical connections environmentally-protected inside.[46]

[46] A full set of engineering diagrams is supplied in Appendix D.

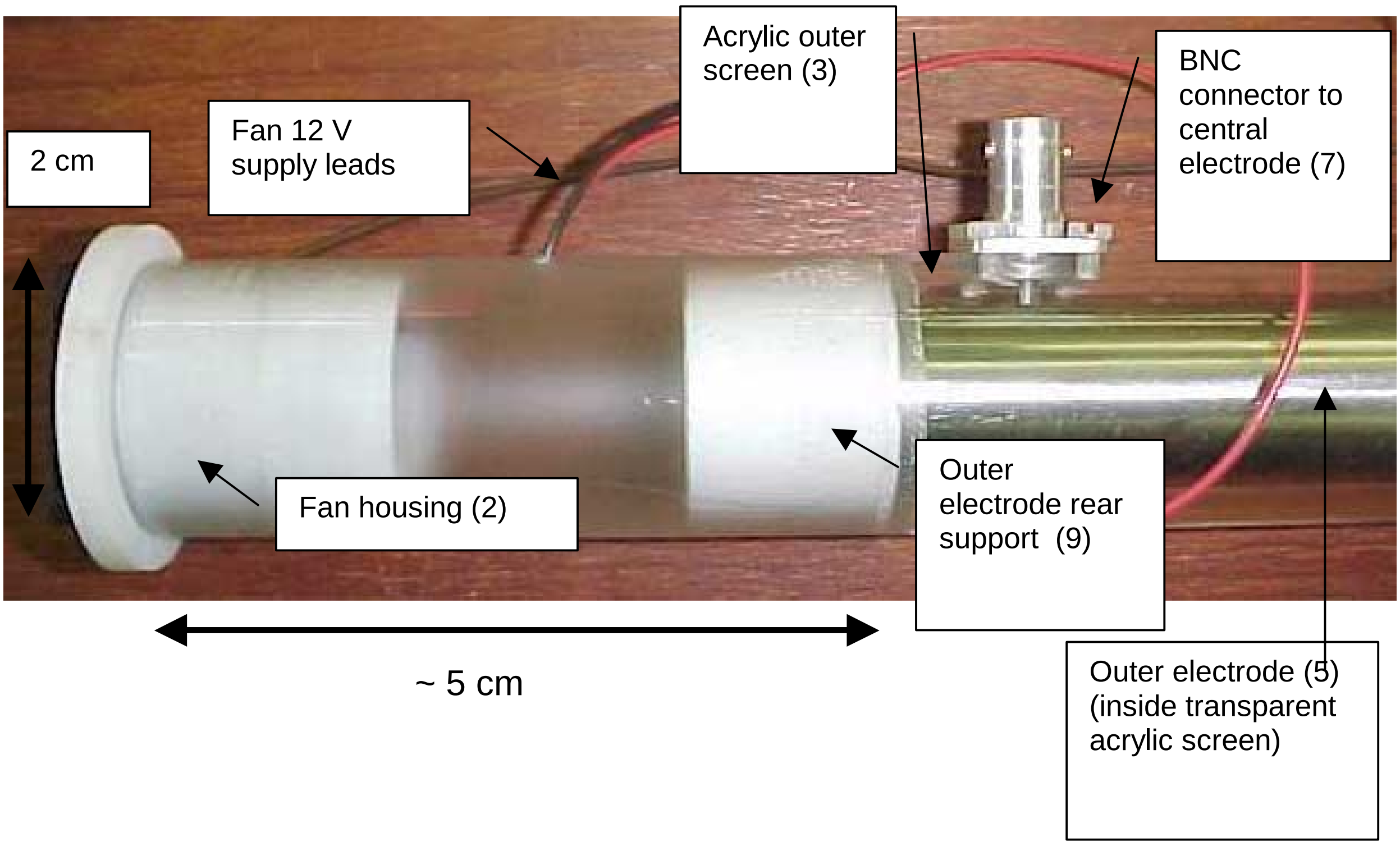

*Figure 5.2 Photograph of one end of the partially constructed new Gerdien tube, before the conducting tape over the acrylic outer screen has been added. (The fan is in place and can just be seen to the left of the fan housing).*

## 5.3 Programmable bias voltage circuitry

Bias voltages for the Gerdien's outer electrode are generated with an 8-bit digital to analogue converter (DAC), programmable with digital signals from the microcontroller. Since the full-scale output of the DAC is limited by its external voltage reference $V_{ref}$, which in this case is well-regulated by its 5 V power supply, an additional high voltage op-amp (IC3 in Figure 5.3) and power supply are required to provide the larger bias voltages necessary for the Gerdien. Power supplies limited the bias voltage range to ± 30 V. However, since high-voltage op-amps and power supplies capable of supplying up to 450 V or greater are available (Horowitz and Hill, 1994; Harrison, 1997b), the bias voltages could in principle be at least an order of magnitude greater, enabling measurements of ions with $\mu > 0.01$ cm$^2$V$^{-1}$s$^{-1}$, or up to a radius of about 16 nm.

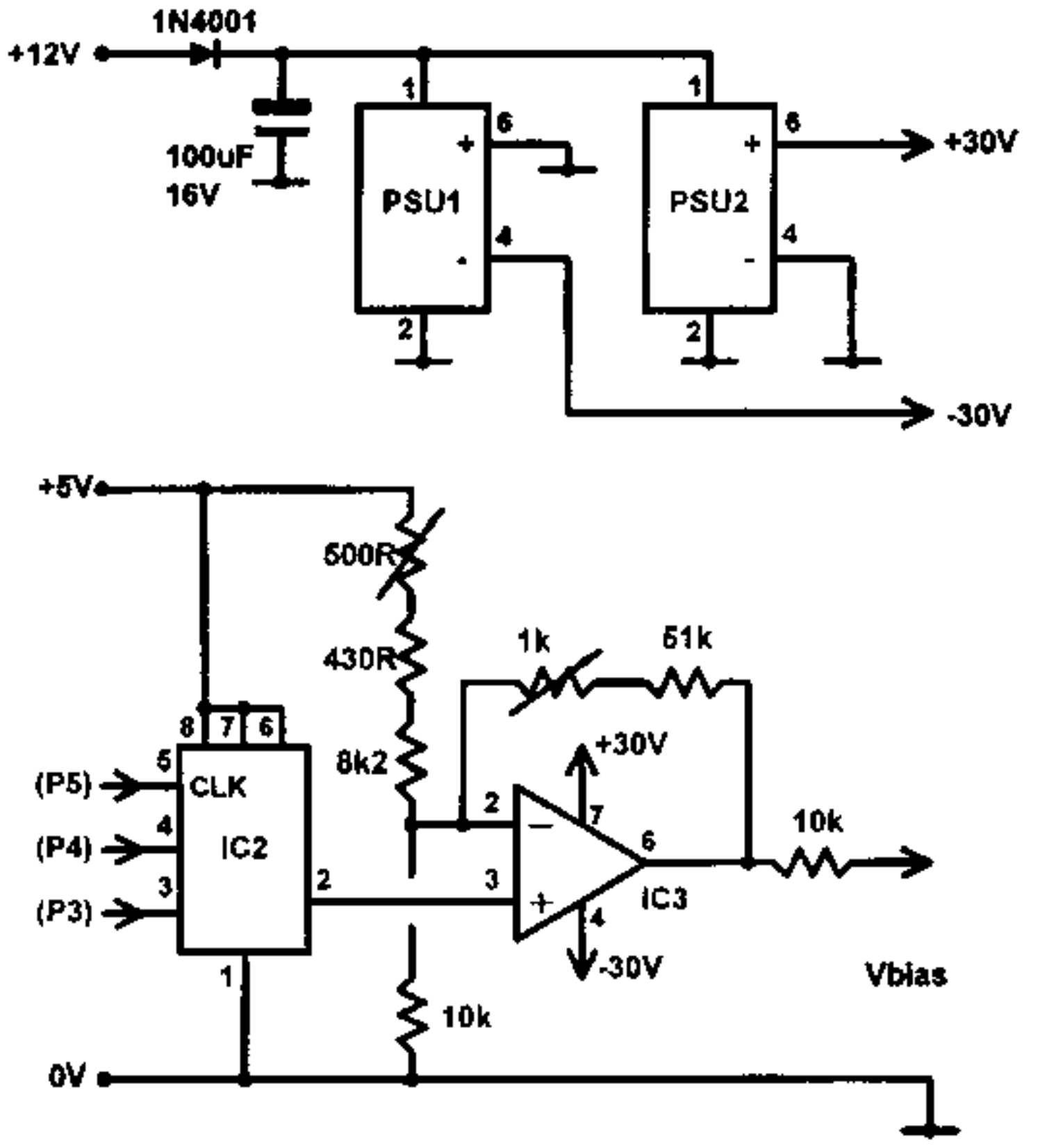

*Figure 5.3 Schematic diagram of the bias voltage generator (from Aplin and Harrison, 2000). PSU 1 and 2 are transformer-isolated 30 V modules (type NMA12155) supplying non-inverting amplifier IC3 (OPA 445). A trimmable voltage offset is applied to IC3 to allow the unipolar generation of the IC2 (MAX 550A) to generate bipolar bias voltages.*

The output voltage of the DAC IC2, *V* is given by

$$V = V_{ref}\left(\frac{n}{256}\right) \quad \text{Eq. 5.3}$$

where $V_{ref}$ was measured as 5.061 ± 0.0005 V, and $n$ is the voltage code supplied serially by the microcontroller. The amplifying stage (IC3) applies a gain of 6.0 to supply the desired bias range of ± 30 V, and an offset is also applied to its inverting input to allow the DAC code to generate bipolar bias voltages. The DAC was calibrated by using the microcontroller to generate a range of voltage codes[47], and measuring the voltage at the output, as shown in Figure 5.4. A least squares fit to the data was then used to relate the DAC code to the output voltage to ± 2 %.

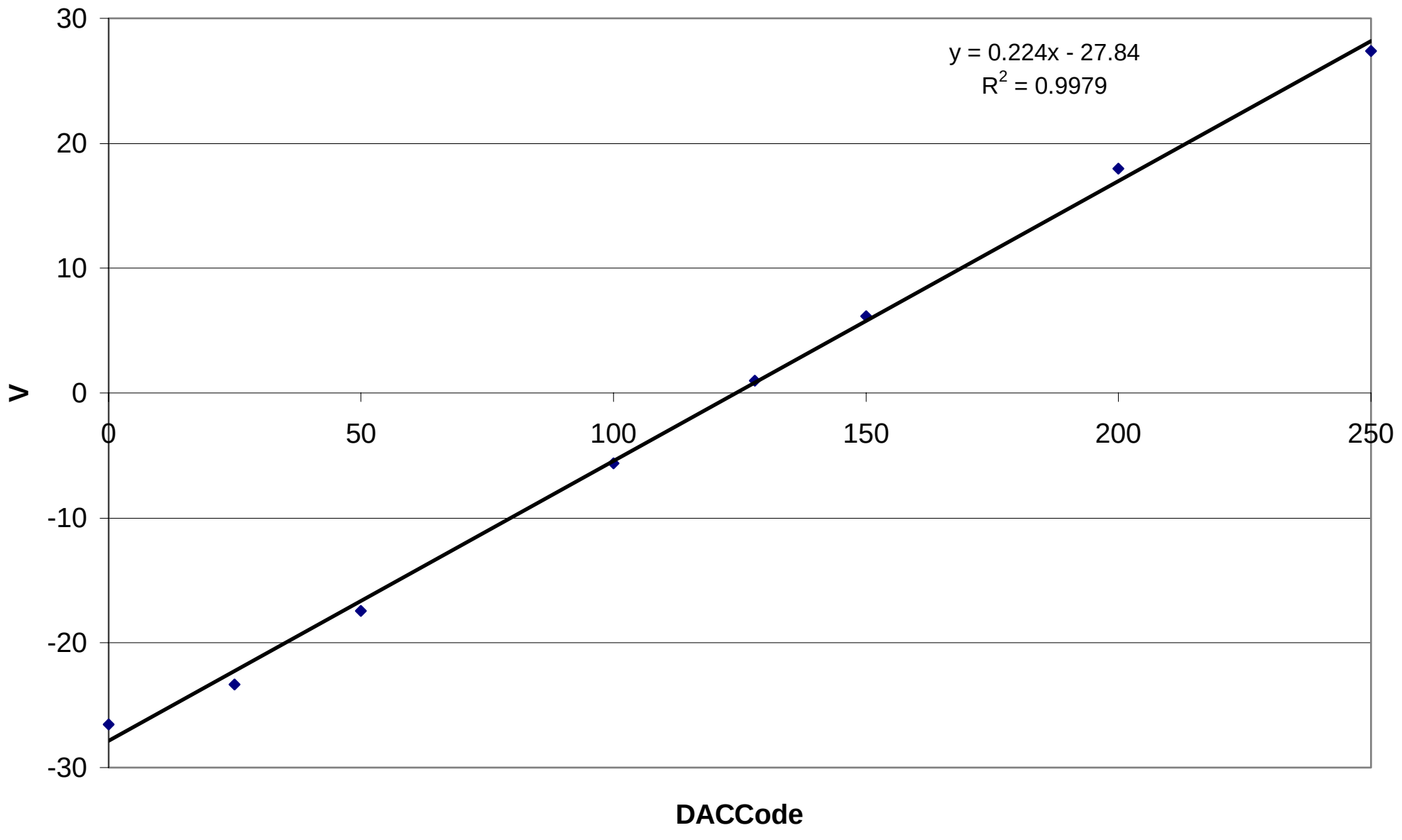

*Figure 5.4 Bias voltage generator calibration, programming the DAC with the microcontroller and measuring the output voltage. Using a least squares linear fit to this data results in an error in the output voltage of ± 2 %, from the standard errors in the fit.*

## 5.4 The multimode electrometer

### 5.4.1 Description

The multimode electrometer (MME) (Harrison and Aplin, 2000c), shown in summary schematic form in Figure 5.6 uses five reed relays and logic circuitry to switch between the modes of voltage follower, picoammeter, $V_{os}$ check, input bias current and electrometer leakage measurement, and charging the central electrode. An overview of the functionality will be given here, and a complete specification is given

[47] The code to initialise and control the DAC is given in Appendix C.

in Appendix A. A photograph of the multimode electrometer can be seen in Figure 5.7.

It is possible to measure the electrometer leakage current (which includes the effect of the input bias current) and input offset voltage by grounding the non-inverting input, as shown in Figure 5.5. In the new multimode electrometer, reed relays are used to measure these quantities.

a) Offset voltage b) Leakage current

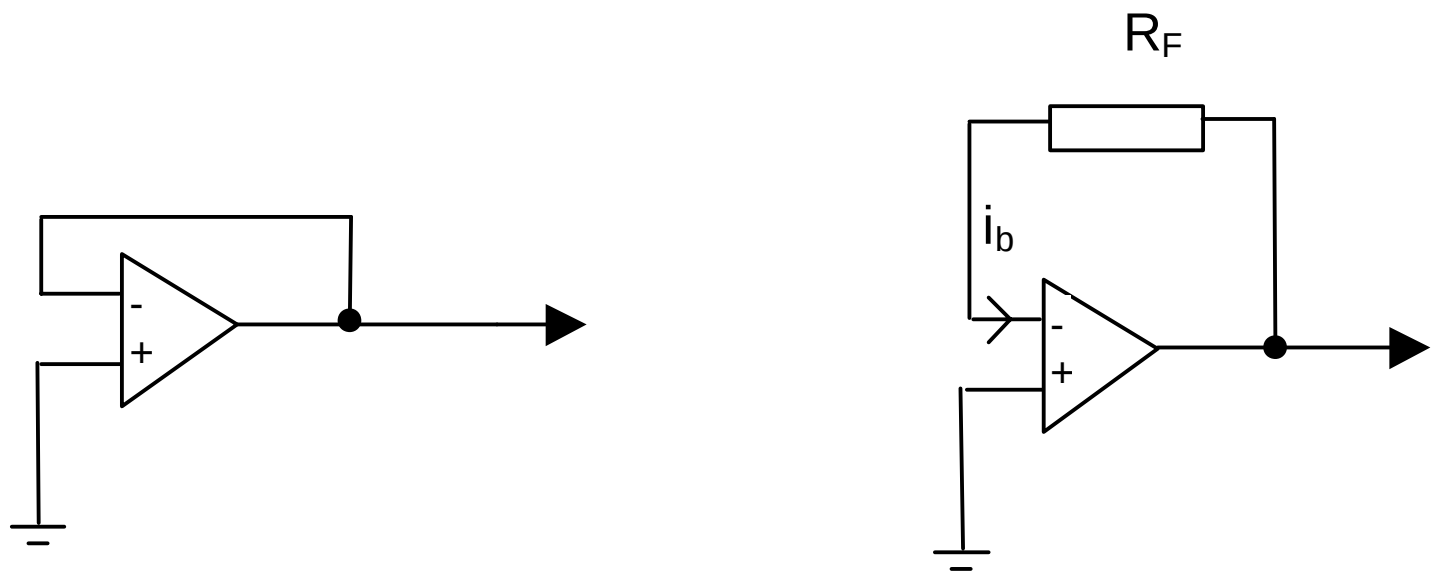

*Figure 5.5 Schematics for measurement of a) input offset voltage $V_{os}$ b) leakage current. (Power supplies are omitted for clarity)*

A voltage reference in the electrometer circuitry, and a fifth reed relay, allows another operating mode to charge the central electrode to a fixed potential, for subsequent voltage decay measurements.

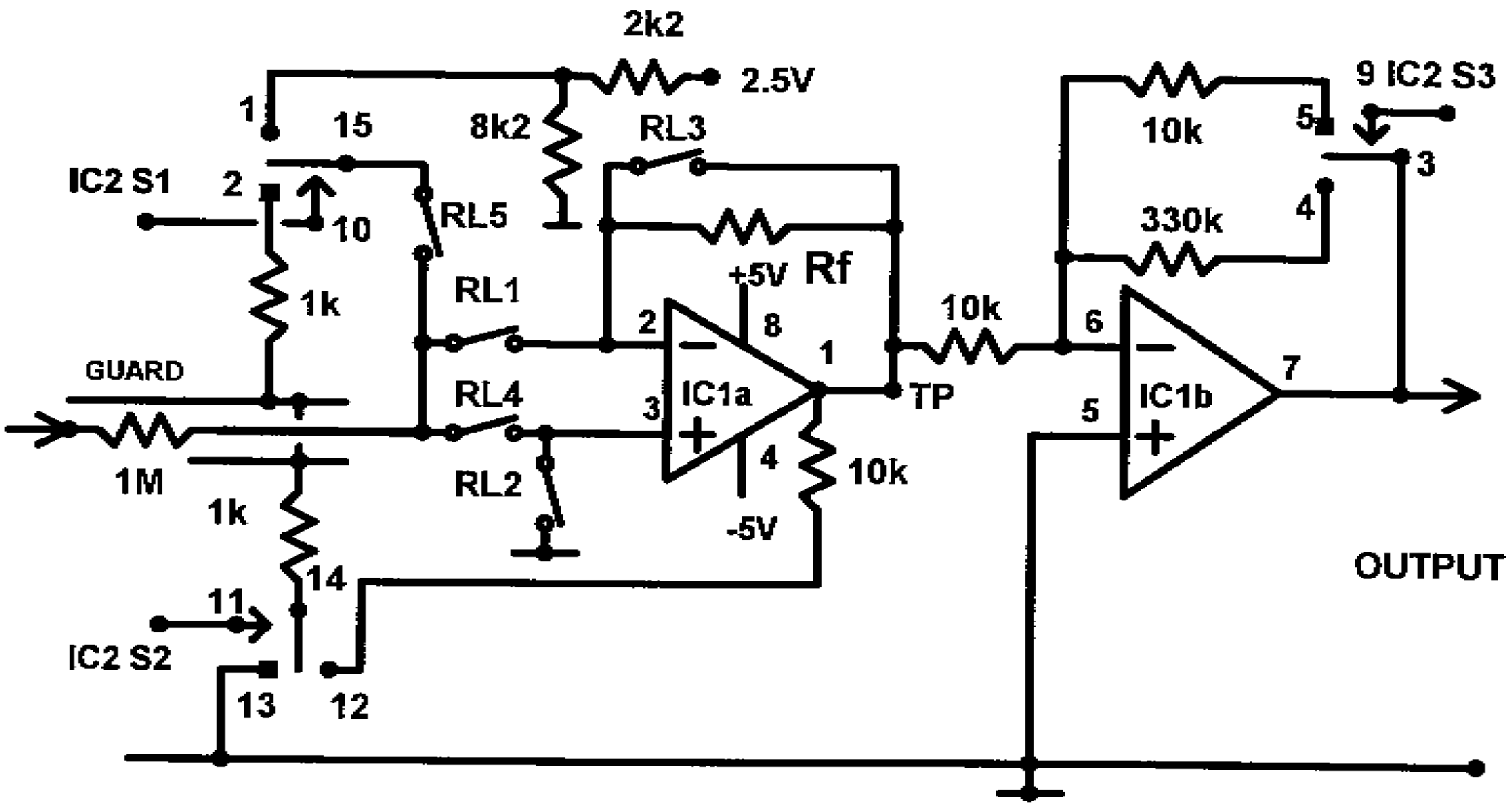

*Figure 5.6 Functional diagram of the switching aspects of the multimode electrometer. IC1a (LMC6042) is switched into voltage follower and current amplifier configurations by reed relay switches RL1 to RL5 with an inverting gain stage of IC1b added.* $R_f$ *is* $10^{12}\Omega$*. MOSFET change-over switches IC2 (4053), are used to guard disconnected reed switch inputs, to minimise leakage. (Square pads on the schematic are used to denote the MOSFET 'off' position and TP is a test point.) From Harrison and Aplin (2000c).*

Three input control lines are decoded by logic circuitry to drive MOSFET switches and, where high impedance switches are required, reed relay coils labelled RL1-5 in Figure 5.6. The reed relay drivers are also connected to LEDs so it is possible to identify the operating mode visually. Switching on RL4 connects the signal to the non-inverting input of the first op-amp, and puts the electrometer into the voltage follower mode. In this mode, the outer screen needs guarding at the input to prevent leakage. This is driven by MOSFET switch S2 in the off position and S1 in the on position, which keeps the outer screen at the same potential as the output of IC1a. An additional leakage-reduction feature uses the guard connection to decrease leakage at the floating side of the open reed relays, controlled by MOSFET switches S1 and S2. RL5 charges the central electrode to 2.0 V, *via* the input connection, before voltage decay measurements can be made. In all the other measurement modes, RL5 is necessarily off, and its open terminal is guarded at the output potential by S1 in the off position. RL3 shorts out the feedback resistor $R_f$ in the follower and $V_{os}$ check modes.

A further inverting, nominally x 33 gain stage, IC1b, is employed to amplify the small signal in the $V_{os}$ and ($V_{os} + i_b R_f$) check modes, which is activated by S3. The gain of this stage is –1 in the measurement mode, as the combination of a high value feedback resistor and unity gain stage has been shown in Section 3.1.3 to have the optimal thermal stability in the picoammeter mode. Therefore, S3 in the off (measuring) position connects a 10 kΩ resistor across IC1b, matching the resistance at the output of IC1a so there is approximately unit gain with sign inversion.

A fifth mode charges up the central electrode to 2.0 V when RL5 is closed, with a 2.5 V REF25Z voltage reference (IC7) and a potential divider. Used in conjunction with the bias voltage generated on the outer electrode, this can charge the electrodes to an effective potential difference of ~ 2.5 ± 30 V for voltage decay measurements.

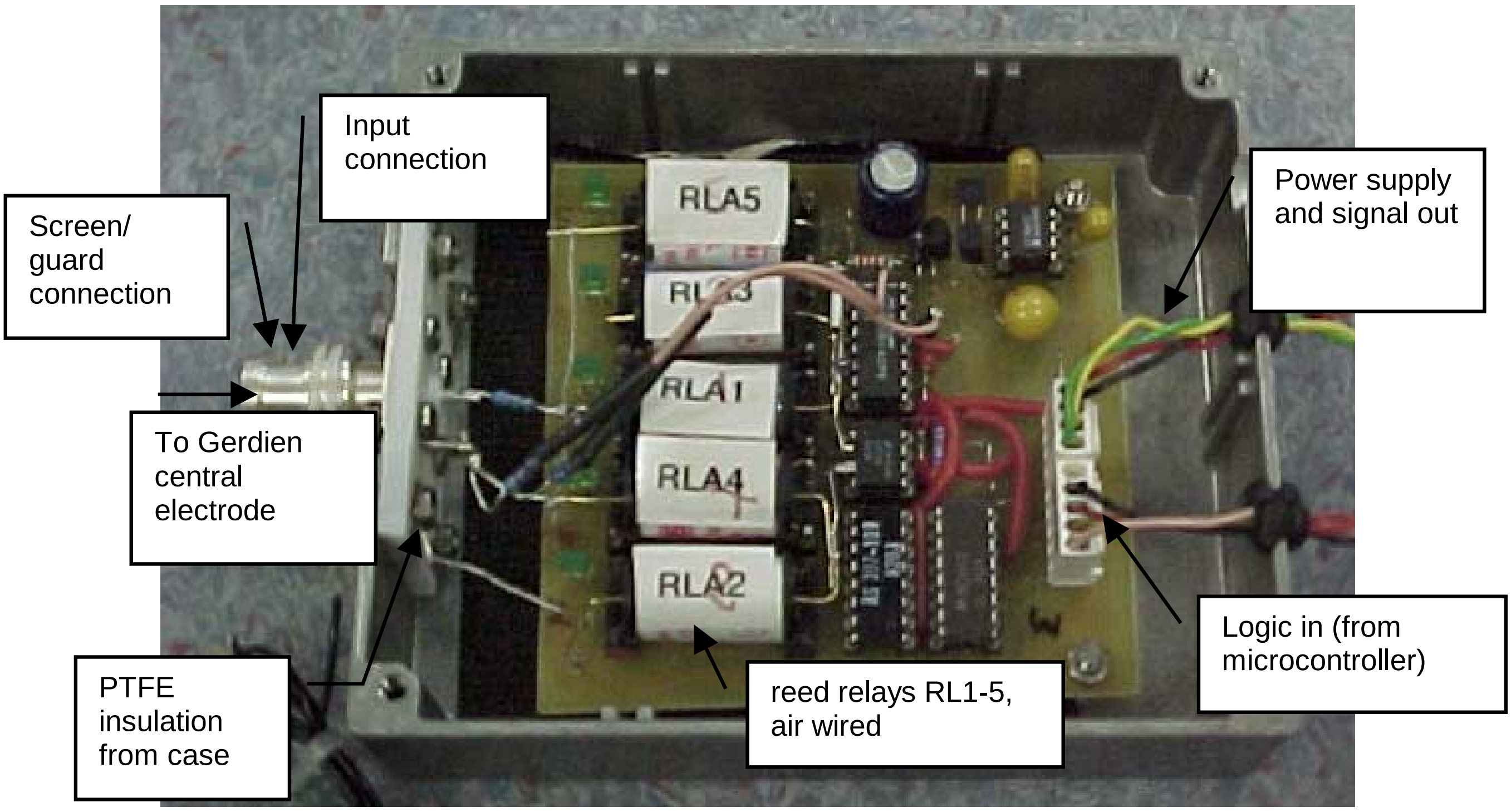

*Figure 5.7 Photograph of the multimode electrometer.*

### 5.4.2 Calibration of the MME

Mode 1: Follower

The follower mode was calibrated by applying a voltage (± 0.05 %) from a millivolt calibrator at the input, and measuring the voltage at the electrometer output. In this mode, the first op-amp operates with unity gain, and the inverting second stage has a gain close to -1. The net effect is therefore a system gain of approximately –1, which can be derived from Figure 5.8.

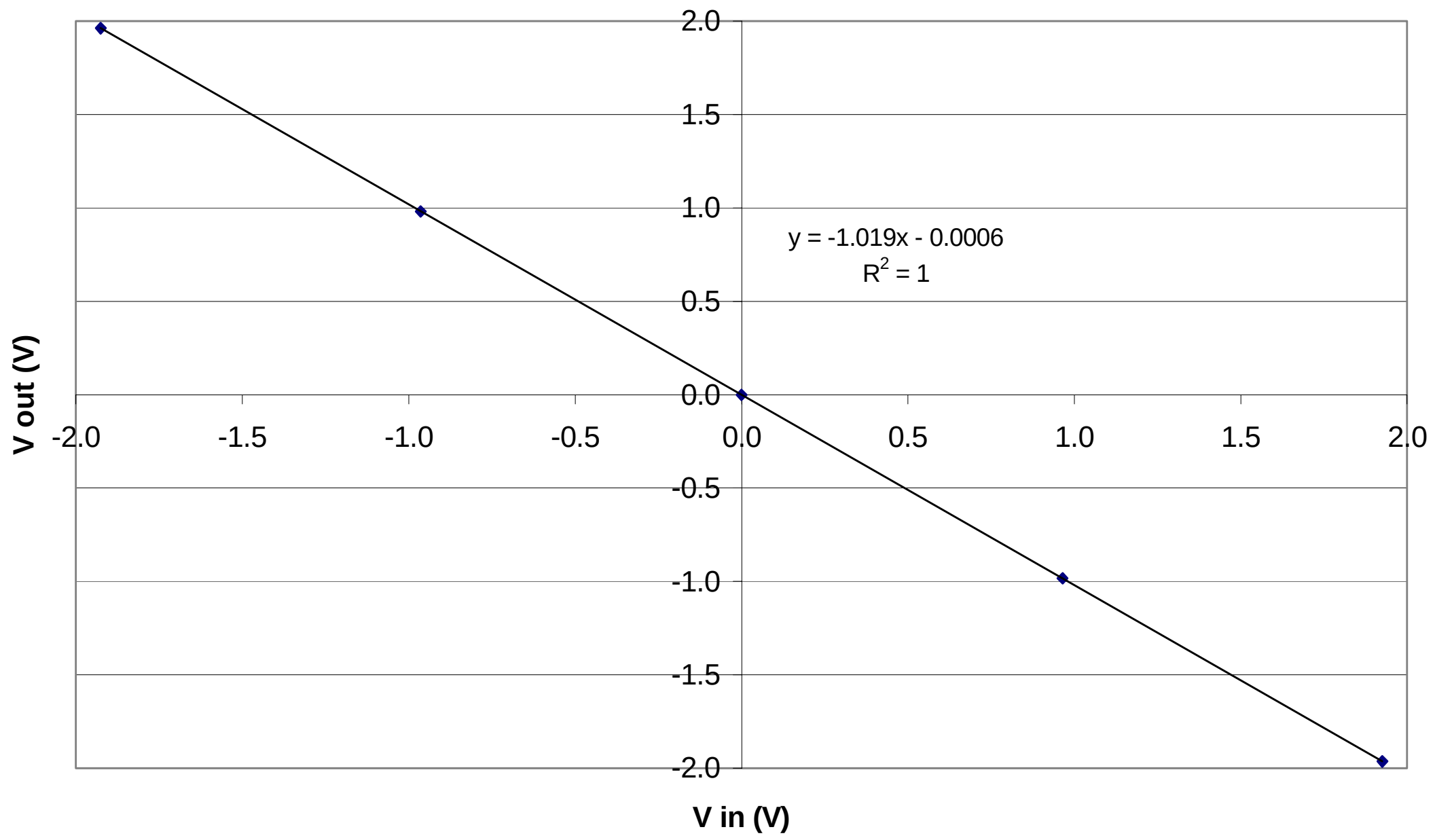

*Figure 5.8 Calibration of the voltage follower mode, for $V_{in}$ applied at the input, and $V_{out}$ measured at the electrometer output.*

The additional inverting gain stage IC1b was calibrated by disconnecting it from the preceding op-amp, and applying a signal from a millivolt calibrator directly to the non-inverting input. The gain was -29.25 with a negligible offset, and typical laboratory values of $V_{os} \sim$ 1mV.

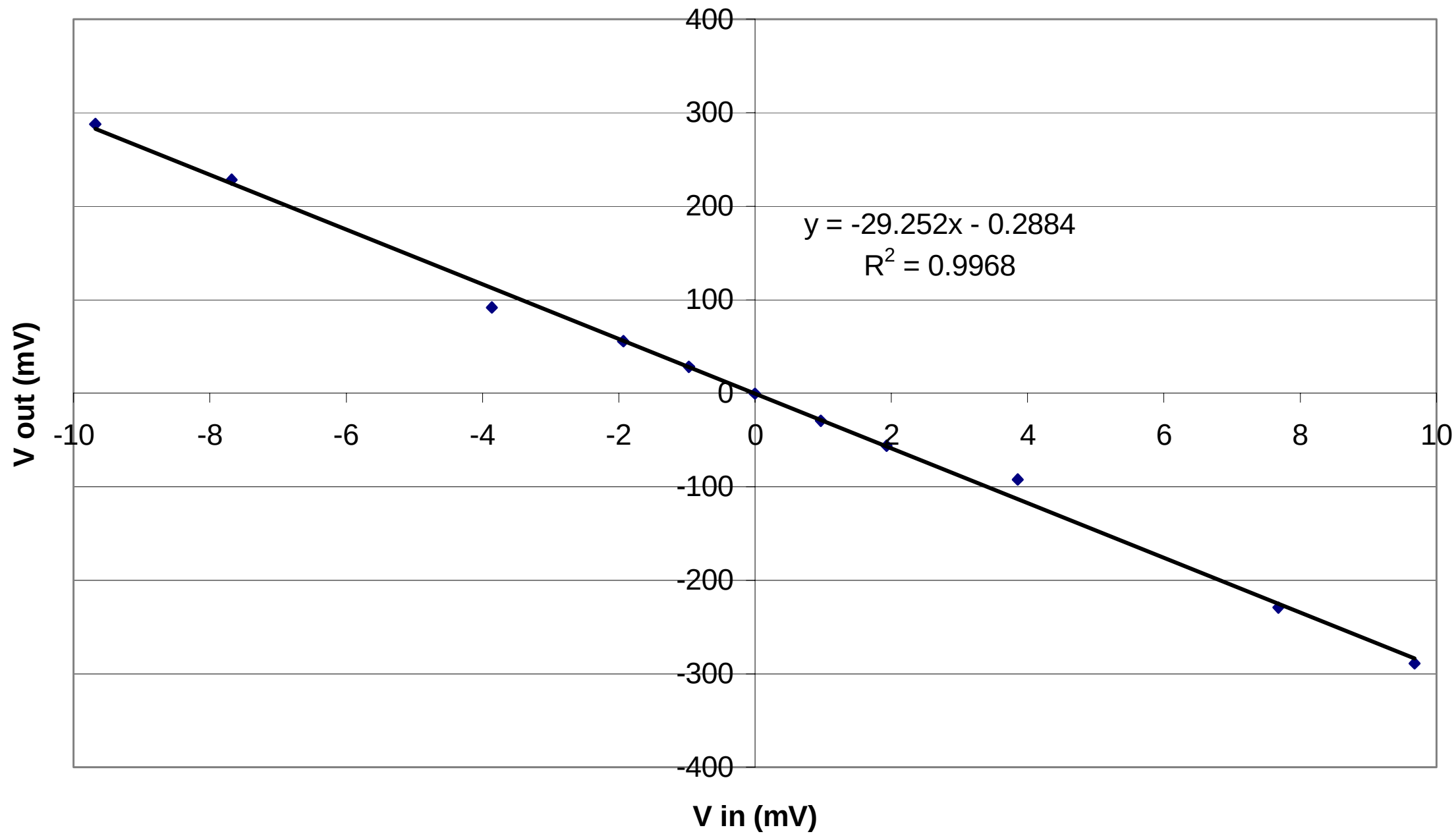

*Figure 5.9 Direct calibration of the second gain stage, with $V_{in}$ applied at the input to the gain stage (with first op-amp disconnected) and $V_{out}$ measured at the electrometer output.*

Mode 2: Current Amplifier

The picoammeter mode was initially calibrated as described in Section 3.2.3, by generating a known current using a millivolt calibrator and a resistor, $R_{cal}$, which must also be calibrated. As in Eq. 3.8, in an i-to-V converter system when there is a secondary stage of gain $G$, the response includes the input offset voltage $V_{os}$ multiplied by $G$. So there is a compromise between using a large magnitude calibration resistor which is difficult to measure directly, or a smaller resistor which introduces large offsets due to magnification of $V_{os}$. In this case both approaches were tried, and are described in Section 5.5.

### 5.4.3 Voltage follower mode

In the voltage decay mode, the charging potential was applied directly to the central electrode via the MME input connection. This was the simplest approach, since the switching could then be controlled directly by the MME. Applying a potential to the outer electrode would require modifications to both the DAC and microcontroller circuitry, and would also be more prone to leakage (the Gerdien condenser central electrode is designed to have ultra-low leakage).

Although the total voltage across the tube is the sum of the bias voltage and the central electrode voltage, the follower only measures the voltage between the central electrode and ground in a maximum range of 2.5 V. It is possible to measure the voltage decay at either electrode because it occurs at an equal rate from both of them. The charging potential $V_f$ was set at + 2 V so that the voltage decay of charge was in the optimal range to be measured. This was achieved by using a 2.5 V reference with a potential divider. A positive electrode charging voltage was chosen because the op-amp's input bias current is much greater for negative input voltages, therefore forcing the input voltage to be positive optimises the measurement.

Measurements in the voltage decay mode are unipolar, resulting from the voltage follower's limiting measurement range of 2.5 V to ground. If $V_b = +30$ V and $V_f = +2$ V, then the exchange of charge due to air ions will not cause the voltage to decay, but will increase it towards 30 V, a change which cannot be detected by the follower. However if $V_b = -30$ V, adding $V_f = +2$ V takes the voltage across the electrodes to above the system ground, which it will decay from to 0 V. Therefore with a positive central electrode voltage, only negative ions can be measured. Electrometer leakage currents are not a function of the bias voltage, therefore it is preferable to have the highest potential difference possible across the electrodes, in order to maximise the signal-current to leakage-current ratio.

## 5.5 Calibration of MME Picoammeter Mode

### 5.5.1 Resistive calibration techniques

$R_{cal}$ nominally 1GΩ

The exact value of a 1GΩ resistor constructed from 10 x (100 ± 5 %) MΩ resistors in series was determined, by measuring the ten 100 MΩ resistors individually with the Keithley 2000 multimeter and summing the values, to be (1.00053 ± 0.00001) GΩ. The response of the MME picoammeter mode to currents generated with a millivolt calibrator, and the (1.00053 ± 0.00001) GΩ resistor, is shown in Figure 5.10.

Finding the value of a 1TΩ ± 20 % calibration resistor

The (1.00053 ± 0.00001) GΩ resistor was used to calibrate another picoammeter with $R_f$ = 4 GΩ (Harrison, 1992 (henceforth H92)), which had a linear ($R^2 = 1$) response. The picoammeter response is known to be

$$V_{out} = -\left[ iR_f - V_{os} \right] G \qquad \textit{Eq. 5.4}$$

which can be rewritten when there is a resistive current source as

$$V_{out} = G \frac{V_{in}}{R_{cal}} R_f - V_{os} G \qquad \textit{Eq. 5.5}$$

Eq. 5.5 can be applied to find the value of 1TΩ ± 20 % resistor, so that it can be used for calibration purposes. It shows that the gradient of a calibration graph of the picoammeter output voltage $V_{out}$, against the voltage used to generate a current with the calibration resistor $R_{cal}$ gives $(G.R_f/R_{cal})$. As the gain can be calculated from the offset of Eq. 5.5, if one of the resistance values in Eq. 5.5 is well-known, then the gradient of Eq. 5.5 can be used to find the other resistance. This method was applied to measure the 1TΩ ± 20 % resistor, by using it as the current-generating resistor $R_{cal}$ with the H92 picoammeter which has $R_f$ = 4 GΩ . The value of the resistor was deduced to be (0.7720 ± 0.0004) TΩ.

$R_{cal}$ nominally 1TΩ

The response of the MME in picoammeter mode to currents generated with a millivolt calibrator using the (0.772 ± 0.0004) TΩ and (1.00053 ± 0.00001) GΩ resistors, was then measured. The results are shown in Figure 5.10.

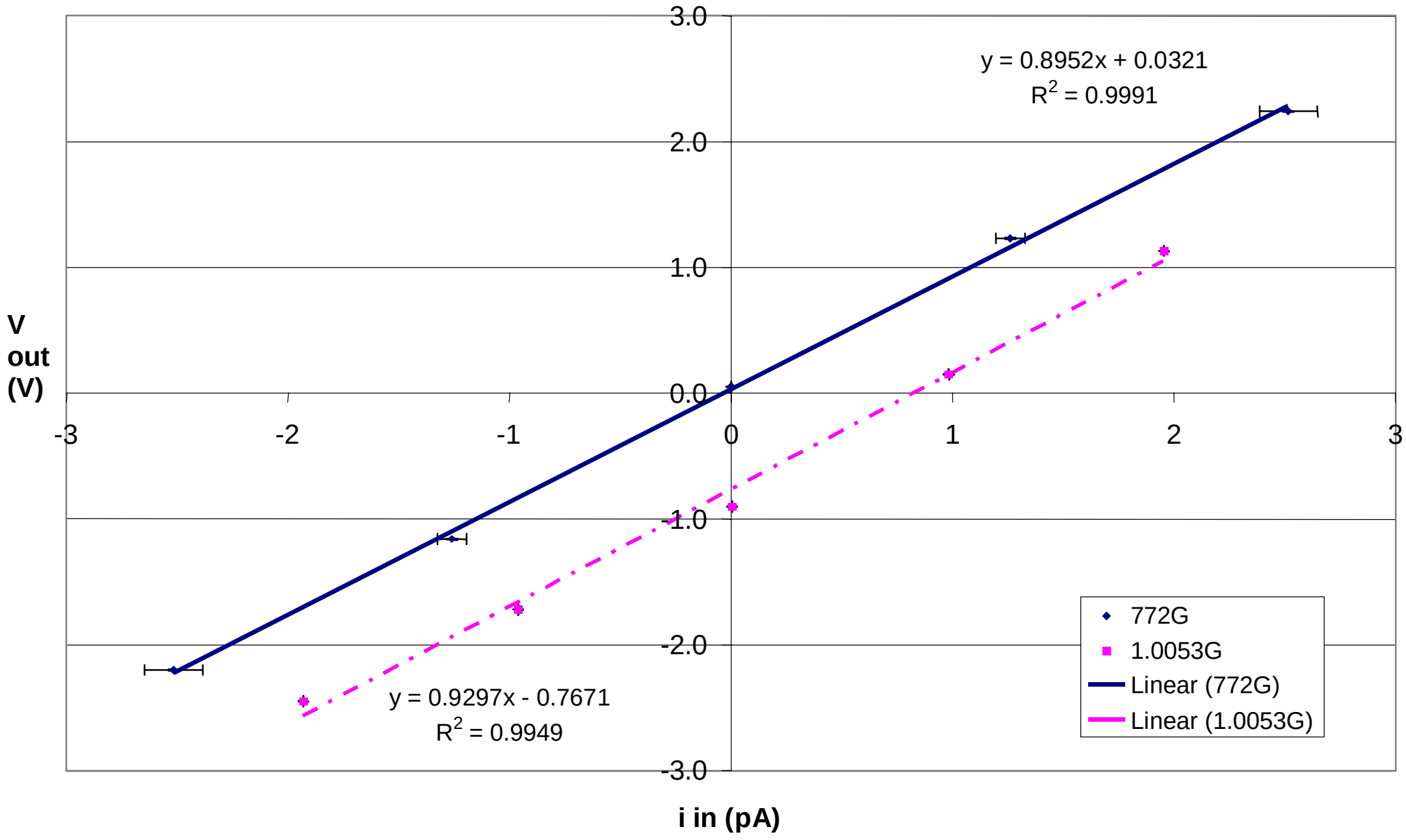

*Figure 5.10 Resistive calibration of the MME in picoammeter mode, using calibration resistors of 772 GΩ and 1 GΩ. The error bars on the abscissa result from the uncertainty in the magnitude of the calibration resistor.*

In Figure 5.10 some inconsistencies can be seen between the two calibration methods. Calibration with $R_{cal}$ = 1.00053 GΩ has a significant offset of -750 mV, in contrast with $R_{cal}$ = 772 GΩ when the offset was only 3 mV. However, the $R_{cal}$ = 772 GΩ calibration was inconsistent with time, and the gain varied by 7 % when the same instrument, which had not been used in the meantime, was recalibrated 9 days later. In Appendix B the system response is derived, and the analysis suggests that significant offsets can result from amplification of the input offset voltage when $R_f >> R_{cal}$. This hypothesis was tested by estimating the magnified offset. $V_{os}$ was measured with the MME in mode 3 (Table A.1), and found to be ~ 1 mV. It is clear that since $V_{os}$ ~ 1 mV, this multiplied by a gain equivalent to $R_f/R_{cal}$. accounts for almost all of the offset in the calibration. It can be concluded that it is unsatisfactory to use 1 GΩ calibration resistors in conjunction with much larger feedback resistors, because the multiplied input voltage offset changes with temperature, and over time, are sufficient to degrade the calibration. This is explicitly given in the overall system response, which is shown in Appendix B as:

$$V_{out} = G\left\{\left[-\left(i + i_b + i_L\right)R_f\right] + V_{os}\left[1 + \frac{R_f}{R_{cal}}\right]\right\} \qquad \textit{Eq. 5.6}$$

The term showing the sum of input, bias $i_b$ and leakage currents $i_L$ in Eq. 5.6 explains the unsatisfactory temporal variation of the gain when the 772 GΩ calibration resistor was used. As explained in Appendix B, both the gain and intercept of the $V_{out}$-$i$ calibration graph will only remain constant if the input bias current, electrometer leakage currents and input offset voltage are invariant, which is unlikely. In conclusion, the method described of resistive calibration described above is inherently unsuitable for picoammeter calibration, because of amplification of the offset voltage and changes in leakage currents.

### 5.5.2 A capacitative calibration technique: the femtoampere current reference

In response to the problems with resistive calibration explained in Section 5.5.1, a different calibration technique was developed based on a capacitative principle, which ensures $R_{cal}$ is large (Harrison and Aplin, 2000a)[48]. It generates a slowly varying voltage ramp (± 52.5 mVs$^{-1}$, over 32 s) using a crystal oscillator. If this signal is presented to a 10 pF polystyrene capacitor, it produces a bipolar current of order ± 500 fA, with its sign depending whether the voltage ramp is on the rising or falling edge. For the calibrator used in this Thesis, the current was measured with the Keithley 6512 electrometer to be ± 470 ± 2 fA over the temperature range -22 to 21.7 ºC, and was also temporally stable. The femtoampere current reference was used to calibrate the MME picoammeter mode, which is shown in Figure 5.11.

[48] The principles of the capacitative calibrator (Harrison and Aplin, 2000a), are described in Appendix B.

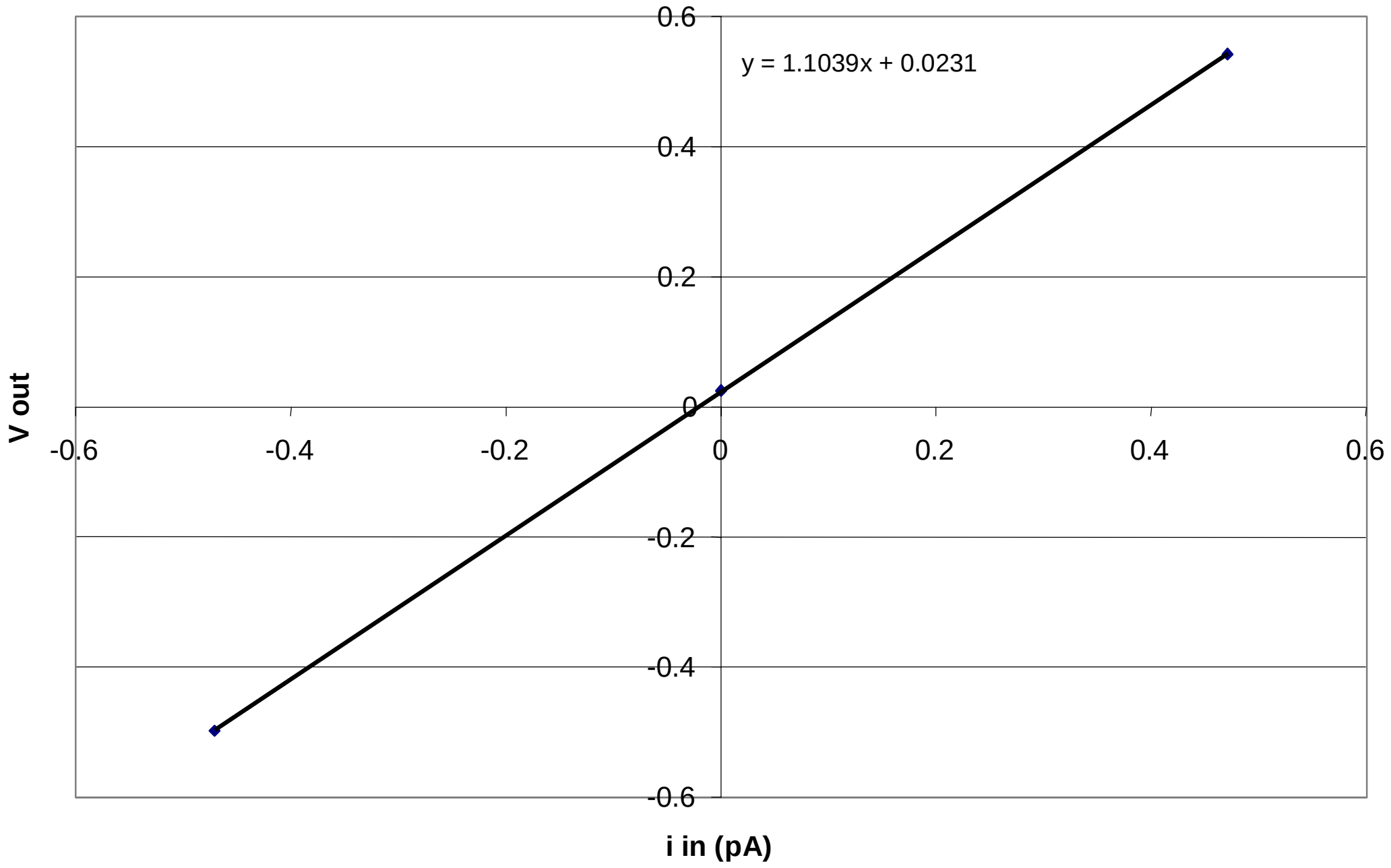

*Figure 5.11 MME calibration in picoammeter mode, with currents generated from the precision voltage ramp connected to a 10 pF capacitor. The offset was measured when the picoammeter input was connected to the screened 10 pF capacitor, with the ramp generator disconnected. The picoammeter gain is (1.104 ± 0.004) V/pA, and the offset is (2.3± 0.1) mV (errors are the standard error in the least squares linear fit).*

## 5.6 Testing of MME measurement modes

### 5.6.1 Error term modes

The input offset voltage was measured at the Reading University Meteorology Field Site on eight days. $V_{os}$ was observed to vary slowly with time, so the averages for each day of measurements were calculated and compared against the average temperature over the same period in Figure 5.12.

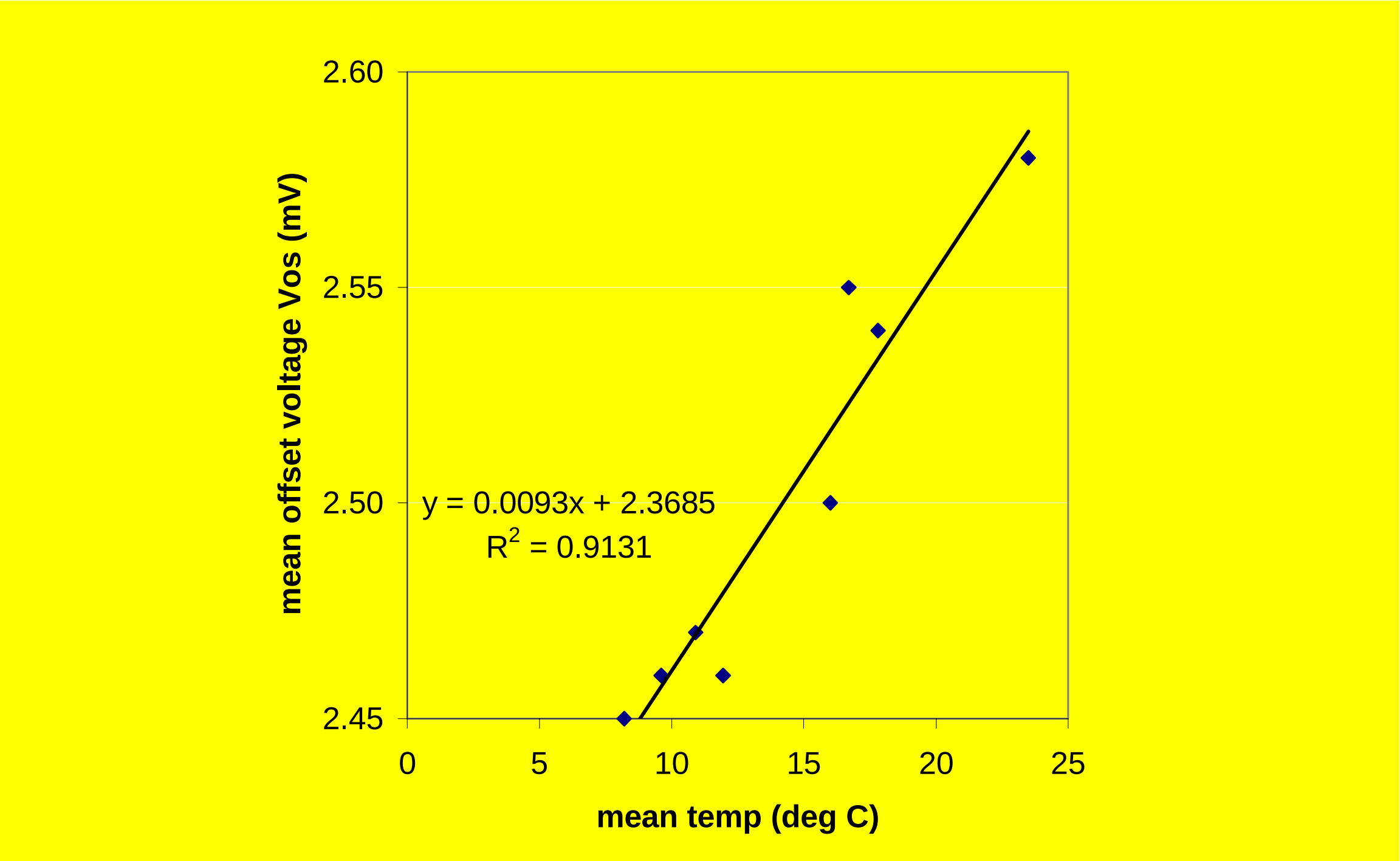

*Figure 5.12 Variation of mean input offset voltage of the MME with mean temperature, over eight diurnal cycles.*

Figure 5.12 shows that $V_{os}$ varies linearly with temperature at a rate of 9.3 µV°C$^{-1}$, and $V_{os}$ typically varies in the range 1-10 µV°C$^{-1}$, (Peyton and Walsh, 1993) with 10 µV°C$^{-1}$ quoted by the manufacturers as typical for the LMC 6042 op-amp used in the MME. The errors introduced due to this source are negligible relative to the ion current, because after calibration the effect of $V_{os}$ included in the picoammeter mode output voltage only represents currents of order atoamperes.

The leakage term $i_L$, found directly using electrometer modes 3 and 4, is composed of the input bias current plus some leakage currents within the MME. The likely origin of the measured leakage currents can be assessed from considering which relays contribute to the leakage in each mode. In mode 4 (leakage current test), RL1, RL3, RL4 and RL5 are open. RL5 is guarded, and RL4 has no current path, so their contributions can be neglected and the leakage must be from RL1, RL3 or leakage across the screen to the input. In mode 3 (current measurement) a similar argument suggests that the leakage current is primarily from RL3 or the screen. This shows that the leakage currents in each mode are slightly different, because the nature of any self-monitoring instrument necessarily requires its configuration to change to measure signal and leakage. Though it appears valid to compensate for some of the errors by this method of switching between modes, it is ultimately unable to completely correct for all electrometer leakages. However, the results of laboratory tests shown in Figure 5.13 imply substantial compensation for the leakage current measured in MME mode 2 *is* relevant, since $i_L$ was strongly temperature dependent, with a variation of -9 fA°C$^{-1}$. At a bias voltage of 20 V, this would contribute an error of 0.4 fSm$^{-1}$°C$^{-1}$ to the conductivity measurement, which is a significant drift.

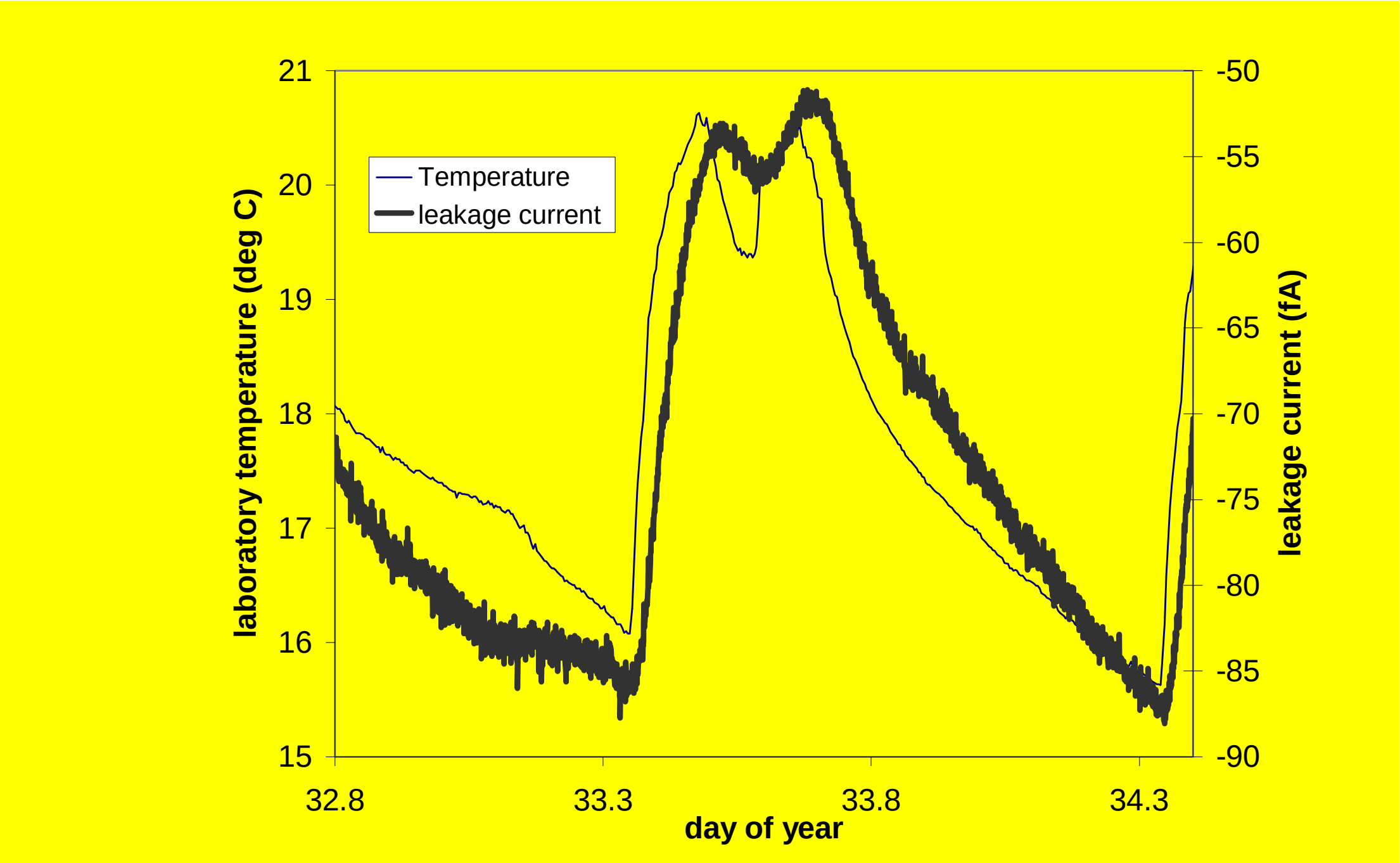

*Figure 5.13 Diurnal cycle of laboratory temperature measured with a platinum resistance thermometer (Harrison and Pedder, 2000) and MME leakage current, found using modes 3 and 4, which is plotted on the right hand axis.*

### 5.6.2 Compensating for leakage currents in the ion current measurement

To justify compensating the ion current measurements for the leakages measured in mode 2, it must be shown that the leakage currents are small with respect to the ion currents, and that they vary more slowly than they are sampled. The time series of $i/i_L$, which shows the relative magnitudes of leakage and signal, was constant with a well-defined mean up to timescales of about two hours, but there was a slow drift, due to the temperature dependence of $i_L$. Over a whole day, a linear fit of $i$ against $i_L$ had $R^2 = 0.41$. This suggests that neither $i$ nor $i_L$ is fluctuating fast enough to reduce the effectiveness of the compensation procedure. Short-term fluctuations in $i_L$ may introduce some uncertainty into the current measurement, because they are both 10 s (1 Hz) averages made with a 5 s pause between them. If $i_L$ is fluctuating rapidly, then subtracting its average value may not be an adequate compensation. $i_L$ was measured in the laboratory at 1Hz for half an hour and the mean 30 s standard deviation was calculated, which is the same time-scale as the compensation. It was 0.5 %, which is comparable to the error in the electrometer calibration.

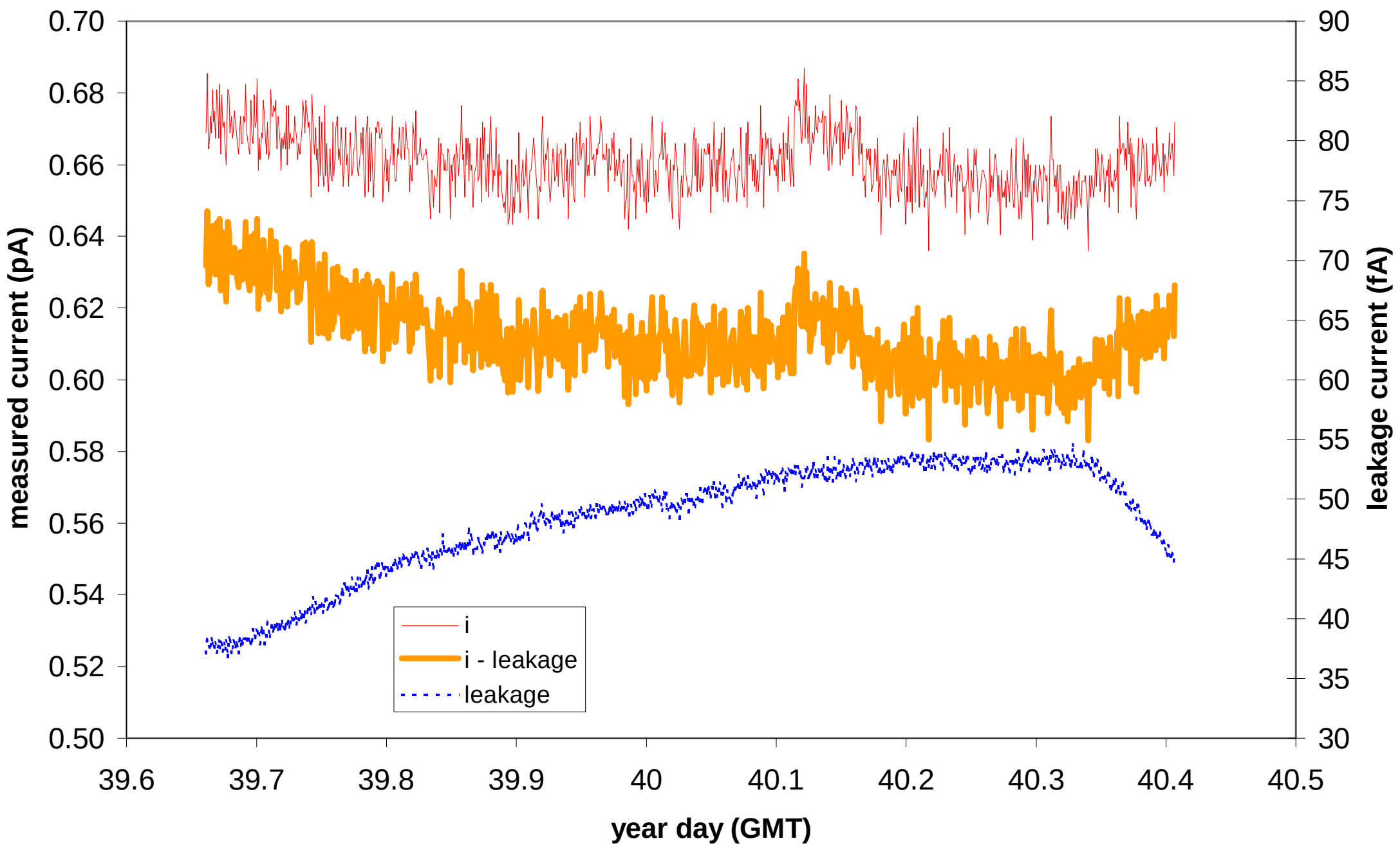

*Figure 5.14 Comparison of measured $i_m$ (thin trace), leakage $i_L$ (thick trace) and ($i_m$ – $i_L$) currents sampled every minute for 18 hours in the laboratory, in response to a resistively generated current of ~ 600 fA. $i_L$ is shown in the blue (dotted) trace on the right-hand axis.*

Figure 5.14 shows that the leakage current is at most 8 % of the total signal, and subtracting it from the measured current brings the mean closer to the reference signal. Here, the reference signal was generated using a 1 TΩ resistor, and is therefore also subject to temperature fluctuations, but since identical calibration and feedback resistors were used for all the current measurements, they could be expected to track together in the same thermal environment. (An absolute calibration technique has been described in Section 5.5.2). The electrometer leakage current can be seen as a source of slowly varying offset, shifting the average value of the signal. When it is compensated for, the average moves but the short-term fluctuations in the signal are diminished. In further data analysis, the $i_L$ contribution has been subtracted from the ion current measurements, as a correction to the mean.

### 5.6.3 Switching transients

In Section 3.1.4 it was shown that one reed relay injects typically 6 fC of charge into the system when closed. It is necessary to allow some time for the system to recover from each transient, so that it does not affect the measurements. However, the comparisons between measurements made in different MME modes, such as the subtraction of $V_{os}$ measured in mode 3 from the results in mode 4 to calculate electrometer leakage currents, necessitate making the measurements as immediately consecutive as possible. The optimal *recovery time* is the time it takes for the exponential decay from the transient to become indistinguishable from the real signal. An empirical investigation to find the optimal recovery time was carried out, by making measurements with the tube sealed but with the voltage switching on, and varying the time before starting to measure after the voltage was changed. This was determined by using the microcontrolled system to start logging with 2 Hz samples immediately after the voltage was switched, and also with a pause before the start of logging. The time series were then examined to see if the samples showed decay characteristics, by fitting a straight line to each set of voltage and time samples and calculating the gradient. If the mean gradient of all the voltage time series was negative, this was used as an indicator that the decay from the transient was continuing to dominate the signal. Recovery times from 0-30 s were tested, and 10 s was the shortest settling time after which the decay from the transient had finished. Hence, 10 s was the recovery time chosen for use in the control program between mode changes.

### 5.6.4 Electrometer control software

Dedicated microcontroller programs were developed for controlling and testing each electrometer mode separately. Source code is given in Appendix C, and Table 5.1 below summarises the function of each program.

| Program Name | Function |
|---|---|
| `ICAL` | Picoammeter (mode 4) calibrator with the femtoampere current source |
| `DECAY` | Charges up central electrode to 2.5 V and logs decay in follower mode (mode 1) at 2 Hz. |
| `IBLOG` | Alternates between modes 3 and 4 and logs them, with a header so they can be separated. |

*Table 5.1 Descriptions of the three programs used to test the MME modes independently.*

## 5.7 Integrated PIMS system

This section describes the integration of the separate components into a complete PIMS system. The capacitance of the Gerdien condenser is determined using the femtoampere reference (Harrison and Aplin, 2000a), and the overall control program is described. Three of these PIMS systems were constructed, and some results from them will be compared in Section 5.7.4.

### 5.7.1 Measurement of capacitance of Gerdien condenser

Since the femtoampere reference (Harrison and Aplin, 2000a) generates a current by passing a voltage ramp through a capacitor, using precision components, it is clear that, alternatively, if the current from a capacitor connected to the voltage ramp is measured accurately, its capacitance can be determined. This technique was employed to calculate the capacitance of the Gerdien condenser, which has classically not been known accurately, as explained in Chapter 2. In the miniature Gerdien condenser there is also a finite capacitance $C_S$ between the outer electrode and screen. This is in addition to the "exposed capacitance" (Arathoon, 1991), which is the effective capacitance of the Gerdien exposed to ions, and is required for conductivity calculation from Current Mode results (Eq. 2.17). Here the exposed capacitance is referred to as $C_G$, between the outer and central electrodes, which determines the ionic charge stored on the electrodes. These capacitances, which are unavoidably present in series, are schematically represented in Figure 5.15.

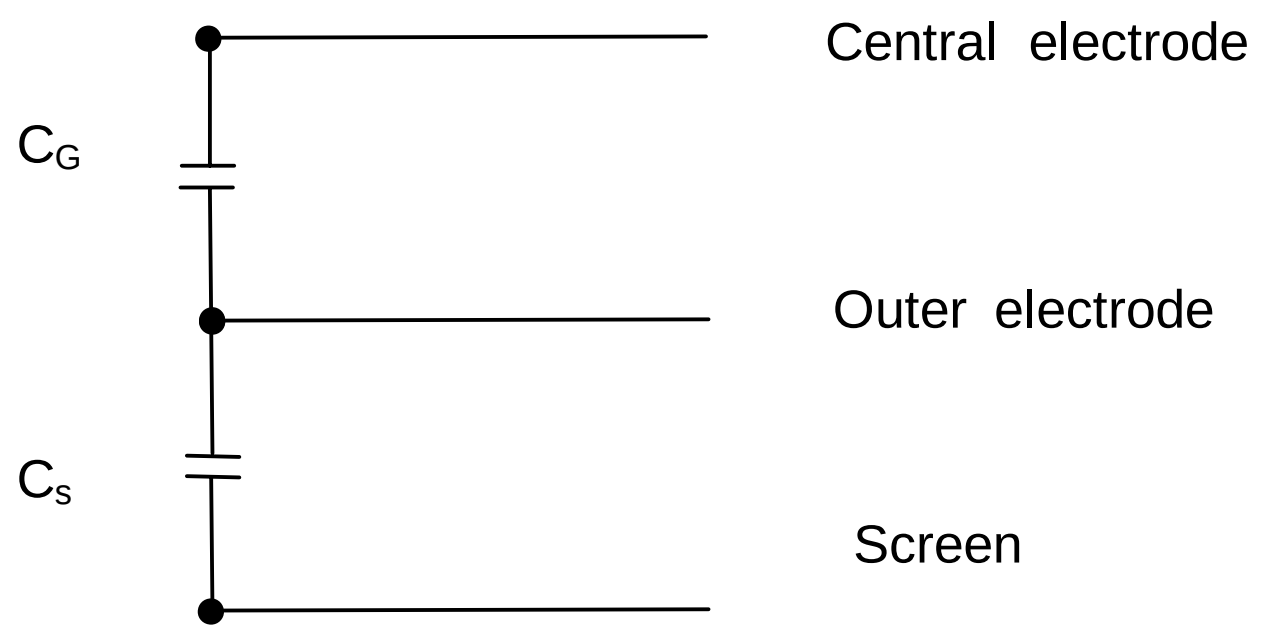

*Figure 5.15 Schematic of the capacitances between the Gerdien condenser electrodes and screen.*

The capacitance $C_S$ is clearly not associated with ion measurements made with the Gerdien condenser, since they only involve a flow of charge to or from the central electrode. It must therefore be discounted in any measurement of capacitance. This was achieved by shorting the outer electrode to the screen, as shown below in Figure 5.16, so $C_S$ cannot contribute to the current measured.

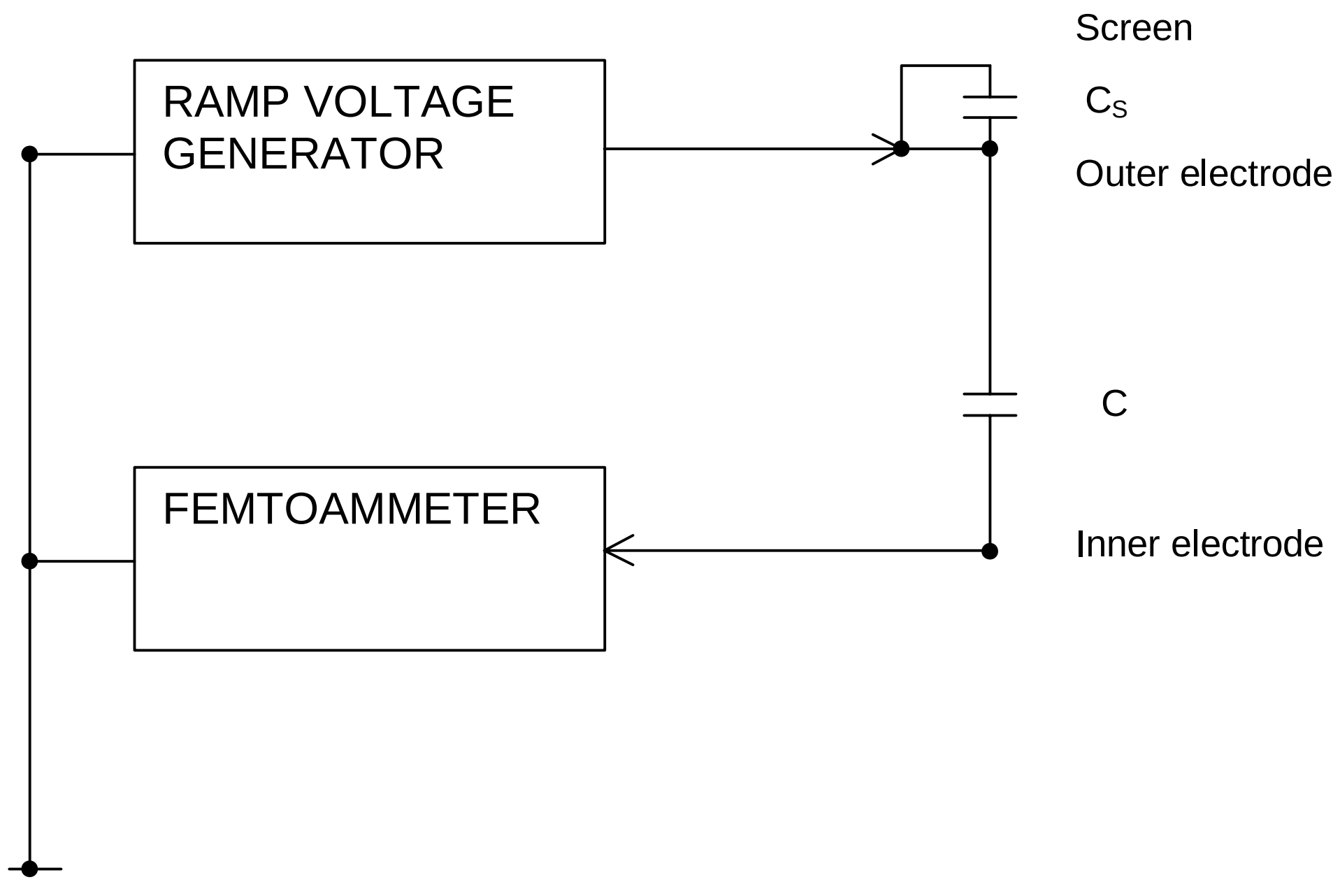

*Figure 5.16 Schematic of the circuit used to measure the Gerdien capacitance.*

The mean magnitude of current, measured with the Keithley 6512 electrometer, was 476 ± 1.5 fA (standard error of 6 cycles, 12 current measurements). Hence, $C_G$ = 9.07 pF, and the standard error of the current above is 0.3 %. Stray capacitance, for example from the connections probably dominates the errors in the capacitance determination; therefore the error in the Gerdien condenser capacitance is estimated as 1 %, 0.09 pF for this Gerdien. This approach has reduced the uncertainty in the current method of conductivity measurement due to the capacitance term by an order of magnitude, compared to other capacitance determination methods, *e.g.* Aplin and Harrison (1999), Smith (1953).

### 5.7.2 Complete control program

The individual control programs for calibration of individual components described above (Section 5.6.4) were merged with the ADC control program described in Chapter 4 to control the whole system. As a variety of measurements were being made, a more sophisticated logging strategy was needed. The sampling rate is a compromise between taking enough samples to average and smooth high-frequency fluctuations, and rapid sampling to improve the voltage decay and current measurement intercomparisons. In the voltage decay mode, the ion current flowing is inferred from the rate of change of potential. Hence, time averaging is not relevant in this case, and the samples should be temporally close to maintain consistency of atmospheric conditions during the sampling period. Therefore, ten measurements of the voltage across the tube were made at 2 Hz.

The sampling characteristics for the other three MME modes were chosen to minimise the standard deviation of the raw measurements, before averaging. This was tested by calculating the variability of measurements at different frequencies, and for varying periods, from a sealed Gerdien in the laboratory at zero bias voltage. Under these conditions, the only variability in the ion concentration should result from background ionisation, therefore the measurements made can be considered equivalent to a background noise level. The minimum variability in the background noise level was found when ten measurements were made at 2 Hz, and these settings were empirically selected for application in the final control program.

The final control program was called *MMESYS**[49], where * represents the PIMS number it was written for[50]. Calibrations of the components differed slightly, and in order to ensure the same ionic mobilities were measured from instrument to instrument, slightly different variables (and timings, discussed in Section 5.7.3) in the programs were required. As well as controlling the different mode switching, MMESYS averages (except in the follower mode) and logs data and sends it serially to the PC. This format was chosen to ease the compensation procedure and separation of results from the Current Measurement and Voltage Decay modes, when the text files are subsequently calibrated using *MS Excel*. A sample set of lines from the PC data file is shown below, and is explained in Table 5.2.

YearDay, Time, I, DACcode1, $(i_b+V_{os})$, $V_{os}$, i

YearDay, Time, V, DACcode1, $V_1 \ldots V_n$

YearDay, Time, I, DACcode2, $(i_b+V_{os})$, $V_{os}$, i

YearDay, Time, V, DACcode2, $V_1 \ldots V_n$

…

YearDay, Time, I, DACcodeN, $(i_b+V_{os})$, $V_{os}$, i

YearDay, Time, V, DACcodeN, $V_1 \ldots V_n$

YearDay, Time, I, DACcode1, $(i_b+V_{os})$, $V_{os}$, i

YearDay, Time, V, DACcode1, $V_1 \ldots V_n$

[49] Source code is in Appendix C.

[50] Use of the name "*MMESYS*" indicates general properties of the program are under discussion. *MMESYS** is used to indicated the aspects in which the individual PIMS programs were different.

| Quantity | Description |
|---|---|
| `Yearday` | Day of year, from the PC clock (0-366) |
| `Time` | Time from the PC clock, hhmmss (0-235959) |
| `I or V` | V denotes voltage decay mode, I denotes current measurement/ error check mode |
| `DACcode` | DAC code to represent bias voltage, where there are N in total ( where N ≤ ~15, limited by the microcontroller memory) |
| $i_b R_f + V_{os}$ | Output in ADC counts, 10s average (0-4095) |
| $V_{os}$ | Output in ADC counts, 10 s average (0-4095) |
| `i` | Output in ADC counts (picoammeter mode), 10 s average (0-4095) |
| $V_1 \ldots V_n$ | n voltage decay measurements, where n = 10 in MMESYS |

*Table 5.2 Format of the serial output logged by the PC.*

### 5.7.3 Timing and Synchronisation

The time of the measurement is derived from the PC clock, which is frequently set manually on the laptops used, at the beginning of each line of logged data. The electrometer leakages and offsets are subsequently subtracted from the current measurement[51], and therefore the three 10 s averages are effectively assumed to be simultaneous with an observation. Section 5.6.2 explains that the ratio of the leakage current to the measured current remains constant for timescales much greater than this.

In the PBASIC microcontroller programming language, timing is controlled by the `PAUSE` ***`milliseconds`*** command, where ***`milliseconds`*** is the pause time expressed in milliseconds ± 1 %. Parallax (1998) explain that the microcontroller BASIC interpreter takes at least 1 ms longer than the `PAUSE` instruction to carry out programming loops including this command. Since the PIMS control programs typically include several such loops, errors are expected in the timing, which were observed when using the microcontrolled picoammeter with the voltage ramp derived current source. This has a precisely-timed ramp time of 32 s (Harrison and Aplin, 2000a), and when trying to make the microcontroller program match this cycle, there was a 15 % error. Processing the voltage decay data involves least squares fitting to voltage as a function of time, and therefore the internal timing of these measurements does need to be known accurately. The voltage decay sampling frequency was nominally 2 Hz, but was measured to be a constant 1.98 Hz with a digital stop clock, on which the timings for data processing were based. (The error in the microcontroller's internal clock was insignificant on the 5 s timescale). The inconsistency between different devices was also investigated.

Three PIMS systems were constructed, and inconsistencies between the microcontrollers' internal clocks meant that the measurements were asynchronous, even when the same program was running on them all. Test programs indicated that the internal timers were tuned slightly differently, so that the PAUSE command produced a pause of a length which was a constant factor different from the same command executed on a different microcontroller. The recovery time between switching the bias voltage was therefore lengthened slightly by an empirically determined factor in the PAUSE statement in *MMESYS**, so that all the microcontrollers had the same program time period. Operating the three microcontrollers from one switchable power supply allowed all the programs to be started simultaneously.

### 5.7.4 Testing the integrated PIMS systems

An important test of the integrated Gerdien system is to compare results from the two ion measurement modes of the MME. The complete system was run at the Reading Meteorology Field Site on 22nd March 2000. For running in atmospheric air, the PIMSs were housed in watertight plastic boxes (28 x 18 x 8 cm), as shown in Figure 5.17. The Gerdien inlets protruded, and the microcontroller/ADC, MME and DAC/power supply circuitry were in three small metal boxes, wired inside the plastic box. There were two Buccaneer connectors on the plastic box, one two-way for power from a 12 V lead-acid battery, and one six-way carrying serial and parallel connections to a laptop PC for logging and programming respectively.

---

[51] Data processing will be discussed in detail in Chapter 6.

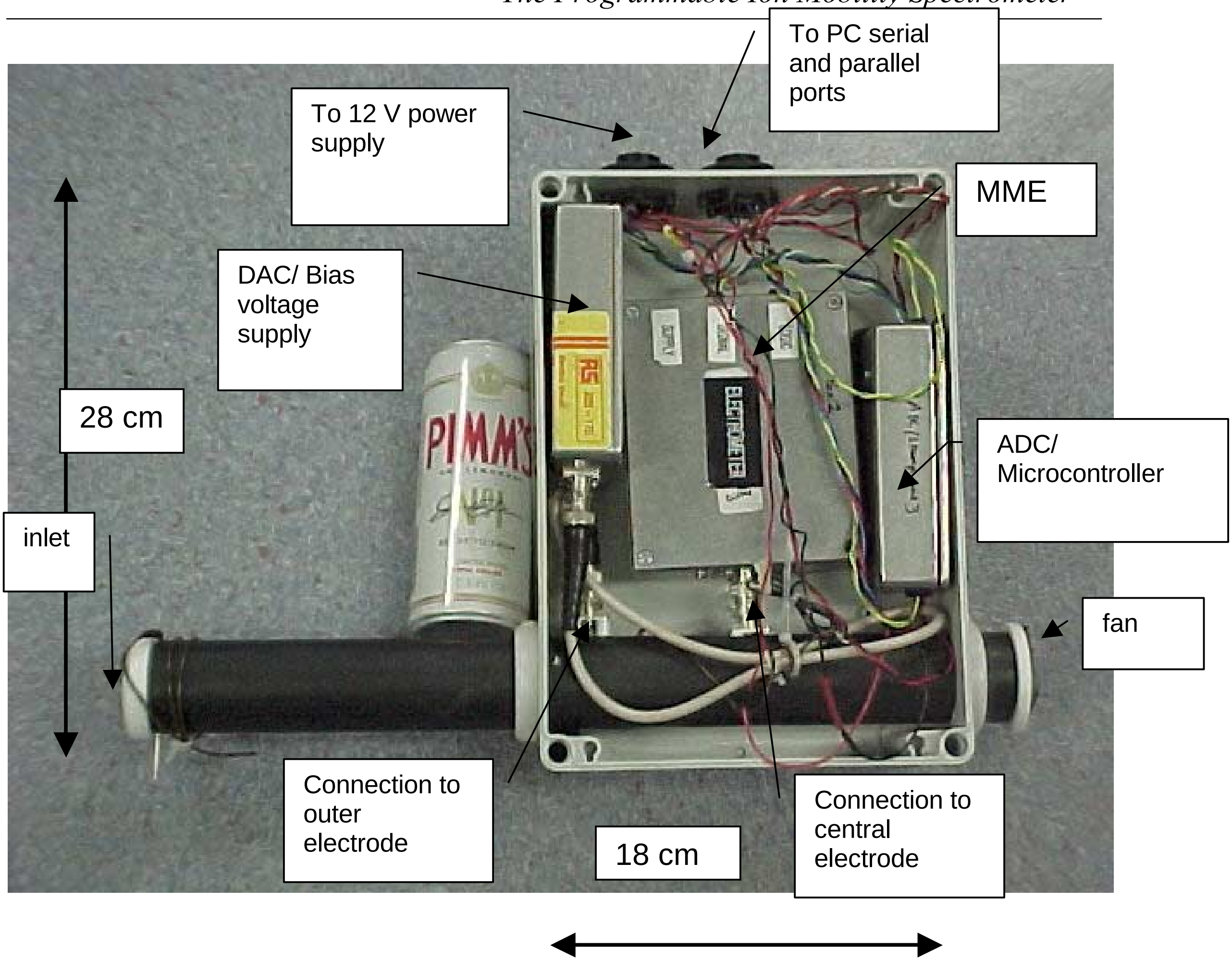

*Figure 5.17 Photograph of the PIMS ready for atmospheric deployment (without lid on box).*

First, PIMS no. 3 was run for approximately two hours measuring conductivity in both current and voltage modes. After PIMS no. 3 was disconnected, PIMS no. 2, a completely different system, was run for a short time at the same location. Meteorological conditions remained effectively constant throughout the test period. Conductivity was calculated from the standard expressions Eq. 2.17 and Eq. 2.4 with $\tau$ calculated from a *Microsoft Excel* exponential fit to the voltage decays[52]. The value of current used in Eq. 2.17 was ($i$-$i_L$).

[52] The details of data processing will be discussed in Chapter 6.

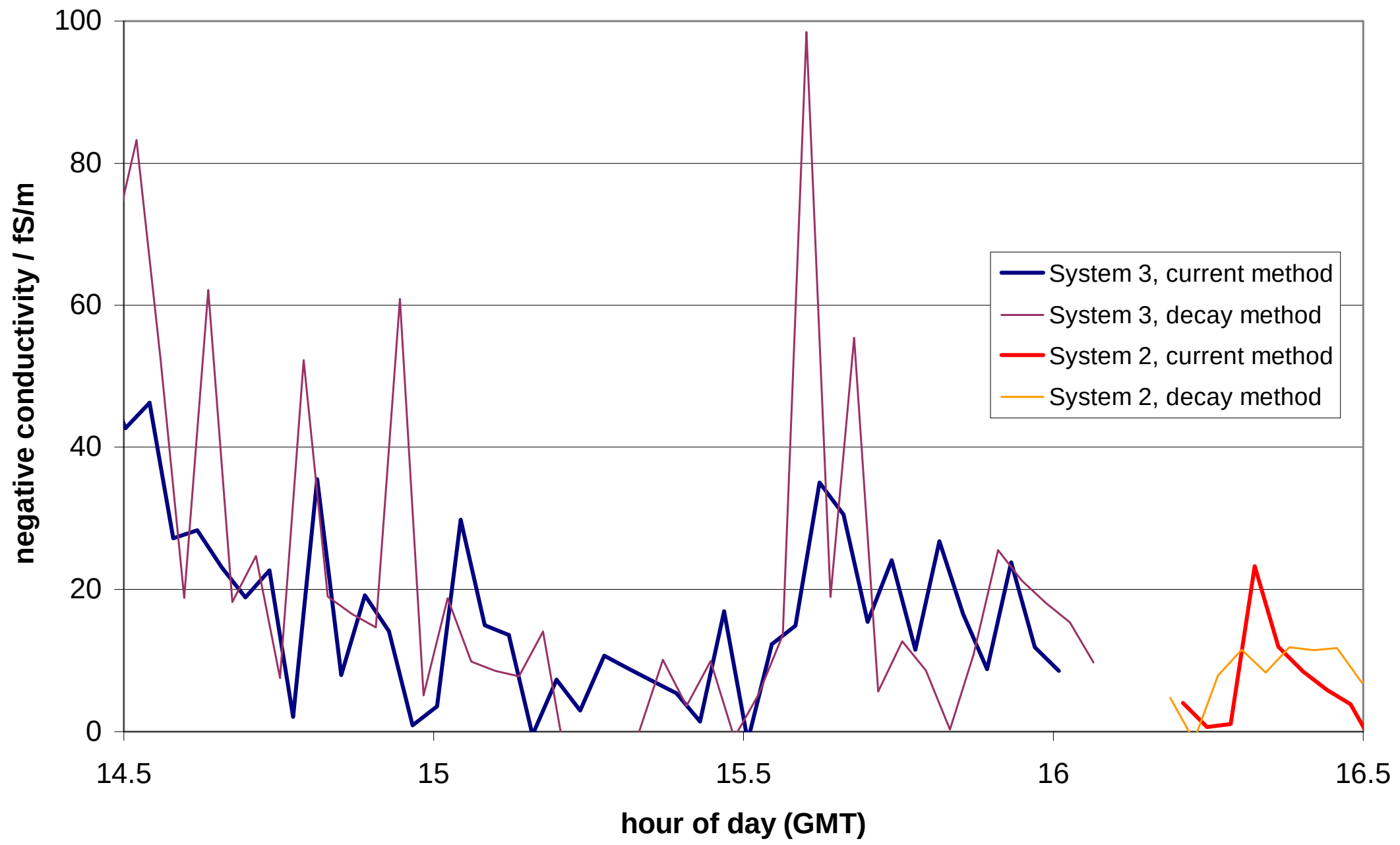

*Figure 5.18 Comparison of negative conductivity calculated with results from both methods, and with PIMS 2 and 3 at Reading on 22nd March 2000.*

There is a general agreement between the conductivity calculated by the two methods, although the absolute magnitudes do vary. PIMS 2 shows comparable mean values. Further analysis is required to explain the disparities in the measurements, but it is clear that the different instruments and techniques show comparable behaviour. All the major peaks are detected by both measurement modes, and the mean value is about 12 $fSm^{-1}$, which conforms to typical fair weather measurements in urban locations (*e.g.* Chalmers, 1967; previous measurements are discussed in Chapter 2).

## 5.8 Conclusions

A programmable ion mobility spectrometer has been designed and tested in this chapter, incorporating a new multimode electrometer and improved Gerdien design. The errors in previous conductivity measurements will have been significantly reduced by determination of the Gerdien capacitance to 1 %, and compensation for temperature-dependent voltage offsets and some leakages. A further innovative feature is the ability to check self-consistency from both current and voltage measurements, results have agreed from two separate PIMS instruments, and such self-calibration of the device appears feasible.

*"... when all conventional physics failed, some property of aerosol could always be invoked to get out of a difficulty, but whether these were cosmetic, or physical in origin is difficult to judge" John Green*[53]

# 6 PIMS atmospheric calibration and testing

In this chapter, the testing of the PIMS instrument in the atmosphere is described in detail. The deployment of three individual PIMS, each with two ion measurement modes, facilitates self-calibration, and these results are discussed in sections 6.4 and 6.5. The PIMS results are also compared to measurements from two independent external instruments, a Geiger counter, which responds to ionising radiation, (Section 6.3) and another Gerdien ion counter of more classical design (section 6.6.1). This chapter also necessarily contains, as a precursor to the results, details of the experimental configuration and data processing, which are outlined in Sections 6.1 and 6.2.

The structure of the calibration and comparison procedures is explained graphically in Figure 6.1. Three PIMS instruments were constructed, and the simplest test is of internal consistency between the two modes on each instrument. The results from each mode can then be compared to other PIMS measurements. The most rigorous calibration procedure is to compare the PIMS results to other instruments measuring ions, a Geiger counter and a Gerdien condenser.

[53] Green J. (1999), *Atmospheric Dynamics*, CUP

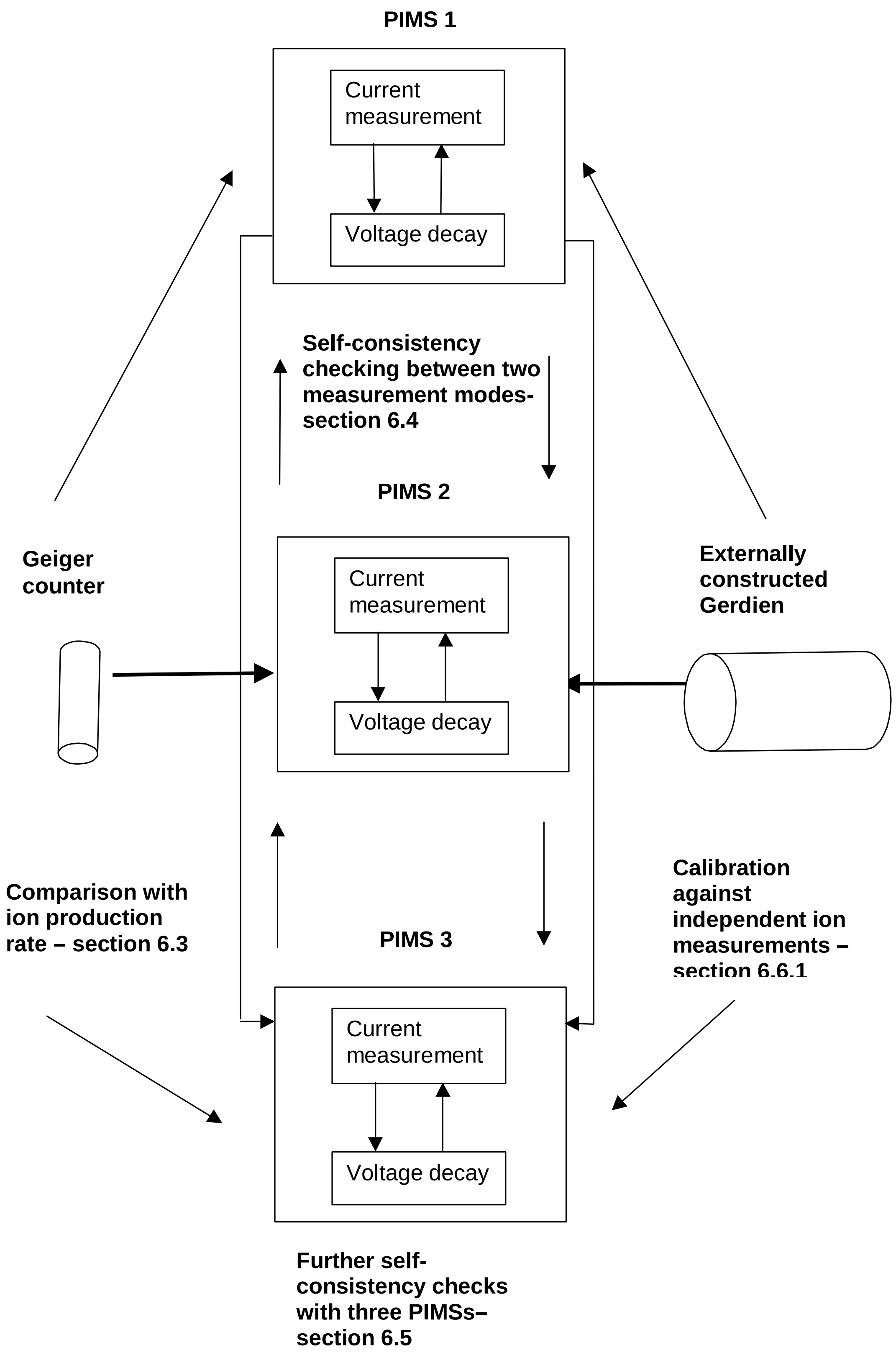

*Figure 6.1 Block diagram showing the approaches to PIMS consistency.*

## 6.1 Description of PIMS atmospheric deployment

Measurements were made at Reading University Meteorology Field Site with the three PIMS instruments mounted next to each other, with the inlets facing south, on a mast at 1.25 m. A photograph of the mast is shown in Figure 6.2. This was the same height the Geiger counter had been at for previous long-term monitoring of local radioactivity at the Field Site. For effective comparison with the Geiger count rate, the PIMS needed to be mounted at the same height. The orientation was chosen so that the average wind velocity would be as close as possible to the wind tunnel test speed of 2 $ms^{-1}$, for which the bias voltage range was selected. (Mean wind speed at the site for 1999 was 2.2 $ms^{-1}$, and direction 197º, so the southerly component is 2.1 $ms^{-1}$, which matches the fan's ventilation rate almost exactly). The inlets were arranged as close as possible, but the plastic PIMS housing kept the minimum separation to about 15 cm; this can be seen in Figure 6.3.

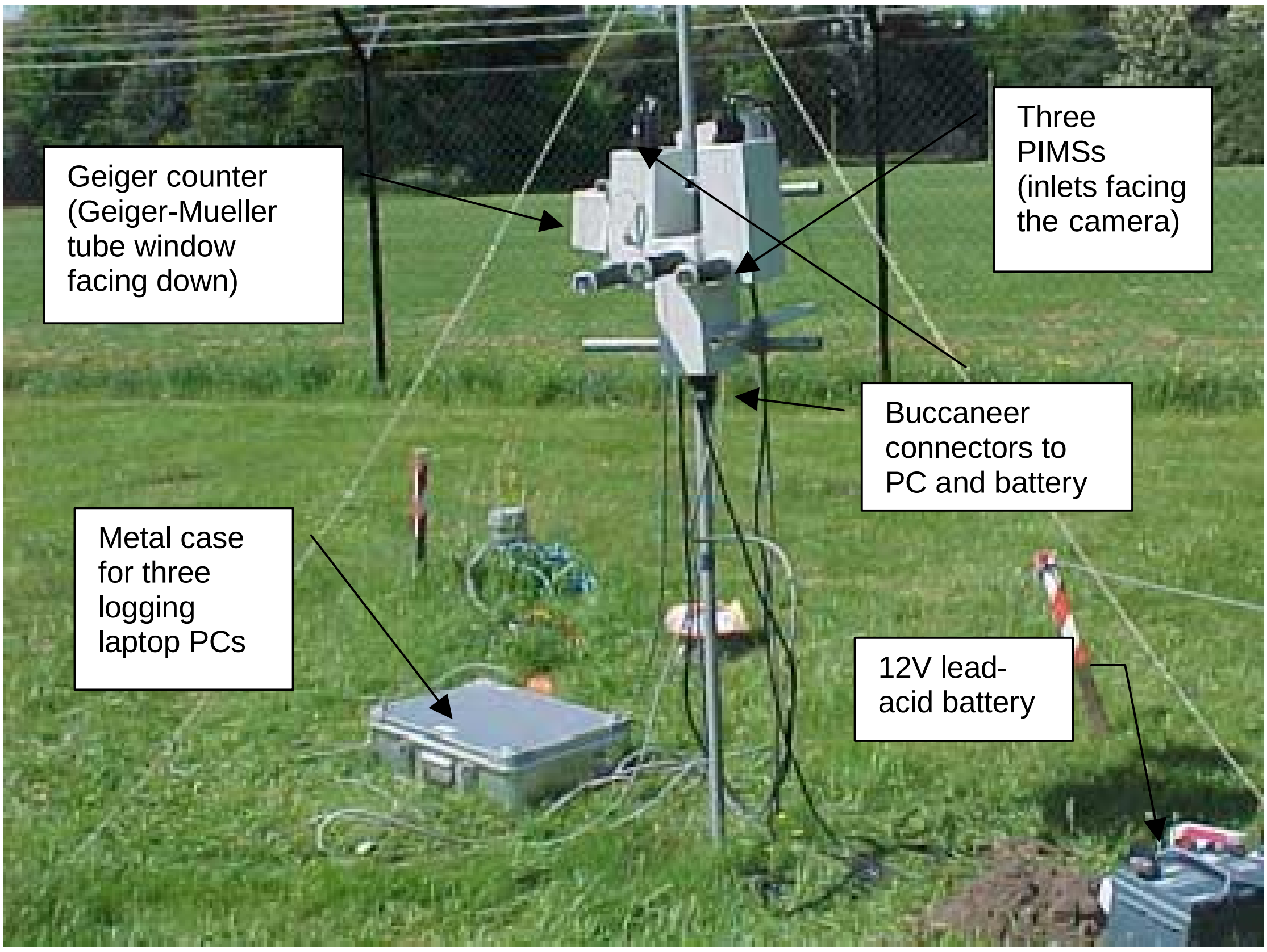

*Figure 6.2 Photograph of the PIMSs running at Reading University Meteorology Field Site.*

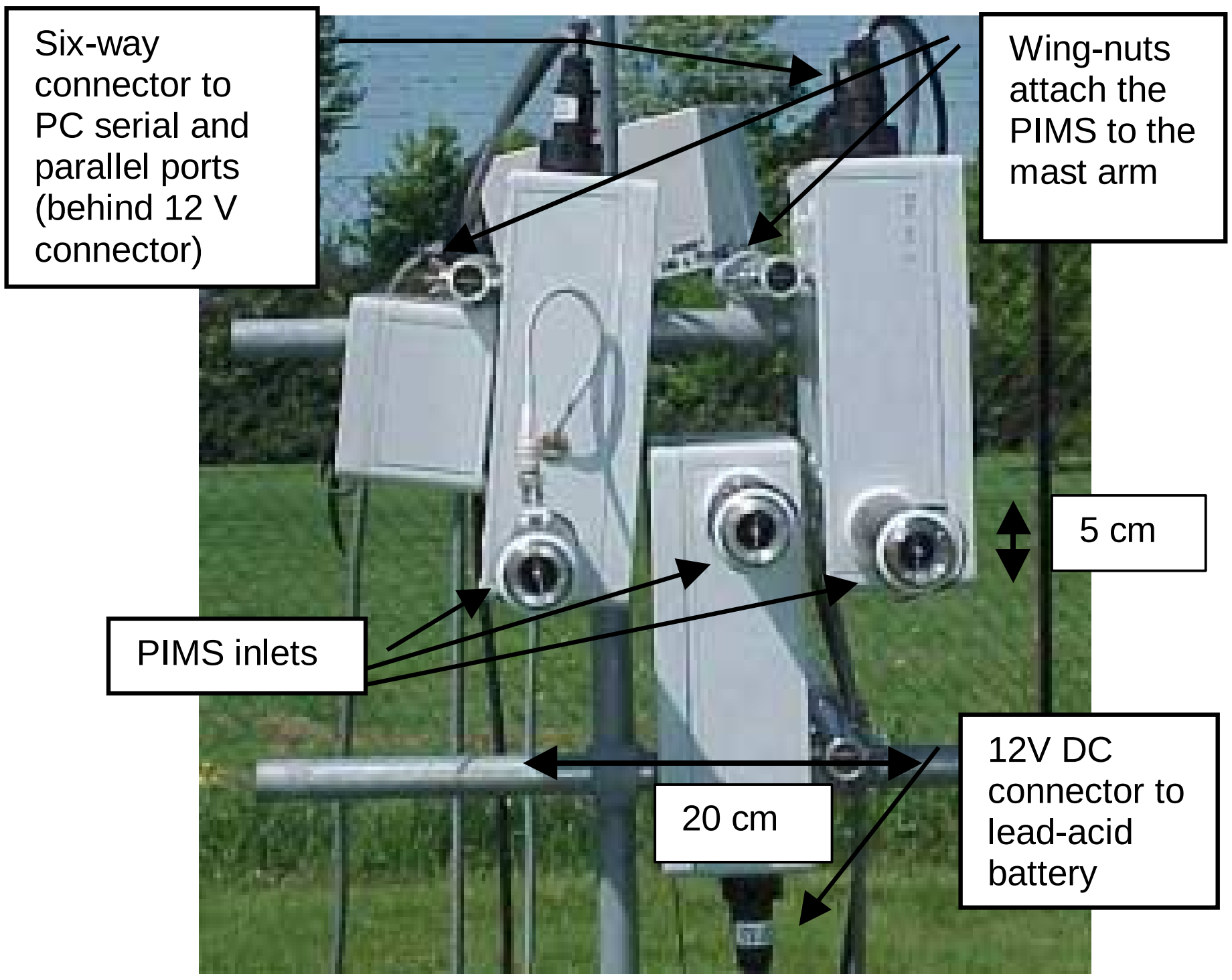

*Figure 6.3 Close-up photograph, taken from the south, showing the PIMS inlets and connections.*

## 6.2 Calibration of raw data

Data is stored in a text file by the logging PC as integer values (ADC counts) from 0 to 4095, representing an output range of ± 2.5 V. Voltage calibration of the ADC converter was described in section 4.2.1; further calibration is then required, depending on the MME mode, to convert the voltage into the correct magnitude of ion current or electrode voltage. The details of this procedure for the two ion measurement modes, and the calculations performed to obtain conductivity values from the measurements, are described below.

### 6.2.1 Current Measurement method

In this mode the measured current $i_{meas}$ is calculated by calibrating the ADC output, as shown in the schematic diagram (Figure 6.4). A voltage offset of 2.5 V has to be added to the bipolar current measurement because the ADC is only unipolar, (see section 4.2.1), therefore 2.5 V must be subtracted from the voltage obtained from the ADC counts before the MME picoammeter calibration can be applied. The electrometer leakage current $i_L$ is calculated separately from the results of the two error checking modes, and is subtracted from $i_{meas}$ to give the compensated current, $i_c$ which is used to calculate conductivity in the conventional manner,

$$\sigma = \frac{\varepsilon_o i_c}{CV} \qquad \textit{Eq. 6.1}$$

for bias voltages $V$ set by the DAC. In Figure 6.4, the entire calibration and compensation procedure is summarised. It begins with the current and error check mode measurements in the integer ADC format stored in the PC text file, and follows the process through to the corrected current used to calculate conductivity $i_c$.

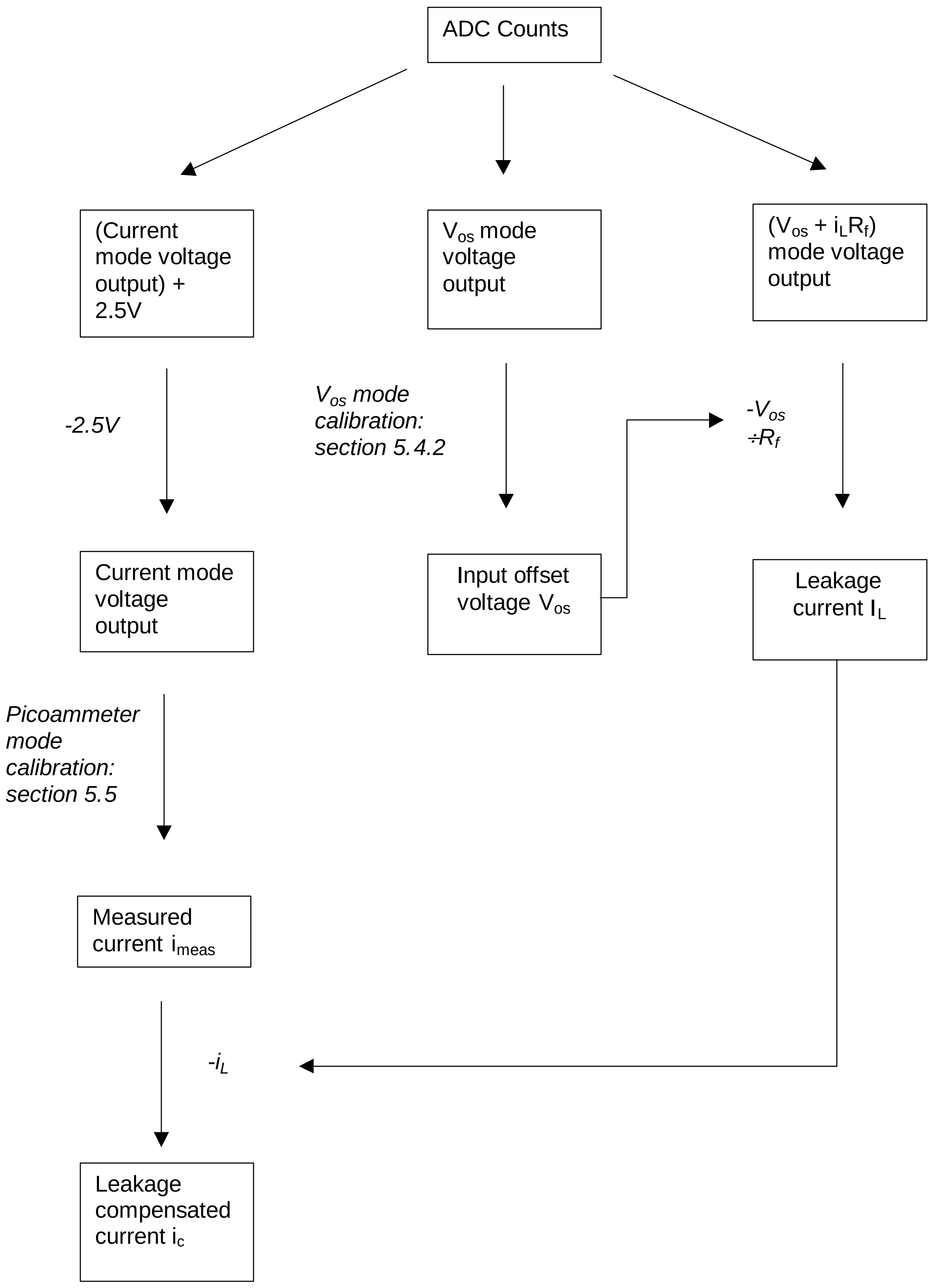

*Figure 6.4 Schematic showing the calibration and correction procedures required to obtain a compensated current measurement. The three modes operate sequentially, for 10 s each. This procedure is repeated for measurements at each bias voltage.*

### 6.2.2 Voltage Decay method

In this mode, repeated electrode voltage measurements are required. Figure 6.5 below outlines the procedure used to calculate the voltage between the Gerdien condenser's central and outer electrodes. Since there is a bias on both electrodes, the potential difference the ions are subject to in the Gerdien is determined by the sum of the bias voltage $V_b$ (potential difference between outer electrode and ground) and $V_f$, the voltage measured by the MME at the central electrode with respect to ground. This will be referred to here as $V_{decay}$,, where $V_f$ is a function of time.

$$V_{decay}(t) = V_b + V_f(t) \qquad \textit{Eq. 6.2.}$$

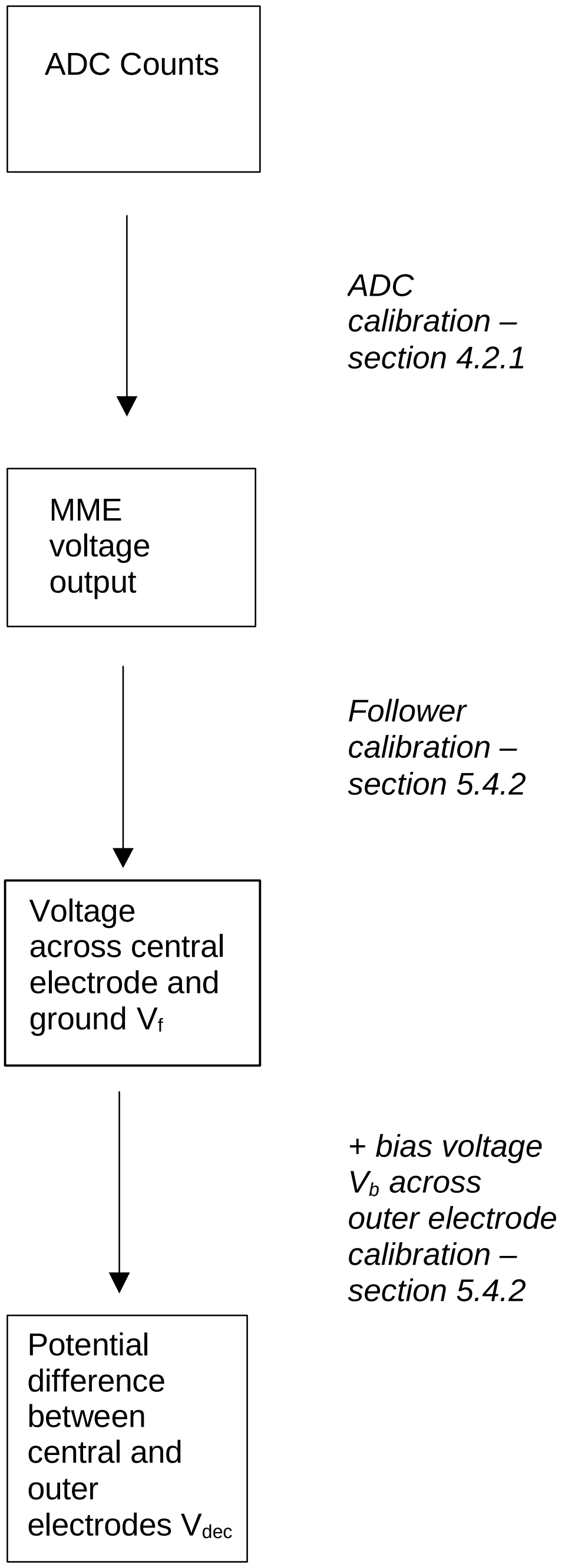

*Figure 6.5 Schematic diagram showing the calibrations used to convert ADC counts stored by the logging PC, to the total potential difference across the outer and central electrodes.*

As explained in Section 2.2.2, it is generally assumed that in the Voltage Decay mode, the potential across the Gerdien electrodes $V_{decay}$ will decay exponentially in a time $t$ such that

$$V_{decay}(t) = V_0 \exp\left(\frac{-t}{\tau}\right)$$ *Eq. 6.3.*

Taking logs, this can be rearranged to give

$$\ln V_{decay} = \frac{-t}{\tau} + \ln V_0$$ *Eq. 6.4,*

and recalling that

$$\sigma = \frac{\varepsilon_o}{\tau}$$ *Eq. 6.5,*

it is clear that the gradient of a graph of ln $V_{decay}$ against time is equivalent to $-\sigma/\varepsilon_0$. This approach was applied to calculate conductivity from the time series of $V_{decay}$. The time of each voltage reading with respect to the start of the measurement (when the reed relay which activates the charging of the central electrode is switched off) is approximately determined using the microcontroller program, and was also calibrated using a digital stopwatch (see section 5.7.3). Since only negative ions can be measured at the central electrode during the active decay from –32.5 V to –30V (see section 5.4.3), all results here refer to negative ion measurements. An *MS Excel* least squares linear fitting worksheet function, *LINEST*, was then applied to the natural logarithm of the magnitude of each voltage reading, with time as the abscissa, starting from a first sample at $t$ = 0.505 s.

*LINEST* was also used to calculate the standard error in fitting the gradient to the points, which can be used to calculate an error in $\sigma$ resulting directly from the least squares regression procedure. The *coefficient of determination*[54] $R^2$, was also determined using the same function. This is a "goodness of fit" parameter from 0 to 1. If $R^2 = 1$, there is no difference between the y-values predicted by the theory and the actual y-value, and decreasing $R^2$ indicates increasing inappropriateness of fitting the regression equation to the data (Johnson and Bhattacharrya, 1996). The fitting procedure described above resulted in a mean $R^2 = 0.65$, which was similar for all the data sets. To test whether this fraction of exponential decays could occur by chance, a Monte Carlo simulation was run to generate 1000 random voltage decay traces. This simulates the input that would be generated by noise. The same exponential curve fitting procedure was applied to the randomly generated decays. Only 0.15 % of the random decay data had $R^2 > 0.65$, which strongly suggests that only a tiny fraction of the exponential decays seen here could occur by chance.

For exponential fitting to voltage decay data, the coefficient of determination is a useful quantity, because, as suggested in Section 2.2.2, an exponential decay may not always be an appropriate method to calculate conductivity. The critical mobility (minimum mobility) of ions contributing to the measurement decreases as the voltage decays, and therefore the number of ions actually being measured varies with decay voltage. This is equivalent to the charge on a capacitor decaying across a variable resistor; thus the theory outlined in Section 2.2.2 leading to Eq. 6.4, may not apply in some cases. The effectiveness of assuming an exponential decay depends on the characteristics of the ion mobility spectrum, which provides information about the number of ions having each mobility[55]. For these reasons, the standard error of the exponential fit is *not* equivalent to the error in the measurement. If the critical mobility were invariant with voltage, there should always be an ideal exponential decay and in this limit, the standard error of the fit would be the true error. The standard error of the exponential fit may therefore be considered as a worst case estimate of the measurement error.

The raw data does not always follow an approximately exponential decay, in some cases the voltage across the electrodes increases, and over the 5 s of sampling the voltage can fluctuate considerably. A sample set of voltage decay time series, measured over one hour from 12:25 to 13:25 on 27 June 2000 is shown below in Figure 6.6. The decay characteristics show substantial variation over this relatively short time, which is common to all the voltage decay measurements. The best and worst exponential fit lines in the hour of measurements are also shown, and span $R^2 = 0.88$ to 0.19. If the voltage across the electrodes increases anomalously, the gradient will be a positive number and the calculated conductivity will be negative. It is consequently inferred that the electrodes are charging up, and the data is rejected. In Figure 6.6, some data sets show both decay and charging characteristics, for example at 12:41 there is an exponential decay for the first three seconds which then changes direction. If an exponential is fitted to the part of the data which is showing decay characteristics, then it has an $R^2$ value of 0.86, but a fit to the whole time series has $R^2 = 0.50$. This will also affect the magnitude of the calculated conductivity, which increases as the gradient of the exponential curve decreases.

[54] MS Excel defines $R^2$ as the square of the Pearson product moment correlation coefficient, $r$, "a dimensionless index that ranges from -1.0 to 1.0 inclusive and reflects the extent of a linear relationship between two data sets". $r = \dfrac{n\left(\sum XY\right) - \left(\sum X\right)\left(\sum Y\right)}{\sqrt{\left[n\sum X^2 - \left(\sum X^2\right)\right]\left[n\sum Y^2 - \left(\sum Y\right)^2\right]}}$

[55] Effects of the ion spectrum on voltage decay data will be discussed in more detail in Chapter 8.

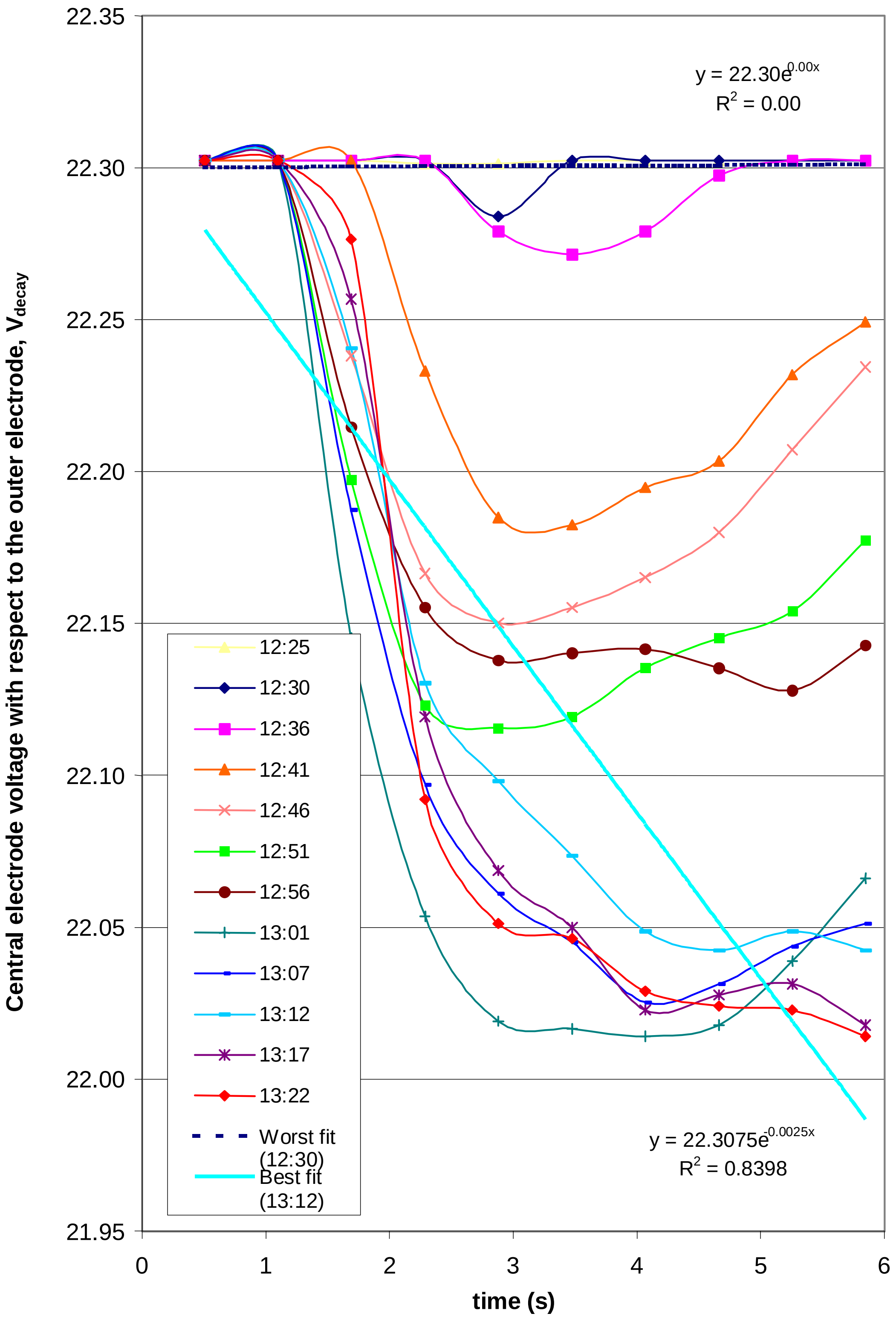

*Figure 6.6 One hour of voltage decay time series measurements, measured with PIMS 2, 27 June 2000. The times in the legend are the time of the measurement (to the nearest minute), and time on the x-axis is seconds from the opening of the relay which charges the central electrode to 2.5 V. $V_b$ = 19.848 V and $V_0$ = 22.35 V.*

There are clearly grounds to reject the data where the voltage increases with time rather than decaying, on the basis that some other physical process rather than unipolar ionic conduction is occurring. There are several possible approaches to filtering the data, for example fitting exponential decays to all sections of the data where the voltage is decreasing with time, and discarding the rest. With this method, sometimes only very few points would be involved in the curve fit, and the number of points would vary. The method of calculation of conductivity would become less rigorous with fewer points in the regression, and individual values would be of varying reliability due to the changing number of points in the fit. The most conservative filtering method would be only to accept data where the voltage decreases throughout the complete time series (*i.e.* every voltage is less than the previous one); however in a 12-hour sample of voltage decay data from 27 June 2000, this would reject 52 % of all the voltage decays. For these reasons, exponential fitting to the complete 5 s of data was instead chosen because it offers an average fit based on results from the whole sampling time. If the average effect is a charging process, *i.e.* $<\Delta V/\Delta t> \; > 0$, the calculated conductivity will be negative, and the data is rejected. In the selected data set, only 1 % of the voltage decay measurements were rejected using this criterion, therefore this method optimises the available data but rejects cases when the central electrode voltage at the end of the sample exceeds its starting value.

## 6.3 Comparisons of PIMS data with ion production rate

In Chapter 3, ion concentrations measured in clean air correlated with the square root of the count rate from an adjacent Geiger counter, which is a proxy for the ion production rate $q$, depending on the aerosol concentration. This relationship is derived from ion-aerosol theory (*e.g.* Clement and Harrison, 1992), explained in Section 2.2.3. In clean air, the ion concentration $n$ varies as $q^{0.5}$,

$$n = \sqrt{\frac{q}{\alpha}} \qquad \textit{Eq. 6.6}$$

and for very polluted air when the attachment of ions to aerosol dominates, and the self-recombination term can be neglected so that $n$ is directly proportional to $q$:

$$n = \frac{q}{\beta Z}$$ *Eq. 6.7.*

Between these two limits, both aerosol attachment and ion recombination terms control ion concentrations. A plot of *n* against *q* should therefore fit a power law with exponent between 0.5 and 1. The ion concentration and Geiger count rate at Reading University Meteorology Field Site were plotted for two fair weather days, 25th February 2000 and 9th April 2000. This is an urban location, with local PM10 mass concentrations of 19 and 18 $\mu gm^{-3}$ for the two periods respectively (National Air Quality Archive (NAQA), 2000[56]). Since typical PM10 concentrations at Mace Head, where the clean air measurements were made in the recombination limit are about 1 $\mu gm^{-3}$, the Reading is data not expected to be represented by Eq. 6.6.

For both of the days studied, the relationship predicted by ion-aerosol theory did not hold, with the ion concentration decreasing slightly as the local radioactivity increased. Fitting a power law to the data did not appear appropriate, since the $R^2$ values were very poor (0.04 and 0.009 respectively). Since this effect was repeatable, and the two days showed similar ion-aerosol characteristics (shown in Table 6.1), this suggests that application of the classical theory is not straightforward for urban air, which was also found by Guo *et al* (1996).

| **Date** | **<$\sigma$> ($fSm^{-1}$)** | **<PM10> ($\mu gm^{-3}$)** | **<$SO_2$>(ppb)** | **<Geiger> (Bq)** |
|---|---|---|---|---|
| 25 Feb 2000 | 12 | 19 | 2.4 | 0.48 |
| 8 April 2000 | 13 | 18 | 2.6 | 0.42 |

*Table 6.1 Comparison of ion and aerosol characteristics for the two days studied. Conductivity and ion production rate were measured at Reading University Meteorology Field Site, and PM10 and $SO_2$ were measured ~ 3 km away (NAQA, 2000)*

Ion-aerosol theory assumes that the aerosol population is monodisperse, and that the ion processes are in equilibrium so that Eq. 2.19, the ion balance equation, is always equal to zero[57]. However, even a very short 1 Hz Geiger time series, such as Figure 6.7, shows that $q$ is extremely variable, and the classical average of 10 ion pairs cm$^{-3}$s$^{-1}$ (Chalmers, 1967) is presumably a very long term average.

[56] http://www.aeat.co.uk/netcen/airqual/

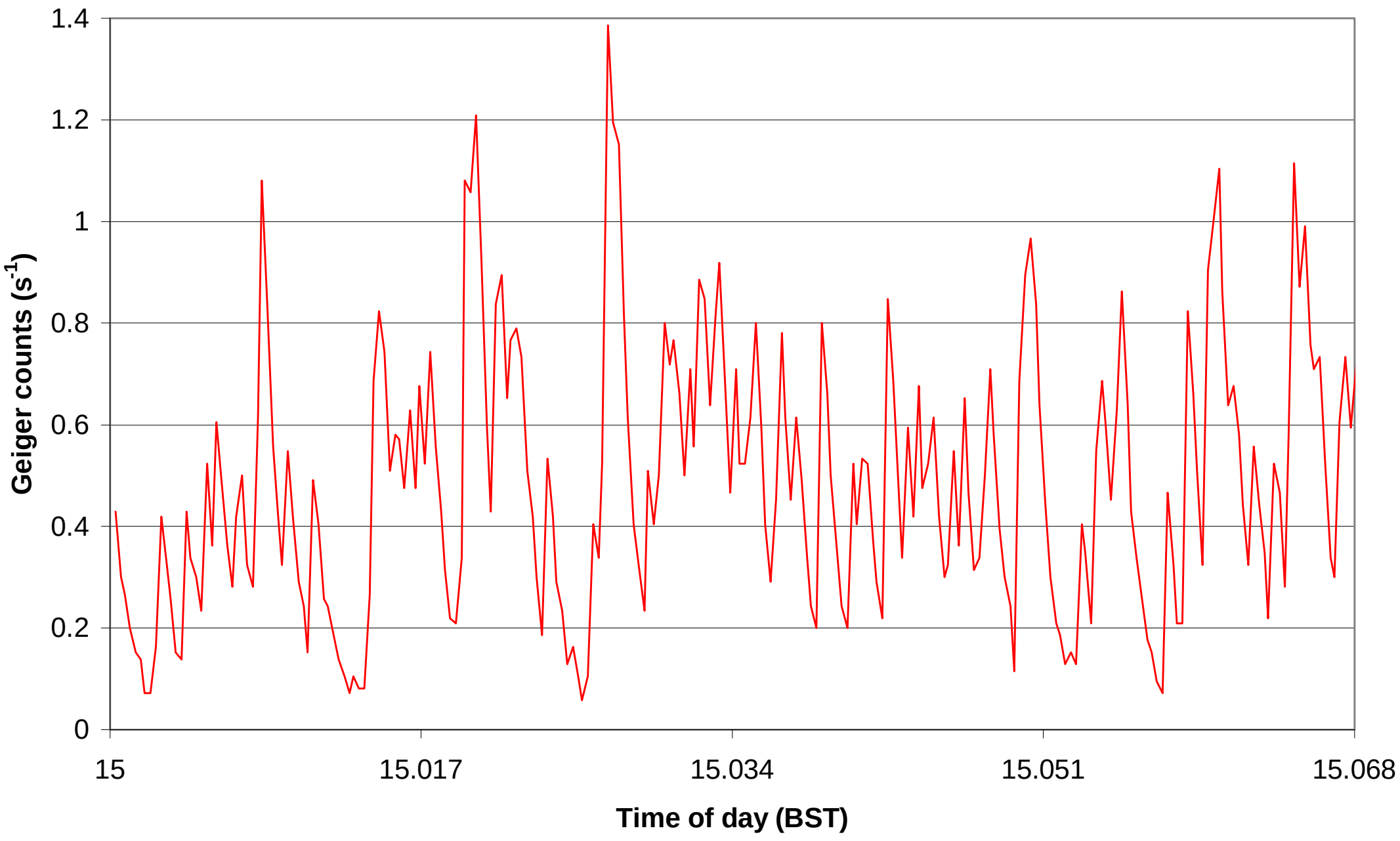

*Figure 6.7 A four minute time series of 1 Hz Geiger counter data from Reading University Meteorology Field Site on 8$^{th}$ April 2000, 1500-1504.*

The great variability of $q$ (an order of magnitude in a few minutes is normal, as in Figure 6.7) perturbs the steady-state assumed by the ion balance equation. This makes it more difficult to use ion-aerosol theory to show that the PIMS is actually responding to ion variations, but its response to large fluctuations in ionisation can be instead be used as a test. Long-term averages could also be used, but the Geiger count rate does not vary in a very wide range, and the ion concentration is also sensitive to other factors (*e.g.* aerosol concentration, meteorological effects), so it would be difficult to distinguish the effect of the ionisation rate alone from the other quantities influencing the ions. Figure 6.8 shows a comparison between the ionisation rate and the ion concentration for a time series of one hour. Negative ions were measured with the PIMS, at a bias voltage of –30 V, and the ion current (which is directly proportional to the ion number concentration) is plotted with the count rate from an adjacent Geiger counter at the same height (1.25 m).

[57] Applying ion-aerosol theory to PIMS data will be discussed further in Chapter 7.

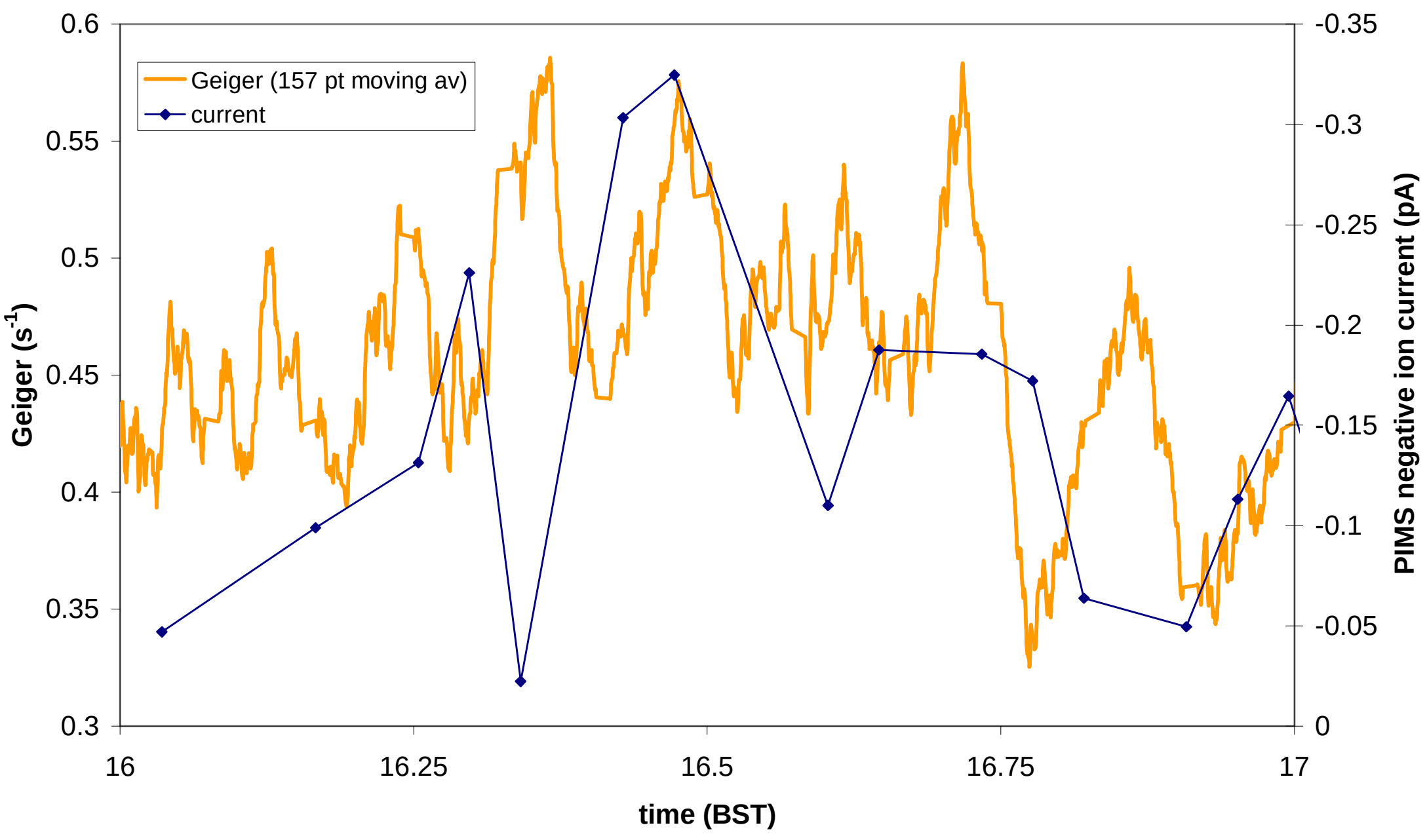

*Figure 6.8 PIMS 3 response to ionisation fluctuations. This is part of a time series measured at Reading on 8th April 2000 from 1500-2100. 1600-1700 is shown here. The thinner dotted trace (RH axis) is the negative ion current, sampled every 157s and filtered according to the criteria discussed above. The thicker trace is a 157 point centred moving average of 1 Hz Geiger counter data.*

Some of the ion peaks, at about 1630, 1645 and 1700 occur at the same time as a peak in the Geiger output, which suggests that the PIMS is able to respond to ionisation fluctuations. The PIMS has now been demonstrated to respond to ionisation transients in urban air on the timescale of minutes, which supports the clean air calibration described in Section 3.3.2.

## 6.4 PIMS ion measurement mode intracomparisons

The two PIMS measurement modes were compared by running both modes on one instrument at the same bias voltage in fair weather conditions, and conductivity calculated from both methods as described in Section 6.1. The two modes frequently showed a good agreement, and an example is given below (Figure 6.9), in which the correlation coefficient of 15-minute averages of the data is 0.51.

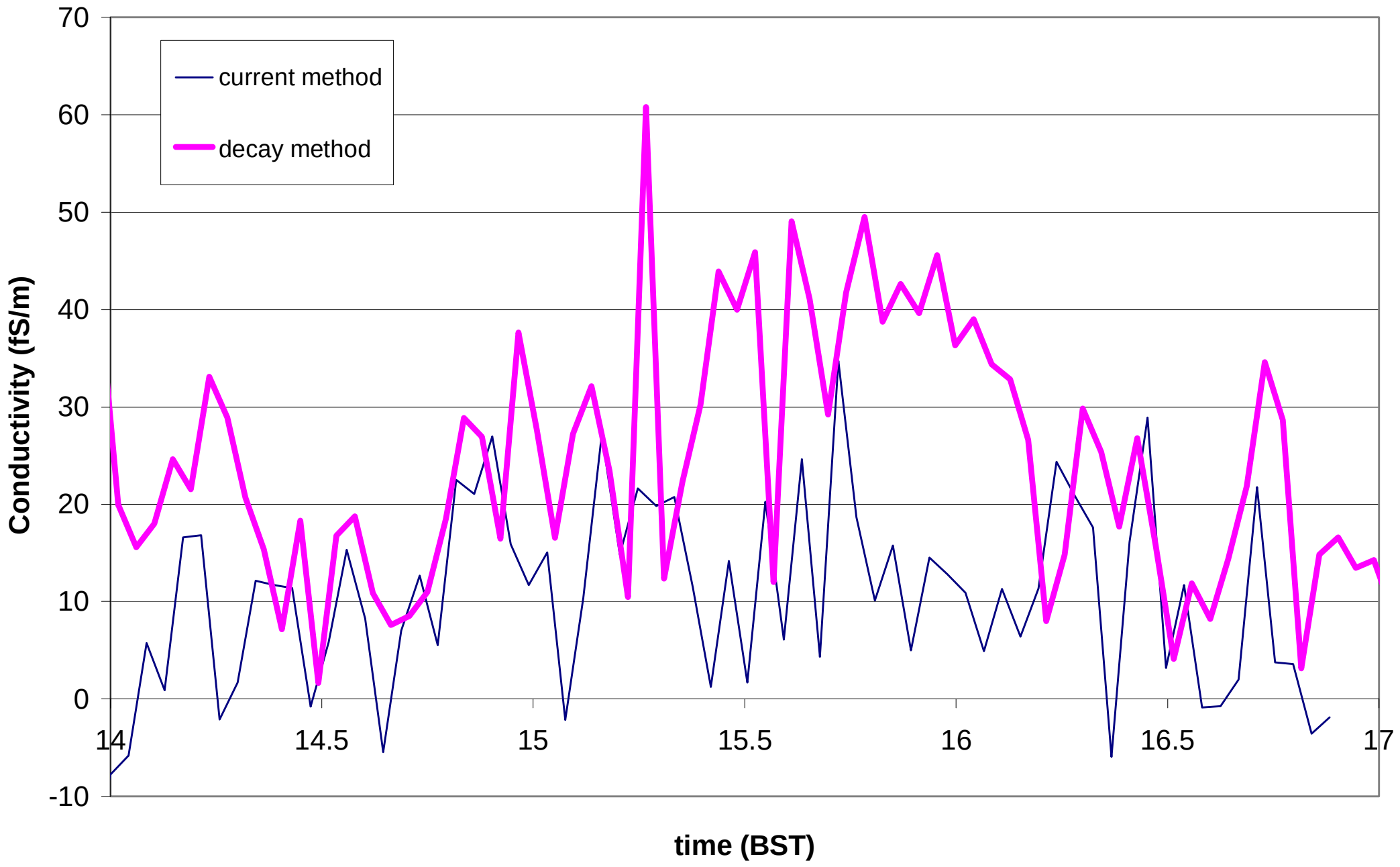

*Figure 6.9 Comparison of the two PIMS measurement modes on 9th April 2000 at Reading. Results are shown from the PIMS 2 instrument from 1400-1700.*

There is generally a correspondence in the trends of the two traces, *e.g.* from 1410-1453, 1536-1648. Although there are some periods when the magnitudes of the two traces agree well, *e.g.* 1424-1453, 1605-1634, their mean values are significantly different. In Figure 6.9, the mean negative conductivity from the Current Method, $\sigma_i$, was 10 fSm$^{-1}$, but from the Voltage Decay mode, $\sigma_v$, it was 18 fSm$^{-1}$. Both these values are acceptable (see review of previous measurements in Section 2.2.3), but the difference between them implies a variable offset or gain in one or both of the measurement modes. As the gain of the MME is known to remain constant for times of order months, and the offset terms show shorter-term variability, for example temperature effects, this disparity is likely to be caused by leakage currents that have not been fully compensated for.

### 6.4.1 Likely origin of conductivity offsets

In the Current Measurement mode, it has been shown (Section 5.6.2) appropriate to apply a correction based on the measured electrometer leakage current $i_L$. In the MME follower mode (the MME schematic is given in Figure 5.6, and a description of the modes is in Appendix A) relays RL1, 5 and 2 are open, as shown in Figure 6.10. RL 2 has no current path and RL5 is guarded, therefore the follower mode leakage current $i_F$ will be the sum of the leakages at the input, from RL2 and the op-amp input bias current. In the leakage check mode, which monitors op-amp bias and electrometer leakage currents, the leakage current $i_L$ is contributed to by RL1, the op-amp input bias current and leakages at the input.

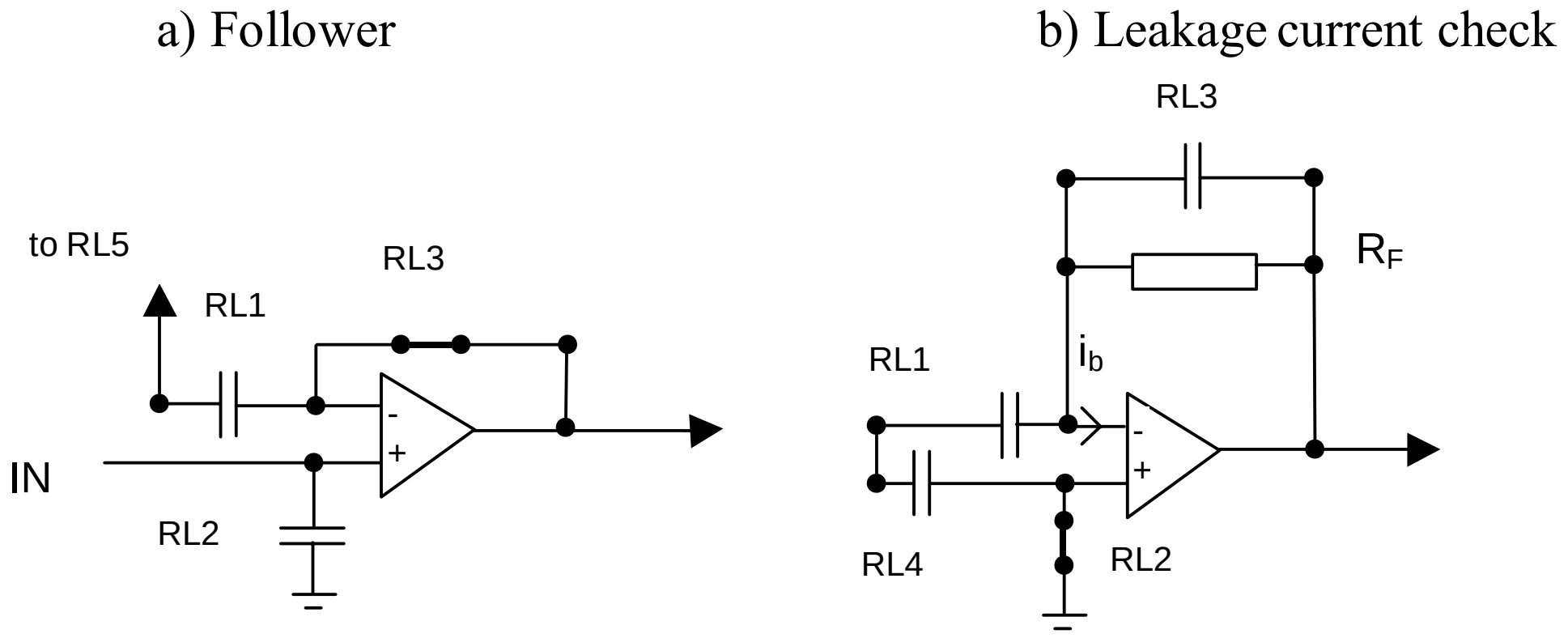

*Figure 6.10 Differences between the MME follower and leakage current measurement configurations a) and b). In this simplified diagram, an open relay is represented as a capacitor, as it transiently injects charges when opening, and a closed relay (thick line) is a direct connection.*

As both $i_F$ and $i_L$ principally originate from leakage across one relay, they are likely to be similar in magnitude. However, in the Voltage Decay mode, current is not measured directly, and has to be inferred from the overall rate of decay. It can therefore be assumed unnecessary to correct for the leakage current in this mode, as long as the leakage can be considered to have a negligible effect on the rate of decay. The contribution to the Voltage Decay mode conductivity $\sigma_v$ from the electrometer leakage current $i_L$ was calculated, and subtracted from $\sigma_v$. For the data shown in Figure 6.9, this changed the conductivity by 2 %, indicating that the electrometer leakage currents are only one-fiftieth the size of the ion signal, which is comparable to the errors introduced from the bias voltage determination in Section 5.3. This shows that the electrometer leakage measurement has no distinguishable effect on $\sigma_v$, given its typical variability. Therefore, no corrections for electrometer leakage currents are applied to the data from the Voltage Decay mode during processing.

### 6.4.2 Negative conductivities

A negative voltage repels negative ions flowing through the Gerdien tube towards the central electrode. When the negative ions reach the central electrode they cause a flow of electrons, and by convention (*e.g.* Bleaney and Bleaney, 1985) this leads to a negative current[58]. Including the sign of the bias voltage, conductivity calculated from Eq. 2.2 should always be a positive value. Hence, currents measured at the central electrode should always be the same sign as the bias voltage, if they are entirely ionic in origin. Yet Figure 6.9 clearly shows that the conductivity is sometimes negative, because $i_c$ was the opposite sign to the bias voltage. The behaviour of the currents when of the opposite sign to the bias still conform to the trend and correlate with $\sigma_v$. This effect is therefore highly likely to be due to an offset imposed by some spurious source of current. This current source can only originate from three places: the electrometer, Gerdien tube itself, or some source of charge external to the PIMS system. Tube leakage currents are caused by insulator degradation, or build-up on the electrode surfaces. Although they probably do contribute some systematic error to the measurements, the effect will be the same for both ion measurement modes, and is unlikely to be the cause of the offset in the current mode. Electrometer leakage is a possible contributor, but in the leakage checking mode, more relays are open than in the current mode, which should exceed the leakage currents that are not compensated for (see Section 5.6.2). To cause the persistent offset in Figure 6.9, the leakage from one relay alone would have to be comparable to the ion current, so the likelihood of this originating from the one relay which is not compensated for in the leakage checking mode is probably small. It is therefore most likely that the offset in the Current Measurement mode results from some external current source, and of a similar magnitude to the ion currents.

### 6.4.3 Temporal inconsistencies

Ion measurements in the two modes are made sequentially and are typically a minute apart. If there are short-term fluctuations in the air passing through the PIMS, the changes in the air being sampled may be detected by consequent measurements in the two modes, which will introduce an error. There are significant turbulent fluctuations in the air of timescales from several minutes to milliseconds (Stull, 1997). As the critical mobility is directly proportional to the wind speed, turbulence will continually modify the fraction of the ion spectrum contributing to the measurement, and may cause disparities between consecutive measurements in different modes. An arbitrarily selected hour of 1 Hz wind data at Reading University Meteorology Field Site, 1100-1200 on 12$^{th}$ June 2000, showed a maximum fluctuation of a factor of 4 on a 30 s time-scale. Since the five minute average wind speed for this period was 3.4 ms$^{-1}$, this corresponds to fluctuations in the range 6.8 – 1.7 ms$^{-1}$. At a typical bias voltage of 20 V, this causes changes in critical mobilities from 1.35 – 0.33 cm$^2$V$^{-1}$s$^{-1}$, spanning about half of the conventionally accepted small ion spectrum as defined by Hõrrak *et al* (1999). This range is for a large fluctuation, but smaller variations in the wind speed will also have some effect on the measurement. Critical mobility variations due to the voltage decay itself will affect the radius of ions contributing to the measurement. However this variation (which is a maximum of 2.5 V) will be insignificant compared with the effect of turbulent fluctuations of wind speed, which is the major factor contributing to the differences between the two measurement modes. Further differences between the modes (as opposed to the differences in measuring them), and the effect of wind fluctuations on the ion spectrum will be discussed in Chapter 7.

[58] An electrical current flows from positive to negative, but the flow of electrons is from negative to positive. Therefore a flow of negative charge is in the opposite direction to the current, so the current is negative.

## 6.5 PIMS systems intercomparisons

Figure 6.11 and Figure 6.12 show the results of conductivity measurements using the different PIMS devices by the two methods on two different fair weather days at Reading.

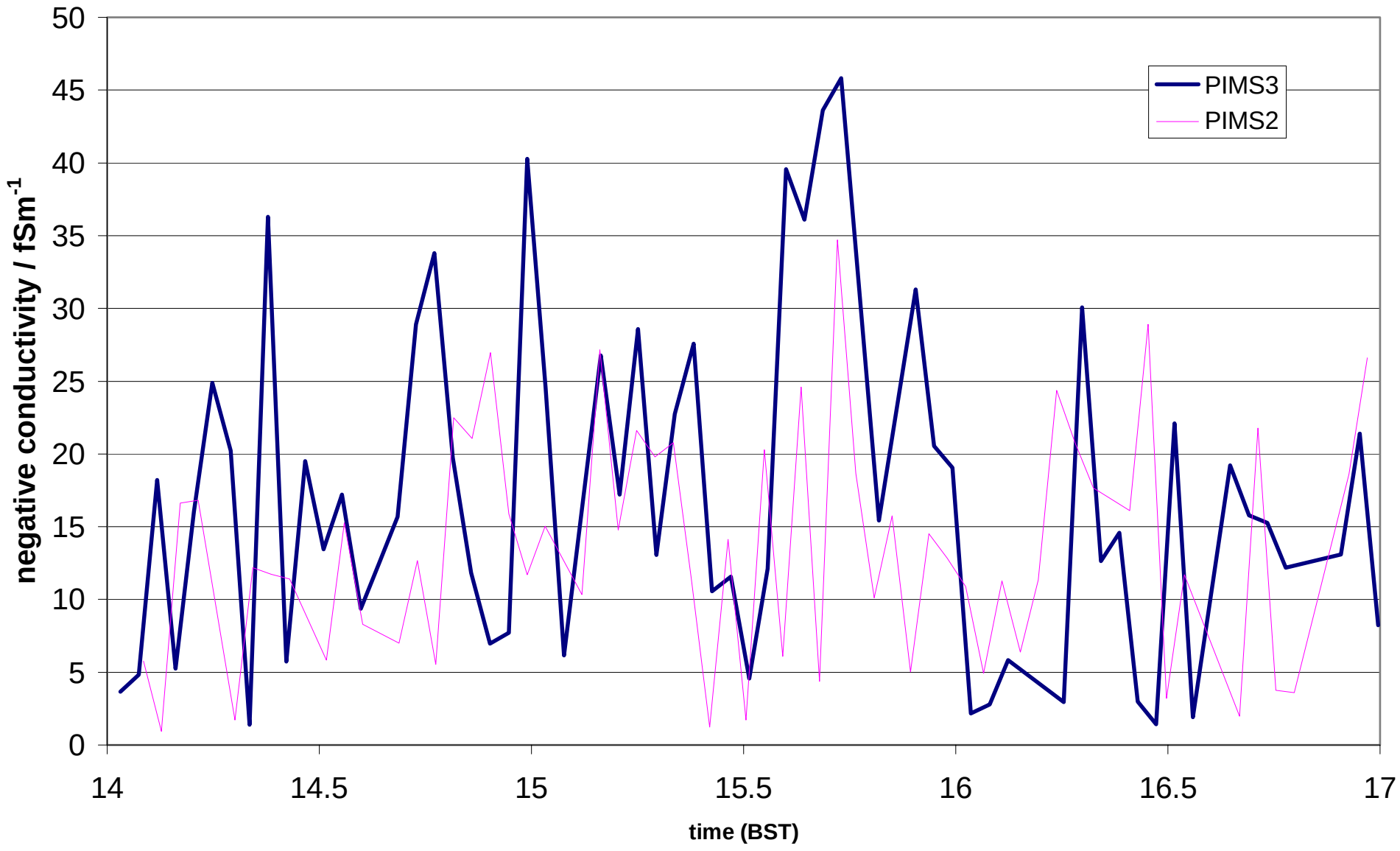

*Figure 6.11 Negative conductivity measured by the current method with PIMS 2 and 3 at $V_b$ = -20 V on 9$^{th}$ April 2000 at Reading. Data shown is a filtered*[59] *sample (12 % of data set discounted) from 1400-1700.*

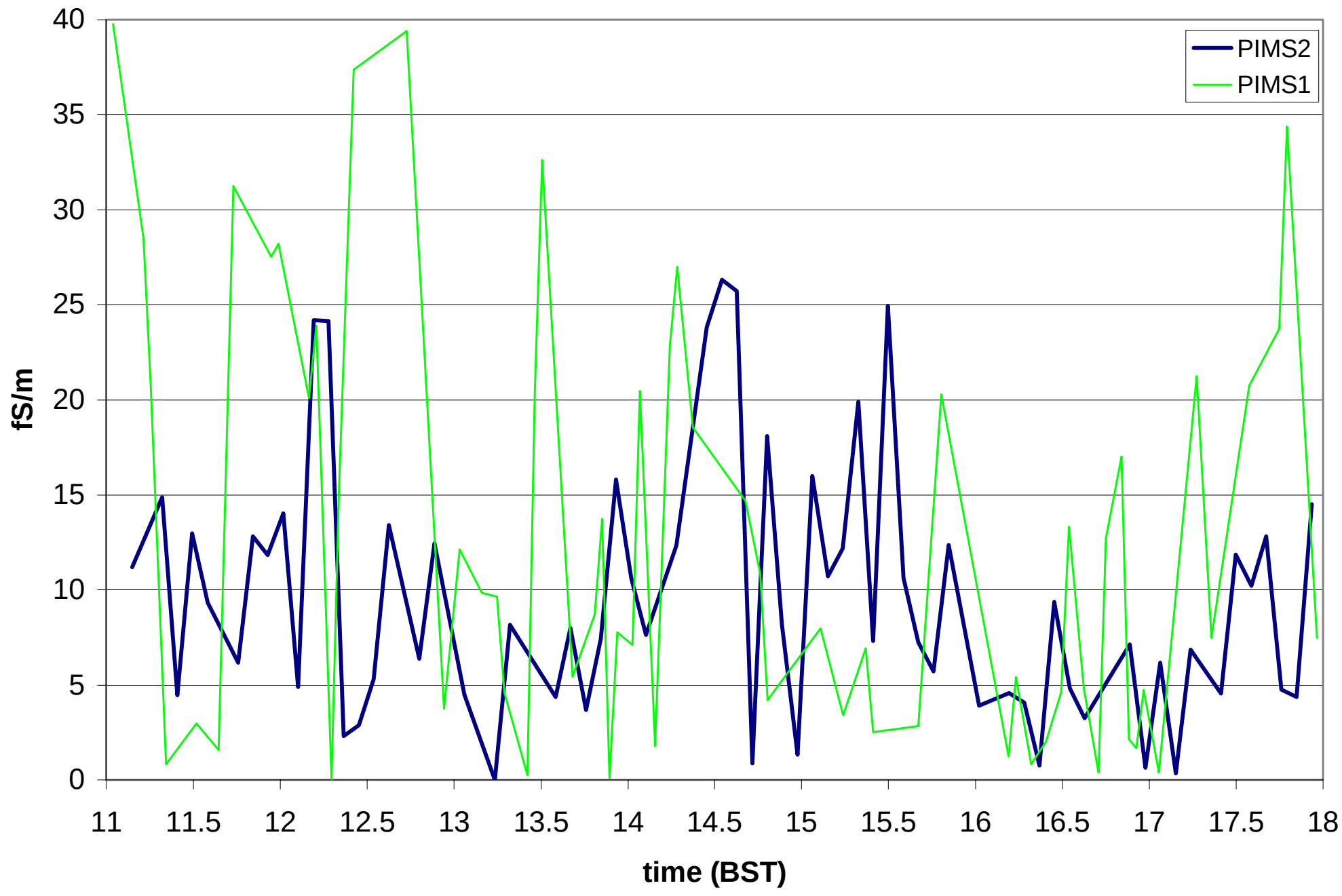

*Figure 6.12 Comparison of negative conductivities measured by the Voltage Decay method with PIMS 1 and 2 at Reading on 18$^{th}$ June 2000. A seven-hour filtered time series (35 % of data set discarded) is shown.*

[59] Filtering criteria are explained in Section 6.7.

In the Current Measurement mode (Figure 6.11), there is an agreement between the two instruments[60]. For the time period shown in Figure 6.11, the correlation coefficient for 15-minute averages of the data was 0.24. However, the averaging period chosen is clearly significant when quantifying the level of agreement, since if the correlation coefficient is calculated for 1400-1530, then the positive correlation is much higher (0.87). The trends in conductivity agree well, with a disparity of 16 % between the mean values. The Voltage Decay mode measurements also concur (15-minute average correlation coefficient of 0.42), with the means separated by 33 %, and the trends tracking each other. The major difference between the measurement of voltage and current with the MME is that the latter approach includes compensation for electrometer leakage currents. It is difficult to precisely evaluate the magnitude of the follower mode electrometer leakage currents, $i_F$ because the relay configurations are significantly different to the leakage current check (see section 6.4), and they are likely to vary with different relays and electronic components. An indication of the difference between the leakage currents from different instruments can be observed by comparing measurements for the same time period from the leakage check mode, which are shown in Figure 6.13.

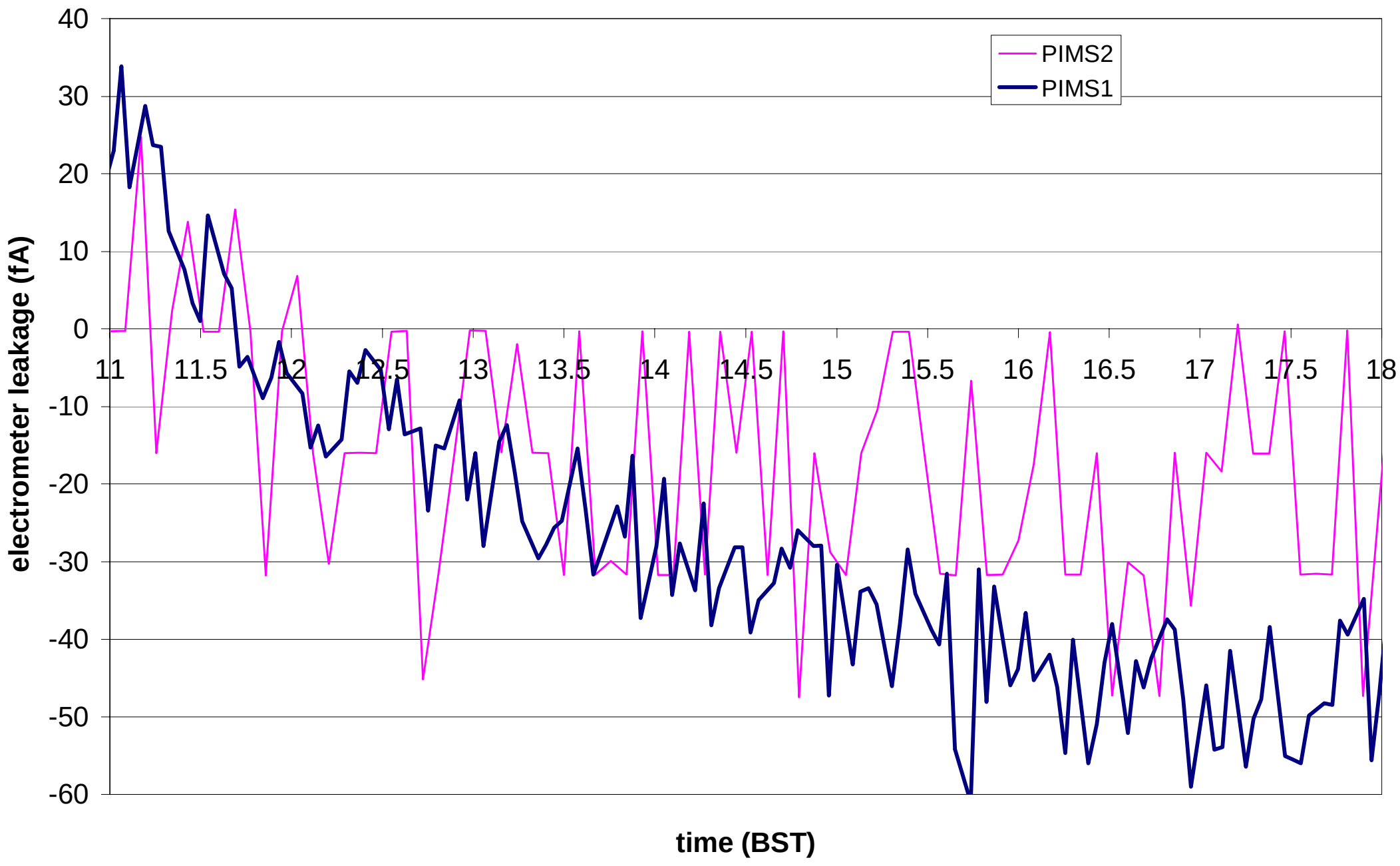

*Figure 6.13 Time series of measured electrometer leakage current $i_L$ for PIMS 1 and 2, on 18th June 2000, found using MME modes 3 and 4.*

[60] Sir John Mason pronounced it to be "excellent" when he visited the Meteorology Department's Experimental Research Laboratory on 23rd June 2000.

The measured electrometer leakage currents varied significantly; for PIMS 1 the mean leakage current was – 26 fA, and for PIMS 2 it was – 16 fA. As mentioned earlier, the measured follower mode leakage current $i_F$ is only contributed to by one of the three reed relays determining $i_L$, so $i_F < i_L$, but $i_F / i_L$ is probably constant, if it is assumed that each reed relay contributes a constant amount of current into the system. So it is not unreasonable to expect voltage decay measurements from two instruments (when $i_F$ is negligible, but not precisely known) to agree less well than current measurements, where $i_L$ is larger, but better-determined and compensated for.

### 6.5.1 Local variations between the PIMSs

As different air, which is constantly fluctuating in turbulent eddies, is sampled by each instrument, this will introduce some errors between the instruments due to local variations. The mean eddy size $l$ can be estimated from von Karman's constant, $\kappa$, at the height of the instruments, $z$ (Stull, 1997),

$$l = \kappa z \qquad \textit{Eq. 6.8.}$$

Since $\kappa = 0.4$ and z = 1.5 m, then the mean eddy size is of comparable scale to the instrument separation, suggesting there may be a significant spatial variation in the wind speed, which may affect the ions sampled by the PIMS.

Leakage across the electrode insulation will vary from instrument to instrument, but the three PIMS are mechanically very similar. Evidence for this is that the Gerdien tubes only varied by 2 % in capacitance, which is an extremely sensitive parameter to changes in dimensions (see Eq. 2.17). The variations in the bias voltage generator have been included in the software, since each PIMS instrument was controlled by a dedicated program with calibration differences accounted for (see Section 5.7.2). The fans ventilating the tubes will also behave slightly differently, but these variations will be small compared to the spatial effects of wind speed fluctuations. This appears the most likely cause for the variability between instruments not attributed to leakage currents.

## 6.6 Comparing PIMS data to other ion measurements

### 6.6.1 Ion measurements made with another design of Gerdien Condenser

Conductivity measurements made with the PIMS were compared to similar measurements made with another Gerdien condenser of more classical design. This design of instrument has been discussed by Blakeslee (1984) and widely applied, *e.g.* by Anderson and Bailey (1991) and Blakeslee and Krider (1992). The instrument used here was developed at the University of Arizona and will consequently be referred to as the "Arizona Gerdien".

The PIMS and the Arizona Gerdien were set up adjacent to each other at Reading University Meteorology Field site on a fair weather day. The response of the Arizona Gerdien is given as 6.2 $fSm^{-1}V^{-1}$ at a bias voltage of 20 V(Blakeslee and Krider, 1992) which (from Eq. 6.1) corresponds to a current amplifier sensitivity of 0.88 $pAV^{-1}$.

A slight difference in the conductivities is to be expected from the two instruments, because of their varying critical mobilities. The Arizona Gerdien is significantly larger than the PIMS (a ~ 5 cm, b ~ 0.5 cm), therefore it is impossible to exactly match the instruments' critical mobilities given the practical limitations to bias voltages and flow rates. In the absence of an ion mobility spectrum for the measurement period, the difference between the measurements cannot be predicted. Primarily for this reason, the Ohmic response of the Arizona Gerdien was measured; an additional benefit is that this is a standard test that the Gerdien condenser is responding to atmospheric ions (Chalmers, 1967). If the Gerdien condenser is in the conductivity measurement régime, the mean conductivity can be calculated from the gradient of the straight line fit to the Ohmic plot in Figure 6.14, *m*, which gives the average output current to bias voltage ratio for the measurement period. Hence, from Eq. 6.1 the conductivity is given by:

$$\sigma = m\frac{\varepsilon_0}{C} \qquad \textit{Eq. 6.9.}$$

Manual current measurements were made across the bias voltage range $V_b = \pm$ 11-106 V. The output was logged on a Gould BS-272 chart recorder. Meanwhile, PIMS 3 measured negative conductivity at 0.5 Hz in the Current Mode. At the end of the PIMS logging period the Arizona Gerdien Ohmic response to negative voltages was tested again. Results from the Ohmic response tests are in Figure 6.14.

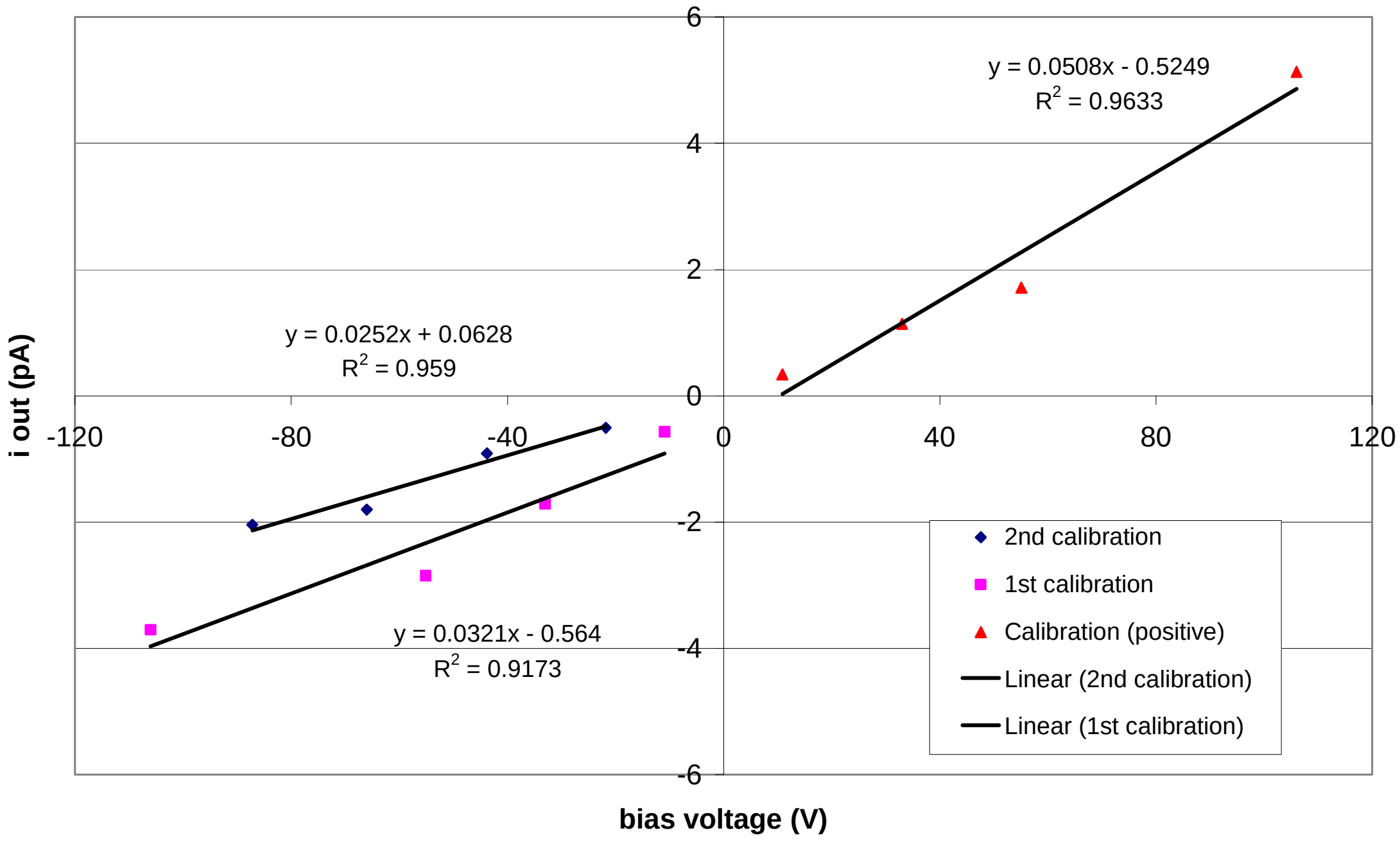

*Figure 6.14 Ohmic response of Arizona Gerdien. The two tests at negative voltages were made approximately an hour apart.*

The linear response shows that the Arizona Gerdien was measuring conductivity at this time. The negative conductivity for the first calibration was 36.0 $fSm^{-1}$, and for the second calibration 45.8 $fSm^{-1}$. The average for the two-hour measurement period is 40.9 $fSm^{-1}$. A sample of the Arizona Gerdien chart recorder output is given below.

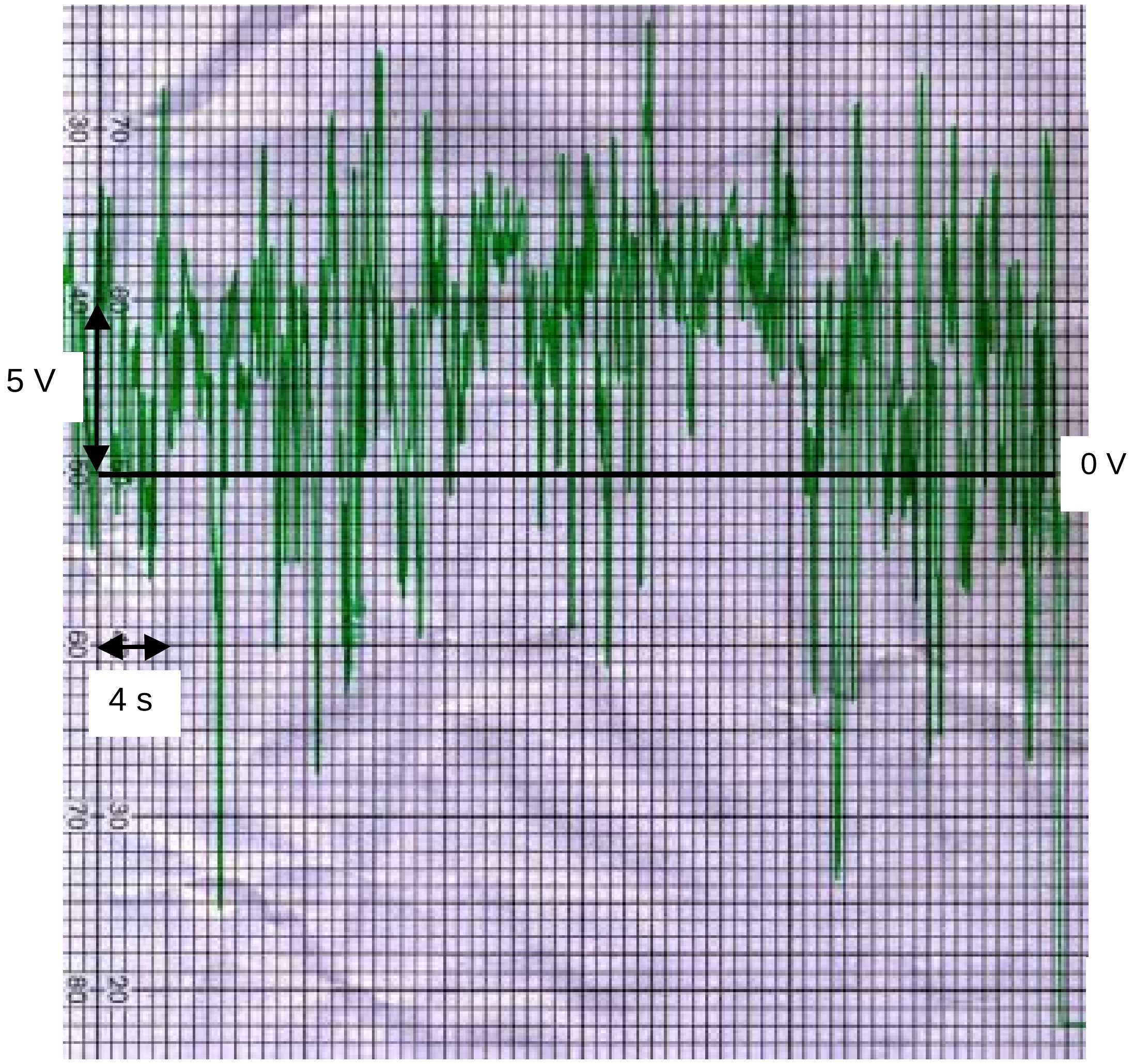

*Figure 6.15 Sample of the Arizona Gerdien output at $V_b$ = -105 V. The chart recorder trace is centred at 0 V, and axis scaling is indicated on the trace.*

The PIMS output (Figure 6.16) appears similar to the Arizona Gerdien trace (Figure 6.15). This suggests that the noise is independent of the instrument, and atmospheric ions may be a fundamentally noisy quantity. The main processes affecting the atmospheric ion concentration: production, attachment and recombination are all constantly fluctuating in rate because of atmospheric turbulence and the stochastic nature of ionisation processes. The air sampled by the instruments is also rapidly changing in its qualities, because of the variability in the ion concentration discussed above, and turbulent eddies affecting the origin of the air sampled[61]. However, the mean conductivities from the two instruments agree well: the PIMS mean conductivity was 41.6 $fSm^{-1}$, which is less than 2 % higher than the average Arizona Gerdien conductivity. It can be concluded that the PIMS can measure conductivity to within 2 % of the conductivity measured by a reference Gerdien condenser.

[61] Variability in the measurements will be discussed further in Chapter 7.

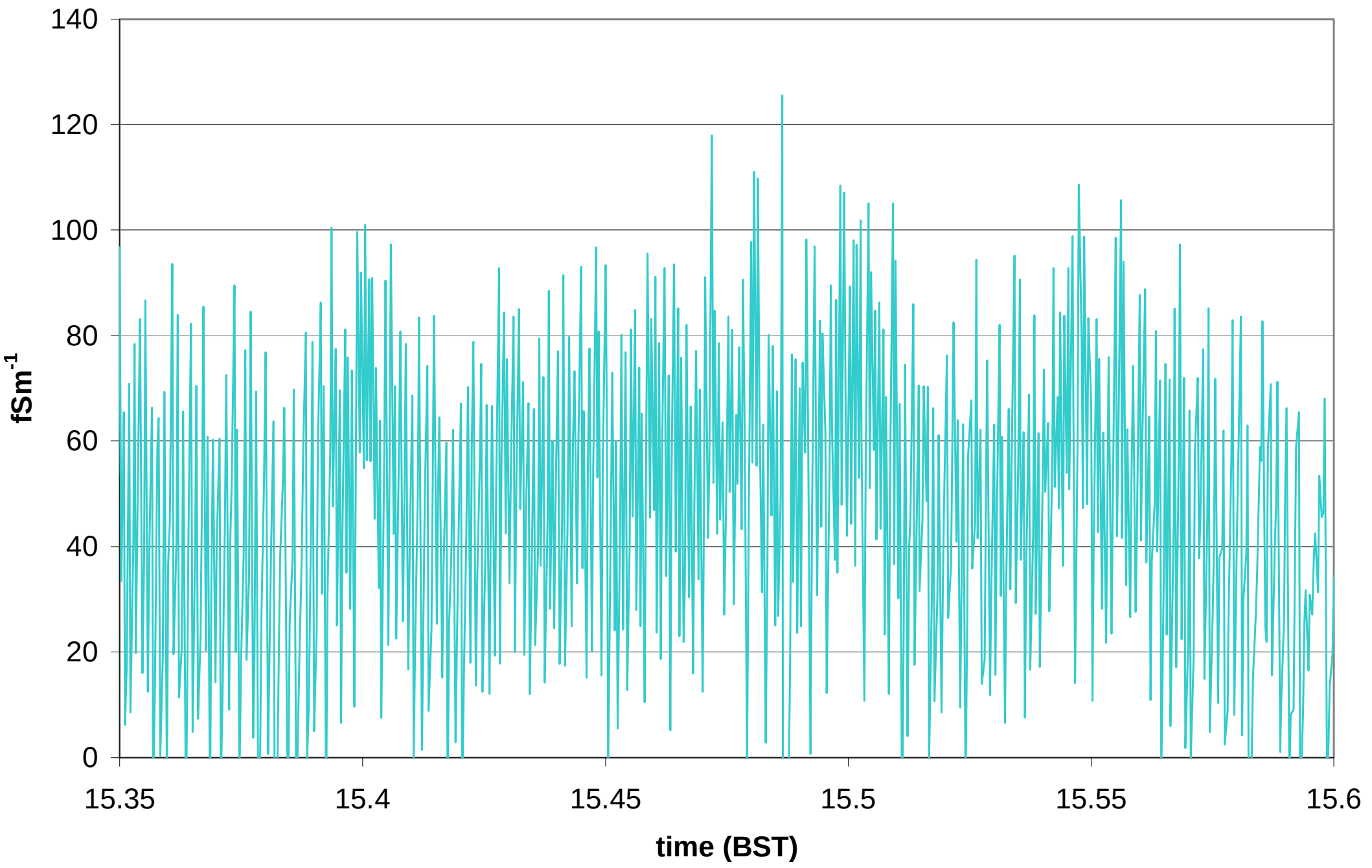

*Figure 6.16 Sample of PIMS 0.5 Hz negative conductivity trace.*

## 6.7 Average conductivities

Ion measurements in the literature are usually presented as average values, although the amount of data averaged is often reasonably small, which suggests that textbook values should not necessarily be equated with reliability. In fact there is significantly more data presented in this Thesis than in much classical work on the subject (*e.g.* Chalmers, 1967) [62]. To compare the PIMS conductivity data from Reading University Meteorology Field Site with such historical averages, data from the continuous measurement campaign from April – June 2000 was averaged. Measurements were only attempted on fair weather days, so that the atmospheric electrical conditions would be relatively unperturbed by the presence of hydrometeors and aerosol (Barlow, 2000). The definition of "fair weather" was based loosely on Reiter's criteria, discussed in Barlow (2000), who concluded that there was a need for redefinition of the term based on micrometeorological considerations. For the results presented here, the major consideration for starting a run of measurements was that there was less than 3/8 cloud cover, which was typically fair weather cumulus.

[62] In the group led by Chalmers at Durham (where I was an undergraduate in the Physics Department, and Atmospheric Electrical work had sadly long since ceased) in the 1960's, two days of consistent atmospheric electrical data was then considered an adequate amount for a PhD. Thesis (Jones, 1999).

The conductivity data were hourly averaged ($<\sigma>$), and then filtered based on the following criteria.

- $<\sigma> > 2i_L$ is a pragmatic treatment of the errors introduced by leakage. The measured leakage current $i_L$ is not a full compensation for the electrometer leakage, because there is one reed relay unaccounted for between the MME current measurement and leakage check modes. Since $i_L$ is mostly leakage from one relay (see discussion in Section 6.4.1), but its sign is not known, $2i_L$ is therefore a conservative estimate for the uncompensated leakage current.
- $<\sigma> < 50\ fSm^{-1}$ excludes measurements that are larger than the theoretical maximum ion concentration, for $q = 10$ cm$^{-3}$s$^{-1}$. In the recombination limit the maximum ion concentration can be easily calculated from Eq. 6.6, which gives an upper limit of 2500 ions cm$^{-3}$. Since

$$\sigma \approx ne\mu \qquad \text{Eq. 6.10,}$$

and the mean mobility of atmospheric small ions is 1.2 cm$^2$V$^{-1}$s$^{-1}$ (*e.g.* Dolezalek, 1974), then $\sigma_{max} \sim 50$ fSm$^{-1}$. This is an absolute maximum upper limit, since it assumes an aerosol-free atmosphere, and any aerosol present (which there clearly is in urban Reading air) reduces the ion concentration.

Conductivity values outside these limits are spurious, and caused by non-ionic current sources dominating the signal. When the conductivity is within the acceptable range, there may still be offsets contributing to the measurement, but the filtering criteria described above only include measurements when the offset is small. Averaging the remaining data will further reduce the error due to random offsets.

Measurements were made with the three PIMS on the mast at 1.25 m (see Figure 6.2) on 24 days in April, May and June 2000 for 292 hours. The total logging time is an average across the three PIMS instruments; they ran for similar lengths of time, but

Although the acquisition of data is now more straightforward, the frequency of suitable fair weather days is unlikely to have changed by orders of magnitude.

this varied slightly as individual PIMS were removed for cleaning and repair. The measurement periods were limited by the weather conditions, for example in April 2000, which was the wettest month since 1756 (Eden, 2000a) fair weather was rare and the instruments were only run for 85 hours. In May, the first half of the month offered some good measurement days (Eden, 2000b), though continual (inaccurate) predictions of thunder meant the logging strategy had to be conservative to avoid possible lightning damage from transients to the circuitry. It is an unfortunate vicissitude of the British climate that fair weather days in the spring and summer, which are favourable for atmospheric electrical measurements, are when the convective activity which leads to thunderstorms is most likely. The apparent need of weather forecasters to alert the public to even the slightest risk of thunder presents a further problem, since instruments designed to measure fair weather quantities (like the PIMS) may not be able to withstand the dramatic changes associated with thunderstorms. Measurements were therefore abandoned when thunder was predicted, which lost considerable logging time during fair weather, as well as the periods when the instruments were being maintained.

After the filtering criteria described above had been applied to the whole set of hourly averages, 50 % of the data were retained. The efficiency of the two modes was different, with 63 % of current mode data retained after filtering and 36 % of the Voltage Decay mode results. The voltage decay data was generally noisier and gave conductivities which were greater in magnitude than those from the current mode. This is summarised in the mean data shown in Table 6.2 below, where the voltage decay mean and standard deviation are greater than in the current mode. The average air conductivity in the literature[63] lies between the means for the two modes at 13 $fSm^{-1}$, but is obviously close to the current mode measurements.

[63] This is calculated from the values given in Table 2.1

| **Measuring Mode** | **Mode efficiency (% data retained)** | **<$\sigma$> (fSm$^{-1}$)** | **<*standard error*> (fSm$^{-1}$)** |
|---|---|---|---|
| Current | 63 | 12 | 3 |
| Voltage | 36 | 19 | 7 |

*Table 6.2 Summary of conductivity measurements from three PIMS running in two modes. Conductivities were averaged across all instruments for the periods when more than two PIMS were running.*

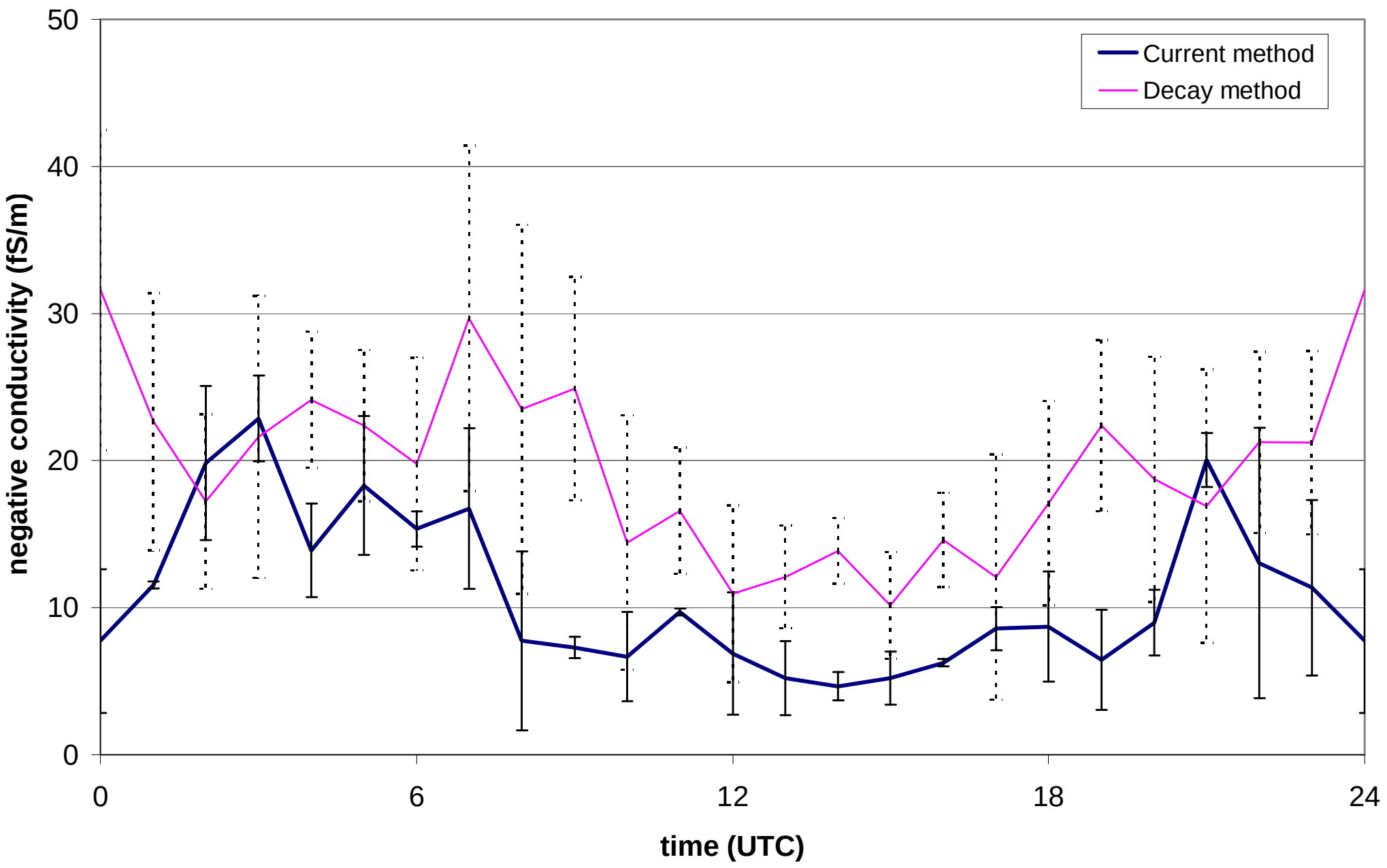

*Figure 6.17 Average diurnal variation in conductivity calculated by the two methods. Error bars are the standard error of the mean for each set of hourly averages, for three instruments.*

Figure 6.17 shows the hourly conductivity averages, taken over 292 hours on 24 days and averaged across the PIMS instruments. The trends in conductivity calculated by the two methods agree well, though the mean values differ. The pattern of this diurnal variation is similar to those in the literature, discussed in Section 2.2.3, with a pre-sunrise peak at 0400-0500, due to a maximum in the ionisation rate caused by thermal stratification. Throughout the day conductivity then decreases, which is traditionally explained by the wind increasing after midday and turbulent mixing dispersing the radioactive gases contributing the majority of the ionisation (Chalmers, 1967). As in Section 2.2.3, studying the trace from a local Geiger counter, which is a proxy for the majority of the ionisation rate, can give some insight into this. The Geiger counter output and the global solar radiation for an example day, 19$^{th}$ October 1999, are shown in Figure 6.18; from midnight to sunrise there is an increasing ionisation rate, which is at its maximum during sunrise. This is probably caused by the stable nocturnal surface layer trapping radioactive gases emitted from the soil (such as radon), which cause most of the ionisation at the surface[64].

[64] Comparison between the ion production rate and the ion concentration is more complicated during the day, because aerosol concentrations and wind speeds are more variable. This has been discussed in Section 2.2.3.

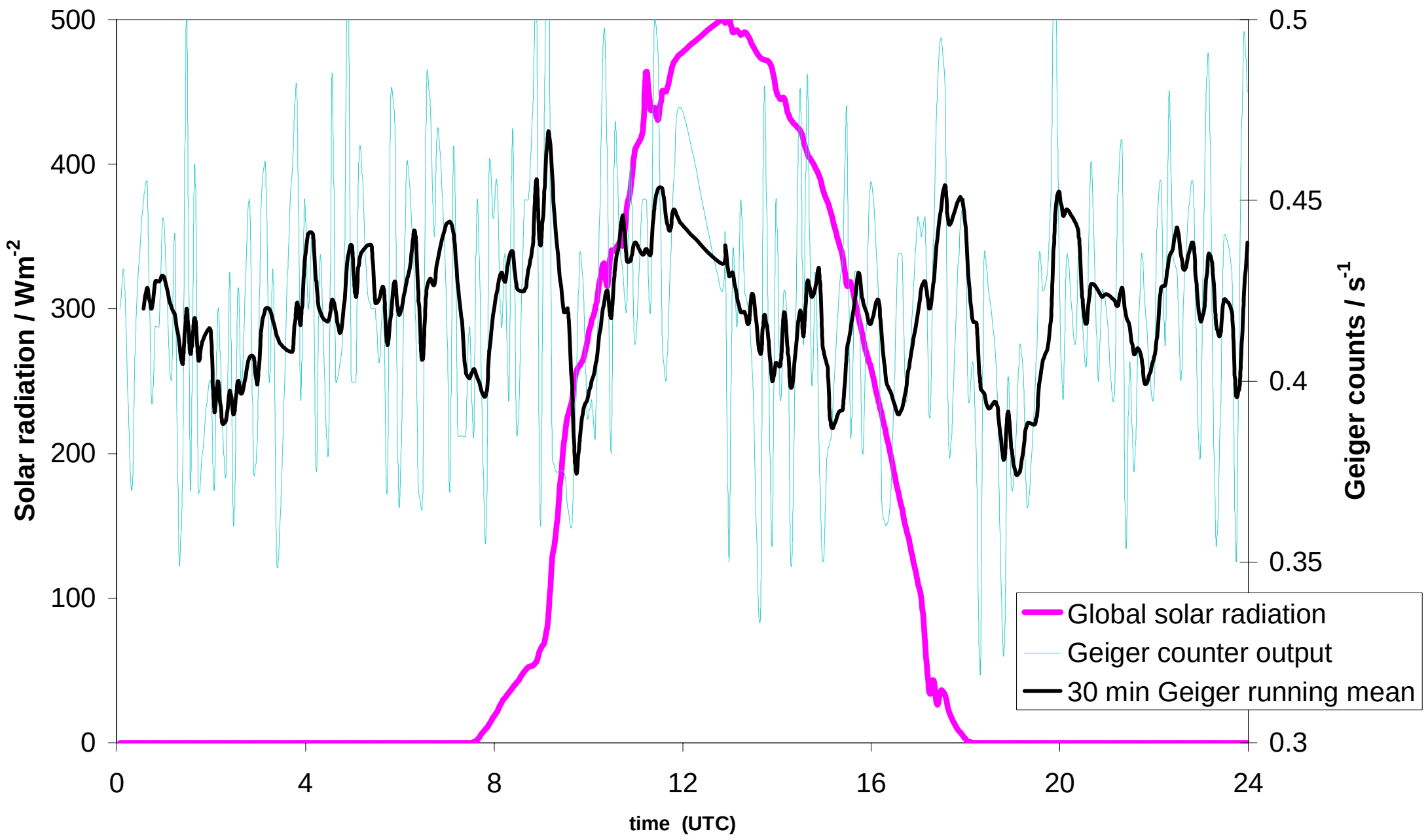

*Figure 6.18 Five minute averages of global solar radiation (left hand axis) and Geiger count rate (right hand axis) for 19th October 1999 at Reading University Meteorology Field Site. A 30 minute running mean of the Geiger counter trace is also shown.*

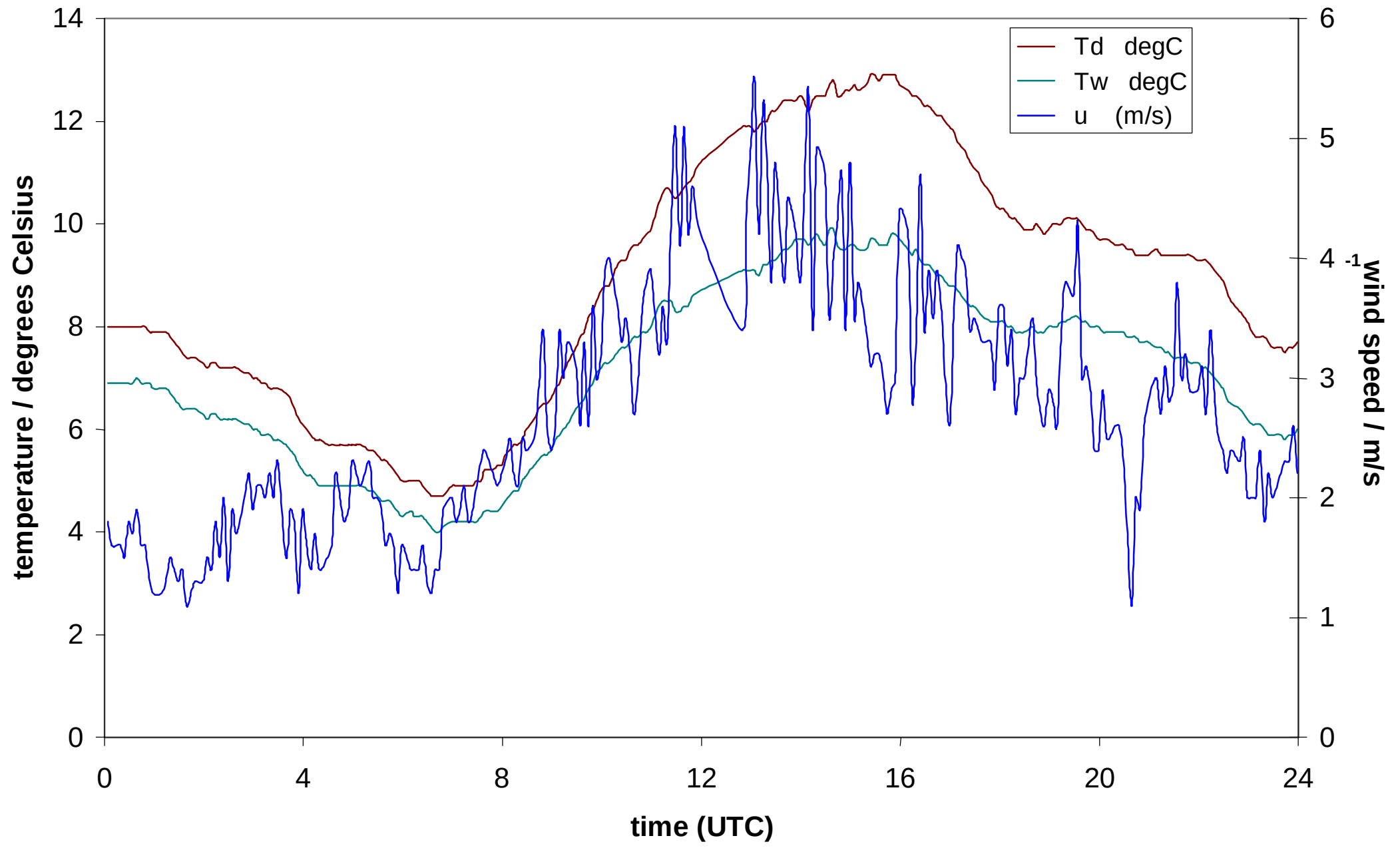

*Figure 6.19 Five minute averages of wet (Tw) and dry-bulb (Td) temperatures at 1 m, and wind speed (u) at 10 m for the same example day, 19th October 1999.*

## 6.8 Conclusions

This chapter has shown that the PIMS devices can measure ions consistently in both the Current Measurement and Voltage Decay modes. The instrument responds to fluctuations in the ionisation rate, and there is good agreement across both modes with the same instrument, and between the three PIMSs. Some of the disagreements between the instruments have been explained in terms of spatial and temporal inconsistencies in the air being sampled. The conductivity trends also agree with those recorded by another Gerdien instrument. The average diurnal variation found over 24 days in a three-month period is concomitant with previous measurements, and many of its variations can be explained in terms of meteorological factors.

*"the fact that an experiment delivers an unexpected answer means simply that you have not asked the question you assume you have asked" C.L. Stong*[65]

# 7 Variability in PIMS measurements

The aim of this chapter is to resolve some uncertainties in the PIMSs data so that a set of operating conditions can be defined during which the PIMS is measuring small ions. In Chapter 6, it was suggested that the presence of charged particulates may have a deleterious effect on the measurements. This is analysed in detail, and other sources of error are investigated to further refine the operating conditions.

## 7.1 Effect of fluctuations in the flow

### 7.1.1 The contribution of intermediate ions to conductivity

The critical mobility is directly proportional to the wind speed (Equation 2.11), so at lower wind speeds the size of ion contributing to the conductivity measurement increases. It can be hypothesised that greater conductivity is observed at night, when there is low wind in the stable nocturnal boundary layer (Stull, 1997), simply because intermediate ions are contributing to the measurement. This was tested using data from a fair weather day, 3 May 2000 when there was a 24-hour set of hourly averaged conductivity data, which itself conformed to the diurnal variation frequently cited in the literature (Section 2.2.3). The maximum radius of ion contributing to the measurement, $r_{max}$, was calculated as in Section 3.2.2. The wind component inside the tube was calculated from Eq. 3.11 and Eq. 3.12. The diurnal variation of negative conductivity and $r_{max}$ are shown in Figure 7.1. It is clear that almost all of the peaks in conductivity coincide with maxima in $r_{max}$, when it is therefore possible that more ions are contributing to the conductivity. For the day chosen, the conductivity is principally due to conventional small ions for the whole day, but at lower wind speeds $r_{max}$ could easily encroach into the intermediate ion size range of r > ~ 1 nm (Hõrrak *et al,* 2000), as also discussed in Section 3.2.2. This is important to recognise, because a significant contribution of intermediate and large ions to the conductivity has been theoretically shown to invalidate the traditional conductivity-aerosol relationship (see Section 2.4) by Dhanorkar and Kamra (1997).

[65] *The Amateur Scientist*, Heinemann, London (1965)

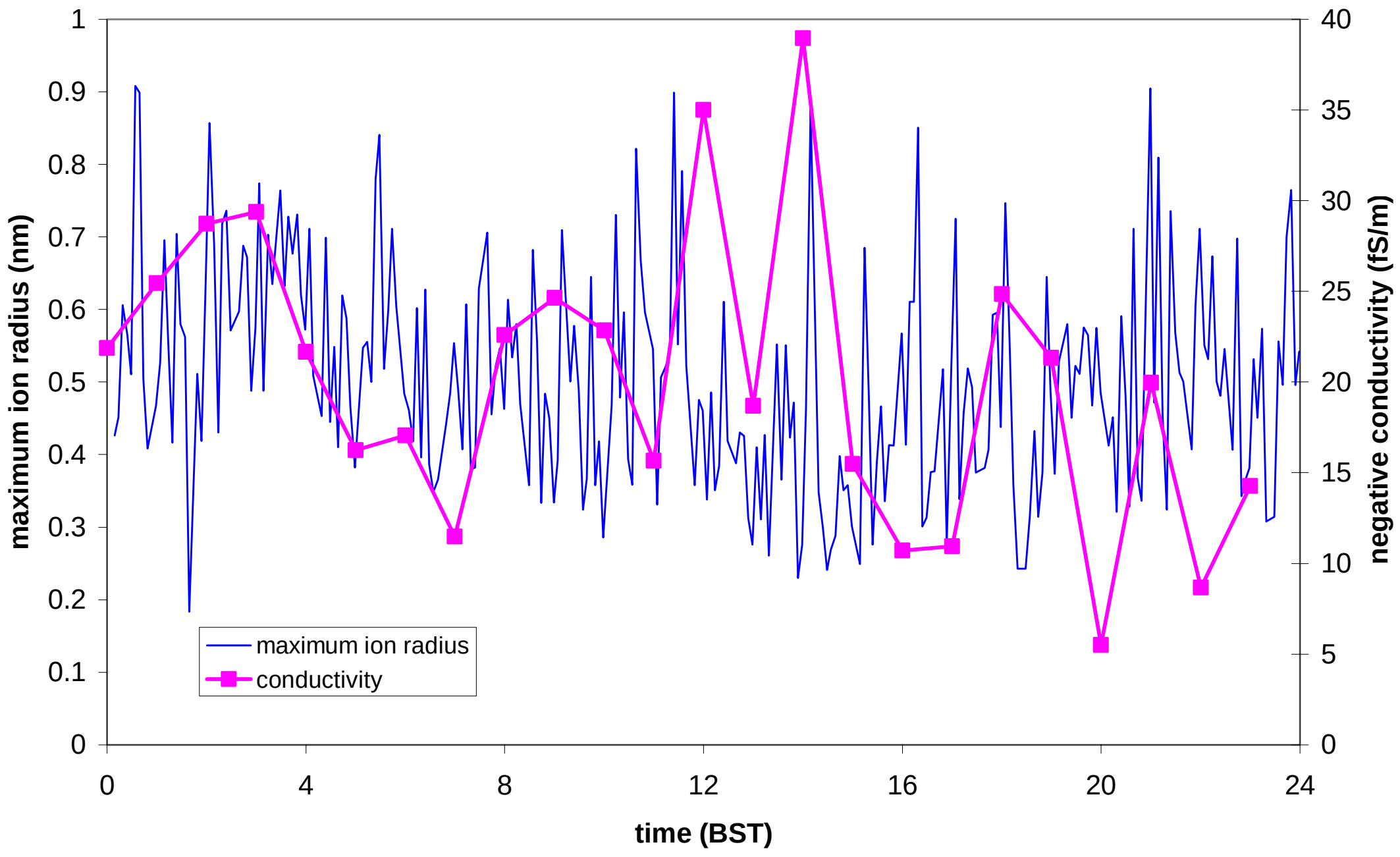

*Figure 7.1 Diurnal variation of conductivity (hourly averages of both methods, for a period when their results agreed) and maximum ion radius (calculated from 5 minute averages of wind component) for 3 May 2000.*

If the measured ion size is not considered, as is conventional, the contribution of larger ions can be misleading. They could make the measurements appear as if they are conforming to the expected patterns, when in reality they are only reflecting the fraction of the ion spectrum contributing to the measurement rather than the ion concentrations. The conventional expression for conductivity (Eq. 2.14) can be rewritten to take account of this

$$\sigma \approx ne\mu + n_i e\mu_i + n_L e\mu_L \qquad \textit{Eq. 7.1,}$$

which separates the contribution of the intermediate and large ions. Ideally, in conductivity measurement, the intermediate and large ions can be ignored when there are few of them because their mobilities are orders of magnitude lower than small ions, and their contribution is negligible. Here it is shown that under certain circumstances, intermediate ions can be measured in such numbers that they do contribute to the conductivity measurement. Classical ion-aerosol theory only holds when the conductivity is entirely due to small ions, because intermediate and large ions are effectively charged aerosol particles (Hõrrak *et al*, 2000), so the spectral separation between them becomes indistinct. Hence, for different conductivity measurements to be comparable, it is necessary to ensure that the conductivity is only contributed to by small ions.

If $r_{max}$ is calculated from wind measurements made near to the ion measurement site, this problem can be alleviated by discarding data where $r_{max}$ falls into the intermediate ion range. 1 Hz wind speed and direction measurements are made continually at 10 m at the Field Site, and five-minute averages from these data were used to rescreen the ion data presented in Section 6.6.2. A further five hours of data were rejected on average for each PIMS, which reduced the total amount of data available for analysis by a further 2 %. The revised diurnal variations are shown in Figure 7.2, where the conductivity traces from the two methods are in better agreement both in magnitude and in trends (this is summarised in Table 7.1). The correlation coefficient between conductivities calculated by the two methods is 0.51, which is improved compared to the correlation when the contribution of intermediate ions was ignored (0.34).

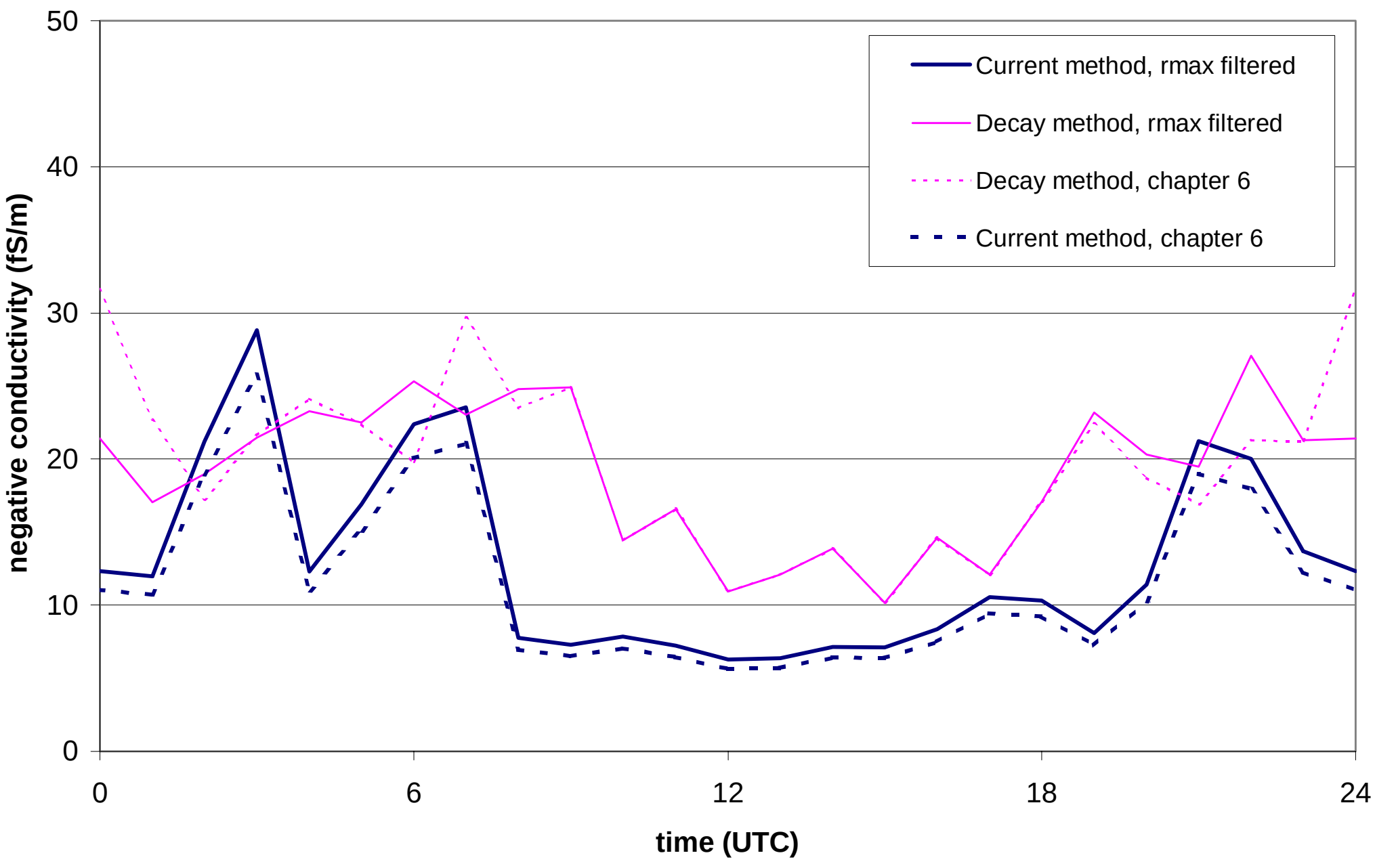

*Figure 7.2 Average diurnal variation of negative conductivity, calculated as in Chapter 6, but with further filtering to remove data where intermediate ions were contributing to the measurement. The averages calculated without excluding* $r_{max}$ *> 1 nm (as in Chapter 6) are shown as dotted lines.*

| **Mode** | **Mode efficiency (% data retained)** | $\langle\sigma\rangle$ **(fSm$^{-1}$)** | <***standard error***> **(fSm$^{-1}$)** |
|---|---|---|---|
| Current | 61 | 13 | 4 |
| Voltage | 35 | 19 | 8 |

*Table 7.1 Mean negative conductivities by the two methods after further filtering to remove the contribution of intermediate ions.*

The majority of changes in the average data compared to Chapter 6 are during the night, when the wind speeds drop and larger ions are therefore more likely to be captured. Since the data from the daytime remains largely unmodified, the poorer agreement between the two modes in the daytime is a consequence of turbulent gusts of wind affecting the ion concentration between consecutive measurements. As the atmospheric surface layer is more stable at night, there is better temporal agreement between measurements made asynchronously. It is clearly possible for intermediate ions to contribute to the conductivity during still periods, particularly at night. I believe, therefore, that this filtering procedure needs to be applied to all conductivity measurements where the Gerdien condenser is susceptible to fluctuations in ventilation. Some filtering of existing data for $r_{max}$ may be necessary with this factor in mind. Reanalysis of historic measurements has been previously shown to be effective at improving conductivity measurements (Gunn, 1960; Andersen and Bailey, 1991)[66].

### 7.1.2 Effect on the fraction of ions measured

Measurement of wind speed and direction, and careful filtering of data should remove the effect of intermediate ions (r > 1 nm (Hõrrak *et al*, 1999)) from analysed ion data, because the times when they are contributing to the measurement can then be identified. However, any change in wind speed causes a variation in the maximum radius of ions being measured, and hence a change in the fraction of the ion spectrum which is being observed. The variation of ion radius with external ventilation is shown below in Figure 7.3. $r_{max}$ is derived from the flow rate in the tube, calculated as in Section 3.2.1. The plot is asymmetrical because of the fan sucking air through the tube, which causes the flow rate in the tube to change depending on the direction of the external wind.

[66] Gunn (1960) corrected conductivity measurements from the Carnegie cruises in the 1930s for errors due to the inlet pipes. Although inlet ducting will damp the effects of turbulent fluctuations in wind

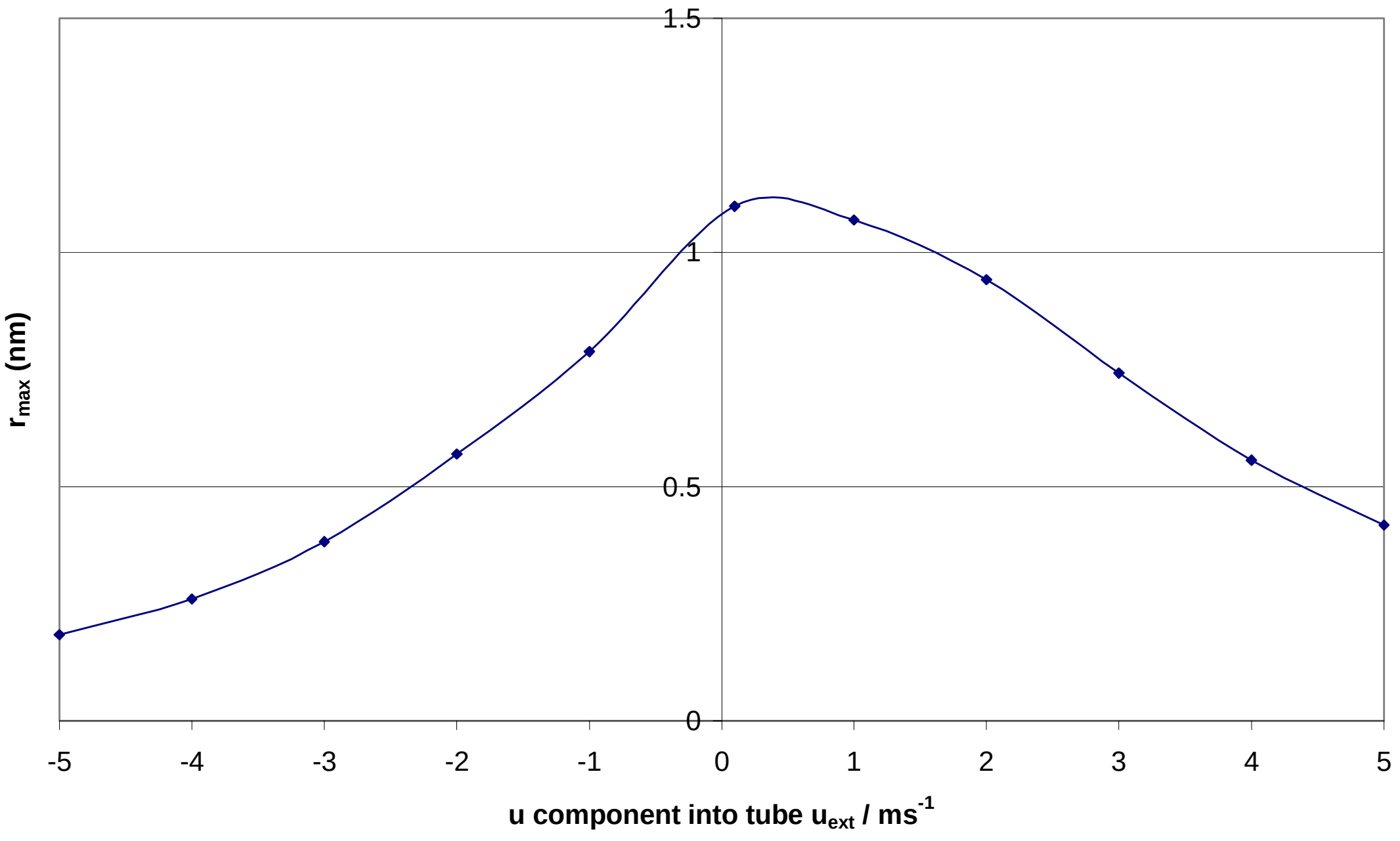

*Figure 7.3 Maximum radius of measured ions with the external ventilation of the Gerdien tube, at a typical bias voltage of 20 V. Flow in the tube is calculated from a polynomial fit to the external flow rate, and radius is calculated from a simple power law fit to approximate Tammet's (1995) expression.*

Figure 7.3 immediately shows that measurements made at very low flow rates permit intermediate ions to contribute to the conductivity, and must be discarded. It also shows that there is a significant variation in the radius of ion contributing to the measurement, even when small ions only are being measured. This will cause some uncertainty, but its magnitude will depend on the characteristics of the ion mobility spectrum. For example, there is a steep change in the radius of the ions measured between flows of 0 and -1 $ms^{-1}$. If there is no variation in the number of ions with radii between 1.1 and 0.8 nm, there will be no effect. If there is a significant difference in the number of ions at the different radii, relatively small fluctuations in the wind speed can cause a large error in the apparent determination of ion concentrations. This error can be called the "spectrum error", and its magnitude can speed, thus reducing the need for the PIMS correction described here, it reduces the instrument's collection efficiency (see Section 2.3).

be estimated from a long-term average negative ion spectrum (Hõrrak *et al*, 2000), which is shown in Figure 2.11. Assuming this as a reference spectrum, the errors in the measurement of negative ion number concentration for 1 $ms^{-1}$ step changes in external ventilation are shown below.

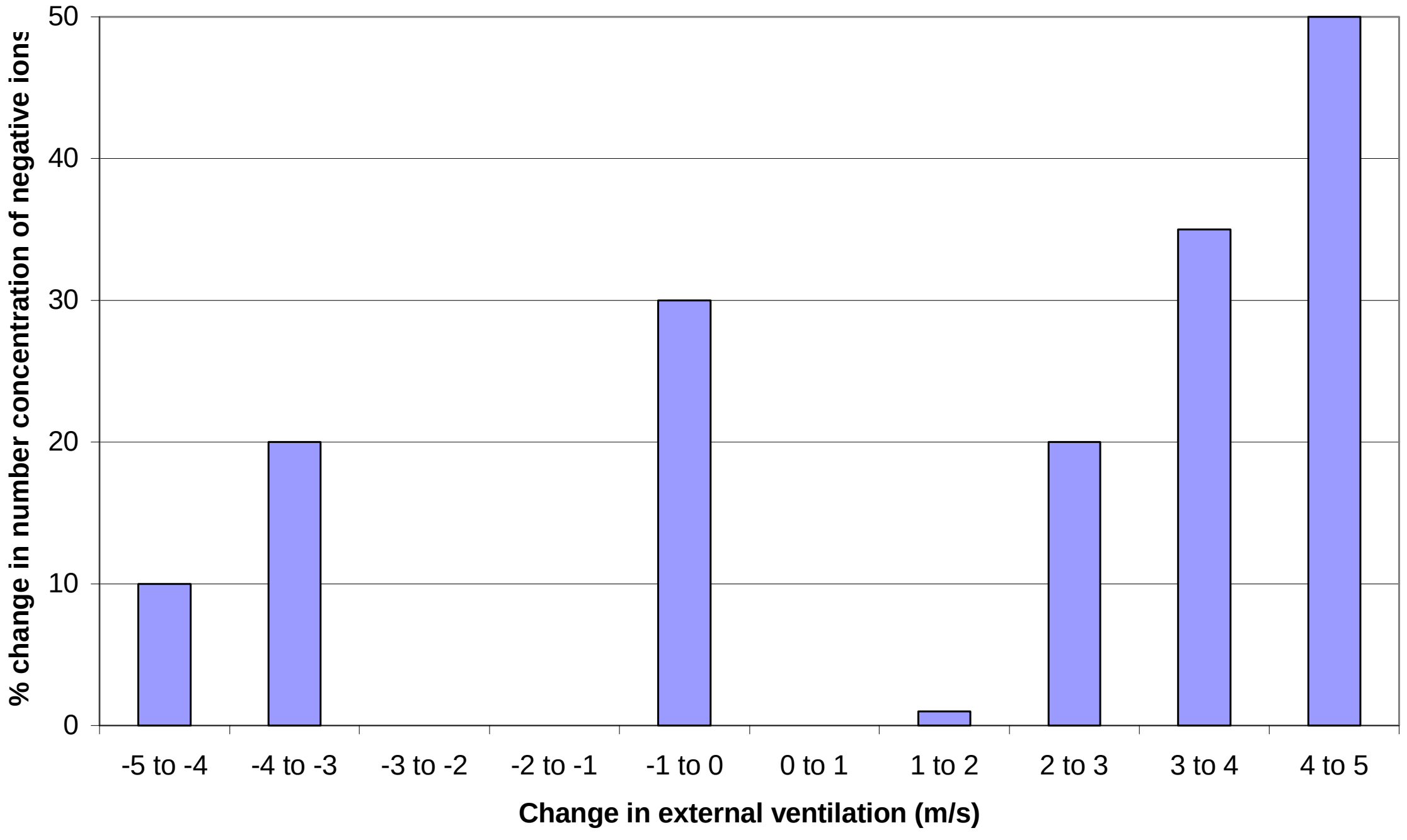

*Figure 7.4 Estimated spectrum error at a bias voltage of 20 V, from 1 $ms^{-1}$ changes in external ventilation.*

Figure 7.4 gives an estimate of the error in the measurement due to the non-uniform nature of the ion mobility spectrum. At higher wind speeds $r_{max}$ is lower, but the PIMS is measuring in a steeper part of the ion mobility distribution so there is a greater variation in the size of ion measured in a 1 $ms^{-1}$ range of wind speeds. The least desirable range of external wind components is from 0 to -1 $ms^{-1}$, because in this range the ions measured change from intermediate to small, and $r_{max}$ is quite sensitive to wind speed. Although the spectrum is flat from 0 – 1 $ms^{-1}$, intermediate ions are making up the majority of the conductivity so this data should be discounted. Conductivity measurements when there is high external ventilation (*i.e.* for the PIMSs orientation used in this Thesis' experiments, strong southerly winds) should be treated with caution, because errors are introduced as the number of ions being measured varies sharply with wind speed. The most favourable wind speeds for conductivity measurement are moderate ones in the range 1-3 ms$^{-1}$, which permit measurement of small ions in a relatively smooth part of the spectrum.

## 7.2 Physical differences between the modes

In the Current Mode, some preliminary averaging is carried out by the microcontroller before the sample is sent to the logging PC. In the Voltage Decay mode, conductivity is calculated from an exponential fit to 10 2 Hz samples. If a transient occurs, for example a charged aerosol particle advected onto one of the Gerdien electrodes, this will cause a perturbation to the electrical charge of the central electrode which is proportional to the charge of the particle. If this occurs whilst current measurements are being made, the transient effect is likely to be smoothed out by averaging, and contributes little to the current. In the Voltage Decay mode, the charge on the central electrode is logged at twice the speed, for half the time[67], so any transient is likely to be measured and sent serially to the PC where it may cause the fluctuations in the "decay" time series shown in Figure 6.6. A fluctuation of 1 V can be estimated to cause the voltage decay trace, which has a maximum range of 2.5 V, to deviate considerably from the expected exponential. This is equivalent to a charged aerosol particle carrying picoCoulombs and impacting on Gerdien electrodes (the Gerdien condenser typically has a capacitance of 10 pF). Aerosol which is not in steady-state, or is in a new region of ionisation, for example particles less than a few minutes old (Harrison, 1992) can be charged highly enough (*i.e.* 10 pC), to cause such perturbations. If one decay measurement lasts for 5 s, then a volume of about 5 x 10$^{-3}$ m$^3$ of air passes through the Gerdien, assuming a flow rate of 2 ms$^{-1}$. As 60 % of the voltage time series have a positive $<\Delta V/\Delta t>$, then this corresponds to about one failure case per 100 m$^3$ of air. This is equivalent to one charged particle per 5 m edge cube of air, which appears a possible explanation for the observed proportion of non-exponential decays. In addition to non-equilibriated aerosol particles, there are numerous other air-borne sources of highly charged particles, for example a bumble-bee (*bombus terrestris*) can carry 50 pC (Harrison, 1997b), and highly-charged particles can be produced by mechanical processes.

[67] The logging strategy has been described in detail in Section 5.6.3

## 7.3 Offset currents measured at zero bias voltage

When there is zero bias voltage $V_b$ across the Gerdien electrodes, the ionic current measured should be negligible, since there is no electric field in the tube. The only ions which will be measured are those which collide with the electrodes by chance due to mechanical motions. MacGorman and Rust (1998) refer to the process of correcting the ion current by subtracting this zero current as "*dynamic zeroing*". They suggest it is the best way to zero the Gerdien system since it accounts for ionic noise in the tube. However, this ionic noise signal is likely to be negligible compared to the contribution also present from larger charged particles. Aerosol particles easily carry several units of electronic charge (Clement and Harrison, 1992), and can be influenced by mechanical forces. So the current when $V_b = 0$ should give an estimate of the noise levels due to the presence of chance ionic and particulate, impaction onto the electrodes.

The current at the central electrode when $V_b = (0 \pm 1)$ V, which will be referred to here as $i_{V=0}$ was measured for 43 hours of fair weather using PIMS 2 at Reading University Meteorology Field Site from 2 - 4 May 2000. The same instrument also measured the currents at $V_b = +25.0$ and $V_b = -19.8$ V. The first observation was that $i_{V=0}$ was comparable in magnitude with currents measured at the higher voltages (here called $i_{V=-20}$ and $i_{V=+25}$), which suggests that it is possible for particles in the tube to cause a significant offset in the measurements. $i_{V=0}$ also showed a close agreement with both $i_{V=-20}$ and $i_{V=+25}$, and a time series of $i_{V=0}$ and $i_{V=+25}$ is plotted below to illustrate this. Since electrical forces seem to have no effect on the current measured, there must be a mechanical process acting in both cases.

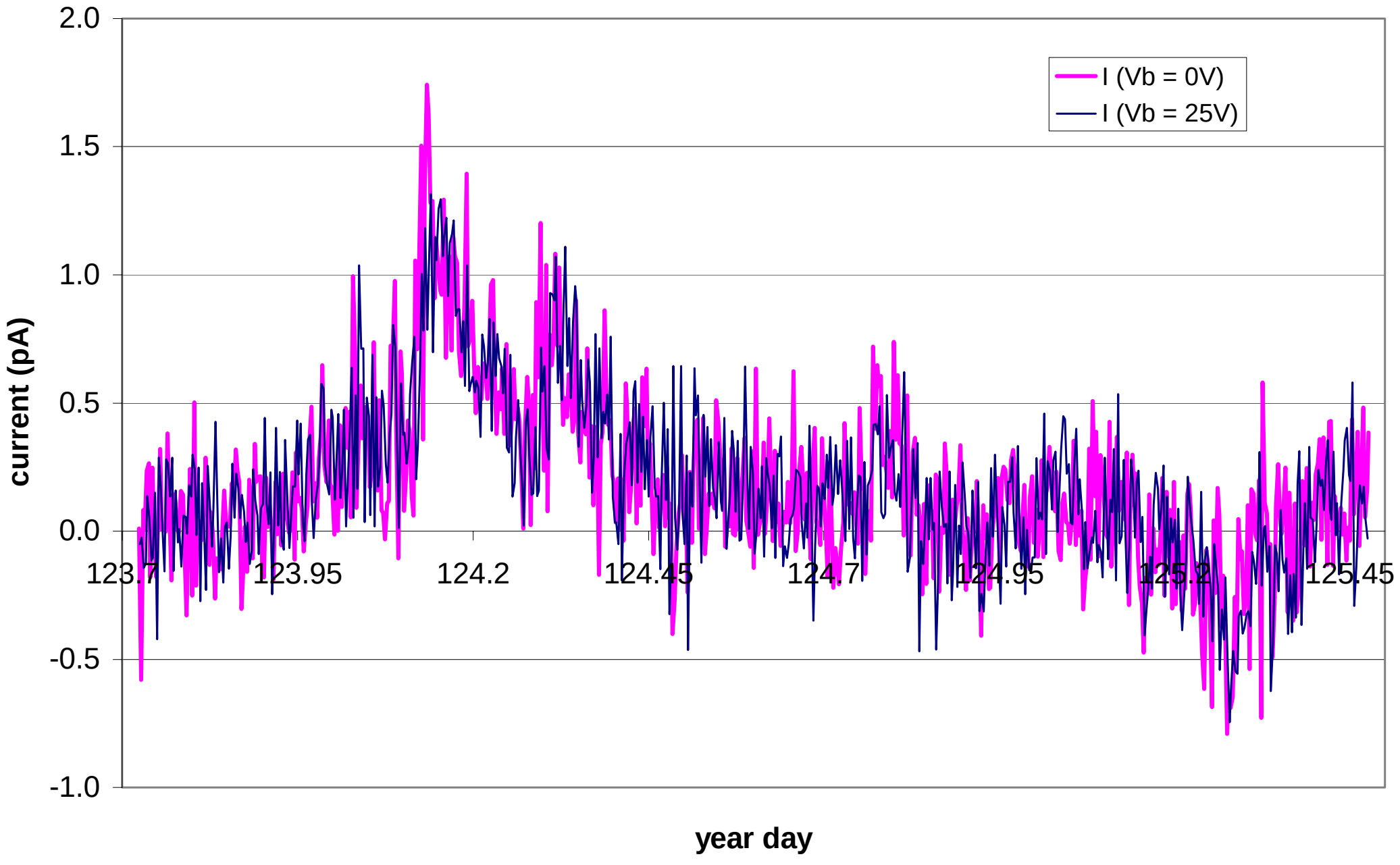

*Figure 7.5 Comparison of the currents measured with PIMS 2 when $V_b = 0$ and $V_b = 25$ V from 1730 2nd May 2000 to 1130 4th May 2000. The x-axis divisions are 6 hours.*

One approach which can be adopted is to investigate the distribution of the noise current. A random source of errors is Gaussian in distribution (Horowitz and Hill, 1994), and if there are other processes affecting $i_{V=0,}$ then the distribution will not be Gaussian. The distribution of $i_{V=0}$ for the same data set plotted in Figure 7.5 was calculated for two contrasting cases. The first case is when $i_{V=0}$ is clearly dominating the signal, defined by the presence of a negative current when the PIMS should only be responding to positive ions. The second case is the average distribution for the whole time series. The distributions are plotted in Figure 7.6 below. For the whole time series, the distribution is approximately Gaussian, although there is a slight bias in the tail towards positive currents. The peak is in the current bin of 0 – 0.2 pA, which shows that the $i_{V=0}$ signal is small for most of the time. The distribution when $i_{V=0}$ is large however, is much less symmetrical, with more frequent negative currents, and less frequent positive currents. The peak is broad and spans larger negative currents than the average distribution. This asymmetry shows that the noise has a non-Gaussian distribution (Berthouex and Brown, 1994) when it dominates the signal at larger bias voltages, compared to the mean distribution of $i_{V=0}$. This suggests that $i_{V=0}$ is not random noise but is symptomatic of a more ordered process, bringing in negative charge.

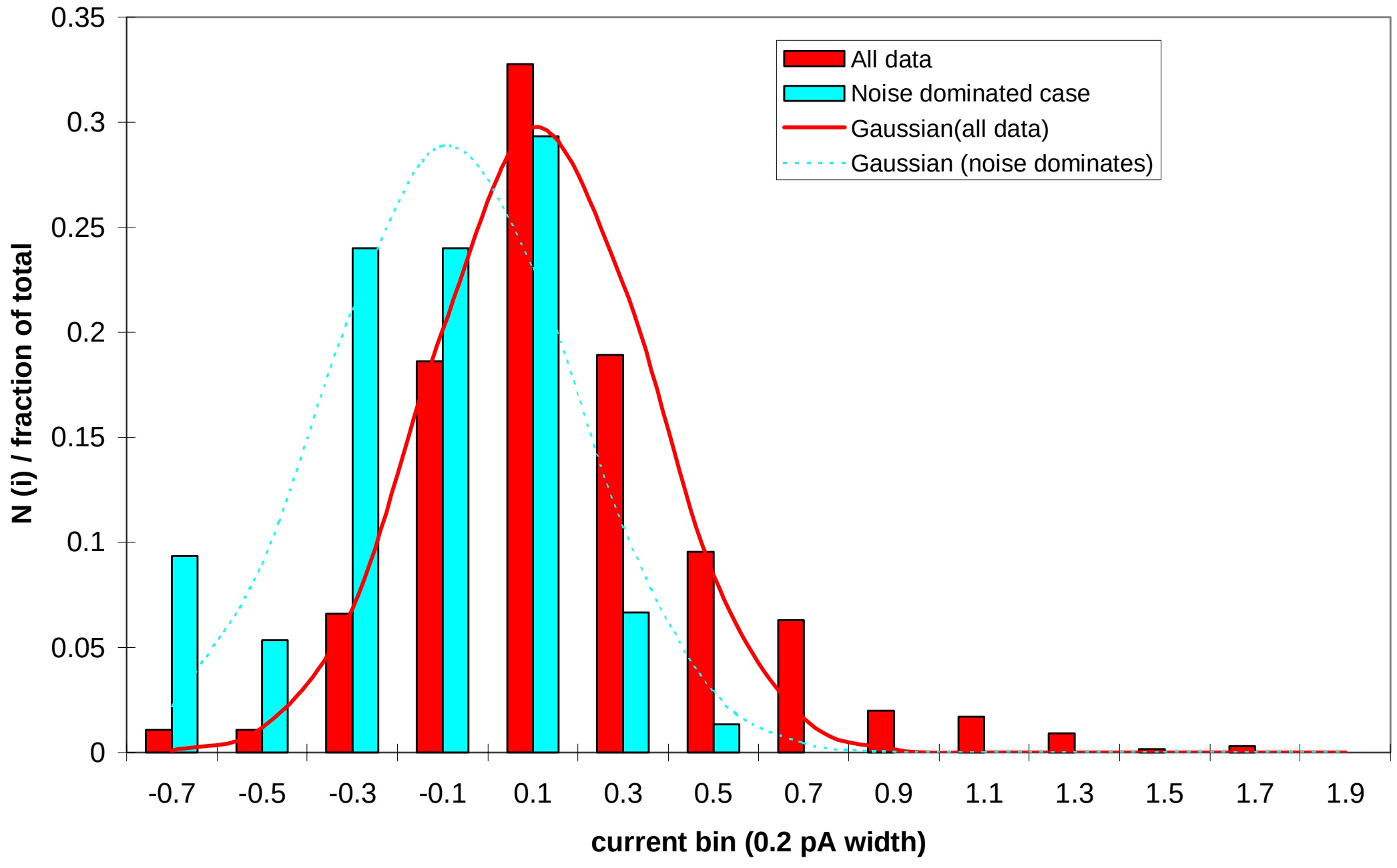

*Figure 7.6 Distribution of currents at $V_b = 0$ for the whole measurement period, and for the period when $i_{V=0}$ is dominating the ion signal. The relative number of occurrences of each current is plotted against the mean current of the 0.2 pA bin, e.g. N(i) at 0.5 pA represents the fraction of readings for which a current between 0.2 and 0.4 pA was measured. Gaussian fits to the distribution are also shown. Correlation coefficients to a Gaussian distribution for the whole data set and the case where noise dominates were 0.98 and 0.95 respectively.*

Having now established that $i_{V=0}$ can be an important source of noise in the measurements, the characteristics of the signal were investigated further. In the period when $i_{V=0}$ was considered to be dominating the signal, it showed a good agreement with the wind component into the PIMS, illustrated in Figure 7.7. This strongly implies that the origin of the dominant negative current is being blown into the tube by the wind.

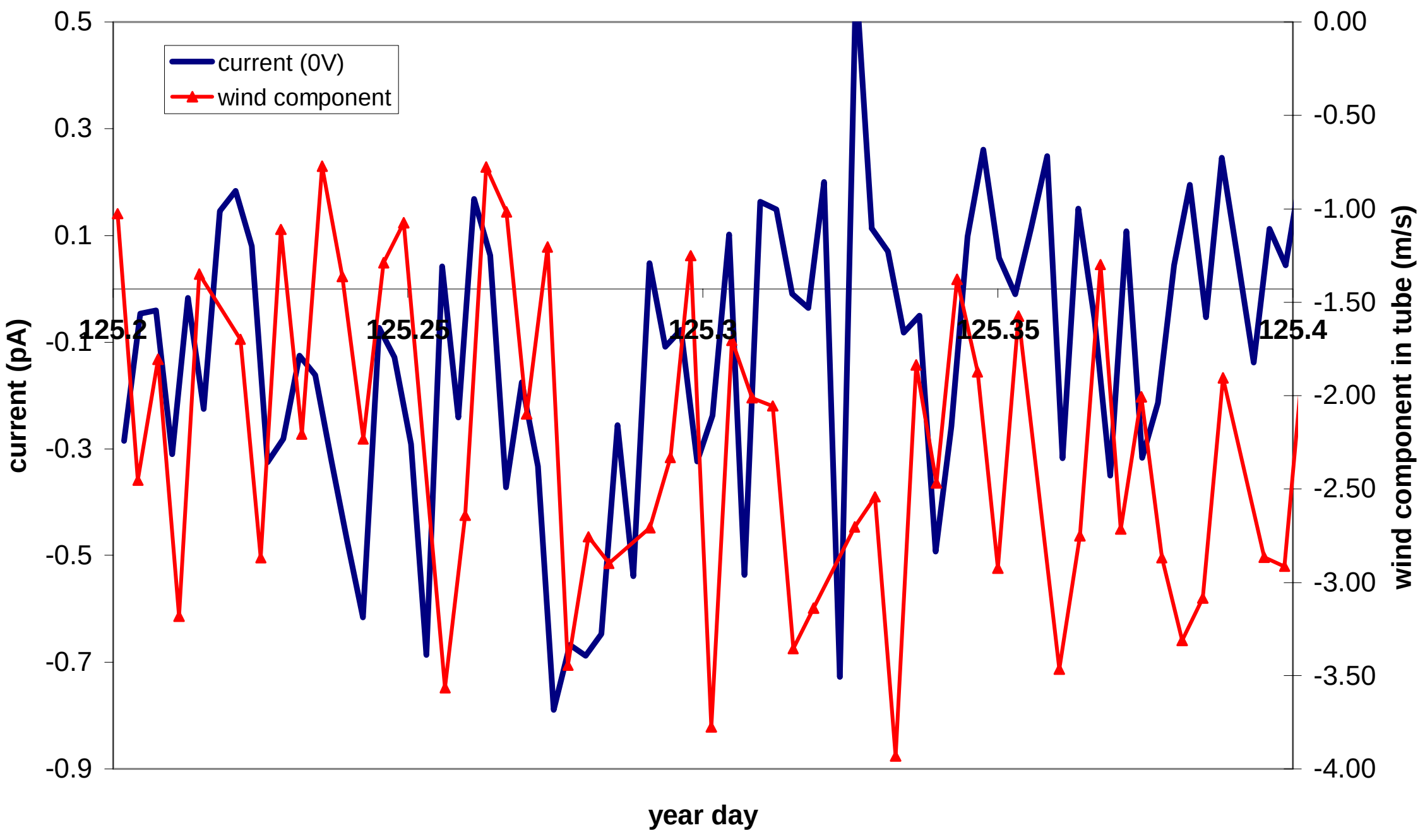

*Figure 7.7 A 4 hour 50 minute section of the $i_{V=0}$ time series (left hand axis) plotted against 5 minute averages of the wind component into the tube (right hand axis).*

A possible origin for the high currents observed at $V_b = 0$ is *particulate space charge*[68] drifting on to the Gerdien condenser electrodes. Space charge was measured at Reading University Meteorology Field Site with an Obolensky-type device[69], on 25th February 2000. Ion measurements with PIMS 1 at a bias voltage of – 30.7 V were also made. A time series of conductivity and the uncalibrated (negative) space charge meter output, plotted in Figure 7.8 below, indicates a positive relationship between the two quantities. This shows that the Gerdien condenser can respond to space charge for at least the measured period. However, the contribution of space charge to the conductivity measurement cannot be determined, as the exact response of the space charge instrument is not known. In order to ascertain the magnitude of the contribution, the partitioning of the space charge between aerosol and ions would also need to be understood. Since conductivity measurement by definition directly

[68] *Space charge* is a generic term for all the free charge per unit volume carried on particles in the air. It is partitioned between the charge carried by aerosol particles and small ions (Chalmers, 1967).

[69] The Obolensky (1925) space charge meter uses a filter to collect the space charge so that the current from it can be measured. Obolensky (1925) used cotton wool, but the instrument fabricated and deployed at Reading uses steel wool as the filter. The tube used was approximately 20 cm long and 2 cm in diameter, ventilated with the same fan as the PIMS at an approximate flow rate of 2 $ms^{-1}$. The

determines small ion concentrations, and therefore includes ionic space charge, if the majority of the space charge on 25th February was ionic, then it is not surprising that the two instruments agree. However, if the space charge were mainly charged aerosol, then the accord between the PIMS and space charge detector might indicate an error in the ion measurements due to the effect of particulate space charge. This is in itself inconclusive, but the relationship between the wind velocity and space charge can be used to estimate whether the space charge is primarily ionic or particulate. If the space charge is primarily particulate it will be influenced by the wind speed, but if it is ionic, there will be a poor relationship between space charge and wind speed because ions are not influenced by mechanical forces. Figure 7.9 is a plot of the space charge detector output with wind speed. It can be seen that there is no relationship between the two quantities, therefore it can be concluded that for this period, the majority of the space charge is ionic.

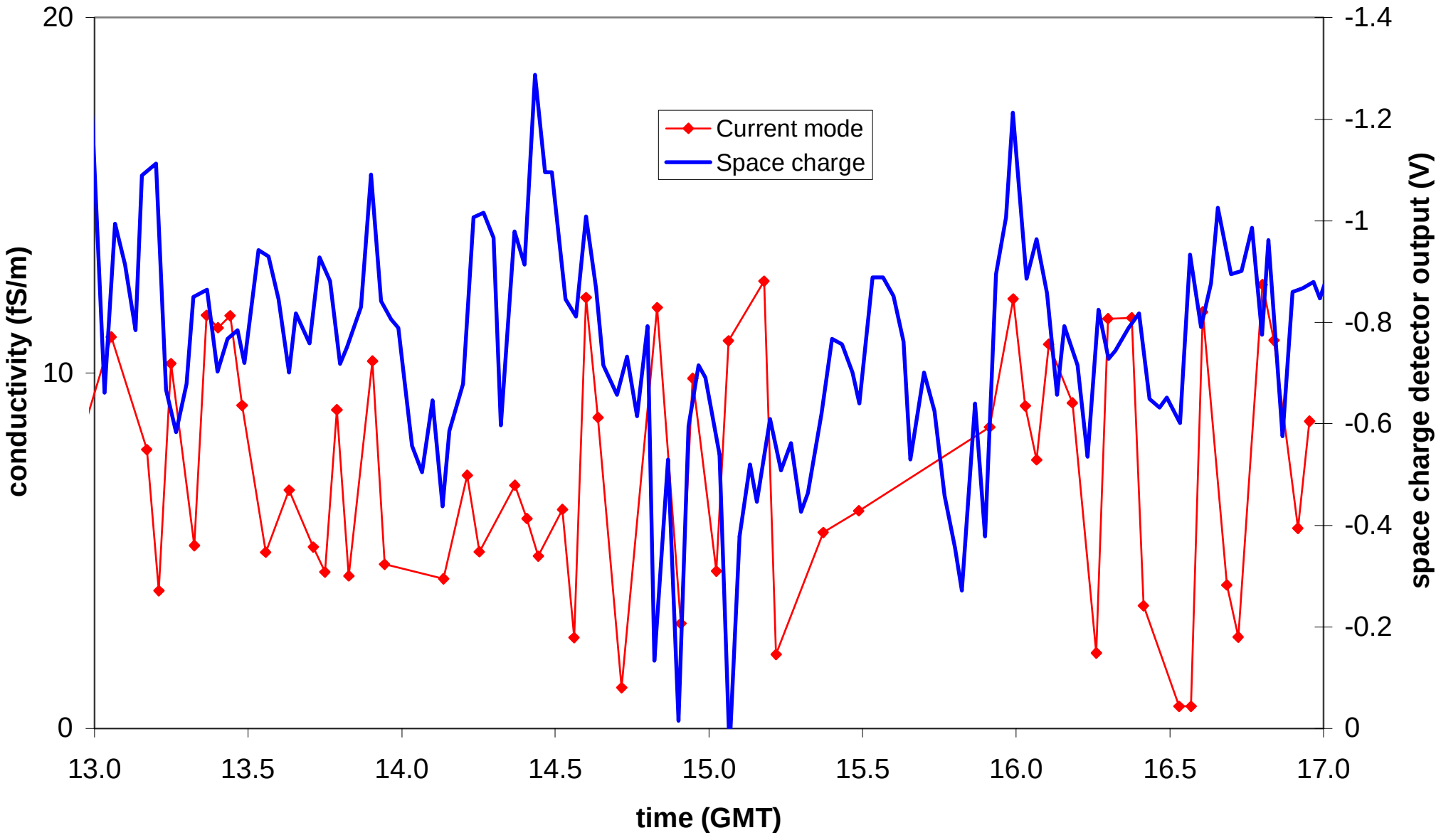

*Figure 7.8 Negative conductivity (left hand axis) measured by the Current Method (conductivity calculated by the two methods agreed) with PIMS 1 on 25th February 2000 at Reading. The uncalibrated space charge detector output, in volts but linearly related to the space charge, is on the right hand axis. The space charge meter was mounted on top of the PIMS with its inlet facing in the same direction.*

---

picoammeter used to measure the current was designed by Harrison (1997a) with an approximate calibration of 1 V/pA.

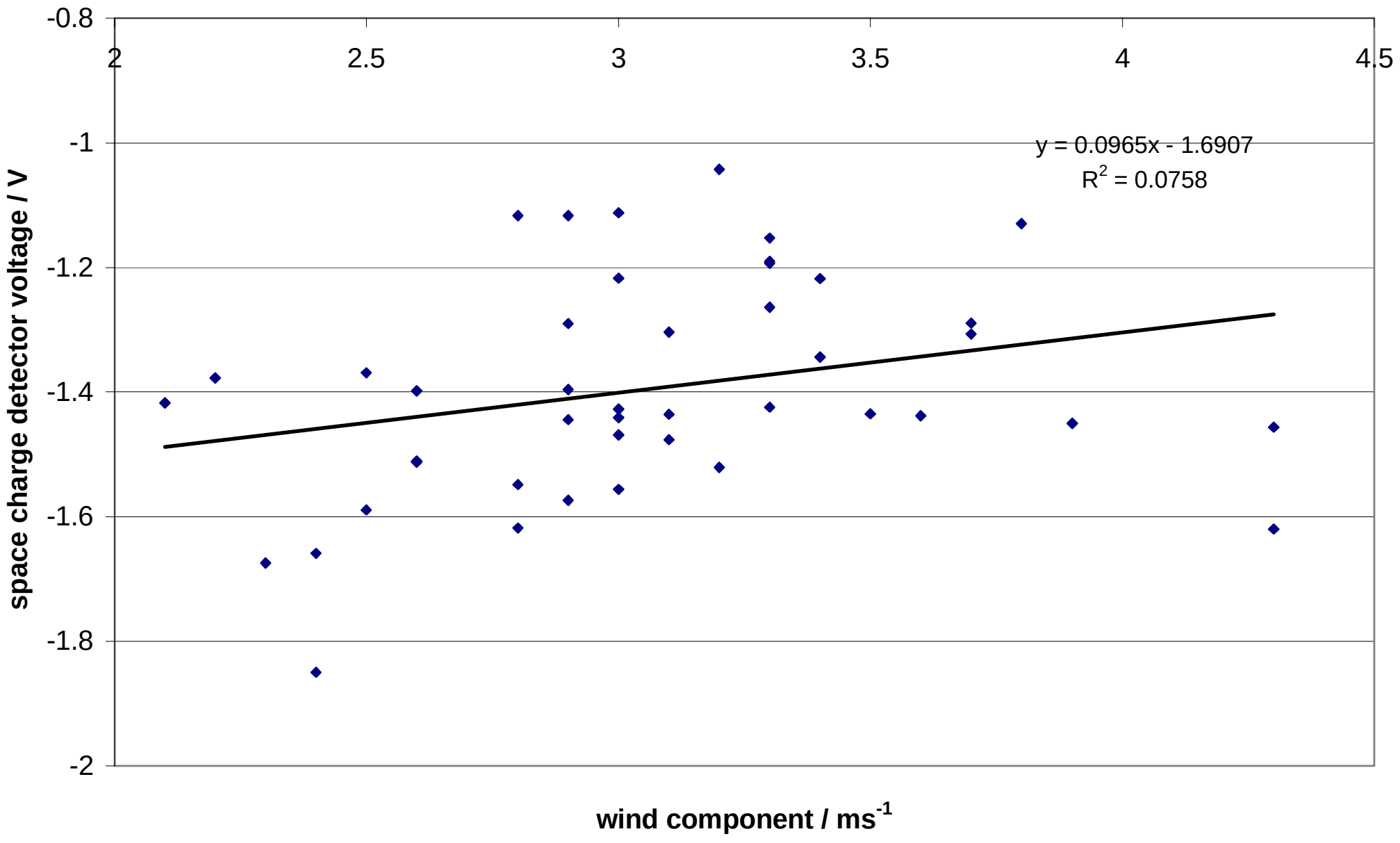

*Figure 7.9 Space charge detector voltage as a function of wind component into the space charge detector, for the same time period as Figure 7.8.*

For the measurements made on 2 - 4 May 2000 at $V_b = 0$, the overall distribution was close to Gaussian and the magnitude was small. This indicates that $i_{V=0}$ is randomly generated noise, the effect of which can therefore be reduced by averaging. However, the same signal as that recorded at $i_{V=0}$ sometimes dominates the measured ion signal at higher bias voltages, and can cause a large enough offset for the measured currents to be the opposite sign to the bias voltage. For these periods $i_{V=0}$ shows a close agreement with the external wind component. It is therefore almost certain that when the current is the opposite sign to $V_b$, the source of the noise causing it (measured at $V_b = 0$) is carried by the wind. On a different day (25th February 2000), space charge was also related to the ion signal (correlation between 15-minute averages was 0.23), in which there were no periods of large offset. Since the space charge was unrelated to the wind speed on this day (Figure 7.9), then it must have been primarily ionic in composition. In the example of the noise dominating the ion signal plotted in Figure 7.7, there is evidence that additional negatively charged particles blown in by the wind are the source of the noise. According to Kamra (1992), negative space charge is carried by the wind, so the results presented here confirm his hypothesis.

The problem of particulate space charge contributing to the conductivity measurement is likely to be worst when there are high aerosol concentrations, as there are lower ion concentrations due to attachment. The combination of a greater proportion of charged aerosol causing more noise, and the associated reduced ion signal will diminish the signal-to-noise ratio. Early radiosonde measurements of voltage decays outside the boundary layer (Hatakayema *et al*, 1958; Venkiteshwaran, 1958) had a high proportion of successful decays, unlike the PIMS measurements at ground level. This supports the concept that the PIMS measurements are perturbed by charged aerosol particles being blown into the Gerdien tube, as aerosol concentrations fall with height (Pruppacher and Klett, 1997). The effect of particulate space charge is thought to be the principal source of noise in the instrument.

In the ion measurements presented above and in Chapter 6, the approach to filtering was to prevent currents with a large non-ionic offset from being considered as small ion measurements. The screening procedure used in Section 6.7 was appropriate for this purpose, because it has excluded measurements when the noise currents were large. A better criterion for data filtering would be to exclude data where the space charge was known to be particulate, which would eliminate all ion measurements containing a contribution from particulate space charge.

## 7.4 Ionic sources of noise

### 7.4.1 Atmospheric ion fluctuations

The main source of ionic noise is fluctuations in the ion concentration itself - here called *ionisation noise*. The ion production rate, measured with a Geiger counter, shows high short-term variability (see Figure 6.7 for an example time series). Fluctuations in the ion concentration are damped by self-recombination and attachment to aerosol, but ion-aerosol timescales to reach steady-state concentrations are of order 100 s (Harrison, 1992). As it takes about one-eighth of a second for an ion to be sucked through the Gerdien at typical flow rates, the effects of a local transient in the ion production rate are likely to be detected before a local steady-state in the tube has been reached. Further evidence for the sensitivity of the instrument to ionic fluctuations has been observed directly by sealing the ends of the Gerdien condenser and measuring the current. On a chart recorder trace, small peaks can be seen, which must be attributed to individual ionisation events in the tube, because the sealed tube prevents new ions entering. These fluctuations probably originate from genuinely stochastic processes such as radioactive decay, and high-energy particles from cosmic rays. However, the sample high-frequency traces shown in Section 6.6.1, when the PIMS was compared to another Gerdien condenser, indicate that this sensitivity is a common feature of all Gerdien-like instruments.

### 7.4.2 Shot noise

Horowitz and Hill (1994) describe the shot noise induced in an electronic system by the discrete nature of the charge carriers. This could be thought directly relevant to the PIMS systems, where the current conceptually results from the effect of individual particles, rather than the continual "flow" of electrons perceived in higher current devices. Shot noise is more important at small currents, such as those measured from the PIMSs, because the fractional effect of the individual charge carriers is greater when the mean current is low. An expression for the current due to shot noise, $i_{shot}$, can be derived from Fourier analysis (Bleaney and Bleaney, 1985)

$$i_{shot} = (2eiB)^{1/2} \qquad \textit{Eq. 7.2}$$

where $i$ is the measured direct current, and $B$ is the bandwidth of the measurement. The average measured PIMSs current is about 300 fA. The bandwidth can be estimated from the RC time constant $\tau$ of the current amplifier

$$\tau = R_F C_F \qquad \textit{Eq. 7.3}$$

with $R_F$ = 1 TΩ and the parasitic capacitance across the resistor $C_F \sim$ 1 pF, Hence $\tau \sim$ 1 s and $B \sim$ 1 Hz. Therefore, $i_{shot} \sim$ 0.3 fA, which is negligible compared to other errors, such as the input bias current of the op-amp.

## 7.5 Conclusions

The effect of wind fluctuations is significant, because at low wind speeds intermediate ions contribute directly to the conductivity measurement. Further screening of data has been shown to remove this effect, improving the agreement between the two measurement modes so that they match almost exactly at night. The poorer agreement between the two modes in the afternoons suggests that the wind is fluctuating sufficiently rapidly for five-minute averages to be an inadequate basis for comparison. This is supported by the evidence presented here, that the external ventilation varies significantly on temporal and length scales comparable with the PIMS measurements. Failure cases have been accounted for by aerosol effects, except for a slight negative offset in the current measurements. This offset may also help to explain the differences in the mean values between the two methods.

Additional sources of uncertainty have been investigated, and drift of particulate space charge onto the Gerdien electrodes appears to be the major source of noise in the data. Fluctuations in the wind speed causing variability in the fraction of the spectrum being measured are also significant at higher flow rates. The PIMS instruments are both self-consistent and reproduce absolute conductivity measurements made in urban air. However, significant filtering is required to remove poor data, when intermediate ions are contributing to the conductivity, and when leakage currents are excessive. Based on the investigations in this chapter, it is now possible to define a detailed set of operating criteria based on instrumental and physical considerations. These are summarised in Table 7.2.

| **Measurement mode** | **Instrumental filtering criteria** | **Physical filtering criteria** |
|---|---|---|
| Current Measurement | $\sigma < \sigma_{max}$ (calculated from recombination limit)<br>$\sigma > -2i_L$ | $u_{ext} > +1$ ms$^{-1}$ or<br>$u_{ext} > -0.5$ ms$^{-1}$<br>Particulate space charge does not contribute to the measurement<br>Favourable $u_{ext}$ to avoid spectrum error: 1-3 ms$^{-1}$ in direction opposing fan flow, or $0-2$ ms$^{-1}$ when aligned with fan flow |
| Voltage Decay | $\left\langle \frac{\Delta V}{\Delta t} \right\rangle < 0$<br>$\sigma < \sigma_{max}$ (calculated from recombination limit) | |

*Table 7.2 Summary of filtering criteria for PIMS conductivity data.*

*"It was in the link between laboratory and sky that Wilson's identification of CN with electrical ions ran into trouble." (Galison, on C.T.R Wilson's article* Atmospheric Electricity*, Nature* ***68****, 102-4, 1903)*

# 8 Observations of ion spectra and ion-aerosol interactions

Atmospheric observations made with the PIMS are described in this chapter, with the aim of showing that the PIMS is a novel instrument capable of obtaining new atmospheric results. A technique to obtain ion mobility spectra by inverting voltage decay data is explained in Section 8.2. Spectra are also calculated from both measurement modes (Sections 8.3.1 and 8.3.2); the two spectra are compared with each other, and another measured spectrum in Section 8.3.3. Ion measurements are analysed with respect to aerosol number concentration in Section 8.4.

## 8.1 Ion spectra

An ion mobility spectrum was defined in Chapter 2 as the distribution of ion number concentration $n(\mu)$ across the range of mobilities $\mu$. It can be readily converted into a size spectrum using the expression derived by Tammet (1995), or by the empirical power law fit to the ion-mobility data (Hõrrak *et al*, 1999) which was described in Chapter 3.

Consider an ion population with a mobility distribution $n(\mu)$: the total number of ions $N$ is given by

$$N = \int_0^\infty \frac{dn}{d\mu} d\mu \qquad \text{Eq. 8.1.}$$

The simplest possible mobility distribution is when all ions have the same mobility, in which case the mobility spectrum is unimodal, as shown in Figure 8.1.

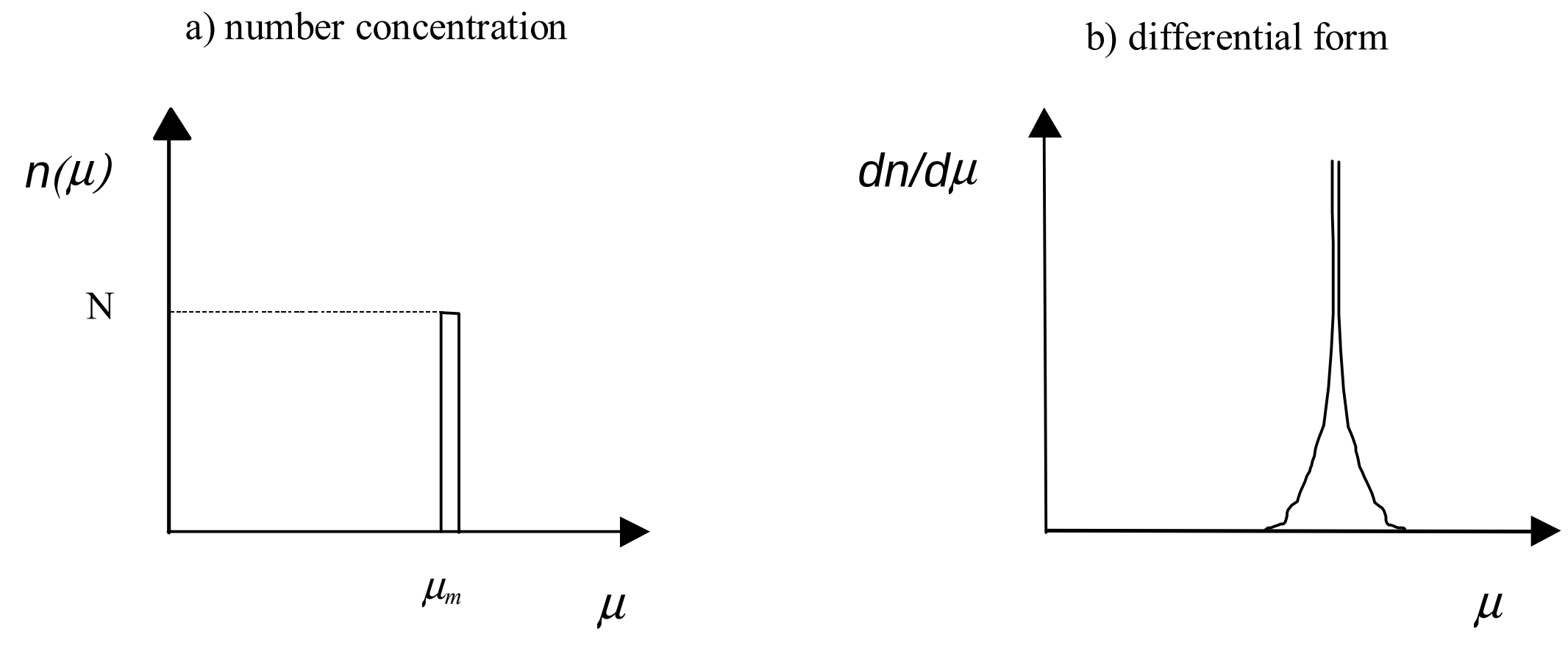

*Figure 8.1 a) The simplest ion mobility spectrum: unimodal, when all ions have the same mobility. b) is the same spectrum expressed in differential form.*

The total number of ions $N$ is given by the area under the spectrum. For the unimodal spectrum, the ionic conductivity is exactly

$$\sigma = eN\mu \qquad \textit{Eq. 8.2.}$$

However, when the number of ions with each mobility changes, the conductivity has to be expressed in integral form. As conductivity by definition is related to the concentration of all atmospheric small ions, the limits of this equation must be chosen to exclude intermediate ions, and therefore have a lower limit of 0.05 $cm^2V^{-1}s^{-1}$. The upper limit is infinity because an infinitely small ion will have infinitely large mobility.

$$\sigma = e \int_{0.05cm^2V^{-1}s^{-1}}^{\infty} n(\mu)d\mu \qquad \textit{Eq. 8.3.}$$

The assumption of an invariant ion mobility spectrum, leading to Eq. 8.2 above, has long been used to estimate ion concentrations from conductivity measurements made with Gerdiens (*e.g.* Shimo *et al*, 1972; Dhanorkar and Kamra, 1993). The assumption has been more fundamentally deployed in the derivation of conductivity from time decay measurements from a Gerdien condenser in the Voltage Decay mode. This expression,

$$\sigma = \frac{\varepsilon_0}{\tau} \qquad \textit{Eq. 8.4}$$

relies on the precept that the conductivity remains constant throughout the measurement period.

Methods to measure and calculate ion mobility spectra have been outlined in Chapter 2, but in summary, they are measured by varying the mobility of ions contributing to the measurement, and then counting the number at the selected mobility. This has typically been effected by varying the distance the ions have to travel to reach the detector, or by varying the electric field the ions are moving in so that one mobility range is selected. Results from these measurements have shown that there is a non-uniform mobility spectrum (*e.g.* Misaki, 1964; Mohnen, 1974; Hõrrak *et al* 2000; some examples of mobility spectra are given in Chapter 2).

## 8.2 Effect of ion mobility spectrum on voltage decays

In the Voltage Decay mode, the changing voltage across the Gerdien electrodes varies the electric field strength in the tube, which has the effect of scanning through a range of mobilities during the measurement. This means that the assumption that a constant number of ions contribute to the measurement at all times is invalidated when there is a non-uniform mobility spectrum. In the special case when all ions have the same mobility, the theory leading to Eq. 8.4 is correct. However, when there is a varying spectrum, the decay cannot be an ideal exponential; its degree of deviation from the exponential depends on the characteristics of the spectrum. Therefore, every voltage decay time series holds spectral information, if it has sufficient resolution, although there may not be a solution to the inversion problem in general.

This is analogous to the classic Gerdien test, the Ohmic response. It is well-known that if a Gerdien is responding correctly to the presence of air ions, a plot of output current against bias voltage will be linear. In reality, varying the bias voltage varies the critical mobility throughout the spectrum, so that if there are species of ions with different mobilities, the plot will deviate slightly from a linear one. However, when the spectrum is unimodal, a Gerdien $i$-$V_b$ plot should be an ideal straight line. The conceptualisation of spectral information concealed within the experimental deviation from the most naïve theoretical predictions has previously been exploited by Nolan (1920) to calculate ion spectra from the "Ohmic" response.

### 8.2.1 Voltage decay for a parameterised mobility spectrum

It is possible to derive an expression which slightly modifies the theory of calculating conductivity from an exponential decay of the voltage across a capacitor (given in Section 2.2.2) to take account of a non-uniform ion mobility spectrum. This starts with the exponential assumption applied in Chapter 6 to compute conductivity from PIMS time series measured in the Voltage Decay mode. Rearrangement of Eq. 6.4 gives the voltage at the central electrode $V_t$ at a time $t$ after the decay is started (by disconnecting the voltage charging up the electrodes), which is related to the conductivity $\sigma$ and the initial voltage $V_0$ by

$$t = -\frac{\varepsilon_0}{\sigma}\ln\frac{V_t}{V_0} \qquad \textit{Eq. 8.5.}$$

When the mobility spectrum is non-uniform, Eq. 8.2 can be substituted for $\sigma$ in Eq. 8.5. The limits of the integral in Eq. 8.3. are equivalent to the mobility range of the measurement; since critical mobility $\mu_c$ and voltage $V$ are inversely related by

$$\mu_c = \frac{ku}{V} \qquad \textit{Eq. 8.6}$$

(see Section 3.1.2) the maximum and minimum critical mobilities $\mu_{ct}$ and $\mu_{c0}$ are measured at $V_t$ and $V_0$ respectively:

$$t = -\frac{\varepsilon_0}{e\int_{\mu_{c0}}^{\mu_{ct}} n(\mu)d\mu}\ln\frac{V_t}{V_0} \qquad \textit{Eq. 8.7.}$$

If a prescribed spectrum is substituted into Eq. 8.7., the form of the voltage decay may be determined from it by calculating $t$ for a set of values of $V_t$. Eq. 8.7. can also be written as Eq. 8.8

$$t = -\frac{\varepsilon_0}{e}\int_{V_0}^{V_t}\frac{1}{n\mu}\frac{dV}{V} \qquad \textit{Eq. 8.8,}$$

which verifies self-consistency with the existing theory, since if the mobility spectrum is unimodal, then inserting $\sigma$ from Eq. 8.2 reduces it to the standard expression for calculating conductivity from exponential decay (Eq. 8.5.). Hence, it is possible to predict the shape of the voltage decay, given a prescribed initial ion spectrum by integrating Eq. 8.7.

To investigate this modification to the exponential theory, a reference ion spectrum was required to substitute in Eq. 8.7.. A Gaussian spectrum was fitted to an existing ion spectrum from 1700 hours of Estonian ion data (Tammet *et al,* 1992). The integral under the Gaussian curve gave a total ion concentration of $N = 127$ cm$^{-3}$, for ions measured in the mobility range 0.36 – 2.84 cm$^2$V$^{-1}$s$^{-1}$. The *decay integral* Eq. 8.7. was solved to give a derived voltage decay time series for the Estonian reference spectrum, using a Turbo Pascal program, *DECINT*[70]. Eq. 8.7. was also solved for a unimodal mobility spectrum, which gave the ideal exponential decay. The characteristics of the unimodal spectrum were prescribed with typical concentrations and mobilities, $N = 100$ cm$^{-3}$, and $\mu = 1$ cm$^2$V$^{-1}$s$^{-1}$ (*e.g.* Chalmers, 1967). A sample voltage range of 10 - 0V was used to simulate the decay. The voltage decay trace calculated when there is a unimodal mobility spectrum was exponential. Therefore a unimodal ion spectrum can be inverted to give the expected exponential decay of potential across the Gerdien electrodes.

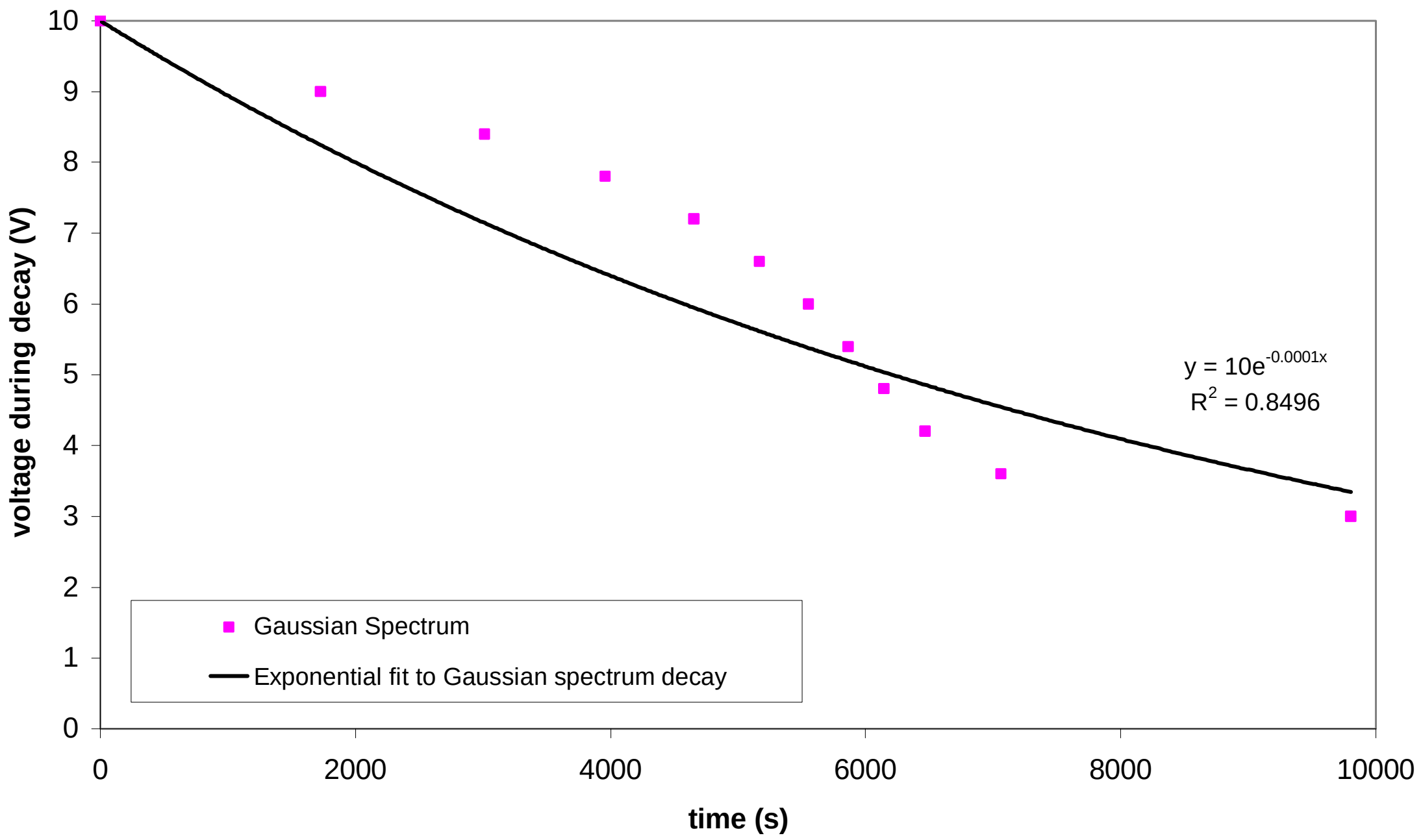

*Figure 8.2 Voltage decay trace derived from Eq. 8.7. assuming a Gaussian spectrum based on measured positive ion data, compared with the exponential fit which would be expected from conventional theory.*

[70] The **dec**ay **int**egral program DECINT (written by R.G. Harrison) numerically integrated Eq. 8.7. using Simpson's Rule for an ion mobility spectrum of any form, which has been numerically parameterised. It evaluates the integral between two values of voltage, which is then repeated

As predicted above, the voltage decay trace calculated from a real (modelled by a Gaussian) ion spectrum is a poorer fit to an exponential decay, though an exponential explains 85 % of the variance in the trace. This shows that when there is a non-uniform mobility spectrum, the voltage decay method of calculating conductivity does not give the ideal exponential. Instead, measurements of voltage against time are clearly able to generate a *quasi-exponential decay.* The ideal exponential decay is of course a special case, which is only expected when all ions have the same mobility and on which conventional theory depends.

### 8.2.2 Inversion problem

Inverting a set of voltage decay measurements to get an ion spectrum numerically is in general impossible, but can be approximated by suitable assumptions. If it is assumed that the characteristics of the ion mobility spectrum vary negligibly in a small mobility range, *i.e.* that the ion spectrum is a superposition of many unimodal distributions, the problem can be simplified. The technique of fitting an exponential and calculating conductivity from Eq. 8.5., and then deducing ion concentration from Eq. 8.2 would then be applicable for small segments of each voltage decay time series (if they are sufficiently small). This *piecewise inversion* approach can be tested by applying the technique to the theoretically derived voltage decay shown above in Figure 8.2; if this assumption is valid, a spectrum calculated in this way should match the original ion spectrum from which the voltage decay was obtained.

A quasi-exponential decay found using *DECINT* with a Gaussian input spectrum was broken down into sets of three consecutive values. If there are ten values of $V_t$ and $t$ from $(V_1,t_1)$ to $(V_{10},t_{10})$, then the first set of three is $(V_{1\text{-}3},t_{1\text{-}3})$, the second $(V_{2\text{-}4},t_{2\text{-}4})$, and so on up to $(V_{10},t_{10})$. An exponential curve fit was then applied to each set of three $(V,t)$ points to calculate the conductivity for this part of the decay. The voltage at the middle of the three points was used to calculate the critical mobility, and from this ion number concentration. The ion spectrum obtained by this technique was then compared to the measured Estonian ion spectrum, shown in Figure 8.3.

---

sequentially. The decay time is found between each pair of voltages and then the values plotted inversely (*i.e.* as voltage as a function of time).

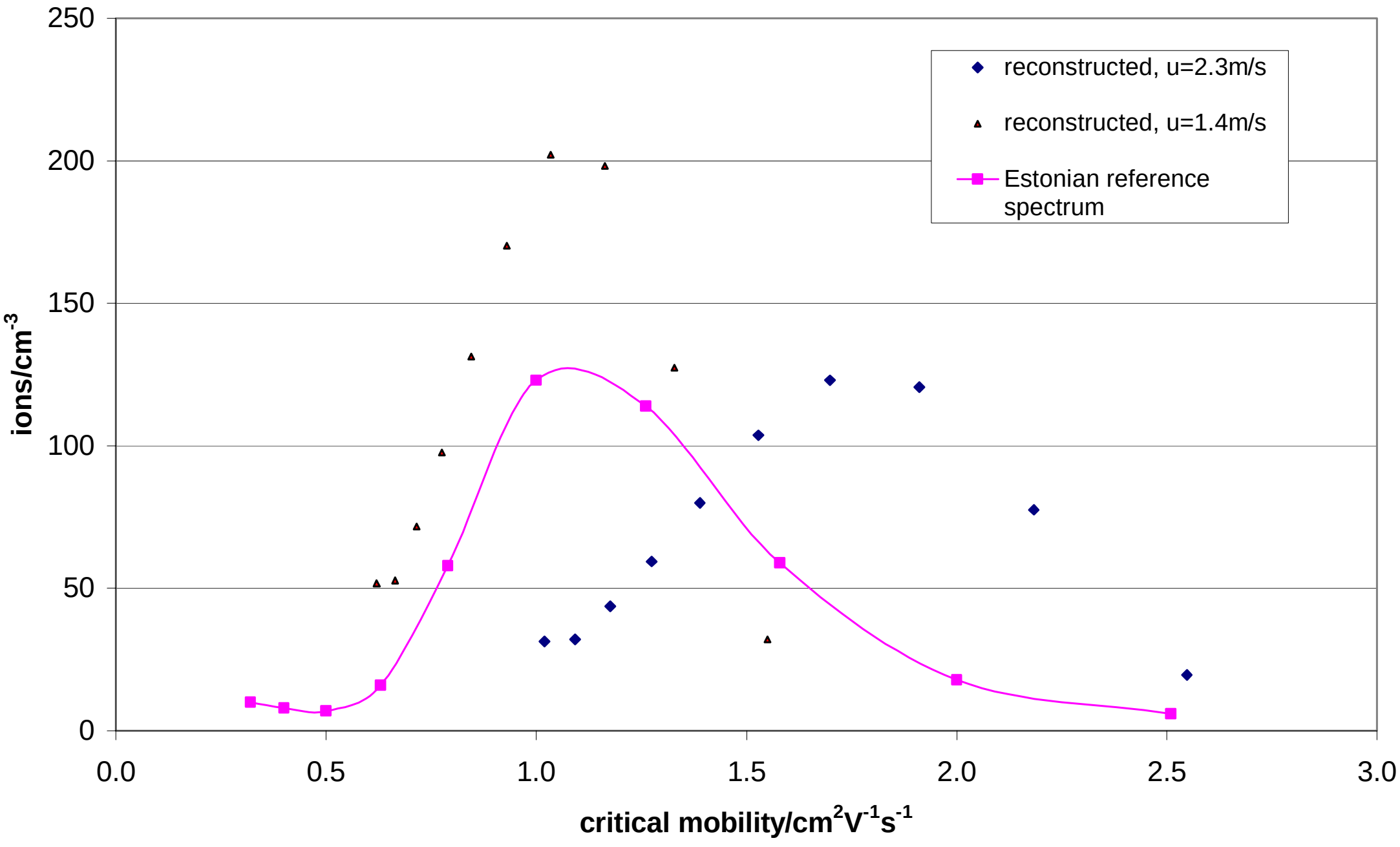

*Figure 8.3 Ion mobility spectra reconstructed from voltage decay time series derived from the reference Estonian ion spectrum. The shape of the reconstructed spectrum depends on the flow speed chosen; two typical speeds are chosen here which best represent the measured spectrum in width and height (u would normally be known from measurements).*

The total number of ions $N$ calculated from integrating a Gaussian fit to the data points was exactly equal between the reconstructed and measured spectra. However, the form of the decay depended on the flow rate $u$, which had to be arbitrarily chosen to calculate the critical mobility and ion concentration. The relationship between $\mu_c$, and $u$ (Eq. 8.2) meant that selecting a faster flow resulted in fewer ions at a higher mobility. In Figure 8.3, spectra at two wind speeds which produced the curve most closely matching the measured spectrum in magnitude and mobility have been chosen.

This has shown that the total ion concentration can be reproduced by fitting a series of exponential decays to small sections of a non-exponential set of measurements of voltage and time. This can be considered a piecewise inversion of Eq. 8.7. The ion spectrum can also be derived in the same way; but the parameters depend on the wind speed. It is also necessary to assume that the ion spectrum changes smoothly and slowly with $\mu$. In the hypothetical test case above, there is no reason to prefer any wind speed over another, so a range of values may be used. However, when deriving ion spectra from real voltage decay data, measured wind speeds will be used so that the critical mobilities can be determined and facilitate comparisons with spectra calculated conventionally from Current Mode measurements.

## 8.3 Measured atmospheric ion spectra

A data set from 1000-1700 on 18$^{th}$ June 2000 was chosen from which to calculate an average ion spectrum by two methods from the two PIMS measurement modes. This day was selected because of the fair weather, no cloud and moderate wind speeds, so that the atmospheric electrical conditions would remain approximately constant over the averaging period. Two PIMS instruments, 1 and 2, were also returning a high proportion of valid data from both modes for the whole period.

### 8.3.1 Current Measurement method

In the Current Measurement mode, the critical mobility is varied by changing the bias voltage across the Gerdien (Eq. 8.6). Since a discrete set of measurements at each mobility is obtained by this method, spectra have to be calculated by differencing the measurements made at two adjacent bias voltages. There also has to be some delay between the measurements made at different bias voltages to allow the electrometer time to recover from the switching-induced transients (see Section 5.6.3). By necessity therefore, this technique requires some temporal or spatial separation of the measurements to be differenced, and associated assumptions about temporal or spatial consistency of the ion spectrum. Differencing measurements made with the same sampling instrument several minutes apart was thought to be inadvisable, given the large fluctuations in ion production rate and wind speed that occur on relatively short timescales (see Chapter 6). Instead, measurements from two PIMSs were used. These were not exactly simultaneous, (despite the efforts to synchronise the instruments described in Section 5.7.3) because of non-linear drift in the individual microcontrollers' timing circuits. Data were only selected for differencing when they were close enough together for the traces of conductivity to be steady to a timescale longer than the differencing timescale, *i.e.* when it can be assumed that the ion spectrum is steady. For this data set, a conservative maximum temporal separation of 2 minutes was selected, as the conductivity trace remained steady on about a 5-minute timescale. (A closer minimum separation would be preferable, but this would reject too much of the data to average the spectra.) The small volume flow rate into the PIMSs (of order $10^{-4}$ $m^3s^{-1}$) also justifies this temporal differencing strategy; the mean electrical conditions in the locality of the PIMSs in fair weather are known not to be changing rapidly, and each instrument only samples a small amount of the air in the vicinity ($10^{-3}$ $m^3$) on the timescales chosen.

$n(\mu)$ is calculated by assuming the ion population remains constant over the time scale of the measurements, so the difference in conductivity between the two measurements is proportional to the number of ions between the two critical mobilities. Data were filtered according to the criteria set in Chapter 7, and the mean critical mobility was calculated from the mean flow in the tube for the measurement period, and the bias voltage. The bias voltage sequences are shown in Table 8.1. The measurements at positive bias voltage from PIMS 2 have not been included in calculation of the spectrum.

| PIMS | $V_{b1}$ | $V_{b2}$ | $V_{b3}$ | $V_{b4}$ | $V_{b5}$ | $V_{b6}$ |
|---|---|---|---|---|---|---|
| 1 | -27.6 | -19.9 | -16.6 | -10.2 | -8.4 | -6.7 |
| 2 | -19.8 | -9.9 | +9.9 | +19.8 | - | - |

*Table 8.1 Sequence of bias voltages used on 18th June 2000, PIMS 1 and 2.*

Eq. 8.9 below was used to calculate the number of ions associated with a mean mobility $<\mu>$, and these points were averaged across the whole measurement period

$$n(\mu) \approx \frac{\Delta\sigma}{e < \mu >} \qquad \textit{Eq. 8.9.}$$

The mean critical mobility was calculated from Eq. 3.1 using the mean of the two bias voltages, and the mean wind component in the tube over the averaging period. Uncertainties were computed as in Section 8.3.2 above, from the standard error of the mean of the flow and the ion concentration. Results are discussed in Section 8.3.3.

### 8.3.2 Voltage Decay method

The ion concentration at a given critical mobility was calculated by deriving the conductivity by exponential fitting to adjacent voltage decay data in each decay. The method was as described above for the reference spectrum, moving sequentially through the time series and fitting exponentials to three points at a time. Hence, each voltage decay time series of ten data points yielded eight spectral points. The flow was calculated using the five-minute averages of measured wind speed and direction, with the actual flow within the tube derived as described in Section 3.2.1. The critical mobility was calculated from Eq. 8.6, using the five-minute average flow and the central voltage of the three points. The voltage across the Gerdien electrodes at the start of the decay was –22.3 V.

The filtering criteria described at the end of Chapter 7 were applied to each three-point data set. After these filtering criteria had been applied, $n(\mu)$ was calculated from Eq. 8.2. The mobilities varied depending on the wind speed at the time of the decay, so they were averaged over the same seven-hour period as well as the ion concentrations. Uncertainties in the ion concentrations and the average mobility were calculated from the standard error of the means of both quantities.

### 8.3.3 Comparison of ion spectra obtained by the two techniques

Negative ion mobility spectra calculated by the two methods described in Sections 8.3.1 and 8.3.2, are shown in Figure 8.5 below.

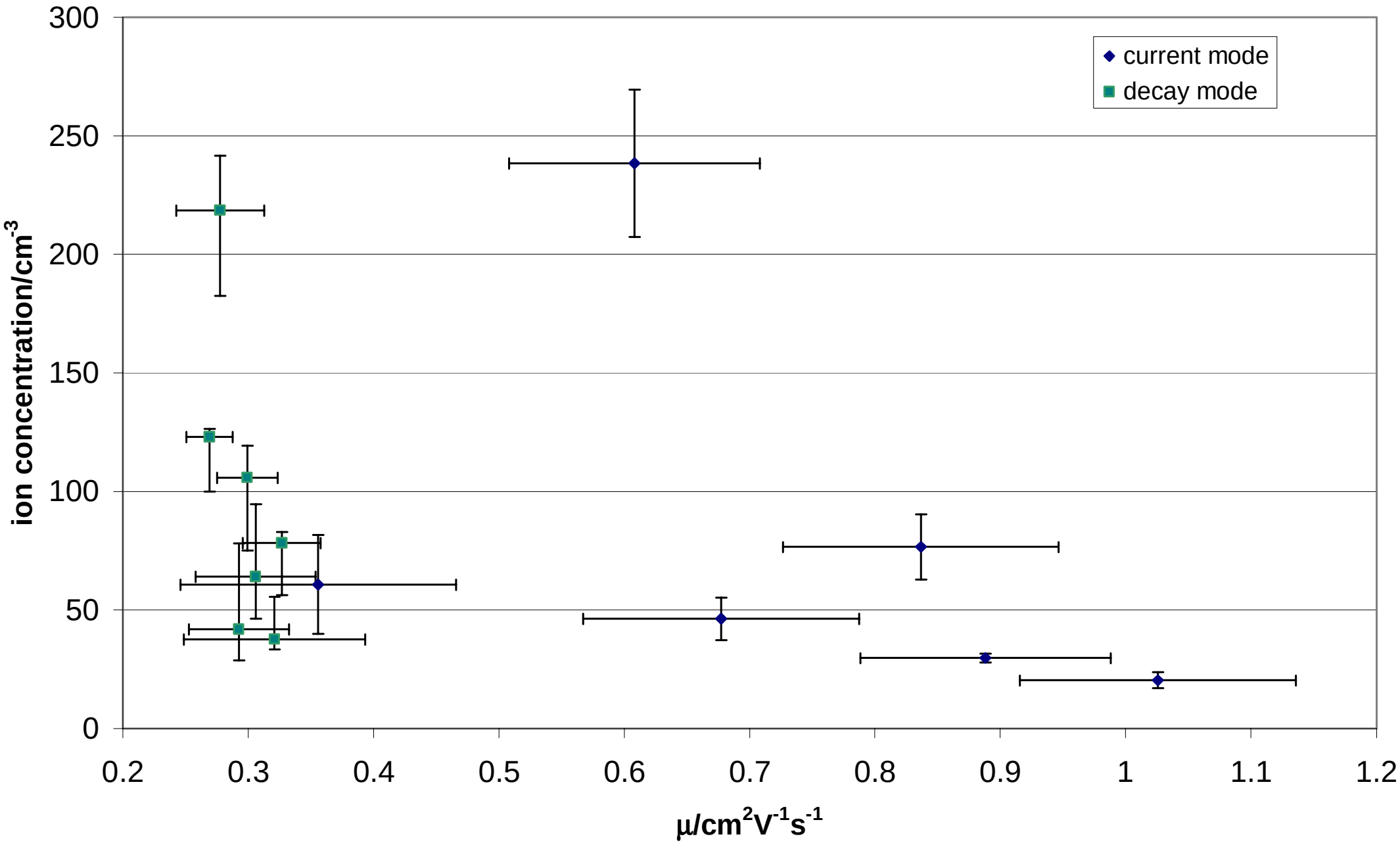

*Figure 8.4 Average negative ion mobility spectra calculated from both current and voltage measurements for 1000-1700 18$^{th}$ June 2000, at Reading University Meteorology Field Site. Error bars are the standard error of the mean.*

The spectra from the Voltage Decay mode are in a smaller mobility range because only a 2.5 V change in voltage at the central electrode can be measured. Additionally, the sampling only lasts for 5 s (for reasons discussed in Chapter 5.7.3), so only a fraction of the complete decay is observed. In the Current Measurement mode, the lower mobility limit is controlled by the bias voltage supply, and the minimum mobility that can be measured ≈ 0.27 cm$^2$V$^{-1}$s$^{-1}$. The spectra are derived from differencing 10 s averages from two instruments separated by up to 2 minutes, whereas in the Voltage Decay mode the spectrum is an average of instantaneous measurements from one instrument in 1.5 s. The assumption that the ion spectrum remains constant over the duration of the measurement may be more difficult to maintain for the current mode measurements. However, this is a more established technique for mobility spectra calculation (Phillips *et al*, 1955; Misaki, 1964; Dhanorkar and Kamra, 1993; Knudsen and Israelsson, 1994). As such, the approximate agreement with this spectrum calculated from the Voltage Decay method is promising.

Conductivity calculated with the average mobility derived from the spectrum is 0.6 ± 0.1 $fSm^{-1}$, which is much lower than expected in urban areas - although it is difficult to define what is "expected" given the plethora of assumptions associated with most conductivity measurements. If conductivity is calculated using a typical value of mobility, 1.7 $cm^2V^{-1}s^{-1}$, it is 2 ± 0.3 $fSm^{-1}$, which is believable for urban areas. It is not appropriate to compare conductivity calculated from the PIMS mobility spectrum with other PIMS conductivity measurements, because the PIMS conductivity measurements discussed previously were calculated assuming the ion spectrum was unimodal.

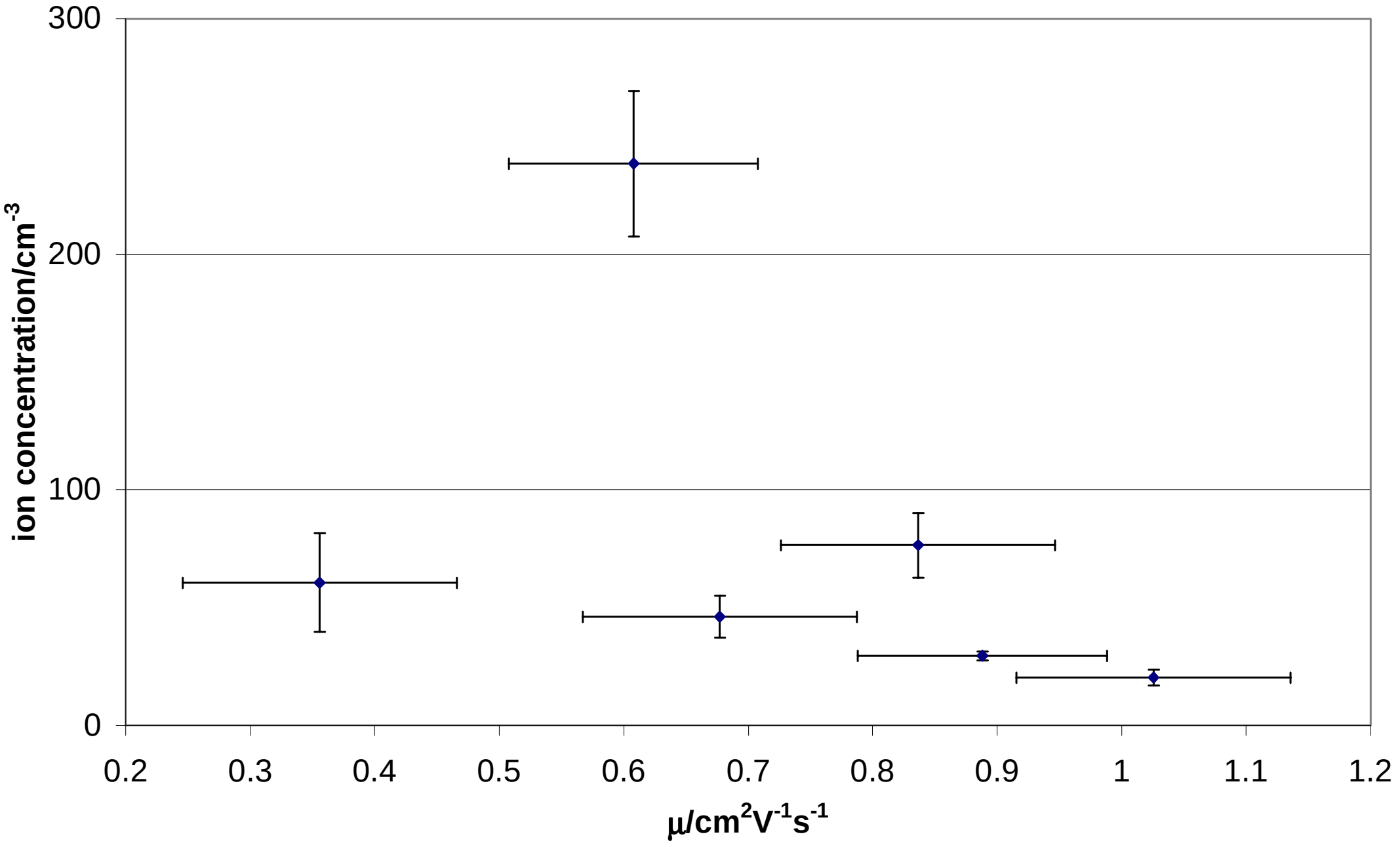

*Figure 8.5 Average negative ion mobility spectrum, calculated from current measurements for 1000-1700 18th June 2000, at Reading University Meteorology Field Site. Error bars are the standard error of the mean.*

Although there is arguably insufficient data to fit a curve to the results shown in Figure 8.5, a Gaussian spectrum was fitted to this data to enable comparisons with the measured Estonian spectrum. Integrating the area under this Gaussian curve showed that the total number of ions is 102 ± 16 $cm^{-3}$, which is a typical value for urban areas (Chalmers, 1967). The Estonian ion spectrum was measured in a rural area, where fewer ions are lost by attachment to aerosol, so $N$ would be expected to be larger than in the air of urban Reading. The spectrum calculated here has an average mobility[71] of about 0.5 $cm^2V^{-1}s^{-1}$, which is at least a factor of two too low for negative ions (*e.g.* Mohnen, 1974; Hõrrak *et al*, 2000). It is not clear why this is, because even at the edges of the (substantial) error bars the mean mobility would only shift by about 0.2 $cm^2V^{-1}s^{-1}$. The ion spectrum measured at Reading covers a relatively small mobility range; the small ion spectrum of Hõrrak *et al* (2000) covered the mobility range 0.3 to 3 $cm^2V^{-1}s^{-1}$. The Gaussian spectrum assumed here is for comparison purposes only, there is currently no scientific theory predicting any particular form for the atmospheric small ion spectrum, and it may be an inappropriate fit to this data. However it is known to be approximately correct, and has a small number (3) of free parameters.

The piecewise inversion of Voltage Decay measurements and the differencing technique for Current Mode measurements have been used to calculate two average negative ion spectra in different mobility ranges for a fair weather day at Reading. Spectra from the two techniques agree approximately in the calculated mobility range. This shows that the piecewise inversion technique is successful in comparison with a more well-established method to calculate ion mobility spectra. The mean mobility from the Reading spectrum is too low, which probably results from measuring a narrower (and smaller in magnitude) range of mobilities than the Estonian reference spectrum. A Gaussian fit to the Reading spectrum calculates a typical value for the total number of ions, which is less than the number calculated in the clean air of rural Estonia, in accordance with ion-aerosol theory.

## 8.4 Relationship between ions and aerosol in Reading air

On 25th February 2000, negative conductivity measurements were made with PIMS 1 at Reading University Meteorology Field Site, at the same time as Condensation Nucleus (CN) observations made manually every two minutes with a Pollak counter (Pedder, 1971). The Pollak counter measures all particles of $r > 3$ nm to 5 % accuracy when operated at an overpressure of 160 mm Hg (Metnieks and Pollak, 1969). This day had the longest time series of coincident conductivity and aerosol number concentration (*Z)* measurements available. Measurements of ions made with the PIMS will be compared to the predictions of classical ion-aerosol theory, in the light of the uncertainties discussed in Chapter 7.

[71] In ion mobility spectrum parlance, the “average” mobility is the most frequently occurring value (*i.e.*

### 8.4.1 Theoretical considerations

The inverse relationship between ions and aerosol was discussed in Section 2.4, and the steady-state solution of Eq. 2.19 is

$$\sigma = e\mu\left(\frac{\sqrt{(\beta Z)^2 + 4\alpha q} - \beta Z}{2\alpha}\right) \qquad \textit{Eq. 8.10.}$$

Because ions attach to aerosol, when the aerosol concentration increases, the ion concentration decreases, and this has led to the concept of using air conductivity as an urban air pollution indicator (again, this has been discussed in Section 2.4). However, the magnitude of the effect is variable. The ion-aerosol attachment coefficient $\beta$ is size-dependent (Gunn, 1954). To get an estimate of the magnitude of the effect to look for on the $25^{th}$ February, Eq. 8.10. has been plotted in Figure 8.6 below for three typical aerosol radii, in each case assuming a monodisperse aerosol population with particles of constant density of $1.77 \times 10^3$ $kgm^{-3}$. Attachment coefficients were calculated using the relationships of Gunn (1954). The mean $Z$, for $25^{th}$ February (which was very well-defined on this day as can be seen from Figure 8.7), has also been indicated on Figure 8.6. It can be seen that whatever the mean aerosol radius, the curve is becoming asymptotic, and the conductivity is insensitive to changes in the aerosol concentration. There is therefore unlikely to be a detectable inverse relationship between conductivity and aerosol concentration at these aerosol concentrations.

what is normally known as the mode). (Mohnen, 1974).

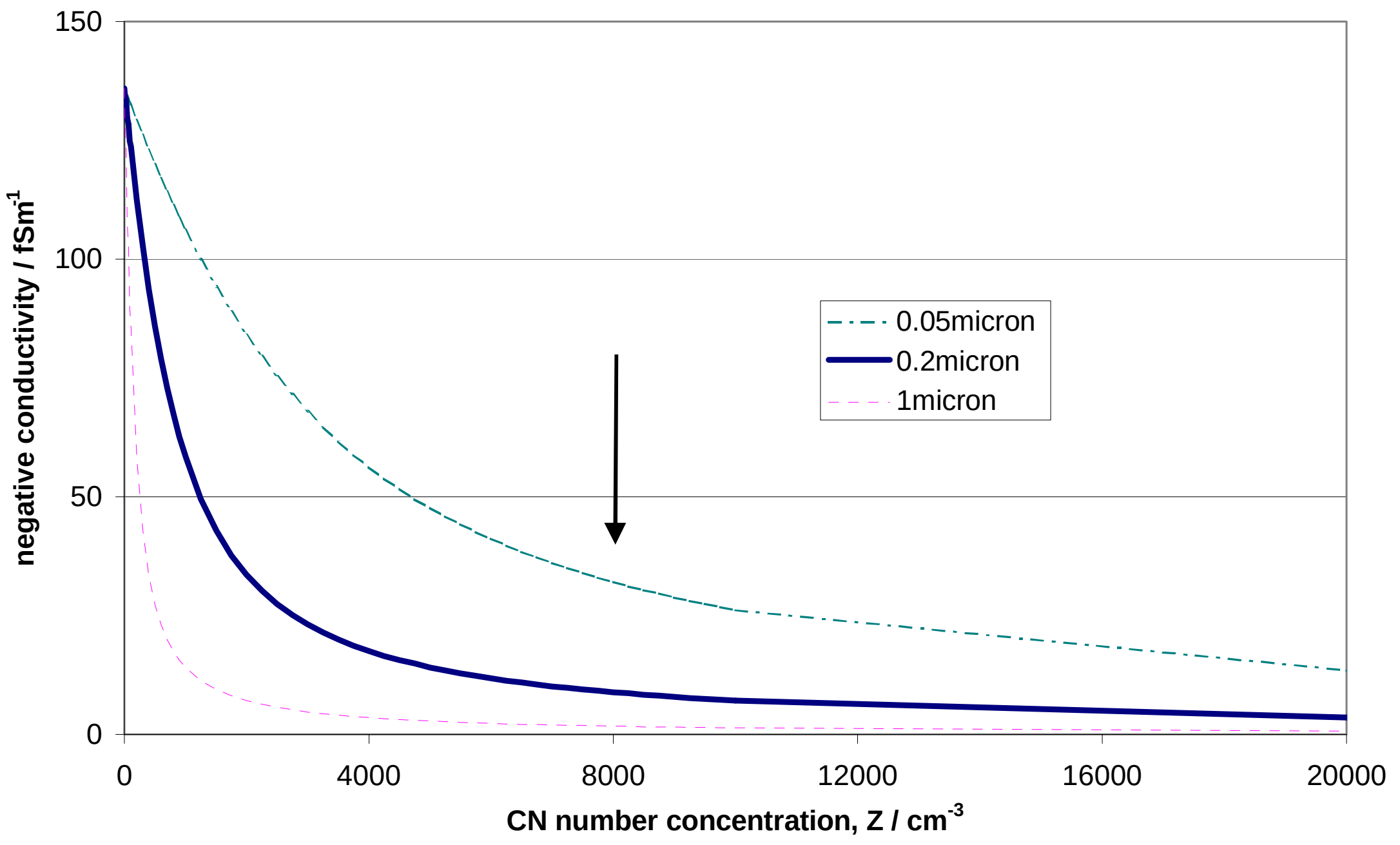

*Figure 8.6 Conductivity as a function of CN number concentration, calculated from Eq. 8.10 for negative ions with attachment coefficients $\beta$ for three different mean aerosol radii. The mean aerosol number concentration on 25th February 2000 is marked with an arrow.*

### 8.4.2 CN and ion measurements

In the time series of CN concentrations and conductivity (Figure 8.7), there is no clear relationship between the two quantities. For some periods there appears to be a positive correlation between the ion and CN concentrations, and for other periods there is an anticorrelation, as conventional theory suggests. These have been indicated with arrows in Figure 8.7 below.

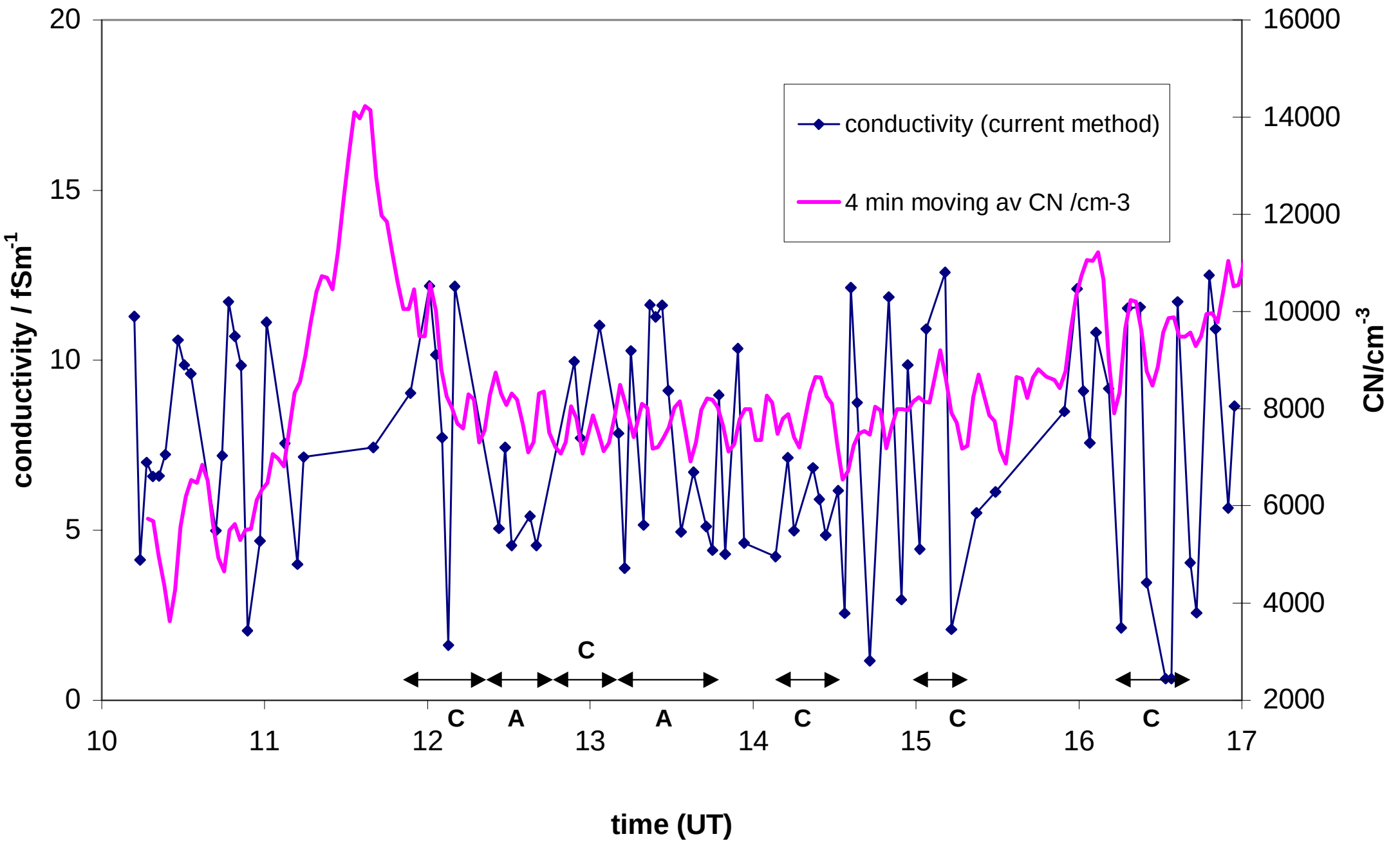

*Figure 8.7 Time series of CN concentration and Current Method filtered negative conductivity (which agreed with the Voltage Decay method conductivity) on 25th February 2000 with PIMS 1 at a bias voltage of –30.7 V. CN concentration was sampled every 2 minutes but a 4 minute centred moving average is shown here to match the mean sampling time for the filtered current method σ measurements (4.5 minutes). Arrows marked with A and C indicate periods of anticorrelation and correlation between σ and Z.*

The changes in régime in Figure 8.7 could be caused by varying radii of ion contributing to the measurement; according to classical ion-aerosol theory, only small ions should be inversely related to the aerosol concentration. If intermediate ions are being measured by the PIMS, due to low wind speeds decreasing the critical mobility, then there may be some overlap between the measuring ranges of the Pollak counter and the PIMS. Accordingly $r_{max}$ was calculated for the periods of correlation and anticorrelation, but there was no difference in the mean radius of ion being measured for the different conditions. This can be seen in Figure 8.8, where the time series of $r_{max}$ is not related to the ion concentration, or its relationship with Z in Figure 8.7.

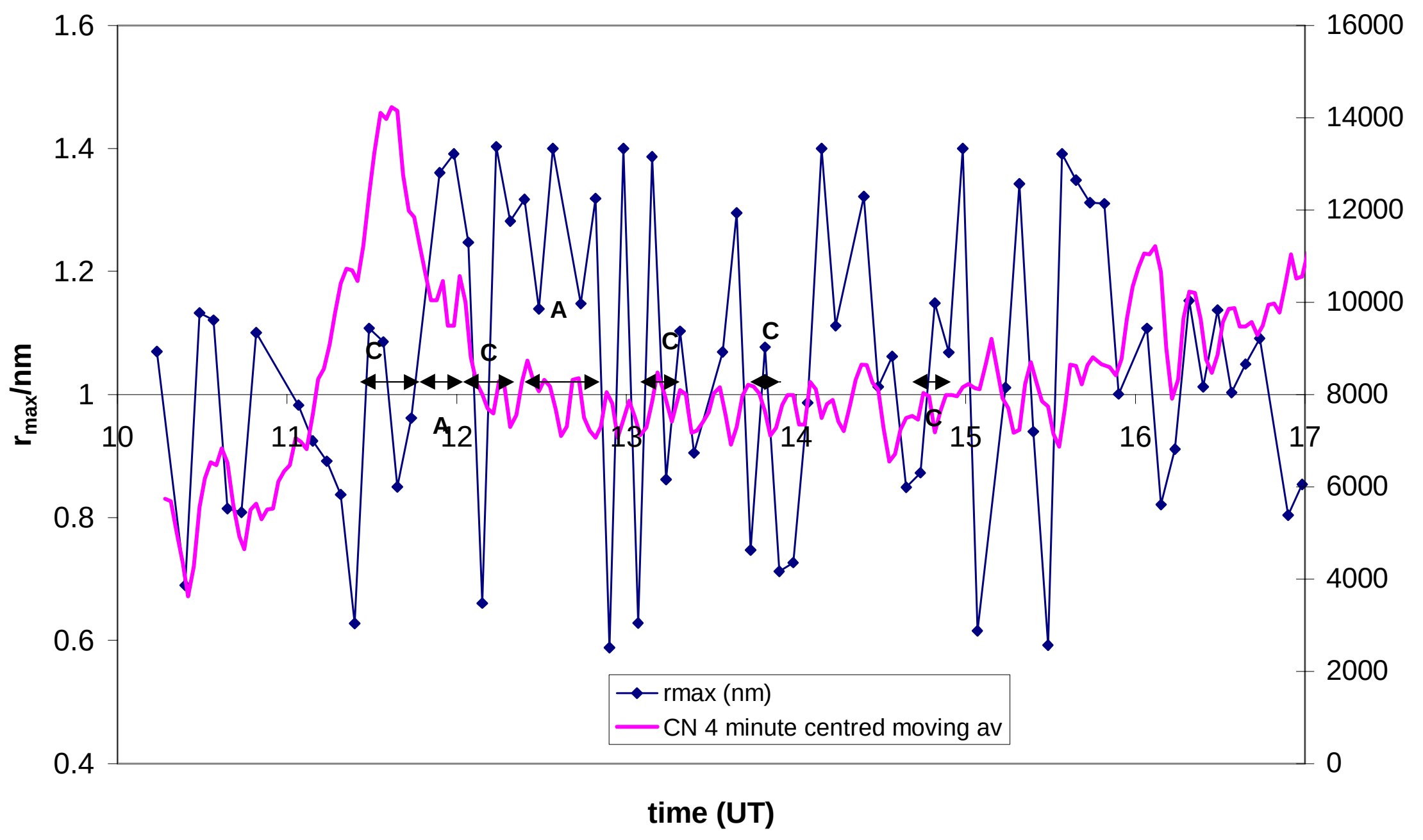

*Figure 8.8 Time series of maximum ion radius detected by the PIMS on 25th February 2000. Periods of correlation and anticorrelation have been marked as in Figure 8.7, and the CN concentration is also indicated on the right-hand axis.*

Space charge was also measured on this day, and the time series showed a good agreement with the conductivity. In Section 7.3, the need to partition the space charge into ionic and particulate components was also discussed. Space charge $\rho$ can be expressed in a simplified form as the sum of the charges carried by the total bipolar ion concentration $(n_+ + n_-)$ and aerosol particles $Z$ each carrying $j$ elementary charges, *i.e.*

$$\rho = e(n_+ + n_-) + ejZ \qquad \textit{Eq. 8.11.}$$

For 25th February, the mean aerosol concentration was 8300 cm$^{-3}$, and the negative ion concentration was about 800 cm$^{-3}$ (assuming a unimodal spectrum). Given that air ions only carry a single charge, it is probable that at these CN concentrations, the space charge was dominated by the particulate fraction. On this day, the space charge was negative for the whole measurement period; negative space charge is commonly caused by charged particulates (Crozier, 1964; Knudsen and Israelsson, 1994; Kamra, 1992). The combination of a high mean CN concentration and problems with applying classical ion-aerosol theory, again suggest that particulate space charge may be perturbing the ion measurements made with the PIMSs, as suggested in Section 7.3.

The agreement seen between the space charge and ion data on this day (plotted in Figure 7.8) also supports this hypothesis. Under these conditions, the variation in $r_{max}$ is insignificant compared with the effect of larger particulates (which are not being influenced by electrical forces) contributing to the conductivity measurement. The conventional relationship between $\sigma$ and $Z$ (Eq. 8.10.) therefore does not apply, to the extent that in a scatter plot of $\sigma$ against $Z$ (Figure 8.9 below), there is a weak positive correlation (with coefficient 0.1) between $\sigma$ and $Z$.

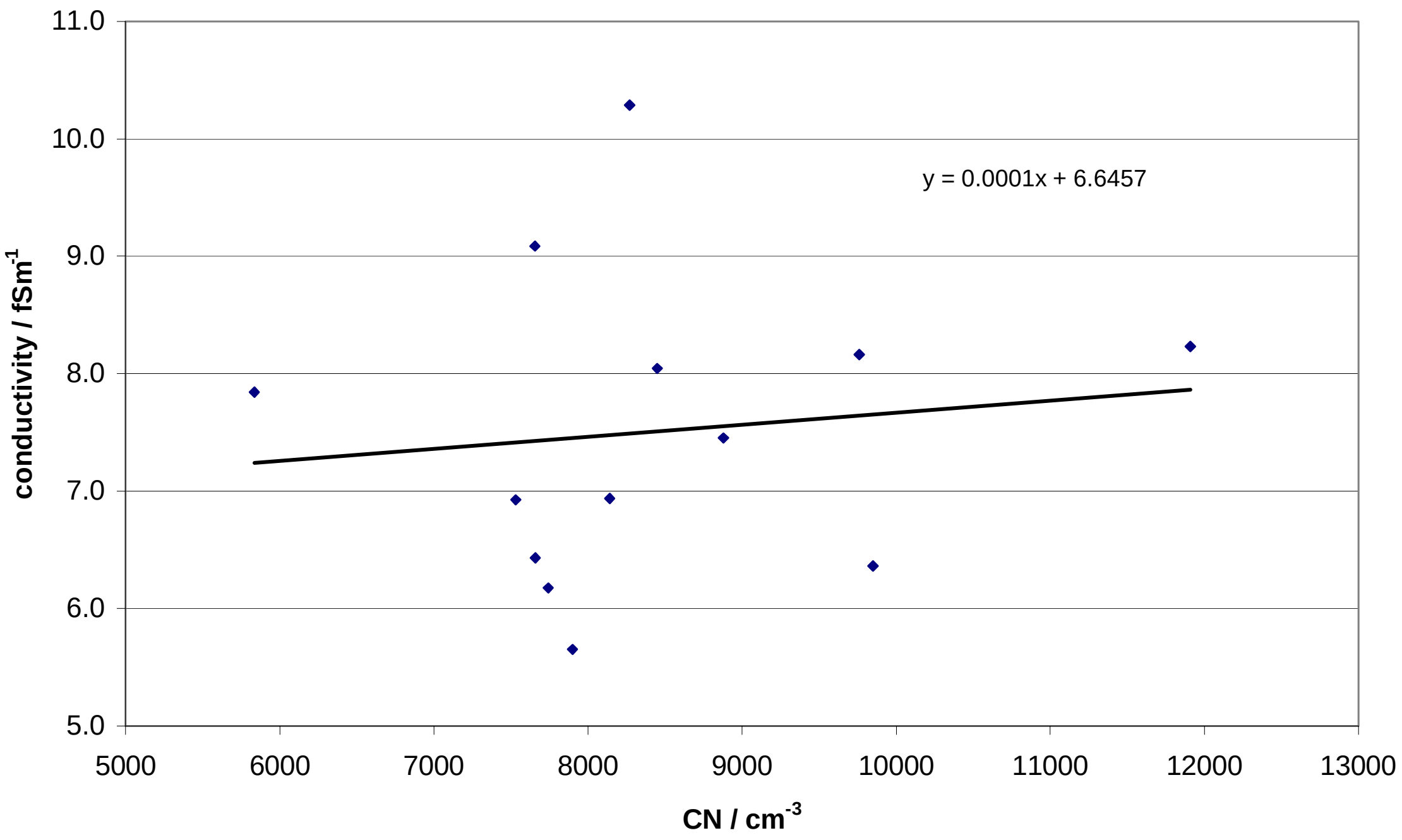

*Figure 8.9 Half-hour averages of negative conductivity and CN concentration on 25th February 2000 between 1000 and 1700 UTC.*

### 8.4.3 Limitations of conventional ion-aerosol theory

According to the theoretically predicted relationship (Figure 8.6) between $\sigma$ and $Z$, $\sigma$ becimes insensitive to changes in $Z$ at high $Z$. $Z$ is also remarkably smooth for ~ 60 % of the time series, so its fluctuations are small and are even less likely to be reflected in the conductivity.

Dhanorkar and Kamra (1997) predicted that conventional ion-aerosol theory does not apply at typical urban aerosol concentrations of about $10^4 – 10^6$ cm$^{-3}$, because intermediate and large ions contribute to the conductivity. The measurements at Reading suggest there was a large proportion of particulate space charge, (with aerosol concentrations ~ $10^4$ cm$^{-3}$) on 25$^{th}$ February, which can mechanically impact on the Gerdien electrodes and perturb the conductivity measurement (as shown in Section 7.4). These results therefore appear to verify the hypothesis of Dhanorkar and Kamra (1997), and show experimentally that particulate space charge can contribute to the conductivity measurement and invalidate conventional theory.

## 8.5 Conclusions

The major result presented in this chapter is a new method to calculate ion mobility spectra based on the Voltage Decay mode results. It has been derived from modifying the classical theory of conductivity calculation from exponential decay of charge from a capacitor through air. It reproduces the classical predictions when there is a unimodal ion mobility spectrum, on which assumption the classical theory is based. The new technique was then applied to calculate an average spectrum for seven hours of data, and spectra were also calculated for the same period by a better-known technique from the Current Measurement mode. Results from the two methods are difficult to compare directly, for reasons outlined in Section 8.3.3, but they are in near agreement in magnitude. Their combined results produce a realistic value for the total ion concentration when a reference spectrum is fitted and compared to a long-term average spectrum. There is some offset in the mean mobility, which may result from fitting the spectrum to a narrower mobility range than the Estonian average spectrum.

The steady-state solution of the ion balance equation (Eq. 2.19) suggests that the response of the small ion population to higher aerosol concentrations is slight. Ion, CN and space charge measurements on 25$^{th}$ February 2000 have shown that this small variation is submerged by the effects of particulate space charge contributing to the measurement. This shows that classical ion-aerosol theory may not be appropriate at typical urban aerosol concentrations, and validates the theoretical predictions of Dhanorkar and Kamra (1997).

*"Pollution in cities produces more and more positive ions…" Cosmopolitan magazine, July 2000*

# 9 Conclusions

*Cosmopolitan's* authority on the physical nature of the world is readily challenged, but the quotation serves to illustrate the confusion surrounding the significance of atmospheric ions. The statement is of course risible to anyone who has some knowledge of the science of atmospheric ions and aerosol. Yet it has become clear in this Thesis that in some circumstances, measurements of atmospheric ions are significantly affected by the presence of aerosol particles. In this context, the *Cosmopolitan* quotation appears more relevant, for aerosol particles can certainly cause a spurious "ion" signal in a Gerdien condenser.

## 9.1 An instrument to measure atmospheric ions

There is a long tradition of atmospheric ion measurements in the last century's history of physics. Many of these measurements have been made with a simple instrument, the Gerdien condenser (Gerdien, 1905), or related variants. Atmospheric ion measurements are again increasingly relevant to modern atmospheric science due to the general interest in aerosols with regard to climate and pollution issues, and the possibility that ions are involved in aerosol particle formation. However, the Gerdien condenser has undergone surprisingly little development, and issues such as its absolute calibration have not been satisfactorily answered in the history of the instrument, with working approximations employed.

Previous measurements of air ions with the Gerdien condenser have been made under a variety of instrumental operating conditions, and it is probably for this reason that they have not yielded conclusive results when compared to each other and to classical ion-aerosol theory. Ion mobility spectrum measurements made with similar instruments are highly necessary for the understanding of air ions, since mobility spectra can resolve ionic growth processes. Ion mobility spectra may also be one of the necessary tools to determine if ions are implicated in ultrafine particle formation.

A miniaturised Gerdien condenser has been designed, in an attempt to avoid many of the inlet errors associated with previous instruments. It has been used to measure air conductivity at different mobilities in two measurement modes. The Current Measurement mode measures an ionic current directly, and the Voltage Decay mode determines the conductivity from change in voltage across the condenser, as ions exchange charge. The device has been developed into a Programmable Ion Mobility Spectrometer, (PIMS) which seems unique in that it can self-calibrate by measuring ions in both modes. Three of these spectrometers have been constructed and tested in atmospheric air for self-consistency, and against independent ion measurements. During the development of the ion mobility spectrometer, issues have arisen which produce uncertainty in its results and preclude certain operating conditions. These findings are general ones, which can be applied to all aspiration ion measurement devices.

## 9.2 The programmable ion mobility spectrometer

The ion spectrometer described in this Thesis is based on the classical Gerdien condenser, but differs from previous instruments in its mechanical construction. It is considerably smaller and incorporates a third electrode to provide electrical screening. Aluminium was found unsatisfactory as a tube material, and stainless steel is preferred. It is designed to be easily dismantled for cleaning and repair. Its capacitance has been calibrated to 1 %, which considerably reduces the sources of uncertainty in conductivity measurement. The ion spectrometer is deployed *in situ*, to avoid inlet errors, and natural ventilation is fan-assisted to prevent saturation. The flow properties inside the tube as a function of the natural ventilation were measured in a wind tunnel and were non-linear. This suggests that using the external ventilation rate alone to calculate critical mobilities in a fan-assisted tube, as has been widely adopted in the past, is inaccurate.

Ions are detected with a multimode electrometer which uses reed relays to switch its functionality between that of a current amplifier and voltage follower, so one electrometer can be used to measure the signals generated in both measurement modes. The multimode electrometer also includes two self-checking modes so that temperature-dependent offsets, such as the op-amp input bias current and electrometer leakage currents can be compensated using software.

The approach of using the same ion spectrometer in the Voltage Decay and Current Measurement modes has been demonstrated to be sound, since results from the two modes agree. These are thought to be the first measurements using both Gerdien measurement modes with one sampling instrument. Comparison of the two modes of the three ion spectrometers constructed was also successful. The three spectrometers have been tested in atmospheric air against independent instruments; a Geiger counter, which measures the ion production rate, and another, reference, Gerdien condenser. In clean air, the ion concentration is associated with the square root of the Geiger count rate, according to theory. In the polluted air of urban Reading a direct relationship between the Geiger count rate and the ion concentration cannot be seen due to the complicating effects of aerosol, but the ion spectrometer still responds to peaks in the Geiger count rate. The trends in conductivity measured by the ion spectrometer and the more classically designed Gerdien reference condenser track each other well, but direct comparisons are difficult owing to the differing sizes of Gerdien tube. Average conductivities from both modes have been calculated from measurements made at Reading on 24 fair weather days in April-June 2000, which reproduce the diurnal variation seen in the literature. This is thought to be the first systematic set of conductivity measurements using the Voltage Decay method in the atmospheric surface layer.

The new, smaller design of Gerdien condenser is advantageous because of its portability and ease of use in the field. For, example operating several such ion spectrometers in an array would be relatively straightforward. The third screening electrode reduces the effect of electromagnetic interference on the measurements, allowing the outer electrode to be operated above ground potential. The smaller design of the Gerdien is advantageous because it requires smaller bias voltages, which can be more precisely controlled for increased accuracy in critical mobility determination. Constructing the ion spectrometer so it can be easily dismantled for cleaning and repair is certainly a worthwhile effort, as all measurements requiring high integrity insulation deteriorate in atmospheric conditions. The multimode electrometer design has enabled the self-calibration hypothesis to be pursued, and its ability to measure voltage offsets and current leakages has improved the accuracy of the ion measurements. Results from the Voltage Decay mode are noisier than those from the Current Measurement mode, perhaps resulting from the effect of highly-charged aerosol particles reaching the electrodes. The increased confidence in the instrument achieved by making measurements in both modes certainly outweighs the heavy filtering which is sometimes required for the Voltage Decay mode data. In practice, the Current Measurement mode will probably be preferred, with Voltage Decay samples also taken to check the measurement.

## 9.3 Experimental findings from the ion spectrometer

### 9.3.1 Ventilation-modulated ion size selection

Varying the flow in the Gerdien tube modifies the *critical mobility*, which is the *minimum* mobility of ion contributing to the conductivity measurement. Mobility is inversely related to ion radius, (the relationship can be expressed in simplified form by a power law) so the concept of critical mobility can be equated to a *maximum* radius of ion which can be detected by the spectrometer. If the flow is sufficiently low, intermediate ions can contribute directly to the measurement of conductivity. This is undesirable because it invalidates the classical indirect theory of ion-aerosol interactions. It was found that the flow in the tube can sometimes become low enough for intermediate ions to be measured at low external ventilation rates, even when the tube is ventilated by a fan. Atmospheric turbulence causes considerable fluctuations in the wind speed and direction on short timescales. This can frequently bring the ion spectrometer into the intermediate ion measurement range.

This can be alleviated by measuring the external wind velocity near to the ion spectrometer, and rejecting the data for the periods when the maximum radius of ion measured is in the intermediate range. This consideration is relevant for any Gerdien measurements where the external ventilation influences the flow rate in the tube. The nature of the flow cannot be confirmed until measurements of the flow in the tube have been made under controlled conditions *e.g.* in a wind tunnel. In the absence of such measurements, any previous measurements where the fan flow rate is comparable to the external wind velocity must be treated with caution. Experiments where the instrument was purely ventilated by the wind will almost certainly have been influenced by this effect, for example Gringel (1978) and Knudsen and Israelsson (1994). Some re-analysis of existing measurements, if wind data is available, may be necessary; if adjacent wind measurements were not made, then the data may be suspect.

### 9.3.2 The effect of space charge

The sum of the charges per unit volume of electrically charged particles in the air (both ionic and particulate) comprises the atmospheric *space charge*. Experiments performed with the ion spectrometer have shown that a significant non-ionic charge can be detected as well as the ionic contribution. The results suggest that particulate space charge can directly contribute to the conductivity when it is advected onto the spectrometer electrodes. However, on other days, only ionic space charge contributes to the conductivity measurement.

Two régimes have therefore been identified, which it is important to differentiate for the following reason. When particulate space charge contributes to the conductivity, charged aerosol particles are measured by the spectrometer as well as small ions. Classical ion-aerosol theory predicts an inverse relationship between conductivity and aerosol number concentration, which when aerosol particles are included in the conductivity measurement, will be invalidated. The particulate space charge can occasionally dominate the ionic signal, invalidating the measurement. The small ion measurement case, of principally ionic space charge, represents the conditions under which ion-aerosol theory should apply.

In the régime when particulate space charge contributes to the conductivity, ion-aerosol theory will not apply, because the direct effect of aerosol on conductivity overrides the inverse relationship predicted by theory. The effect is exacerbated by the insensitivity of the conductivity to changes in the aerosol concentration at high aerosol concentrations. This is experimental confirmation of the theoretical predictions made by Dhanorkar and Kamra (1997), of the failure of the inverse conductivity-aerosol relationship at high aerosol concentrations.

### 9.3.3 Validity of air conductivity as a pollution indicator

Practical implications of the corroboration of Dhanorkar and Kamra's (1997) predictions are that it is clearly unfeasible to use conductivity as an urban air pollution indicator, except in relatively clean air, with a low aerosol concentration. Even if the contribution of particulate space charge to the conductivity measurement is particularly large with the small instrument used in this Thesis (although there is no scientific reason to suppose it might be), the insensitivity of the conductivity to aerosol at high aerosol conditions implies that the combination of these two factors still synergise in invalidating the inverse relationship. This may help to explain the large scatter in attempts to observe an inverse conductivity-aerosol relationship (Guo *et al*, 1996). The direct use of space charge, rather than ions, as a pollution indicator may therefore be more appropriate in urban air.

The utility of using conductivity to infer local aerosol concentrations in clean air may be doubtful, but clean-air conductivity measurements could perhaps still be used to infer a global change in aerosol levels. The existence of this secular effect is still unresolved despite oceanic conductivity measurements adding a new point to the plot of conductivity against time every few years (the most recent being Kamra and Deshpande (1995)). However, meteorological effects and to a lesser extent ion variations may cause local variability. Variations in the operating conditions of the Gerdien, for example the different critical mobilities, may also cause differences between the conductivity measurements. If clean-air conductivity measurements are to have any chance of observing secular changes, there need to be several standard instruments recording conductivity at different clean air locations for some time (*e.g.* several months every five years), so that changes in the long-term average can be detected.

## 9.4 The importance of the ion mobility spectrum

The existence of a spectrum of atmospheric small ions has been known for many years, *e.g.* Torreson (1949). However, the effect of the spectrum has usually been ignored in conductivity measurements; the derivation of conductivity from Voltage Decay measurements assumes a unimodal spectrum, so that charge on an electrode decays exponentially at a rate dependent on the air conductivity. This approximation is probably a consequence of the history of electrometry: in the early years of ion measurements the only form of electrometer available was a gold-leaf electroscope (*e.g.* Zeleny, 1898). This could only be used to determine an overall rate of decay, and could not make discrete measurements of the change of voltage. A theoretical background to support ion measurements was being developed concurrently, and assumed a unimodal ion spectrum because the instruments to measure an ion mobility spectrum did not yet exist. Even when the existence of the ion spectrum became well-known, the unimodal theory developed in the early 1900s has been glibly repeated to the present day, with no modification arising from the improved observations of the ion mobility spectrum. The theory clearly needs to account for a mobility spectrum, which, in general, will cause a departure from an exponential decay.

In this Thesis, the consequences of considering the ion mobility spectrum in conductivity measurement have been investigated. The exponential relationship for the rate of decay of charge from the Gerdien electrodes through air assumes there is a unimodal ion mobility spectrum. Several sets of measurements have shown that the mean spectrum is clearly not unimodal (Torreson, 1949; Misaki, 1964; Hõrrak *et al*, 2000), and in this case there is a quasi-exponential decay, because the voltage change during the decay affects the critical mobility of the measurement.

The concept of a change in voltage varying the critical mobility of the measurement is a simple one, and clear from use of the Current Mode. Swann (1914) described the Gerdien condenser as an instrument which is designed to be run in the Voltage Decay mode, but recent publications do not refer to the Voltage Decay mode as an alternative means of conductivity measurement at all (*e.g.* MacGorman and Rust, 1998). Only three papers seem to have been published using the Voltage Decay technique in the last fifty years (Venkiteshwaran, 1958; Hatakayema *et al*, 1958; Rosen *et al*, 1982), and all of these described voltage decay measurements on radiosonde-borne Gerdiens, which was presumably for simplicity. There do not appear to be any systematic conductivity measurements by the Voltage Decay method in the boundary layer. Another, related, explanation is because only measurements in clean air, out of the boundary layer, appear to have been made. This is virtually devoid of the trace gases that react to produce different atmospheric ion species. The free tropospheric ion spectrum might therefore be much more unimodal in nature than in the surface layer. Voltage Decay measurements in the upper troposphere might be expected to conform to more ideal exponentials than have been observed in the surface layer. Venkiteshwaran (1958) shows some sample voltage decay traces, which are ideal exponential decays, supporting this suggestion.

### 9.4.1 Derivation of the quasi-exponential decay

An expression has been derived for the form of the decay when there is a non-unimodal spectrum. This is similar to the traditional exponential decay, except that it includes an integral to include the effect of the changing voltage on the mobilities of ions being measured. The equation is no longer amenable to analytical solution, and requires numerical techniques to yield a quasi-exponential voltage decay curve from a prescribed ion spectrum.

### 9.4.2 Calculating the ion mobility spectrum from Voltage Decay measurements

A further technique was developed to obtain a spectrum by piecewise inversion, from the Voltage Decay mode time series. This involves approximating small steps of the voltage decay time series with exponential decays. The technique was applied to a theoretically derived voltage decay time series generated from a Gaussian ion spectrum, and successfully inverted the voltage decay series to recover the original Gaussian spectrum. It was then applied to a set of voltage decay measurements made with the ion spectrometer to obtain a spectrum is a novel technique, although over a restricted range. Nolan (1920) classified ion categories from the “quasi-linear” Ohmic response resulting from the existence of the mobility spectrum, which is a similar conceptualisation to the quasi-exponential decay. However, the ion spectrum does not appear to have been calculated from Voltage Decay measurements before.

### 9.4.3 Programmable capabilities of the spectrometer

The ion mobility spectrometer developed is fully programmable; it is operated by a microcontroller, which can be programmed in a variant of BASIC from a PC. This enables it to be run as a spectrometer in the Current Mode by varying the bias voltage, so that the mobility of ions contributing to the measurement can be changed.

Conductivity measurements from two adjacent PIMS instruments running at different bias voltages in the Current Mode have been used to calculate ion spectra. The calculated ion concentrations and mobilities were similar to those obtained for the same day from the Voltage Decay mode, reinforcing the technique used to calculate the spectrum from piecewise inversion. The Current Mode ion spectrum agreed with a reference spectrum obtained from long-term ion measurements.

### 9.4.4 Relevance of the ion mobility spectrum to conductivity measurements

The ion mobility spectrum is highly relevant to conductivity measurement for two main reasons. The first reason is the calculation of conductivity from the Voltage Decay mode. The exponential model is inappropriate, except when the mobility spectrum is unimodal. The most accurate way to calculate conductivity from the Voltage Decay mode would be to calculate the spectrum for the mobilities scanned through by the voltage decay. When this is known, the total number of ions can be found and used to calculate the conductivity within that mobility range.

Conductivity calculated from discrete measurements in both modes is clearly dependent on the fraction of the mobility spectrum contributing to the measurement (*i.e.* the instantaneous flow rate and bias voltage). Ideally therefore, conductivity should be an integrated measure of the ion number concentration, calculated from the total number of ions in the mobility spectrum. This requires parallel measurements of conductivity and ion spectra, which the PIMS is able to provide. In the absence of an ion mobility spectrometer, the critical mobility at which the measurement is made should be quoted, with some indication of its variability, since it is inappropriate to compare conductivity measurements directly if their critical mobilities are significantly different.

## 9.5 Atmospheric measurements

The PIMS instrument was designed to make fair weather measurements of conductivity and ion spectra in the surface layer. It is fundamentally unsuited to measuring in precipitation, as are Gerdien devices in general, because water inside the tube disturbs the measurement. Perhaps surprisingly, conductivity has not been reported to vary very much from fair weather conditions in the vicinity of a thunderstorm (Blakeslee and Krider, 1992; Bailey *et al*, 1999). The PIMS could therefore be deployed in disturbed weather, providing it was protected from precipitation. However, addition of circuitry to protect the control system from transients would be advisable.

In considering whether the PIMS would be suitable for measurements at altitude, it is necessary to consider whether the electrometer can measure the higher conductivities due to increased ionisation by cosmic rays. In the Current Mode the multimode electrometer can measure conductivities up to 100 $fSm^{-1}$, and the maximum conductivity measurable in the Voltage Decay mode is 200 $fSm^{-1}$. Conductivity reaches these levels at about 10 km (Woessner *et al* 1958; Gringel, 1978) so the PIMS in its present form would be able to measure up to about this height, needing little modification to be flown on an aeroplane. As it is, the PIMS is too heavy to be suitable for radiosonde or gondola ascents, and it was designed specifically to send data serially to a nearby PC. The idea of a microcontrolled spectrometer with specially-designed circuitry and power supplies would be a novel one, and might be able to obtain the first ion spectra out of the surface layer.

## 9.6 Further work

Further research is needed into the effect of space charge onto the conductivity measurements. To predict the occurrence of the régime dominated by particulate space charge described in Section 9.3.2, the space charge needs to be partitioned into ionic and particulate components. This is a further motivation for study of the turbulent transport of space charge, begun by Barlow (2000). A measurement campaign comprising ion and aerosol number concentration, ion production rate, meteorological and space charge measurements in clean and polluted air should provide insight into the conditions in which the PIMS instrument can and cannot be known to respond solely to small ions. Bipolar ion measurements near the ground would facilitate investigations of the electrode effect, and such a campaign might also supply more details about the efficacy of space charge as an urban pollution indicator.

There is a need for more ion mobility spectra, particularly at high time resolution, to investigate the properties of the spectrum in general, and to attempt to resolve ion nucleation processes. Shifts in the spectrum towards larger ions have been observed on timescales of order half a day (Hõrrak *et al*, 1998a). Ions have been observed to form particles at high radioactivity, there is strong laboratory evidence that particles can be formed from ions at natural levels of radiation (Vohra *et al*, 1984) and modelling studies have predicted that particles are formed from ion-ion recombination (Turco *et al*, 1998; Yu and Turco, 2000). However there are no direct observations of ion growth leading to ultrafine particles in the atmosphere. To resolve this, ion mobility spectra associated with ultrafine particle measurements would be required. The Voltage Decay mode of the PIMS is clearly capable of resolving high time resolution mobility spectra: 1 spectrum every 5 s, or more. The programmable capability of the PIMS also allows voltages to be selected over a wider range of interest, to calculate spectra from the Current Measurement mode.

The piecewise inversion technique could readily be applied to simple Voltage Decay Gerdien measurements made on radiosondes, to investigate the ion spectrum in the free troposphere. These new measurements would provide similar information about the ion population and particle formation as described above, but applied to the troposphere where cloud formation occurs. They would perhaps give some more direct atmospheric answers to the microphysical questions raised in the controversial debate over cosmic ray ionisation and cloud formation.

Some improvements to the instrumentation developed in this Thesis would help to achieve these aims. Sensing the wind speed within the tube, and electronically modifying the bias voltage so that it remains constant, thus smoothing the effect of wind fluctuations is the most obvious next step. Measurements with an array of PIMS would be useful in determining local variability in ion concentrations. Installing a more powerful microcontroller (these are readily available) would permit further sophistication in automation of the measurements. Synchronisation could be improved by addition of further timing circuitry, which is commercially obtainable. The microcontroller could be used to sense a mobility range of interest, for example where the spectrum is changing most rapidly. The bias voltage could then be automatically changed to concentrate on measuring in the desired mobility range. Additionally, an investigation of sampling issues, and characterisation of the response spectrum would be desirable to improve the fraction of data retained from the PIMS in fair weather.

# Appendix A

# The multimode electrometer

This appendix provides a complete description of the MME (constructed by S.R. Tames), which is summarised in Section 5.4. A full circuit diagram is given here which includes the logic and power supply aspects. The logic table to operate the MME modes (designed by J.R. Knight) is also provided. A photograph of the MME is Figure 5.7 in the main text.

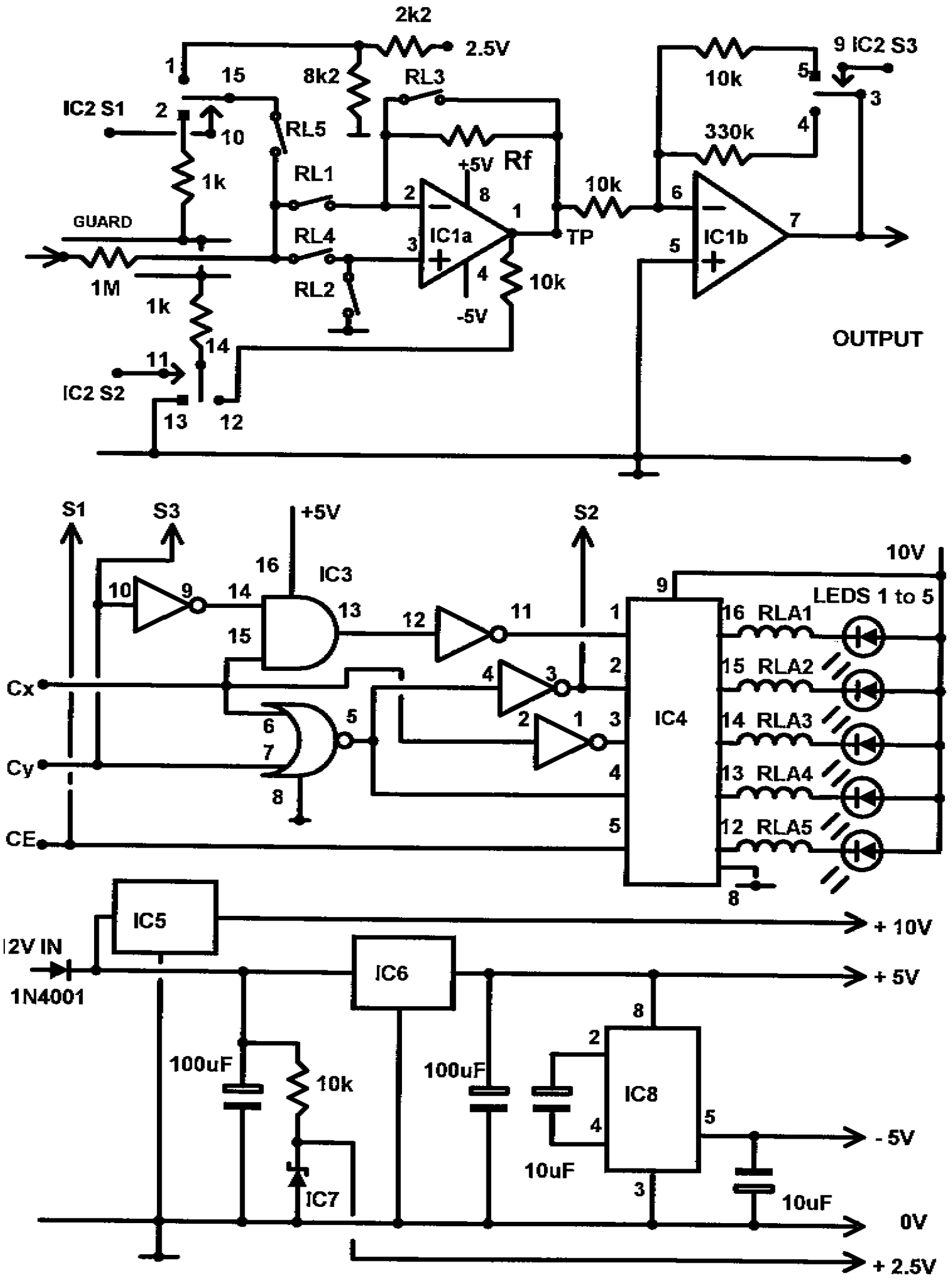

*Figure A.1 Complete schematic of the multimode electrometer. IC1a (LMC6042) is switched into voltage follower and current amplifier configurations by reed relay switches RL1 to RL5 with an inverting gain stage of IC1b added. $R_f$ is $10^{12}\Omega$. MOSFET change-over switches (IC2 4053), are used to guard floating inputs. (Square pads on the schematic are used to denote the MOSFET 'off' position in Table 1 and TP is a test point.) IC3 (4572) decodes the control lines to operate the MOSFET switches, using IC4 (ULA2003) to drive the reed relay activator coils RLA1 to RLA5. The power supply derives +10V, ±5 and +2.5V from a 12V lead-acid battery using three terminal regulators IC5 (78L10) and IC6 (78L05), voltage reference IC7 (REF25Z) and voltage inverter IC8 (7661).*

| OPERATING MODE | | INPUT CONTROL LINES | | | REED SWITCH ACTIVATIONS | | | | | MOSFET SWITCHES | | |
|---|---|---|---|---|---|---|---|---|---|---|---|---|
| | | | | | | | | | | S1 | S2 | S3 |
| | | | | | RL1 | RL2 | RL3 | RL4 | RL5 | | | |
| | | $C_x$ | $C_y$ | $C_E$ | | | | | | | | |
| 1 | **Follower** | 0 | 0 | 0 | | | ON | ON | | | ON | ON |
| 2 | **Picoammeter** | 1 | 0 | 0 | ON | ON | | | | ON | ON | ON |
| 3 | **system $V_{os}$ check** | 0 | 1 | 0 | | ON | ON | | | ON | ON | |
| 4 | **$V_{os} + i_b R_f$** | 1 | 1 | 0 | | ON | | | | ON | ON | |
| 5 | **Charge Electrode** | 0 | 0 | 1 | *supplies 2.0V to input* | | | | ON | | | |

*Table A.1 Relationship between mode setting, input control lines and switching elements*

# Appendix B

# Calibration of feedback current amplifiers

The content of this appendix is taken from an internal technical memorandum (TechNote number 304) produced at the Department of Meteorology, University of Reading.

## B.1 Feedback current amplifiers

Electrometer Current Amplifiers allow the measurement of ultra-small currents, which are typically in the range 10 fA to 10 pA for atmospheric ion measurements. A feedback current amplifier uses a low input-bias current opamp as the amplifying element, with negative feedback applied using a large (~GΩ to TΩ) resistor. This is frequently called a *transresistance* configuration: it is a current-to-voltage converter. The high-value resistor is usually only of a nominal value, and therefore a calibration is required for precise measurement of current to be obtained.

Calibration requires either a known source of small currents, or a scaling argument based on determining the resistor ratios in bridge circuits using known fixed resistors to derive the value of the large, unknown feedback resistor. Systems based on scaling techniques have already been described (Harrison 1995, 1997a), and it is the calibration based on small current generations which are discussed further here.

### B.1.1 Resistive generation of small currents

Currents can be generated using a voltage source applied through a series resistor $R_{cal}$ to the current amplifier. The feedback current amplifier input is a virtual earth, therefore the whole potential difference of the voltage source is applied across the resistor and the current generated can be calculated *if* the value of the resistor $R_{cal}$ is known. Its resistance will have to be measured by a different method, which usually means direct measurement using a high value ohmmeter, or by fabricating the resistor from series connection of resistors which are either precision values or small enough to be directly measured.[72]

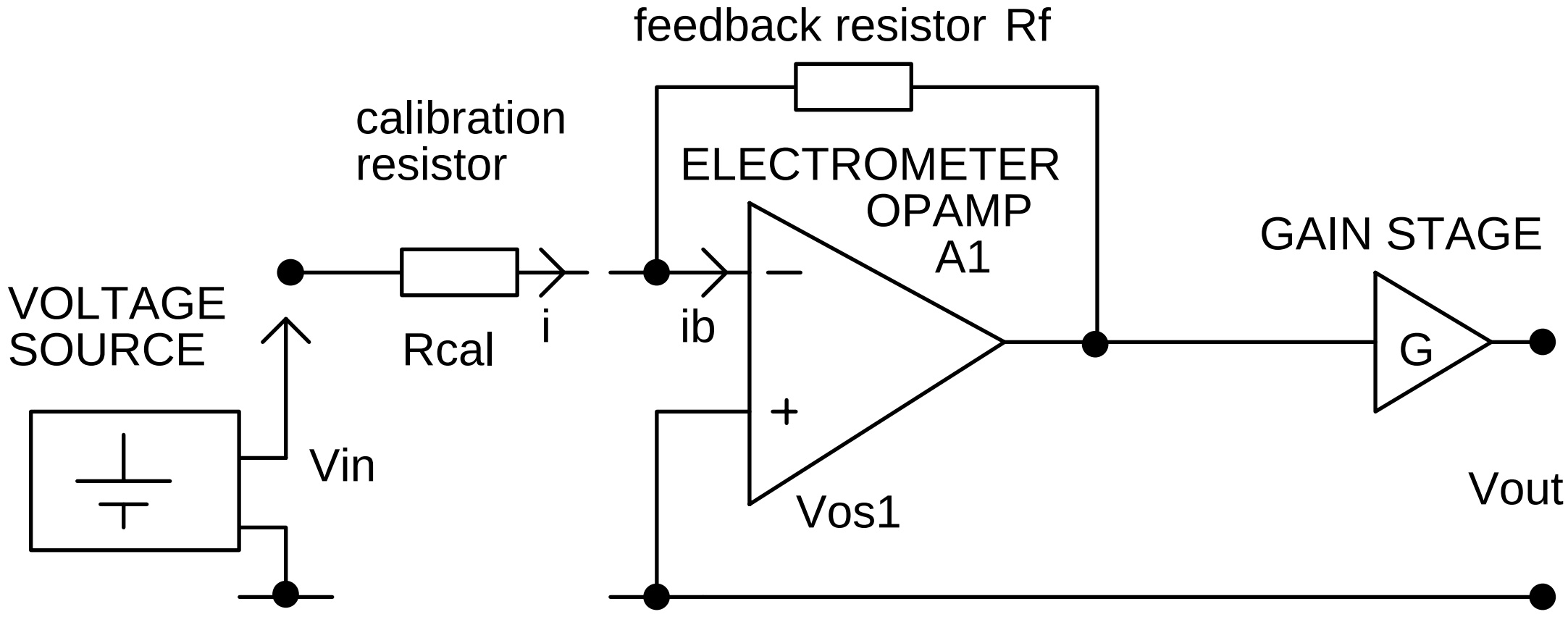

*Figure B.1 Resistive generation of calibration current* i.

The circuit configuration will be some variant of a resistive current source generating a known current $i$, a feedback electrometer opamp, and, possibly, a gain stage (as in Figure A.1). Calibration essentially requires the determination of $R_f$, as the response is linear if leakage terms are not significant and any voltage dependence of the resistance $R_f$ is negligible. The current-to-voltage stage A1 is *inverting*, so that a current flowing *into* the input leads to a *negative* voltage at the output of the opamp A1. If there is a subsequent gain stage, it may have an inverting or non-inverting response. An stage with an adjustable gain of nominally G ~ −1 is useful, as it allows the response of the overall system to preserve the sign of the input current, and the gain can be trimmed to give an exact value, *e.g.* of 1V per pA, when $R_f$ has been determined by calibration.

Assuming an ideal system (all inputs screened, voltages steady, input-bias current $i_b$ and any additional leakage currents at the input $i_L$), the output voltage will be

[72] An alternative is to construct a preliminary feedback current amplifier using the intended current source resistor $R_{cal}$ as the feedback resistor, but working at *larger* current range (μA-nA), for which currents are more easily generated with conventional, calibrated components. This is effectively a variation on the scaling method already mentioned.

$$V_{out} = G\left\{\left[-(i + i_b + i_L)R_f\right] + V_{os1}\left[1 + \frac{R_f}{R_{cal}}\right]\right\} \qquad \textit{Eq. B.1,}$$

where $V_{os1}$ is the offset voltage[73] of amplifier A1, typically a few millivolts. This term is variable and therefore usually unknown: it is subject to thermal and temporal drifts.

Eq. B.1 shows that it is desirable for $R_{cal}$ to be at least comparable with, or preferably greater than $R_f$, so that the right hand square bracket term (amplification of A1's offset voltage) can be neglected. However this is not always possible given the restricted range of high-value resistors available for $R_f$ and $R_{cal}$[74]. In that case, an alternative is to fit a straight line to the response of $V_{out}$ against $V_{in}$, which will have a gradient

$$-G\frac{R_f}{R_{cal}} \qquad \textit{Eq. B.2}$$

This can be determined with essentially any well-defined value of $R_{cal}$.

The intercept, however, is given by

$$G\left\{\left[-(i_b + i_L)R_f\right] + V_{os1}\left[1 + \frac{R_f}{R_{cal}}\right]\right\} \qquad \textit{Eq. B.3,}$$

which will only remain constant from one calibration to the next if $i_b$, $i_L$ and $V_{os1}$ all remain constant. This is only likely to be the case for short time intervals, using *exactly* the same combination of resistors and connectors, or if it is established that $V_{os1}$ is steady and $i_L << i$.

In practice, use of such a "gradient-only" calibration for atmospheric ion measurements will not cause significant difficulty, as the source resistance of ion counters is exceedingly high. The ion counter can be seen as a resistive voltage source, with $R_{cal} >> R_f$ and therefore no amplification of the A1 offset voltage occurs. However it is important to realise that the offset present in the calibration of the Current Amplifier (if $R_{cal} << R_f$) will not be present in the measurement using the ion counter.

---

[73] voltage required at A1's *input* to make its *output* voltage zero

### B.1.2 Capacitative generation of small currents

The need to provide fixed small currents with which to test Current Amplifiers is complicated by the variable offset terms in Eq. B.1, due to finite calibration resistance. Any device operating with $R_{cal} << R_f$ will not generate a well-defined response from a Current Amplifier because of the offset term. An alternative method of generating a fixed current is capacitative, applying a voltage which is steadily changing to a capacitor.

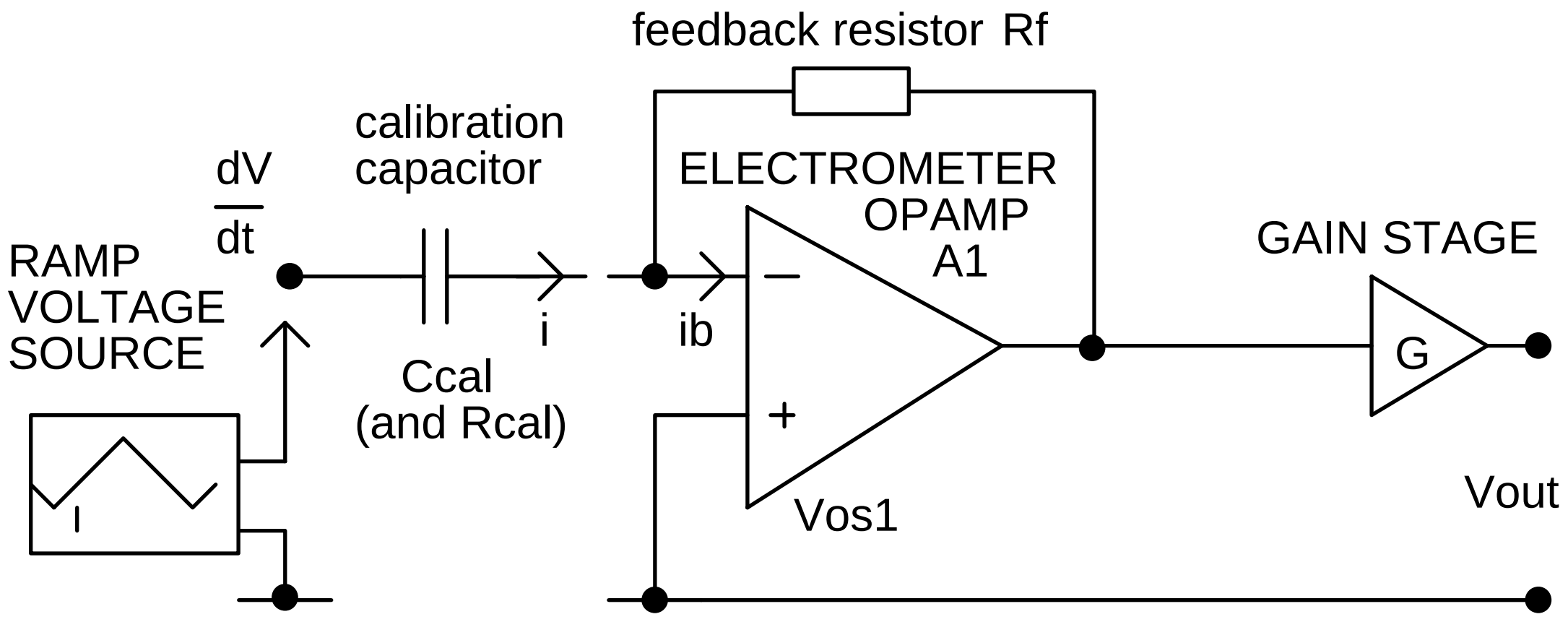

*Figure B.2 Capacitative generation of calibration current* i.

This method generates the reference current by differentiation, using a precise, slowly-varying linear ramp voltage, presented to a low value high-quality capacitor. The current generated *i* is proportional to the rate of change of voltage *V*(*t*), *i.e.*

$$i = C_{cal}\frac{dV}{dt} \qquad \text{Eq. B.4}$$

There is actually a resistive component to any practical capacitor, so that the current generated will vary with time according to

$$i(t) = C_{cal}\frac{dV}{dt} + \frac{V(t)}{R_{cal}} \qquad \text{Eq. B.5}$$

if the resistive generation of current is comparable with the capacitative generation. This can be minimised using a small ramp voltage. With a 50 $mVs^{-1}$ ramp voltage, and 10 pF polystyrene capacitor gave a *steady* current during the ramp, to 1 fA.

[74] The maximum value of resistor generally available is 1TΩ, which will give a 1V per pA response from an electrometer. If this sensitivity is required, calibration with a significantly larger $R_{cal}$ resistor is not practical.

Consequently the resistive contribution to the current is small, and $R_{cal}$ due to the capacitor can therefore be estimated as ≥1 PΩ.

The properties and stability of the ramp generator become the important requirements in the capacitive method, together with any AC response of the electrometer as the source and feedback *impedances* which are significant. The source impedance at high frequencies, $Z_{cal}$ however, will be substantially lower than $R_{cal}$[75], and therefore this approach is most ideally used with a 'slow' (low frequency) electrometer to minimise any noise gain.

### B.1.3 Practical implementation of capacitative technique

A voltage ramp generator has been described for this application (Harrison and Aplin, 2000a), in which the rising and falling periods of the voltage ramp can be used to generate symmetrical positive and negative calibration currents, Figure B.3. With a suitable capacitor connected, it becomes a Current Reference, readily arranged to generate ± 0.5 pA.

The reference current is generated by differentiation, using a slowly increasing and decreasing linear ramp voltage, presented to a low value polystyrene capacitor. The reference current is only available during the duration of the ramp, but symmetrical bipolar currents can be derived alternately, from the rising and falling edges, allowing two point calibration. Measurements need to be made a little *after* the ramp begins, to avoid transients. The output current can be deduced to reasonable accuracy (~2%) solely by using precision components, although the accuracy may be improved by direct measurement of the current using a standardised electrometer.

The circuit operates in the following manner. A stabilised 5.00 V supply is generated using voltage reference IC1, and a 2.50 V reference derived from it using equal precision resistors and follower IC4b. A crystal oscillator (IC2) is used to generate a stable and symmetrical 15.625 mHz square wave at the output of IC3. The MOSFET output stage of IC3 was found to deliver this square wave with precise amplitudes to within 1 mV of the 5.00 V supply and ground. The square wave is used as the input to an integrator based on C1 and IC4a, via R1, with the integrator centred at 2.50 V. If IC4 is chosen to have a negligible input bias current, the 5.00 V square wave will cause equal bipolar integrator input currents, and associated rising and falling voltage ramps at IC4a output. The final output current is generated via the differentiator comprising C2 (the calibration capacitor $C_{cal}$) and the input stage of the Current Amplifier connected.

[75] $Z_{cal}$ is the parallel combination of $R_{cal}$ and $X_{cal}$, the capacitative reactance.

The reference current $i$ is given by

$$i = C_2 \frac{dV}{dt} = \frac{C_2}{C_1}\frac{\Delta V}{R_1} \qquad \textit{Eq. B.6,}$$

where $\Delta V$ is the integrator input voltage (±2.50 V). C1, C2 and R1 are thermally stable, close-tolerance components. The choice of C2 is based on a compromise between a sufficiently large value to minimise the effect of unknown stray capacitances (with the associated thermal and temporal stability of the connection geometry) and a small value to maximise the source impedance and reduce the noise gain of the electrometer at high frequencies. In practise the first constraint was found to be the more important parameter, as the noise generated on the ramp voltage was small and the electrometer could be operated in a low frequency mode.

The integrator can be reset using the MOSFET switches in IC5, which are set by default in the RUN (*i.e.* current reference operating) condition. In the STOP state the oscillator and counters are reset, and the output of IC4 is pulled to ground via IC5c. This effectively eliminates leakage across the output capacitor C2 in the STOP state, allowing the Current Reference to be permanently connected to a slow response Current Amplifier, without any additional high impedance switching.

For the component values used, the rate of change of ramp voltage was measured using a chart recorder to be ±52.5 $mVs^{-1}$, with the ramp starting and finishing at 2.5 V. Tests using a Keithley 6512 electrometer showed the output current to be of magnitude 470 ± 2 fA, alternately positive and negative for 32 s each. Measurements were made 5 s after the ramp transition. The Current Reference was found to be stable to ± 2 fA for slowly-varying temperatures over the range −22 to 21.7 °C. The bipolar output currents also remained symmetrical to within the accuracy of measurement over this range, although sudden thermal shocks to the output capacitor C2 during initial cooling led to some transient asymmetric currents.

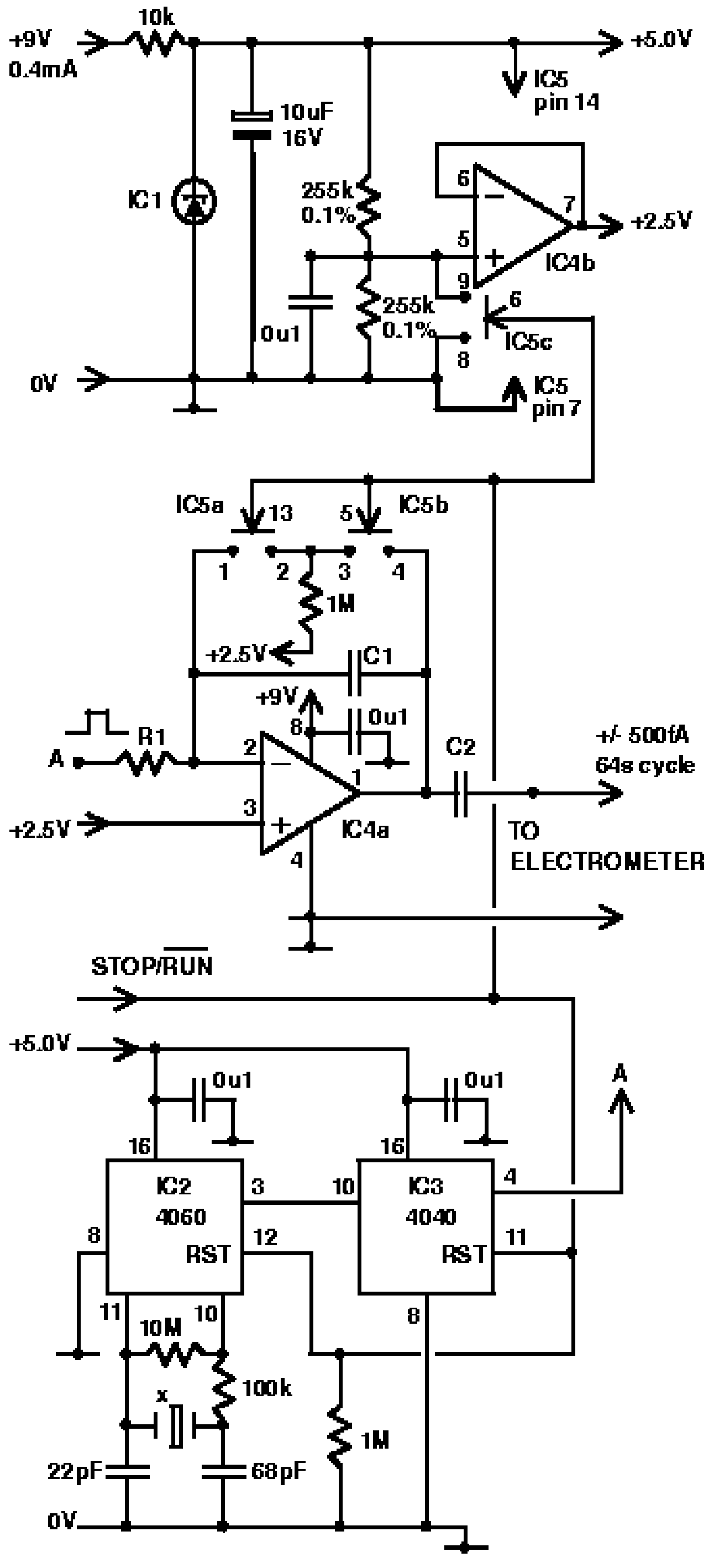

*Figure B.3 Schematic of the 500 fA Current Reference circuit. IC1 REF50Z 5 V ±30ppm/°C voltage reference; IC2 4060 oscillator divider; IC3 4040 binary divider; IC4 LTC1047 chopper stabilised dual opamp; IC5 4066 quad MOSFET switch; R1 10 MΩ ±1% ±50ppm/°C metal film; C1 4.7 μF ±1% polyester layer ±200ppm/°C; C2 10 pF polystyrene ±80ppm/°C; X 32768 Hz quartz crystal.*

# Appendix C

# Program source code

Listings for the most important microcontroller programs are given below. They are written in PBASIC (Parallax, 1998), which is a variant of BASIC with extra commands to control the microcontroller (*e.g.* setting the I/O pin states).

Source code for a program written in Turbo Pascal, DECINT, which performed the decay integral in Chapter 8 is also given.

## C.1 Chapter 4

The program GNOCLOCK interfaces a 12-bit A to D converter, and switches between two critical mobilities using reed relays to change the bias voltage.

```
‘GNOCLOCK.BAS
‘Karen Aplin 02/02/99
dirs=%01100000                 ' makes pins 5 and 6 (connected to
                               relays)outputs
symbol      CS    = 2          ' ADC Chip select (Low True)
symbol      CLK   = 3          ' ADC Clock
symbol      Din   = pin4       ' Data input pin number
symbol      DatIn = 4          ‘ ADC Data Input
symbol      Dout  = pin4       ' Data output pin number
symbol      DatOut= 4          ' ADC Data Output
symbol      ADbits= b1         ' Loop counter used to receive bits
symbol      ADRes = w1         ' 12-bit ADC conversion result
symbol      sglDif= 1          ' 1 -> Single-ended, two-channel mode.
symbol      msbf  = 1          ' Output 0s after data transfer is
                               complete
symbol      oddSign     = bit0      ' Channel number

Init:
      high CS                  ' Deactivate the ADC to begin.

switch:                        ' Main part of program
      serout 1,N300,(90)       ' outputs "Z" to PC serially to
      identify serout 1,N300,(10)   mode
      serout 1,N300,(13)
      pin5=1
      pin6=1                   ' Both relays on (shorts out Rf to
                               prevent transients)
      pause 10000              ' waits for 10s
      gosub log
      serout 1,N300,(95)       ' outputs "-" to PC serially
      serout 1,N300,(10)
      serout 1,N300,(13)
      pins=0                   ' Both relays off – switches on
                               negative Vb and turns off shorting
                               relay
```

```
      pause 120000            ' waits for 2 min
      gosub log               ' reading 1
      pause 6000              ' waits for 1 s
      gosub log               ' reading 2
      pause 6000              ' waits for 1 s
      gosub log               ' reading 3
      pause 6000              ' waits for 1 s
      gosub log               ' reading 4
      pause 6000              ' waits for 1 s
      gosub log               ' reading 5
      pause 6000              ' waits for 1 s
      gosub log               ' reading 6
      serout 1,N300,(43)      ' outputs "+" to PC
      serout 1,N300,(10)
      serout 1,N300,(13)
      pin6=0
      pin5=1                  ' Switches on positive Vb
      pause 1800000           ' waits for 3 min
      gosub log               ' reading 1
      pause 6000              ' waits for 1 s
      gosub log               ' reading 2
      pause 6000              ' waits for 1 s
      gosub log               ' reading 3
      pause 6000              ' waits for 1 s
      gosub log               ' reading 4
      pause 6000              ' waits for 1 s
      gosub log               ' reading 5
      pause 6000              ' waits for 1 s
      gosub log               ' reading 6
      goto switch             ' repeats whole process forever
'*******************************************************************
log:                          ' logging subroutine
      oddSign = 0
      gosub Convert
      serout 1,N300,(#ADRes)  ' outputs ADC reading to PC
      serout 1,N300,(10)
      serout 1,N300,(13)
      return                  ' returns to switching loop

Convert:                      ' ADC Converter subroutine
      low CLK                 ' Low clock--output on rising edge.
      high DatOut             ' Switch DIO to output high (start
bit).
      low CS                  ' Activate the 1298.
      pulsout CLK,5           ' Send start bit.
      let Dout = sglDif       ' First setup bit.
      pulsout CLK,5           ' Send bit.
      let Dout = oddSign
      pulsout CLK,5           ' Send bit.
      let pin4 = msbf         ' Final setup bit.(total 3 bit input)
      pulsout CLK,5           ' Send bit.
      input DatIn
      let ADRes = 0           ' Clear old ADC result.
      for ADbits = 1 to 13    ' Get null bit + 12 data bits.
      let ADRes = ADRes*2+Din ' Shift AD left, add new data bit.
      pulsout CLK,5           ' Clock next data bit in.
      next                    ' Get next data bit.
      high CS                 ' Turn off the ADC
      return                  ' Return to program.
```

## C.2 Chapter 5

This section contains the programs developed to control the PIMS system. Section C.2.1 gives the routine written to control the DAC, and the remainder of the programs control the multimode electrometer. The programs used separately to test the modes are given first, and followed by MMESYS, the overall PIMS control program. The MME switching logic is given in Appendix A.

### C.2.1 DAC driver

The DAC driver routine was written by J.R. Knight. It stores the DAC voltage codes in the microcontroller memory, and after the DAC is initialised, it reads through the binary DAC code to send it bit by bit to the DAC.

```
‘JRK2.BAS
'DAC driver routine JRK 25/05/99 simple-data version
dirs=%11111111                ' makes all pins outputs
symbol      CSDAC = 0         ' DAC Chip Select
symbol      CLK   = 3         ' ADC/DAC Clock
symbol      Din   = pin4
symbol      stp   = b2
symbol      stp2  = b3
symbol      data  = b4
symbol      tem   = b5
symbol      mask  = b6
symbol      vsteps      = 2   ' (number of voltage steps – 1)

read 255,data                 ' reads out eeprom space available
debug data


init:   eeprom 0,(0,128,255)  ' stores voltage codes in EEPROM
      high csdac
      low clk

loop:                         ‘ main part of DAC driver
      for stp = 0 to vsteps   ‘ steps through sequence of voltage
codes
      low csdac
      data = 9                ' control byte needed to initialise DAC
      gosub send
      read stp,data           ‘ Reads voltage code into memory
      debug data
      gosub send
      high csdac

pause 5000
                              ' Additional functions of the program
                              can be added here.
next
goto loop                     ‘ Starts sequence again
```

```
send:                         ‘ Produces a control byte to send to
DAC
      mask=128
      for stp2= 0 to 7        ‘ Loop which moves along the bits in
                              the control byte one by one
      tem=mask & data
      din=0
      if tem=0 then pulse
      din=1

pulse:                        ‘ Sends 1 digital pulse to DAC
      pulsout CLK,500

      mask=mask/2             ‘ Moves one step along the binary
                              sequence to be sent to DAC
      next
      return                  ‘ Returns to program
```

### C.2.2 MME test programs

IBLOG switches the MME into leakage check mode and input offset voltage measurement mode. It then logs the output serially.

```
' error check mode logging program (IBLOG.BAS)
' Karen Aplin 25/01/00
dirs=%11110111                'makes all pins except 4 outputs
symbol      CE    = pin7      'MME logic control lines
symbol      CY    = pin5
symbol      CX    = pin6
symbol      CSADC = 2         ' ADC Chip select (Low True)
symbol      CLK   = 3         ' ADC/DAC Clock
symbol      Din   = pin4
symbol      Dout  = pin4
symbol      Dpin  = 4
symbol      ADRes = w1        ' 12-bit ADC conversion result
symbol      sglDif= 0         ‘ ADC in differential mode
symbol      msbf  = 1
symbol      oddSign= 0
symbol      ADBits  = b4
symbol      hh    = b8        ' variable controlling number of
samples

main:                         ‘ main part of program – steps through
                              error check modes
      gosub ib
      gosub log
      gosub Vos
      gosub log
      goto main

ib:                           ‘ MME leakage current mode
      CE = 0                  ‘ Logic control to select mode
      CY = 1
      CX = 1
      serout 1,N300,("ib",44) ‘ Sends character to denote mode to
                              logging PC serially
      pause 1000
      return
```

```
Vos:                          ' MME input offset voltage mode
      CE = 0
      CY = 1
      CX = 0
      serout 1,N300,("Vos",44)
      pause 1000
      return

log:  for hh = 1 to 29        ' logging loop
      pause 5000              ' pause length determines sampling
                              frequency
      gosub convert
      serout 1,N300,(#ADRes,44)' output reading to PC
      next
      pause 5000              ' recovery time before last reading
      gosub convert
      serout 1,N300,(#ADRes,13,10)'last reading with carriage return
      return

convert:                      ' ADC Driver
      low CLK                 ' Low clock--output on rising edge.
      high Dpin               ' Switch DIO to output high (start
bit).
      low CSADC               ' Activate the 1298.
      pulsout CLK,5           ' Send start bit.
      let Dout = sglDif       ' First setup bit.
      pulsout CLK,5           ' Send bit.
      let Dout = oddSign
      pulsout CLK,5           ' Send bit.
      let Dout = msbf         ' Final setup bit.(total 3 bit input)
      pulsout CLK,5           ' Send bit.
      input Dpin
      let ADRes = 0           ' Clear old ADC result.
      for ADbits = 1 to 13    ' Get null bit + 12 data bits.
      let ADRes = ADRes*2+Din ' Shift AD left, add new data bit.
      pulsout CLK,50          ' Clock next data bit in.
      next                    ' Get next data bit.
      high CSADC              ' Turn off the ADC
      output Dpin             ' makes D output again
      return                  ' Return to program.
```

ICAL sets the MME into current measurement mode and calculates average currents which are sent to the PC serially. It is designed to calibrate the MME picoammeter function with the femtoampere current reference reference (Harrison and Aplin, 2000a).

```
' ical.BAS
' Karen Aplin 25/01/00
dirs=%11110111                'makes all pins except 4 outputs
symbol      CE    = pin7      'MME logic control
symbol      CY    = pin5
symbol      CX    = pin6
symbol      CSADC = 2         ' ADC Chip select (Low True)
symbol      CLK   = 3         ' ADC/DAC Clock
symbol      Din   = pin4
```

```
symbol      Dout  = pin4
symbol      Dpin  = 4
symbol      Datasum = w3      'Word variable storing the sum of
samples
                               to average
symbol      ADRes = w1        ' 12-bit ADC conversion result
symbol      sglDif= 0         ' ADC in differential mode
symbol      msbf  = 1
symbol      oddSign     = 0
symbol      ADBits  = b4
symbol      hh    = b8        ' no of samples to average over
symbol      CSDAC = 0         ' DAC Chip Select
symbol      stp   = b2
symbol      stp2  = b3
symbol      data  = b4
symbol      tem   = b5
symbol      mask  = b6
symbol      vsteps = 2        ' (voltage steps) - 1

      CE = 0                  ' sets MME into picoammeter mode
      CX = 1
      CY = 0

main:  gosub log              ' main loop
      goto main

log:
      Datasum = 0             ' Resets sum of samples to zero
      for hh = 1 to 5         ' Number of points to average over
      gosub convert
      Datasum = Datasum + Adres ' Adds the instantaneous reading to
the
                              total
      pause 1000              ' sampling frequency (1000 = 1 s)
      debug Datasum
      next
      Datasum=Datasum/5       ' calculates the average value
      debug "mean",#Datasum   ' shows mean on screen
      serout 1,N300,(#Datasum)' outputs reading serially to PC
      return

convert:                      ' ADC driver
      low CLK                 ' Low clock--output on rising edge.
      high Dpin               ' Switch DIO to output high (start
bit).
      low CSADC               ' Activate the 1298.
      pulsout CLK,5           ' Send start bit.
      let Dout = sglDif       ' First setup bit.
      pulsout CLK,5           ' Send bit.
      let Dout = oddSign
      pulsout CLK,5           ' Send bit.
      let Dout = msbf         ' Final setup bit.(total 3 bit input)
      pulsout CLK,5           ' Send bit.
      input Dpin
      let ADRes = 0           ' Clear old ADC result.
      for ADbits = 1 to 13    ' Get null bit + 12 data bits.
      let ADRes = ADRes*2+Din ' Shift AD left, add new data bit.
      pulsout CLK,50          ' Clock next data bit in.
      next                    ' Get next data bit.
      high CSADC              ' Turn off the ADC
      output Dpin             ' makes D output again
```

```
    return                  ' Return to program.
```

DECAY tests the MME follower mode, and the voltage decay of the central electrode. It charges up the central electrode, switches it off and makes 2 Hz measurements of the voltage decay in follower mode. The bias voltage can also be varied cyclically.

```
' Multimode electrometer decay mode test program (decay.BAS)
' Karen Aplin 15/02/00
dirs=%11110101              ' makes pins 0,2,4-7 outputs
symbol      CSADC = 2       ' ADC Chip select (Low True)
symbol      CLK   = 3       ' ADC/DAC Clock
symbol      Din   = pin4
symbol      Dout  = pin4
symbol      Dpin  = 4
symbol      data  = b0
symbol      ADRes = w1      ' 12-bit ADC conversion result
symbol      sglDif      = 0
symbol      msbf  = 1
symbol      oddSign     = 0
symbol      ADBits  = b4
symbol      hh    = b8
symbol      CE    = pin7    ' MME logic
symbol      CY    = pin5
symbol      CX    = pin6
symbol      ADComp      = w5
symbol      CSDAC = 0       ' DAC Chip Select
symbol      stp   = b2      ' DAC controller variables
symbol      stp2  = b3
symbol      tem   = b5
symbol      mask  = b6

      let stp = 1
      eeprom 0,(0,0,0)        ' stores voltage codes
      high csdac
        low clk
      low csdac
      data = 9  'control byte
      gosub send
      read stp,data
      debug data
      gosub send
      high csdac

main:                         ' main part of program
      CE = 1                  ' MME logic to charge up central
                              electrode
      CY = 0
      CX = 0
      debug "Charged",cr      ' screen message
      pause 100
      CE = 0                  ' switches off central electrode
                              voltage and puts MME into follower mode
      pause 100
      serout 1,N300,(10,13)   ' starts new line in logging file
      for hh = 1 to 30        ' defines how many points to be
                              measured in decay
```

```
      gosub log
      next
      goto main

log:                          ' subroutine to take a measurement,
                              display it on the screen and send it to
                              the PC
      gosub convert
      serout 1,N300,(#ADRes,44)' output reading to PC
      debug ADRes,cr
      pause 1000
      return

convert:                      ' ADC driver
      low CLK                 ' Low clock--output on rising edge.
      high Dpin               ' Switch DIO to output high (start
bit).
      low CSADC               ' Activate the 1298.
      pulsout CLK,5           ' Send start bit.
      let Dout = sglDif       ' First setup bit.
      pulsout CLK,5           ' Send bit.
      let Dout = oddSign
      pulsout CLK,5           ' Send bit.
      let Dout = msbf         ' Final setup bit.(total 3 bit input)
      pulsout CLK,5           ' Send bit.
      input Dpin
      let ADRes = 0           ' Clear old ADC result.
      for ADbits = 1 to 13    ' Get null bit + 12 data bits.
      let ADRes = ADRes*2+Din ' Shift AD left, add new data bit.
      pulsout CLK,50          ' Clock next data bit in.
      next                    ' Get next data bit.
      high CSADC              ' Turn off the ADC
      output Dpin             ' makes D output again
      return                  ' Return to program.

send: mask=128                ' DAC driver
      for stp2= 0 to 7
      tem=mask & data
      din=0
      if tem=0 then pulse
      din=1
pulse:      pulsout CLK,500
      mask=mask/2
      next
      return
```

ADCAV switches between current measurement and error checking modes at a range of bias voltages. It then calculates an average measurement and sends it to the PC serially, so that the three measurements are in the same line in the output file.

```
' ADCAV.BAS - Logs and serially outputs averages
' Karen Aplin 25/01/00
dirs=%11110111                'makes all pins except 4 outputs
symbol      CE    = pin7      'MME logic
symbol      CY    = pin5
symbol      CX    = pin6
```

```
symbol      CSADC = 2          ' ADC Chip select (Low True)
symbol      CLK   = 3          ' ADC/DAC Clock
symbol      Din   = pin4
symbol      Dout  = pin4
symbol      Dpin  = 4
symbol      Datasum = w3
symbol      ADRes = w1         ' 12-bit ADC conversion result
symbol      sglDif= 0          ' ADC in differential mode
symbol      msbf  = 1
symbol      oddSign= 0
symbol      ADBits  = b4
symbol      hh    = b8         ' used in loop: no of points to average
                               over
symbol      CSDAC = 0          ' DAC Chip Select
symbol      CLK   = 3          ' ADC/DAC Clock
symbol      Din   = pin4
symbol      stp   = b2
symbol      stp2  = b3
symbol      data  = b4
symbol      tem   = b5
symbol      mask  = b6
symbol      vsteps = 2         'voltage steps minus one


main:                          ‘ main loop
      eeprom 0,(0,128,255)     ' reads voltage codes into memory
      for stp = 0 to vsteps
      low csdac
      data = 9                 ' DAC control byte
      gosub send
      read stp,data
      debug data
      gosub send
      high csdac
      low clk
      gosub ib                 ‘ leakage check mode
      gosub log                ‘ logs data
      gosub Vos                ‘ input offset voltage check mode
      gosub log
      gosub current            ‘ current measurement mode
      gosub log
      goto main

ib:
      CE = 0                   ‘ switches MME to leakage check mode
      CY = 1
      CX = 1
      debug "ib"
      serout 1,N300,(44)       ‘ sends comma to PC logging file
      pause 10000              ‘ 10s recovery time from transient
      return
Vos:
      CE = 0
      CY = 1
      CX = 0
      debug "Vos"
      serout 1,N300,(44)
      pause 10000
      return

current:
```

```
      CE = 0
      CX = 1
      CY = 0
      debug "current"
      serout 1,N300,(10,13)   ‘ sends carriage return and line feed
                              to PC logging file
      pause 10000
      return


log:                          ‘ averaging/logging routine
      Datasum = 0
      for hh = 1 to 10
      gosub convert
      Datasum = Datasum + ADres
      pause 1000
      debug Datasum
      next
      Datasum=Datasum/10
      debug "mean",#Datasum
      serout 1,N300,(#Datasum)' output reading to PC
      return

convert:                      ‘ ADC driver
      low CLK                 ' Low clock--output on rising edge.
      high Dpin               ' Switch DIO to output high (start
bit).
      low CSADC               ' Activate the 1298.
      pulsout CLK,5           ' Send start bit.
      let Dout = sglDif       ' First setup bit.
      pulsout CLK,5           ' Send bit.
      let Dout = oddSign
      pulsout CLK,5           ' Send bit.
      let Dout = msbf         ' Final setup bit.(total 3 bit input)
      pulsout CLK,5           ' Send bit.
      input Dpin
      let ADRes = 0           ' Clear old ADC result.
      for ADbits = 1 to 13    ' Get null bit + 12 data bits.
      let ADRes = ADRes*2+Din ' Shift AD left, add new data bit.
      pulsout CLK,50          ' Clock next data bit in.
      next                    ' Get next data bit.
      high CSADC              ' Turn off the ADC
      output Dpin             ' makes D output again
      return                  ' Return to program.

send: mask=128                ‘ DAC driver
      for stp2= 0 to 7
      tem=mask & data
      din=0
      if tem=0 then pulse
      din=1
pulse:      pulsout CLK,500
      mask=mask/2
      next
      return
```

MMESYS is the final program used to operate the PIMS in all its modes (as given in Table A.1), and take measurements of voltage decays and current samples. MMESYS2 is shown here; the only difference between the programs for the different

PIMS is in the microcontroller timings and DAC calibrations. It is essentially a combination of ADCAV and DECAY.

```
'MMESYS2.BAS uses MME all modes, logs and serially outputs averages
'Karen Aplin 23/02/00
dirs=%11110111                  'makes all pins except 4 outputs
symbol      CE    = pin7        'MME logic
symbol      CY    = pin5
symbol      CX    = pin6
symbol      CSADC = 2           ' ADC Chip select (Low True)
symbol      CLK   = 3           ' ADC/DAC Clock
symbol      Din   = pin4
symbol      Dout  = pin4
symbol      Dpin  = 4
symbol      Datasum = w5
symbol      ADRes = w1          ' 12-bit ADC conversion result
symbol      sglDif= 0
symbol      msbf  = 1
symbol      oddSign = 0
symbol      ADBits  = b4
symbol      CSDAC = 0           ' DAC Chip Select
symbol      stp   = b1
symbol      stp2  = b9
symbol      data  = b7
symbol      tem   = b5
symbol      mask  = b6          'DAC control and counting loop variable
symbol      vsteps = 9          '(voltage steps-1)

      eeprom 0,(0,55,0,65,0,75,0,85,0,106)' stores voltage codes

main:
      for stp = 0 to vsteps     ' voltage switching for i measurements
      low csdac
      data = 9                  ' DAC control byte
      gosub send
      read stp,data
      debug data,cr
serout 1,N300,(10,13,"I",44,#data) ‘ starts new line and sends “I” to
                              PC to denote averaging measurement mode
      gosub send
      low clk
      gosub ib
      gosub Vos
      gosub current
      gosub decay
footer:     next
      goto main

ib:                             'leakage current check mode
      CE = 0
      CY = 1
      CX = 1
      'debug "leakage",cr
      gosub logav

Vos:                            'input offset voltage check mode
      CE = 0
      CY = 1
```

```
      CX = 0
      'debug "Vos",cr
      gosub logav

current:                     'current measurement mode
      CE = 0
      CX = 1
      CY = 0
      'debug "current",cr
      gosub logav

decay:                       ' subroutine to charge up central
                             electrode and log the voltage as it
                             decays from it
      CE = 1                 'charges central electrode
      CY = 0
      CX = 0
      pause 100
      CE = 0
      'debug "decay",cr
      pause 100   'decay mode
      serout 1,N300,(10,13,"V",44,#data)' starts new line of data for
                             voltage decay measurements
      for mask = 1 to 10
      gosub convert
      serout 1,N300,(44,#ADRes) ' data points are comma-separated
      debug #ADRes
      pause 500
      next
      goto footer

logav:                       ' averaging routine for current
                             measurement and error check modes
      serout 1,N300,(44)     ' data points are comma-separated
      pause 10000
      Datasum = 0
      for mask = 1 to 10
      gosub convert
      Datasum = Datasum + ADres
      pause 1000
      debug Datasum
      next
      Datasum=Datasum/10
      'debug "mean",#Datasum
      serout 1,N300,(#Datasum)' output reading to PC
      return

convert:
      low CLK                ' Low clock--output on rising edge.
      high Dpin              ' Switch DIO to output high (start
bit).
      low CSADC              ' Activate the 1298.
      pulsout CLK,5          ' Send start bit.
      let Dout = sglDif      ' First setup bit.
      pulsout CLK,5          ' Send bit.
      let Dout = oddSign
      pulsout CLK,5          ' Send bit.
      let Dout = msbf        ' Final setup bit.(total 3 bit input)
      pulsout CLK,5          ' Send bit.
      input Dpin
      let ADRes = 0          ' Clear old ADC result.
```

```
      for ADbits = 1 to 13    ' Get null bit + 12 data bits.
      let ADRes = ADRes*2+Din ' Shift AD left, add new data bit.
      pulsout CLK,50          ' Clock next data bit in.
      next                    ' Get next data bit.
      high CSADC              ' Turn off the ADC
      output Dpin             ' makes D output again
      return                  ' Return to program.

send:                         ‘ DAC driver
      mask=128
      for stp2= 0 to 7
      tem=mask & data
      din=0
      if tem=0 then pulse
      din=1
pulse:      pulsout CLK,500
      mask=mask/2
      next
      return
```

## C.3 Chapter 8

DECINT was written by R.G. Harrison and performs the numerical integration to simulate a voltage decay with a predefined spectrum.

```
Program DECINT; {DECayINTegral}
{ program to perform numerical integrations to calculate the decay
 of voltage across electrodes of Gerdien tube
 (1) in the simplest case (single ion mobility) this is an
exponential
 (2) with an ion mobility spectrum, the decay deviates from an
exponential
 which is found by integrating over the mobility spectrum up to the
critical
 mobility determined by the instantaneous voltage generated from the
decay
 (3) mobility spectrum is assumed to be Gaussian (fitting Estonian
data)
 (4)integration performed using quality-controlled Simpson's rule

 Giles Harrison March 2000}

type integer=longint;

const
 {tube parameters}
 a=11.0;    {mm}
 b=2.0;           {mm}
 L=0.25;    {m}
 u=2.0;           {m.s-1}
 epsilon  = 8.841941e-12;  {permittivity of free space   [Farads per
metre]}
 Qe      =1.6021892e-19;  {modulus of electronic charge
[Coulombs]}

var
 timesum,V0,VT,V1,V2,T12,upperMu,lowerMu,MobInt,VoltInt,IonSum:real;
 TrapzdIt: integer;
```

```
 Npoints,i:integer;
 fout:text;

{1. function to compute critical mobility from tube parameters, as a
function of the applied bias voltage V}
function MuCrit(V:real):real; {in cm2.v-1.s-1}
begin
MuCrit:= ((sqr(a)-sqr(b))*1.0e-6*ln(a/b) * u)/(2.0*V*L)/1.0e-4;
end;

{2. Integrand for ion mobility spectrum, generated by Gaussian fit to
Estonian data}
                   FUNCTION func1(X:EXTENDED):EXTENDED;
                  const
                  a=130.22902;
                  b=1.4791917;
                  c=0.39179526;
                  var
                  y:real;
                   BEGIN      {integral for (positive) ion spectrum}
                  {fitted to Gaussian curve}
                  y:=100.0*   {to convert cm-based units to SI}
                        10.0*a * exp( -sqr(x-b)/ (2.0*sqr(c) )  );
                 func1:= y;
                   END;


{implementation of Simpson's rule, as coded in "Numerical Recipes in
Pascal"
which calls the Trapezium rule first}

PROCEDURE trapzd1(a,b: real;
               VAR s: real;
                   n: integer);
VAR
   j: integer;
   x,tnm,sum,del: real;
BEGIN
   IF n = 1 THEN BEGIN
      s := 0.5*(b-a)*(func1(a)+func1(b));
      TrapzdIt := 1
   END
   ELSE BEGIN
      tnm := TrapzdIt;
      del := (b-a)/tnm;
      x := a+0.5*del;
      sum := 0.0;
      FOR j := 1 TO TrapzdIt DO BEGIN
         sum := sum+func1(x);
         x := x+del
      END;
      s := 0.5*(s+(b-a)*sum/tnm);
      TrapzdIt := 2*TrapzdIt
   END
END;

PROCEDURE qsimp1(a,b: real;
              VAR s: real);
LABEL 99;
CONST
   eps = 1.0e-3;
```

```
   jmax = 20;
VAR
   j: integer;
   st,ost,os: real;
BEGIN
   ost := -1.0e30;
   os := -1.0e30;
   FOR j := 1 TO jmax DO BEGIN
      trapzd1(a,b,st,j);
      s := (4.0*st-ost)/3.0;
      IF abs(s-os) < eps*abs(os) THEN GOTO 99;
      os := s;
      ost := st
   END;
   writeln ('pause in QSIMP1- too many steps');
   readln;
99:
END;


{3. find time taken for the voltage decay (which requires the
integration over
the mobility spectrum up to the critical mobility at each voltage) }
function decaytime(V0,VT:real):real;
begin
upperMu:=MuCrit(VT);
lowerMu:=MuCrit(V0);
writeln(lowerMu,' ', upperMu);
qsimp1(lowerMu,upperMu,MobInt);           {mobility spectrum}
VoltInt:=ln(VT/V0);
decaytime:=-(epsilon/Qe)* (VoltInt/MobInt) ;
end;



BEGIN {main program}

{open output file file, comma-separated for Excel}
assign(fout,'c:\temp\decay.csv');
rewrite(fout);
writeln(fout,'volts,timediff,time');

{compute time taken T12 for decay from V1 to V2 ultimately, by
evaluating the
decay integral from V1 down to VT over ion mobility spectrum
-evaluated at 10 points in the decay}
V1:=10; V2:=1;
Npoints:=10;
writeln(V1:3:2,',0');
writeln(fout,V1:3:2,',0');
timesum:=0;
for i:=Npoints downto 1 do
      begin
      VT:=i*( (V1-V2)/Npoints);
      V0:=((i+1)*(V1-V2))/Npoints;
{     writeln(V0:3:2,' ',VT:3:2);}
      T12:=decaytime(V0,VT);
      timesum:=timesum+T12;
      writeln(VT:3:2,',',T12:16,',',timesum:16);
      writeln(fout,VT:3:2,',',T12:16,',',timesum:16);
      end;
```

```
close(fout);      {close output file}
END.
```

# Appendix D
# Gerdien Engineering Diagrams

This appendix contains the engineering diagrams for the new design of stainless steel Gerdien described in Chapter 5. The Gerdien and diagrams were designed and prepared by S. D. Gill and A.G. Lomas. The diagrams are supplied in a folder in the inside back cover of the Thesis. A summary of their content is given in Table D.1 below. (Figure 5.2 is a photograph of the Gerdien in construction.)

| **Drawing Title** | **Description** |
|---|---|
| Gerdien tube | Outside of the tube showing BNCs and end of central electrode |
| Gerdien tube assembly – full detail | As above but shows internal components as dotted lines |
| Gerdien tube – sectioned | Shows two cross-sections of the tube<br>AA: longitudinal, showing central electrode detail<br>BB: transect through central electrode connector |
| Gerdien tube – part sectioned | As *Gerdien tube* but shows central electrode inside the tube. |
| Gerdien tube – components | Shows individual components. Modified and given in main text as Figure 5.1. |

# Appendix E

# The development of an integrated scientific perception of atmospheric particulates during the twentieth century

## E.1 Introduction

Interest in the electrical properties of the air is not new and probably began as a result of curiosity about spectacular natural phenomena like the aurorae and lightning. Yet successful measurements and understanding of air-borne particles, which contribute to these phenomena, are clearly products of the science of the twentieth century. The reasons for the surge of interest in the field from 1898 onwards are examined and suggest that a complex combination of scientific circumstances played a crucial rôle, motivating and facilitating research into air ions and their electrical properties. The development of the subject is followed, towards the "modern" approach of the present day, with air ions generally considered as part of the continuum of atmospheric aerosol particles. Finally, a case study of the relationship between atmospheric small ion and aerosol concentration acts as a supporting example, and relates historical scientific discoveries to contemporary work (*e.g.* Guo *et al.*, 1996).

### E.1.1 Air ions in context

Today the study of air ions is based primarily upon the idea of their continual interaction with other parts of the atmospheric aerosol spectrum, but their special contribution to atmospheric electricity is also recognised. This article traces the rôle of air ions and their relevance to different aspects of twentieth century physics, within a broad historical context.

The route to our current perception of air ion behaviour can be characterised by the development of two focal areas. Firstly, there was a need for a semantic framework to describe and answer problems effectively. The unification of the initially separate studies of "dust" and air ions was also a crucial step in the progression towards the contemporary approach.

### E.1.2 Background: The discovery of the electron

The first publication directly related to air ions seems to be by Zeleny in 1897 and research into the subject became very popular in the early years of this century. In a thorough and well-researched article, Flagan (1998) suggests that the problem of charge dissipation in the atmosphere was enough to motivate the plethora of papers published from 1898-1930. This problem had been known for many years, and could ostensibly only be investigated after the discovery of the electron in 1897 by J.J. Thomson explained the true nature of the ion. Yet it seems unlikely that this would be a pressing enough problem for so many workers to become involved, especially when there was so much new work to do in physics at the turn of the century. In addition, much of the research into air ions clearly evolved from experiments initially unrelated to atmospheric science, see for example Rutherford (1897) and McClelland (1898).

Ions were first identified by Faraday in his $19^{th}$ century work on electrochemistry, and were known to be produced by some sort of breakdown of molecules in an electric field. Helmholtz developed this work to define the electric charge on the atom as the finite quantity of electricity carried by all ions. Stoney then used the word "electron" for the first time to describe a fundamental amount of electricity (Robotti, 1995). It is well known that J.J. Thomson identified the constituents of cathode rays as the first sub-atomic particles, which he called "corpuscles", in 1897. As the idea of the atom breaking up in any way had been inconceivable since Democritus, who first used the term, Thomson lacked sufficient evidence to immediately identify the corpuscle with a fundamental unit of both electricity and matter. Robotti discusses Thomson's attempts to measure the charge of the corpuscles independently. It was only when this measurement had been completed, that the corpuscle could be recognised as the electron. According to Robotti, Thomson based the whole research programme at the Cavendish Laboratory around this aim. It is fortunate for physics, and meteorology in particular, that he did so. Ernest Rutherford and C.T.R. Wilson were taken on as research students in the 1890s. Although they became famous for work outside this field, (Wilson's being particularly important to meteorology) both men produced valuable work related to ions in the air during this period at the Cavendish Laboratory.

One of the areas investigated by Thomson was the conduction of electricity through gases. Coulomb in 1795 was the first person to notice that the atmosphere is slightly electrically conductive, finding that an insulated gold leaf electrometer would discharge slowly in air. Roentgen's timely discovery of X-rays (in 1896) was another circumstance that abetted research into ions in gases. It was quickly found that "Roentgen rays" made gases electrically conductive by ionisation. This probably appealed to Thomson because of the potential for measuring the charge of the corpuscles that were produced. The *Concise Dictionary of Scientific Biography*[76] (1981) explains that Thomson took a new approach to the problem of conduction of electricity through gases, describing ions in terms of experimentally defined parameters (such as mobility) when previously, only the dynamics of molecular collisions were considered (proving inadequate to explain the problem). This could well derive from his desire to measure the charge on the corpuscle without reference to pre-existing particles. The arrival of the experimentalist C.T.R. Wilson at the laboratory permitted further progress with some theoretical work of Thomson's from 1893, relating to the effect of charged particles on condensation. Roentgen rays were found to increase the condensation rate in a saturated chamber, and these results, combined with Thomson's relating to ionisation increasing the electrical conductivity of gases, led them to surmise that ions acted as cloud condensation nucleii in supersaturated conditions not found in the atmosphere (Robotti, 1995).

This work did not lead to the direct measurement of the charge of the corpuscle, but it provided a method for $e$ to be measured using the photoelectric effect. Thomson could then present his revolutionary idea, that the electron was fundamental, both as a unit of electrical charge *and* within the atom. Robotti summarises, "Thomson's measurement of the charge of the corpuscle meant that … all electrical phenomena would have to be re-examined". This is absolutely fundamental to the history of air ion measurements. Thomson's preliminary work had provided a conceptual and experimental framework within which atmospheric electricity could be considered, and his eventual result provided motivation for further work in all aspects of electricity.

---

[76] *Concise Dictionary of Scientific Biography* (1981), Scribner

## E.2 Development of the study of air ions

One way to define science is as a progressive activity which moves from one theory to another (*e.g.* Kuhn, 1967), with later theories generally thought of as closer approximations to the truth. One does not have to subscribe to any particular philosophical approach to scientific progress to appreciate that every new framework of scientific thought brings with it a new terminology. The acknowledgement and ultimately, application of the new language by scientists working in similar fields signifies the true acceptance of the exemplar. Thomson's work on gaseous ions was hardly revolutionary on the same scale as the discovery of the electron, more a new approach which had evolved from his efforts to solve a different problem. Yet it did require a new terminology, and the development of the "20$^{th}$ century" approach to ion measurements can be traced by the introduction of Thomson's new terms into the scientific vernacular.

Consider the concept of electrical mobility as defined by Thomson in his book *The Conduction of Electricity Through Gases*, which was first published in 1903.

$$\mu = \frac{v_d}{E} \qquad \textit{Eq. E.1}$$

Mobility, $\mu$, was a powerful parameter to describe the motion of ions in an electric field because it relates the drift velocity attained by the charged particle, $v_d$ to $\underline{E}$, the magnitude of the electric field. This assumes the Stokes régime, which is when the electrostatic forces acting on the particle balance the drag forces. After the work of Wellisch (1909) and Cunningham (1910) electrical mobility could also be related to particle radius and mass.

Zeleny (1898) discovered that negative ions needed a larger potential difference to be deflected by the same amount as the positive ions. This implies that the negative ions have a higher velocity (and hence mobility) than positive ions. Zeleny reported these results apparently without appreciating the concept of mobility as defined by Thomson. In his papers from 1898 and 1900, he refers continually to the velocity of ions in an electrical field, and his tables of results are headed with, "velocity of ions in an electric field of 1 $Vm^{-1}$". According to Flagan (1998), Zeleny was responsible for the definition of critical mobility in the cylindrical condenser, that is, the minimum mobility of ion that would hit the plate T and contribute to the electrometer deflection.

With hindsight this is true, but Zeleny only defined a critical ionic *velocity*, indicating that by 1900 the importance of mobility as a parameter had not been directly recognised.

In 1900 C.T.R. Wilson described some experiments by Elster and Geitel related to the rate of discharge of an electrified body in air. Using Thomson's value for the charge on the ions, the ionisation rate was calculated at 20 ions $cm^{-3}s^{-1}$ for each sign. Wilson's terminology in this paper is interesting. He was still working at the Cavendish Laboratory at this time and had begun his scientific career under Thomson, so Wilson, if anybody, would use the rhetoric developed by his supervisor. Yet he is still referring to the charge on the ion, and not explicitly equating it to what is now called the charge on the electron. However, as Thomson himself continued to refer to the electron as the corpuscle occasionally until the 1920s, this is possibly understandable. It appears that the concept of mobility was not outlined clearly until *Conduction of Electricity through Gases*, which was published in 1903. This is supported by the terminology used by Thomson's students; Rutherford (1897) only refers to ionic velocity and not mobility. The concept of mobility (if not a full understanding) seems to have been recognised by 1909, as Wellisch referred to it continually in a paper entitled, *On the mobilities of ions produced by Rontgen rays in gases and vapours*. Even though Wellisch first derived the relationship between mobility $\mu$ and mass $m$,

$$\mu \propto \frac{1}{\sqrt{m}} \qquad \textit{Eq. E.2}$$

he incorrectly used the units of velocity when referring to mobility. By 1914 Swann was aware of the critical mobility of a cylindrical condenser, but still discussed mobility indirectly, as a mass-velocity relationship. Pollack (1915) used the new term explicitly when presenting results pointing to the existence of a larger, low mobility ion in the air.

## E.3 From "dust" to "aerosol"

Aitken (1880) was one of the first scientists to study atmospheric aerosol, covering many of the most important issues in aerosol science today in one short paper. He suggested that "infinitesimally small and invisible particles" were necessary to create droplets in clouds, because "burned sulphur" had been shown to be a good fog producer. Aitken also proposed the following sources of fine atmospheric dust: dried spray from the ocean (salt was invoked separately), meteoric matter and combustion products. In the 1890s Thomson and Wilson showed ions could also act as cloud condensation nucleii. The first direct link between dust and ions is credited to Rutherford (1897), who explained that dust particles reduced the conductivity of the air because of the small ions attaching to them. In 1923 Nolan presented the results of experiments showing that ions were surrounded by water molecules. This must have provoked analogies between the smallest (ions) and largest particles (cloud droplets) then known to exist in the atmosphere. Langevin (in Flagan, 1998) also furthered the relationship between small ions and aerosol by discovering large ions. The large ion concentration was found to correlate with the aerosol concentration, and was anticorrelated with ionic variations, perhaps placing large ions between aerosol and smaller ions in some broad spectrum. The third edition of Thomson's book (1928) summarises contemporary work on ions in gases and their relationship to aerosol. He interprets some very low mobility ions found by Langevin in Paris in 1908 as "charged particles of dust" and relates them obliquely to pollution: "the number of them [is] much greater in large towns than in the open country".

The above summary of aerosol science in the first quarter of the century shows the basis of the "continuum" view of atmospheric aerosol, and the growing perception of aerosol as a product of anthropogenic activity. Yet nobody had directly referred to particulates as pollutants. Aitken (1880) even suggested that sulphur's combustion products might somehow deodorise London's polluted streets. By the 1940s and 1950s scientists were becoming more aware of the magnitude and variety of particulate matter emissions. The relevance and potential harmfulness of small particles suspended in the air, which were becoming known as "aerosol" had also been noted (Hewson, 1951). The emergence of this perception of aerosol pollution was concurrent with the massive and uncontrolled growth of industrialisation and car usage in the Western world (Elsom, 1992).

## E.4 Small ions and particulate pollution

A useful way to follow the progressive "integration" of initially discrete studies of small ions and atmospheric particulate matter is to consider the direct relationship between them. The direct effect that particulates have on the air is to reduce its electrical conductivity. The number of small ions, which cause the air to be slightly conductive, decreases as the aerosol number concentration increases. This is because of attachment of the ions to the much larger aerosol particles. This effect was first noticed in the atmosphere by Wait (1946) following conductivity measurements over the oceans on the cruises of the research ship *Carnegie* (Chalmers, 1967) from 1912 onwards. Rutherford (1897) had been the first to publish the idea that larger particles reduce the number of small ions, but it took many years for this idea to be formalised from its first inception (Rutherford, 1897). The relationship between aerosol concentration and air conductivity has only recently been experimentally investigated with the aim of improving aerosol measurements (Aplin *et al.*, 1998). The progression of separate measurements of air ions (e.g. Rutherford, 1897) and aerosol (Aitken, 1890) of the late $19^{th}$ century towards their integration in the late $20^{th}$ century (*e.g.* Guo *et al.*, 1996) is a useful indicator of the change in perceptions of atmospheric particles.

Rutherford's initial attempts to describe ionic equilibrium in the air did not include any effects from ions recombining with aerosol, although he clearly understood the physical processes involved:

> *"Since the dust-particles are very large compared with the ions, an ion is more likely to strike against the dust-particle, and give up its charge to it or to adhere to the surface... and the rate of loss of conductivity is much more rapid than if the loss of conductivity were due to collisions between the ions themselves".*

No attempt was made to express the attachment of small ions to larger particles mathematically until Nolan *et al.* (1925). It is unclear why the Cambridge physicists all neglected this effect in their equations, though they were well aware of it. (It is possible that their outlook was so focussed upon investigating ions in the laboratory, that they thought attachment to particles in the air was unimportant or unnecessary.) In considering the equilibrium case, where the ion production rate equals the recombination rate, the only loss term expressed in the equation described self-attachment (ions recombining with others of the opposite sign). The equation used by Thomson, Rutherford and Wilson in many papers on this topic was:

$$q = \alpha n^2 \qquad \textit{Eq. E.3}$$

Here, $q$ is the rate of ion production, $n$ is the ion number concentration and $\alpha$ is an self-attachment coefficient. Nolan *et al.* (1925) added an extra term to account for ions combining with ions and larger particles. The rate of attachment is proportional to the number of ions and the number of aerosol particles. The equation here is given in the form used today (Nolan *et al.* used a slightly different notation).

$$q = \alpha n^2 + \beta n Z \qquad \textit{Eq. E.4}$$

In Eq. E.4, $Z$ is the monodisperse aerosol number concentration and $\beta$ is an attachment coefficient. It took some time for this coefficient to be described theoretically, possibly because of the complexity of the physics underlying the attraction between a small charged particle and a larger uncharged one. Various attempts were made to find $\beta$ experimentally, but they varied over an order of magnitude (Wait and Parkinson, 1951). $\beta$ is dependent on the size of the ions and aerosol particles measured, so further measurements made in different locations and using different techniques added to the confusion. Bricard (1949) appears to have developed the first working method for calculating $\beta$. After Bricard, a number of other workers approached the problem; some of this is reviewed in Harrison (1992).

Acceptance of the idea of using the properties of the smallest particles in the air to monitor the largest ones appears to have required the simultaneous development of three branches of this field. Air conductivity measurements had to be carried out in a variety of locations before it was realised that the urban areas showed much lower conductivity than the rural ones. Long term clean air measurements of atmospheric electrical parameters in the remote oceans led to the perception that clean air conductivity could be used as a secular pollution indicator (Wait, 1946). An awareness of increasing particulate pollution levels, combined with a rigorous approach to aerosol attachment processes completed the knowledge needed to bring the relationship between atmospheric electrical processes and aerosol to the state they are in today.

## E.5 Conclusions

There has been a kind of paradigm shift (Kuhn, 1967) in the science of atmospheric particles during the twentieth century. Aitken started to investigate aerosol particles in the 1880s, but understanding of the smallest particles lacked motivation or a theoretical basis. Discovery of the electron provided justification for the existence of the atmospheric ion, plus terminological and experimental frameworks to understand the ions and their properties. Following this, measurements of air ions combined with an increase in pollution necessitated development of theories explaining the interaction of these particles. This established the methodology for the integrated way scientists have studied atmospheric particulates in the second half of this century. Without the progression elucidated above, there would not be the understanding of the atmospheric aerosol continuum and its contemporary relevance for climate change, pollution issues and atmospheric electricity.

# References

Adachi M., Ishida T., Kim T.O., Okuyama K. (1996), Experimental evaluation of ion-induced nucleation in nanometre-aerosol formation by α-radiolysis of $SO_2/H_2O/N_2$ mixtures, *Colloids and Surfaces* A **109**, 39-48

Aitken J. (1880), On dust, fogs and clouds, *Nature*, **22**, 195

Aitken J., (1890), On the number of dust particles in the atmosphere, *Trans. Roy. Soc. Edin.*, **35**, 1-19

Anderson R. V. and Bailey J.C. (1991), Errors in the Gerdien measurement of atmospheric electrical conductivity, *Meteorol. Atm. Phys*, **46**, 101-112

Aplin K.L. and Harrison R.G. (1999), The interaction between air ions and aerosol particles in the atmosphere, In: Taylor D.M. (ed.), *Proceedings of 10th International Electrostatics Conference*, Cambridge, 28th-31st March 1999, Institute of Physics Conference Series **163**, 411-414

Aplin K.L. and Harrison R.G. (2000), A computer-controlled Gerdien atmospheric ion counter, *Rev. Sci. Instrum.*, **71**, 8, 3037-3041

Aplin K.L., Harrison R.G. and Wilkinson S. (1998), An electrical method of urban pollution measurement, *J. Aerosol Sci.*, **29**, S1, 869-870

Arathoon J. (1991), *Atmospheric Electricity Investigations: the Measurement of Atmospheric Conductivity*, BSc. Dissertation, University of Lancaster

Atkins P.W. (1989), *Physical Chemistry,* 4th edition*,* Oxford University Press

Bailey J.C., Blakeslee R.J., and Driscoll K.T. (1999), Evidence for the absence of conductivity variations above thunderstorms, In: Christian H.J. (ed.), *Proceedings 11th International Conference on Atmospheric Electricity*, Guntersville, Alabama 7th-11th June 1999 (NASA/CP-1999-209261), 646-649

Barlow J.F. (2000), *Turbulent transfer of space charge in the atmospheric surface layer*, PhD Thesis, University of Reading.

Berthouex, P.M. and Brown, L.C. (1994), *Statistics for Environmental Engineers*, CRC Press, Boca Raton

Blakeslee R.J. (1984), *The electric current densities beneath thunderstorms*, PhD Thesis, University of Arizona

Blakeslee R.J. and Krider E.P. (1992), Ground level measurements of air conductivities under Florida thunderstorms, *J. Geophys. Res*., **97**, D12, 12947-12951

Bleaney B.I. and Bleaney, B. (1985), *Electricity and magnetism*, 3rd edition, Oxford University Press

Bricard P.J. (1949), L'equilibre ionique de la basse atmosphere, *J. Geophys. Res*., **54**, 39-52

Bricard J., Billard F. and Madelaine G. (1968), Formation and evolution of nuclei of condensation that appear in air initially free of aerosols, *J.Geophys Res* **73**, 14, 4487-4496

Brownlee J.N. (1973), *A fast mobility spectrometer for atmospheric ions*, PhD Thesis, University of Auckland

Brownlee J.N. (1975a), A new approach to atmospheric-ion mobility spectrometry, *J. Terr. Atm. Phys*, **37**, 1139-1144

Brownlee J.N. (1975b), Some measurements of atmospheric ion mobility spectra and conductivity at Auckland, *J. Terr. Atm. Phys*, **37**, 1145-1149

Castleman A.G. (1982), In: Schryer D.R., *Heterogeneous atmospheric chemistry*, AGU, Washington

Chalmers J.A. (1967), *Atmospheric Electricity*, 2nd edition, Pergamon Press, Oxford

Clement C.F. and Harrison R.G. (1992), The charging of radioactive aerosols, *J. Aerosol Sci* **23**, 5, 481-504

Coulomb C.A. (1795), *Premier mémoire sur l'electricité et le magnetisme*, Histoire de l'Académie Royale des Sciences, Paris

Crozier W.D. (1964), The electric field of a New Mexico dust devil, *J. Geophys Res*., **69**, 5427-5429

Cunningham E. (1910), On the velocity of steady fall of spherical particles through fluid medium, *Proc. Roy. Soc. A*., **83**, 357-365

Dhanorkar S. and Kamra A.K. (1992), Relation between electrical conductivity and presence of intermediate and large ions in the lower atmosphere, *J. Geophys. Res.*, **97**, 20345-20360

Dhanorkar S. and Kamra A.K (1993), Diurnal variations of the mobility spectrum of ions and size distributions of aerosol in the atmosphere, *J. Geophys. Res.*, **98**, D2, 2639-2650

Dhanorkar S. and Kamra A.K. (1997), Calculation of electrical conductivity from ion-aerosol balance equations, *J. Geophys Res* **102**, D25, 30147-30159

Diamond G.L., Iribarne J.V., Corr D.J. (1985), Ion-induced nucleation from sulphur dioxide, *J. Aerosol Sci*., **16**, 1, 43-55

Dolezalek H. (ed.) (1974), *Electrical processes in atmospheres*, Springer Verlag, Darmstadt

Duffin W.J. (1980), *Electricity and magnetism*, 3rd edition, McGraw Hill, London

Eden P. (2000a), Weather Log, *Weather*, **55**, 6

Eden P. (2000b), Weather Log, *Weather*, **55**, 7

Eisele F.L. (1988), First tandem mass spectrometric measurements of tropospheric ions, *J. Geophys. Res*., **93**, 716

Eisele F.L. (1989), Natural and transmission line produced positive ions, *J. Geophys. Res*., **94**, 6309

Elsom D.M. (1992), *Atmospheric Pollution*, 2nd edition, Blackwell, Oxford

Emi H., Shintani E., Namiki N. and Otani Y. (1998), Measurement of the ion mobility distribution at a new mobility analyser with separation in axial direction to the flow *J.Aerosol Sci.* **29**, S1, S1247-S1248

Flagan R.C. (1998), History of electrical aerosol measurements, *Aerosol Science and Technology*, **28**, 301-380

Gerdien H. (1905), Demonstration eines Apparates zur absoluten Messung der elektrischen Leitfähigkeit der Luft, *Phys. Zeitung*, **6**, 800-801

Gringel W. (1978), *Untersuchungen zur elektrischen Luftleitfähigkeit unter Berücksichtigung der Sonnenaktivität und der Aerosolteilchenkonzentration bis 35km Höhe*, PhD Thesis, University of Tübingen

Gringel W., Rosen J.M., Hofman D.J. (1983), Electrical structure from 0 to 30 kilometers, In: Krider E.P. and Roble R.G. (eds), *The Earth's Electrical Environment*, National Academy Press, Washington D.C.

Gunn R. (1960), The electrical conductivity and electric field intensity over the North Atlantic and its bearing on changes in the world-wide pollution of the free atmosphere, In: *The American University*, Washington DC., 49-55

Guo Y., Barthakur N., Bhartendu (1996), Using atmospheric electrical conductivity as an urban air pollution indicator, *J. Geophys. Res*., **101**, 9197-9203

Handloser J.S. (1959), *Health physics instrumentation,* Pergamon Press, Oxford

Harrison R.G. (1992), *Aerosol charging and radioactivity*, PhD. Thesis, University of London

Harrison R.G. (1995), A portable picoammeter for atmospheric electrical use In: Cunningham S.A.(ed.), *Proceedings of 9th International Electrostatics Conference*, York, April 1995, Institute of Physics Conference series **143**, 223-226

Harrison R.G. (1997a), A noise-rejecting current amplifier for surface atmospheric ion flux measurements, *Rev. Sci. Instrum.,* **68**, 9, 563-3565

Harrison R.G. (1997b), An antenna electrometer system for atmospheric electrical measurements, *Rev. Sci. Instrum.,* **68**, 3, 1599-1603

Harrison R.G. (2000), Cloud formation and the possible significance of charge for atmospheric condensation and ice nuclei, (in press, *Space Science Reviews*)

Harrison R.G. and Aplin K.L. (2000a), Femtoampere current reference stable over atmospheric temperatures, *Rev. Sci. Instrum.,* 7**1**, 8, 3231-3232

Harrison R.G. and Aplin K.L. (2000b) (On Yu F. and Turco R.P., Ultrafine aerosol formation via ion-mediated nucleation *Geophys. Res. Lett.*, **27**, 883-886, 2000), *Geophys. Res. Lett.* **27***,* 13, July 1st 2000

Harrison R.G. and Aplin K.L. (2000c), A multimode electrometer for atmospheric ion measurements, accepted for publication in *Rev. Sci. Instrum.*

Harrison R.G. and Aplin K.L. (2001), Atmospheric electricity, In: MacCracken M., Munn E., Perry J. (eds), *Encyclopaedia of Global Environmental Change*, Wiley, Chichester (to appear)

Harrison R.G. and Pedder M.A. (2000), A fine wire thermometer for micrometeorology (submitted to *Rev. Sci. Instrum.*)

Hatakeyama H., Kobayashi J., Kitaoka T., Uchikawa K. (1958), A radiosonde instrument for the measurement of atmospheric electricity and its flight results, In: Smith L.G. (ed.), *Recent advances in atmospheric electricity*, Pergamon Press, Oxford

Hewson E. (1951), *Atmospheric Pollution* In: Malone T.F. (ed.), *Compendium of Meteorology*, American Meteorological Society

Higazi K.A. and Chalmers J.A. (1966), Measurements of atmospheric electrical conductivity near the ground, *J. Atm. Terr. Phys*., **25**, 327-330

Hogg A.R. (1939), The conduction of electricity in the lowest levels of the atmosphere, *Mem. Commonw. Solar. Obs. Australia*, **7**

Horowitz P. and Hill W. (1994), *The art of electronics*, 2nd edition, Cambridge University Press

Hõrrak U., Salm. J., Tammet. H. (1998a), Bursts of intermediate ions in atmospheric air, *J. Geophys. Res*, **103**, 13909-13915

Hõrrak U., Mirme S., Salm J., Tamm E., Tammet H. (1998b), , Air ion measurements as a source of information about atmospheric aerosols, *Atmospheric Research*, **46**, 233-242

Hõrrak U., Salm J., Tammet H. (1999), Classification of natural air ions near the ground, In: Christian H.J. (ed), *Proc. 11th Int. Conf. on Atmospheric Electricity, (NASA/CP-1999-209261)*, Alabama, 7th-11th June 1999, 618-621

Israël H. (1971), *Atmospheric Electricity*, 2nd edition, Israeli Program for Scientific Translations, Jerusalem

Johnson R.A and Bhattacharryya G.K. (1996), *Statistics: principles and methods*, 3rd edition, Wiley, Chichester

Jones C. D. (1999), Personal communication

Jorgensen T. B. and Hansen A.W. (2000), Comments on "Variation of cosmic ray flux and global cloud coverage-a missing link in solar-climate relationships", *J. Atmos. Solar-Terrestrial Phys*., **62**, 73-77

Kamra A.K. (1982), Fair weather space charge distribution in the lowest 2 m of the atmosphere, *J. Geophys. Res*., **87**, C6, 4257-4263

Kamra A.K. and Deshpande C.G (1995), Possible secular change and land-to-ocean extension of air pollution from measurements of atmospheric electrical conductivity over the Bay of Bengal, *J. Geophys. Res*., **100**, 7105-7110

Keesee R.G. and Castleman A.W. (1985), Ions and cluster ions, experimental studies and atmospheric observations, *J. Geophys. Res*, **90**, 5885-5890

Keithley (1992), *Low Level Measurements*, 4th edition, Keithley Instruments Ltd.

Kirkby J. *et al* (2000), *A study of the link between cosmic rays and clouds with a cloud chamber at the CERN PS,* European Centre for Nuclear Research (CERN) internal document, proposal SPSC/P317, SPSC 2000-021

Knudsen E. and Israelsson S. (1994), Mobility spectra of ions in the electrode effect layer, *J. Geophys Res*., **99**, D5, 10709-10712

Kuhn T.S. (1967), *The structure of scientific revolutions*, 2nd edition, Chicago University Press

Kulmala M., Pirjola L., and Mäkelä J. M. (2000), Stable sulphate clusters as a source of new atmospheric particles, *Nature*, **404**, 66-68

Laaksonen A. and Kulmala M. (1991), Homogeneous heteromolecular nucleation of sulphuric acid and water vapours in the stratospheric conditions: a theoretical study of the effect of hydrate interaction, *Journal of Aerosol Science*, **22**, 6, 779-787

MacGorman D.R. and Rust W.D. (1998), *The electrical nature of storms*, Oxford University Press

Mäkelä J.M., Aalto P., Jokinen V., Pohja T., Nissinen A., Palmroth S., Markkanen T., Seitsonen K., Lihavainen H., Kulmala M. (1997), Observations of ultrafine aerosol particle formation and growth in boreal forest, *Geophys. Res. Lett*., **24**, 10, 1219-1222

Mäkelä J.M. (1992), *Studies on irradiation induced aerosol particle formation processes in air containing sulphur dioxide*, PhD. Thesis, University of Helsinki

Mason B.J., *Physics of clouds*, Pergamon, Oxford (1971)

Matisen R. Miller F., Tammet H., and Salm J., Air ion counters and spectrometers designed in Tartu University, In: *Air ions and electrical aerosol analysis*, University of Tartu, Estonia (1992)

McClelland J.A. (1898), On the conductivity of the hot gases from flames, *Phil. Mag.,* **46**, 29-42

Metnieks, A.L. and Pollak, L.W. (1969), *Instructions for use of photo-electric condensation nucleus counters.* Communications of the Dublin Institute for Advance Studies, Series D, Geophysical Bulletin No 16. Dublin

Misaki M. (1964), Mobility spectrums [sic] of large ions in the New Mexico semidesert, *J. Geophys. Res*., **69**, 16, 3309-3318

Mohnen V.A. (1974), Formation, nature and mobility of ions of atmospheric importance In: Dolezalek H. (ed.), *Electrical processes in atmospheres*, Springer Verlag, Darmstadt

Moody N.F. (1981), Measurement of ion-induced noise in atmospheric counting chambers, and a novel chamber for its elimination, *Int. J. Biometeor*., **25**, 331-339.

Moody N.F. (1984), Design and construction of an improved, portable air-ion counter, *Int. J. Biometeor*., **28**, 169-184

Moore C.B. and Vonnegut B. (1988), Measurements of the electrical conductivities of air over hot water, *J. Atmos. Sci.*, **45**, 5, 885-890

Nagato K. and Ogawa T. (1998), Evolution of tropospheric ions observed by an ion mobility spectrometer with a drift tube, *J. Geophys. Res*, **103**, 13917-13925

Ney E.P. (1959), Cosmic radiation and the weather, *Nature*, **183**, 451-452

Nolan J.J. (1920), Ions produced in air by radio-active bodies, *Proc. Roy. Irish Acad. A*., **35**, 38-45

Nolan J.J. (1923), Ionic mobilities in air and hydrogen, *Proc. Roy. Irish Acad. A*., **36**, 74-92

Nolan J.J., Boylan R.K. and de Sachy G.P. (1925), The equilibrium of ionisation in the atmosphere, *Proc. Roy. Irish Acad. A*., **37**,1-12

O'Dowd C.D, Smith M.H., Lowe J.A., Harrison R.M., Davison B., Hewitt C.N. (1996), New particle formation in the marine environment *In*: Kulmala M. and Wagner P.E., *Proc. 14th Int. Conf. on Nucleation and Atmospheric Aerosols*, 925-928

Obolensky W.N. (1925), Über elektrische Ladung in der Atmosphäre, *Ann. Phys. Lpz.*, 77, 644-666

Parallax Inc. (1998), *BASIC Stamp® Manual*, Version 1.9

Pauletti D., Schirripa Spagnolo G. (1989), Atmospheric electricity in a rural site and its possible correlation with pollution: a preliminary study, *Atmospheric Environment*, **23**, 1607-1611

Pedder M.A. (1971), Measurement of size and diffusion characteristics of aerosols with particle sizes less then 0.01μm using the Pollak condensation nucleus counter, *J Phys. D.* **4**, 531-538

Peyton A.J. and Walsh V. (1993), *Analog electronics with op-amps*, Cambridge University Press

Phillips B.B., Allee P.A., Pales J.C., Woessner R.H. (1955), An experimental analysis of the effect of air pollution on the conductivity and ion balance of the atmosphere, *J. Geophys. Res.*, **60**, 3, 289-296

Pollack J.A. (1915), A new type of ion in the air, *Phil. Mag.*, **29**, 636-646

Pruppacher H.R. and Klett J.D. (1997), *Microphysics of Clouds and Precipitation*, 2$^{nd}$ edition, Kluwer Academic Press, Dordrecht

Pudovkin M.I. and Veretenko S.V. (1995), Cloudiness decreases associated with Forbush-decreases of galactic cosmic rays, *J. Atmos. Terr. Phys.*, **75**, 1349-1355

Raes. F., Janssens A. and Eggermont G. (1985), A synergism between ultraviolet and gamma radiation in producing aerosol particles from $SO_2$-$H_2SO_4$ laden atmospheres, *Atmos. Env.*, **19**, 7, 1069-1073

Raes F., Janssens A. and Van Dingenen R. (1986), The role of ion-induced aerosol formation in the lower atmosphere, *J. Aerosol Sci.* **17**, 3, 466-470

Retalis D., Pitta A and Psallidas P. (1991), The conductivity of the air and other electrical parameters in relation to meteorological elements and air pollution in Athens, *Meteorol. Atmos. Phys.*, **46**, 197-204

Robotti S. (1995), J.J. Thomson at the Cavendish Laboratory: the history of an electric charge measurement, *Annals of Science*, **52**, 265-284

Rosen J.M. *et al.,* (1982), Results of an international workshop on atmospheric electrical measurements, *J. Geophys. Res.* **87**, 1219-1227

Rutherford E. (1897), The velocity and rate of recombination of the ions in gases exposed to Rontgen radiation, *Phil. Mag*., **44**, 422-440

Schonland B.F.J. (1953), *Atmospheric electricity*, 2nd edition, Methuen, London

Shimo M., Ikebe Y., Nakayama T., Kawano M. (1972), Measurement of small ions and condensation nuclei over the sea near the land, *Pure and Applied Geophysics*, **100**, 109-122

Smith L.G. (1953), On the Calibration of Conductivity Meters, *Rev. Sci. Inst*., **24**, 998

Stull R.B. (1997), *An introduction to boundary layer meteorology*, Kluwer Academic Press, Dordrecht

Svensmark H. and Friis-Christensen E. (1997), Variations of cosmic ray flux and global cloud coverage – a missing link in solar-climate relationships *J Atmos Solar-Terrestrial Phys,* **59**, 1225-1232

Swann W.F.G. (1914), The theory of electrical dispersion into the free atmosphere, with a discussion of the theory of the Gerdien conductivity apparatus, and of the theory of the collection of radioactive deposit by a charged conductor, *J. Terr. Mag. Atmos. Elect*., **19**, 81-92

Tammet H. (1995), Size and mobility of atmospheric particles, clusters and ions *J. Aerosol Sci.,* **26**, 3, 459-475

Tammet H. (1992), Comparison of model distributions of aerosol particles sizes, In: *Air ions and electrical aerosol analysi*s, University of Tartu, Estonia

Tammet H., Iher H. and Salm J. (1992), Spectrum of atmospheric ions in the mobility range 0.32-3.2 $cm^2/(V.s)$, In: *Air ions and electrical aerosol analysi*s, University of Tartu, Estonia, 35-49

Thomson J.J. (1928), *Conduction of electricity through gases*, 3rd edition, Cambridge University Press

Thuillard M. (1995), Electric mobility measurements of small ions in the temperature range –20 - -40°C at constant relative humidity of 87%, *J. Aerosol Sci*, **26**, 219-225

Torreson O.W. (1949), Instruments used in observations of atmospheric electricity, In: Fleming J.A. (ed.), *Terrestrial Magnetism and Electricity*, 2nd edition, Dover, New York

Tritton D. (1988), *Physical fluid dynamics*, Oxford University Press

TSI Incorporated (1994), *Model 8250 DustTrak Aerosol Monitor Operation and Service Manual*, TSI Inc., St. Paul

Turco R.P., Zhao Jing-Xia, and Yu F. (1998), A new source of tropospheric aerosols: Ion-ion recombination *Geophys. Re.s Lett*, **25**, 5, 635-638

Van der Hage J.C.H. and de Bruin T.F (1999), The atmospheric electric fog effect, In: Christian H.J. (ed.), *Proceedings 11th International Conference on Atmospheric Electricity*, Guntersville, Alabama 7th-11th June 1999 (NASA/CP-1999-209261), 595-597

Verzar F. and Evans (1953), *Arch. Meterol. Geophys*., **5**, 372

Venkiteshwaran S.P. (1958), Measurement of the electrical potential gradient and conductivity by radiosonde at Poona, India, In: Smith L.G. (ed.), *Recent advances in atmospheric electricity*, Pergamon Press, Oxford

Vohra K.G., Subba Ramu M.C. and Muraleedharan T.S. (1984), An experimental study of the role of radon and its daughters in the conversion of sulphur dioxide into aerosol particles in the atmosphere, *Atmos. Env*., **18**, 8, 1653-1656

Wåhlin L. (1986), *Atmospheric electrostatics*, Research Studies Press, Letchworth

Wait G.R. (1946), Some experiments relating to the electrical conductivity of the lower atmosphere*, J. Washington. Acad. Sci*., **36**, 321-343

Wait G.R., Parkinson W.D. (1951), *Ions in the atmosphere*, In: *Compendium of Meteorology*, Malone T.F. (ed.), American Meteorological Society

Warneck P. (1987), *Chemistry of the natural atmosphere*, Academic Press, London

Wellisch E.M. (1909), On the mobilities of ions produced by Rontgen rays in gases and vapours, *Phil. Trans. A*, **209**, 249-279

Wilson C.T.R. (1897), *Proc. Camb. Phil. Soc.*, **9**, 333-336

Wilson C.T.R. (1899), On the condensation nuclei produced in gases by the action of Rontgen rays, Uranium rays, Ultra-violet light, and other agents, *Phil. Trans. A.*, 403-453

Wilson C.T.R. (1900), On the leakage of electricity through dust-free air, *Proc. Camb. Phil. Soc.*, **11**, 32

Wilson C.T.R. (1901), On the ionisation of atmospheric air, *Proc. Roy. Soc.*, **68**, 151

Woessner R.H., Cobb W.E., Gunn H. (1958), Simultaneous measurements of the positive and negative light-ion conductivities to 26 km, *J. Geophys. Res.*, **63**, 171-180

Yu F., Turco R.P. (1997), The role of ions in the formation and evolution of particles in aircraft plumes, *Geophys. Res. Lett.*, **24**, 15, 1927-1930

Yu F. and Turco R.P. (2000), Ultrafine aerosol formation via ion-mediated nucleation, *Geophys. Res. Lett.*, **27**, 883-886

Zeleny J. (1898), On the ratio of the velocities of the two ions produced in gases by Rontgen radiation and on some related phenomena*, Phil. Mag., **46**, 120-154*

Zeleny J. (1900), The velocity of ions produced in gases by Rontgen rays, *Phil. Trans. A.*, **195**, 193-234